\documentclass{elsarticle}

\usepackage{graphicx}

\usepackage{latexsym,amsmath,amssymb}
\usepackage{xcolor}

\newcommand{\be}{\begin{equation}}
\newcommand{\ee}{\end{equation}}
\newcommand{\ba}{\begin{align}}
\newcommand{\ea}{\end{align}}
\newcommand{\bea}{\begin{eqnarray}}
\newcommand{\eea}{\end{eqnarray}}

\begin{document}

\begin{frontmatter}



\title{Review of the No-Boundary Wave Function}


\author{Jean-Luc Lehners}

\address{Max--Planck--Institute for Gravitational Physics (Albert--Einstein--Institute) \\ 14476 Potsdam, Germany}

\ead{jlehners@aei.mpg.de}

\begin{abstract}

When the universe is treated as a quantum system, it is described by a wave function. This wave function is a function not only of the matter fields, but also of spacetime. The no-boundary proposal is the idea that the wave function should be calculated by summing over geometries that have no boundary to the past, and over regular matter configurations on these geometries. Accordingly, the universe is finite, self-contained and the big bang singularity is avoided. Moreover, given a dynamical theory, the no-boundary proposal provides probabilities for various solutions of the theory. In this sense it provides a quantum theory of initial conditions.

This review starts with a general overview of the framework of quantum cosmology, describing both the canonical and path integral approaches, and their interpretations. After recalling several heuristic motivations for the no-boundary proposal, its consequences are illustrated with simple examples, mainly in the context of cosmic inflation. We review how to include perturbations, assess the classicality of spacetime and how probabilities may be derived. A special emphasis is given to explicit implementations in minisuperspace, to observational consequences, and to the relationship of the no-boundary wave function with string theory. At each stage, the required analytic and numerical techniques are explained in detail, including the Picard-Lefschetz approach to oscillating integrals. 

\end{abstract}

\end{frontmatter}

\newpage

\tableofcontents{}

\newpage

\section{Introduction}

The principles of quantum theory are the most basic physical principles uncovered to date. They have been tested over the past century in numerous experiments, and form the basis of modern electromagnetic and nuclear technology. What has not been achieved so far is a direct experimental verification of the quantisation of the fourth fundamental force, namely gravity. The reasons for this are easily understood to lie in the relative weakness of the gravitational force compared to electromagnetic and nuclear interactions. 

So even though we are not compelled by experiment to assume that gravity is quantised, there are good arguments that indicate that it must be. One such argument, based on the internal consistency of the laws of physics, will be provided in the next section. Once we contend that gravity is indeed quantised, and that the laws of quantum theory apply to all interactions, it is clear that the entire universe must also be considered to be a quantum system. As such, it must be described by a wave function. This wave function is then not only a function of the matter fields, but also of spacetime itself. But how should one calculate the wave function of the universe?

Note that if the wave function includes a description of space and time, then it will necessarily also tie in with the question of boundary and initial conditions of the universe, since these are the conditions at the ends of space and time. In other words, a wave function of the universe links together the laws of dynamical evolution and the specification of initial conditions. This is in stark contrast with a laboratory setting, where initial conditions are carefully prepared by the experimenter. 

But then what should the conditions be at the ends of space and time? Here J. Hartle and S. Hawking made a suggestion that is as radical as it is elegant \cite{Hawking:1981gb,Hartle:1983ai}: they proposed that there should be no such ends! In other words, they proposed that space and time should have no boundary to our past. And if there is no boundary, then there is no need for further boundary conditions -- hence (the hope is that) this fully specifies the initial conditions for our universe, and it indicates how to calculate the wave function of the universe.

The idea that the universe is entirely self-contained, both in space and in time, sounds almost like a tautology. What it means is that if there was a boundary, then one would have to specify conditions at that boundary and they would link to an evolution on the other side of the boundary -- put differently, if there was a boundary then outside information would be required. But there is more: in ordinary classical general relativity, the no-boundary condition is plainly impossible. This is because under reasonable assumptions on the matter content of the universe, the celebrated singularity theorems \cite{Hawking:1970zqf} imply that in our past there must have existed a curvature singularity, the big bang. Such a singularity would be an end point of spacetime. Therefore one cannot impose the no-boundary condition in classical physics. It is a true quantum gravitational condition. And it eliminates the big bang singularity. As will be discussed in detail, the no-boundary condition in fact has similarities with quantum tunnelling. Thus one possible interpretation of the no-boundary proposal is that it eliminates the big bang singularity, and replaces it by describing the origin of the universe as a tunnelling event from nothing, {\it i.e.} from the absence of not just matter, but also the absence of spacetime.  

\begin{figure}[ht]
	\centering
	\includegraphics[width=0.75\textwidth]{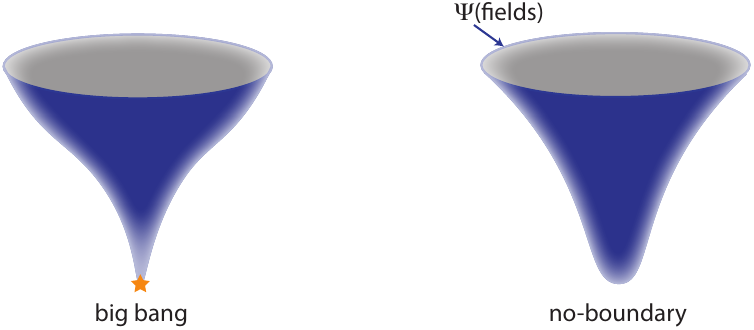}
	\caption{{\it Left panel:} A cartoon of the evolution of a classical, expanding universe with closed $1$-dimensional spatial sections. Space is horizontal and time vertical. Going back in time, one reaches a singularity, the big bang. {\it Right panel:} By contrast, the no-boundary proposal suggests that one should consider geometries that are (smoothly) rounded off in the past, and that contain no boundary there. Such a condition automatically eliminates the big bang.}
	\label{fig:cartoon}
\end{figure}

A cartoon of the no-boundary idea is shown in Fig.~\ref{fig:cartoon}. The figure should make it clear that no-boundary geometries, though they have no boundary to the past, do actually have a boundary (but a single one): this boundary may be thought of as a spatial slice of our universe, either a current slice or a spatial slice in the early universe. The no-boundary wave function then has as arguments the field values (or momenta) on this ``final'' hypersurface. It is not a transition amplitude in the usual sense, since this would depend on field values or momenta on two separate hypersurfaces. However, one may at least heuristically think of it as a transition from nothing to today.

It should be clear by now that the no-boundary proposal assumes a fully quantum view of spacetime: actual spacetime does not exist when not measured, {\it i.e.} when not in interaction with either itself or matter. It is the interactions between the different constituents of the universe that result in our perception of classical spacetime, and of large scale classical laws of evolution. And vice versa, going back in time towards the putative big bang, one will necessarily encounter departures from the classical evolution. A significant part of this review will be concerned with making these ideas, which may sound rather vague at first, more precise.

There is an immediate concern that should be mentioned from the outset. Let us assume we are given a wave function for the universe. Then it will imply probabilities for different histories of the universe, or one could say that it will provide probabilities for different universes. Yet we live in a single universe, so how can we talk about probabilities for different universes? It should be said that this issue is not fully resolved. Certainly, inside of the universe probabilities for the outcomes of experiments make sense, and we will show in some detail how this comes about. But probabilities are inherently linked to a classical notion of time (think about any operational definition of probabilities). And in quantum cosmology, time itself is not {\it a priori} classical, it is not fixed from outside. There is no overarching concept of spacetime in which universes reside -- rather, each universe has its own spacetime. Hence it may not make sense to compare probabilities for different universes. Nevertheless the common practice has been to do just that \cite{Halliwell:1989myn,Hartle:2008ng}, and to ask, given a set of dynamical laws, whether the no-boundary wave function implies that a universe like ours is typical or highly unlikely. We will return to this question at various points, and mention ways in which the interpretation of the wave function might be elucidated. Let us emphasise though that even within a given universe the no-boundary wave function can make post- and predictions for observations. 

Besides such conceptual issues, the first challenge in fact is to find a convincing mathematical implementation of the no-boundary proposal. This will be a central focus of this review, and it is probably also the area in which most progress has been made in recent years. As we will see, this is directly linked to an understanding of gravitational path integrals quite generally, and many connections with string theory have emerged. These connections are now providing a two-way flow of insights: string theory provides clues regarding the mathematical implementation of the no-boundary proposal, and the no-boundary wave function provides a framework in which to address some of the landscape/swampland-related puzzles of string theory. These connections provide promising starting points for further research, and one aim of this review is to provide the necessary background for undertaking such research.

In brief, the outline of the review is as follows: we will start with a general overview of quantum cosmology in section~\ref{sec:qc}. This will serve to introduce the basic concepts one is led to, both in the canonical and path integral quantisation schemes. Section~\ref{sec:nb} then introduces the no-boundary proposal, starting from heuristic motivations and building up in mathematical rigour. Both analytic and numerical methods for finding no-boundary solutions will be described in detail. The crucial question of how to extract probabilities from the no-boundary wave function, and to derive predictions for observations, will be tackled in section~\ref{sec:ob}. More recent topics, and in particular the above-mentioned connections with string theory, will be the topic of section~\ref{sec:st}. To end, we will list a number of outstanding questions in the discussion section~\ref{sec:di}. Finally, three appendices fill in a few details about canonical and path integral quantisation,  as well as some perhaps lesser known mathematical methods, in particular Picard-Lefschetz theory. This review aims to be self-contained, but it assumes prior knowledge of general relativity, quantum mechanics, and the basics of cosmology and string theory, with the latter only being required in section~\ref{sec:st}.

{\it Notation and conventions:} Greek indices run over space and time, lower case Latin indices only over space, while capital Latin indices run over fields. For Lorentzian metrics, the signature is taken to be mostly plus, and the conventions for gravity are $R^\lambda{}_{\mu\sigma\nu} = \partial_\sigma \Gamma^\lambda_{\mu\nu} - \partial_\nu \Gamma^\lambda_{\mu\sigma}+\Gamma^\tau_{\mu\nu}\Gamma^\lambda_{\tau\sigma}-\Gamma^\tau_{\mu\sigma}\Gamma^\lambda_{\nu\tau}$ with $R_{\mu\nu}=R^\lambda{}_{\mu\lambda\nu}.$ We set $8\pi G=1,$ but mostly retain $\hbar$ explicitly.


\section{Quantum Cosmology} \label{sec:qc}

Quantum cosmology is, simply put, the application of quantum principles to the whole universe. No less. Despite this highly ambitious mission statement, this task is, at least to some extent, already within reach of current physics. In practice, there are two main simplifications that are helpful: the first is based on the observational fact that the early universe was comparatively simple, in the sense that it was spatially highly isotropic and homogeneous. This suggests an approximation scheme where one starts with geometries and matter configurations that are exactly isotropic and homogeneous on spatial slices, and then includes perturbations as a small correction. A second simplification that is typically made use of is the semi-classical approximation, i.e. one considers an expansion in Planck's constant $\hbar$ (or, rather, in the dimensionless inverse Planck mass) and works to leading or first sub-leading order. Of course, at the end of the calculations, one has to check whether the approximations used were justified. As we will see, with these approximations at hand, one may actually get quite far in describing quantum effects in the early universe. 

\begin{figure}[ht]
	\centering
	\includegraphics[width=0.8\textwidth]{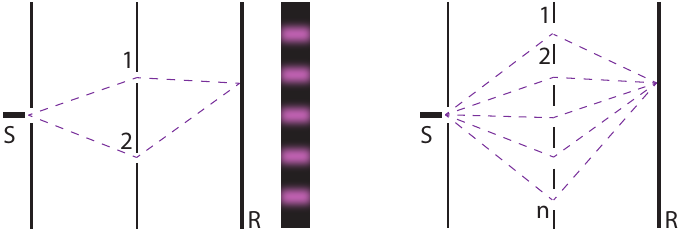}
	\caption{{\it Left panel:} The double slit experiment: a source $\textrm{S}$ emits particles that are detected on a recording surface $\textrm{R}.$ When two slits are present in the intermediate screen, an interference pattern is observed, as indicated on the right. {\it Right panel:} By opening successively more slits, one arrives at the picture that the amplitude should be calculated by summing over all possible paths.}
	\label{fig:doubleslit}
\end{figure}

\subsection{The Need for Quantum Gravity}

We will start by recalling the famous double slit experiment. This experiment highlights the most striking features of quantum mechanics. Moreover, it provides a direct motivation for the path integral and, as we will argue, the quantisation of gravity. Fig.~\ref{fig:doubleslit} shows the setup. Consider a source $\textrm{S}$ sending particles, e.g. electrons, through slits $1$ and $2,$ and a recording plane $\textrm{R}$ that measures the impact of the particles. We call the flux of particles measured at $\textrm{R},$ or equivalently the probability of detecting particles, $P_1$ when only slit $1$ is open, $P_2$ when only slit $2$ is open, and $P$ when both slits are open. Classically, we expect $P=P_1 + P_2.$ However, this is famously not what is observed. Rather, the probability distribution at $R$ when both slits are open is given by a formula that is akin to having waves emanating from $\textrm{S},$ passing through the slits, and {\it interfering} before reaching $\textrm{R},$
\begin{align}
    P = |\Phi_1 + \Phi_2|^2\,,
\end{align}
leading to the interference fringes shown in the figure. The crucial point is that this formula remains valid even when the source is dimmed so much that only a single particle is emitted during a time interval that is at least as long as the crossing time. This experiment then directly illustrates the superposition principle: amplitudes that correspond to the same outcome, but to different evolutions, are summed. The two possible trajectories of the particle interfere with each other.

Above we did not specify how the amplitude is numerically determined. One guideline is that the amplitude should recover classical physics when $\hbar$ is small. First note that $\hbar$ has dimensions of action, namely $[$energy$] \times [$time$].$ Second, recall that classical solutions are given by extrema of the action $S$ ($\delta S = 0$). Then one may guess that the amplitude should be an oscillating function of $S/\hbar.$ The appropriate choice turns out to be \cite{feynman2010quantum}
\begin{align}
    \Phi = e^{\frac{i}{\hbar}S}\,.
\end{align}
With this choice, whenever $S/\hbar$ is large, nearby trajectories cancel each other out, as $e^{iS/\hbar}$ varies rapidly, except near extrema where $\delta S =0.$ That the exponent is linear in the action can be derived from the additional requirement that the amplitude leads to the correct composition property of paths. With the choice above, for two paths in succession, with actions $S_1, S_2,$ we get a combined amplitude that factorises,  $e^{\frac{i}{\hbar}(S_1+S_2)}= e^{\frac{i}{\hbar}S_1}e^{\frac{i}{\hbar}S_2},$ as it should. Moreover, one may show \cite{feynman2010quantum} that for this choice one recovers the standard Schr\"{o}dinger equation by splitting up paths into small intervals. 

Now it is just a small step to the path integral. This step is illustrated in the right panel of Fig.~\ref{fig:doubleslit}. Imagine adding additional slits to the screen, say $n$ slits in total. Then the amplitude will be given by a sum over trajectories passing through all of the slits,
\begin{align}
    \Phi = \sum_{i=1}^{i=n} e^{\frac{i}{\hbar}S_i}\,,
\end{align}
and new interference patters will be generated. In fact, we could have added more screens in between the source and the recording plane, and then opened slits in all of these intervening screens. By imagining a dense array of screens, with infinitesimal slits, one arrives at the conclusion that the amplitude should be given by the sum over all possible trajectories, each trajectory being weighted by a phase that is the action divided by $\hbar,$
\begin{align}
    \Phi = \int {\cal D}x \, e^{\frac{i}{\hbar}S[x]}\,,
\end{align}
where $x$ denotes the position of the particle and ${\cal D}x$ is an appropriate measure over paths. This is Feynman's path integral approach to quantum theory \cite{feynman2010quantum}. 
In fact, what we have just described is the \emph{propagator} $\Phi.$ Due to Heisenberg's uncertainty relation, we cannot in general assume a definite initial position of the particle -- rather, we should consider an initial wave function $\psi_0.$ $\Phi$ then propagates this wave function to a later time $t$, according to
\begin{align}
\psi_t(x) = \int \Phi \, \psi_0(x_0)  \mathrm{d}x_0\,.    
\end{align}
That is to say, for each part of the initial wave function one considers a sum over all possible trajectories to a later configuration. A useful example of an initial wave function is given by a Gaussian wave packet centered at $x_i$ with spread $\sigma$ and with momentum $p_i,$ 
\begin{align}
    \psi_0(x_0) = \frac{1}{\pi^{1/4}\sigma^{1/2}}\, e^{-\frac{(x_0-x_i)^2}{2\sigma^2}+\frac{i}{\hbar}p_i x_0}\,.
\end{align}
The state is normalised such that $\int_{-\infty}^{+\infty} |\psi_0|^2 \mathrm{d}x_0=1.$ The limit of a pure position state is then achieved by considering $\sigma \to 0.$ This yields a delta function in position, and hence corresponds to a non-normalisable state, but nevertheless can be a useful idealisation to consider. Above, we implicitly assumed such an idealisation when saying that the particles were emitted at the source location~$\textrm{S}$. 

We just saw that the double slit experiment suggests an approach to quantum theory in which one sums over all possible particle paths. In extending these arguments to field theories, one would reach the conclusion that one should integrate over all possible matter configurations. But what about gravity?

Let us go back briefly to the case with just two slits. Then the amplitude is well approximated by the contributions from the saddle points
\begin{align}
    \Phi \approx {\cal N} \left( e^{\frac{i}{\hbar}S_{saddle,1}} + e^{\frac{i}{\hbar}S_{saddle,2}} \right)\,,
\end{align}
where ${\cal N}$ is a normalisation factor. Here $S_{saddle,1}$ corresponds to the classical action for the trajectory that passes through slit $1,$ and similarly for slit $2.$ When gravity is taken into account, this means that it is the action for the trajectory of particle $1$ including the backreaction on the spacetime, as it is the action of a classical solution to the full equations of motion. In other words, the amplitude is a superposition not just of two particle paths, but of two spacetimes that each contain a particle path (for a similar argument, see \cite{Feynman:1996kb}). 

Note that it would not make sense to assume that there is just one underlying spacetime: if we imagine adding more slits, then the backreaction due to the mass of the particle taking various paths would add up. Considering multiple slits, finely spaced, one could make the backreaction arbitrarily large, an absurd conclusion considering that the limit of infinitely many slits corresponds to having no screens at all, i.e. corresponds to free propagation. Thus we certainly cannot assume that the various trajectories together lead to a combined gravitational effect. It makes equally little sense to assume that there is a fixed background spacetime, maybe given by the ``average'' backreaction. After all, classical motions of particles follow geodesics, which depend on the curvature of the spacetime. And this curvature depends on the trajectory itself, and hence must be different for the two trajectories we are considering. Basically, the gravitational force exerted on other objects would be ill defined if there was a fixed background spacetime.

The only interpretation that makes sense is that we have a superposition of spacetimes-with-matter. In other words, conceptually it makes sense to think of the double slit experiment as a quantum gravity experiment. It is only the relative weakness of the gravitational force which implies that one can ignore the gravitational backreaction in laboratory experiments (so far). But conceptually, requiring internal consistency of the physical laws implies that gravity and spacetime must be quantised along with matter. 

\begin{figure}[ht]
	\centering
	\includegraphics[width=0.8\textwidth]{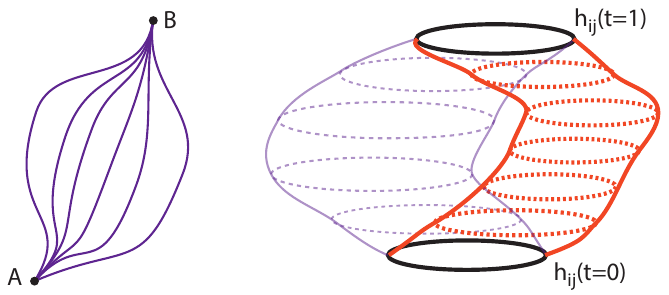}
	\caption{{\it Left panel:} when calculating the amplitude for a particle to propagate from $A$ to $B,$ we sum over all paths that connect the end points. {\it Right panel:} the analogous situation in quantum cosmology is to sum over all geometries and matter configurations that connect two $3-$dimensional boundary surfaces (here illustrated as $1-$dimensional circles located at coordinate values $t=0,1$) with metrics $h_{ij}$. In this picture, the connecting geometries are cylinders, and only two are shown so as not to overload the picture. Each is its own spacetime, and one should not think of them as existing in an ambient spacetime.}
	\label{fig:boundaries}
\end{figure}

The path integral then provides an obvious framework to attempt such a quantisation: it simply implies that we must also sum over geometries, so that heuristically a wave function $\Psi$ should be calculated as
\begin{align}
    \Psi = \int {\cal D}g_{\mu\nu}{\cal D}\phi \, e^{\frac{i}{\hbar}S[g_{\mu\nu},\phi]}\Psi_0\,,\label{eq:piheuristic}
\end{align}
where $\Psi_0$ describes an initial spatial geometry and matter configuration, see Fig. \ref{fig:boundaries} for an illustration. One can also think of $\Psi_0$ as specifying boundary conditions. In the following, the rather schematic formula above will be made much more precise\footnote{The path integral quantisation of gravity can however not represent the full theory of quantum gravity. This is because gravity is not renormalisable, which implies that when loop corrections are taken into account effective higher order curvature terms are generated. Only when the spacetime curvature remains well below the Planck scale can such corrections be ignored. In the applications we will consider this will always be the case.}.

\subsection{An Example: Transitions in a de Sitter Universe} \label{sec:ex}

We have just given arguments for why gravity should be quantised, and how one is naturally led to a path integral quantisation scheme. To illustrate this approach, it is useful to study an example. We will specialise to the simplest, yet physically relevant, setting, namely that of gravity coupled 
 to a (positive) cosmological constant $\Lambda$, with action (setting $8\pi G = 1$)
\begin{eqnarray}
S = \int_{\cal M} \mathrm{d} ^4 x  \sqrt{-g} \left( \frac{R}{2} - \Lambda \right)  + \int_{\partial \cal M} \mathrm{d}^3 y \sqrt{h}K \,. \label{LorentzianAction}
\end{eqnarray}
This theory is a first approximation to the current, dark energy dominated, universe, and it may also be a good approximation to the early universe, if inflation took place. The action is integrated over a $4-$manifold ${\cal M},$ with $3-$dimensional boundaries $\partial {\cal M}$ and induced $3-$metric $h_{ij}.$ If one wants to keep the metric fixed on the boundaries, then one has to include the Gibbons-Hawking-York (GHY) \cite{York:1972sj,Gibbons:1976ue} surface term, which involves the determinant $h=det(h_{ij})$ and the trace of the extrinsic curvature $K,$ defined in \eqref{eq:extrinsic}. It has the effect of eliminating second derivatives in the action. The GHY term thus allows for so-called Dirichlet boundary conditions, where one keeps the boundary metric fixed. If one does not include any surface term, one obtains a Neumann condition instead, meaning that one can keep the momentum conjugate to the metric fixed on the boundary \cite{Krishnan:2016mcj}. This will become clear in the example below. 

At first, we will restrict our analysis to the simplified context of  closed Friedmann-Lema\^{i}tre-Roberston-Walker (FLRW) universes with metric
\begin{eqnarray} \label{eq:LorentzianMetric}
\mathrm{d}s ^2 =  - \tilde{N}^2(t) \mathrm{d}t^2 + a ^2(t) \mathrm{d} \Omega _3 ^2 \;,
\end{eqnarray}
where $\tilde{N}$ is the lapse function and $\mathrm{d}\Omega_3^2$ the metric on the unit three-sphere\footnote{Explicitly, one can take $\mathrm{d}\Omega_3^2 = \mathrm{d}\chi^2 + \sin^2\chi \left( \mathrm{d}\theta^2 + \sin^2\theta \, \mathrm{d}\phi^2\right),$ with $0 \leq (\chi, \theta) \leq \pi,$ $0 \leq \phi \leq 2\pi.$ The unit $3-$sphere has volume $2\pi^2.$}. This symmetry reduced setting is an example of {\it minisuperspace}. Not only does this reduction greatly simplify the analysis, it is also expected to be a good approximation when describing the universe on the largest scales, and at early times. The reason for taking the spatial section to be closed is that this prevents a divergent volume integral. One could for instance also take flat sections, and assume the topology of a torus. 

With these choices, the action simplifies to
\begin{eqnarray}
S = 2 \pi ^2 \int_0^1 \mathrm{d} t \tilde{N} \left( - 3 a \frac{ \dot{a} ^2 }{\tilde{N}^2} + 3 a -  a ^3 \Lambda \right) \;, \label{eq:miniact}
\end{eqnarray}
where an overdot stands for a derivative w.r.t. the time coordinate. We are free to choose the time coordinate such that the integral interpolates between fixed $3-$geometries at coordinate values $t=0,1;$ the proper time elapsed is then determined by the lapse $\tilde{N}$. Note that the kinetic term for the scale factor $a$ is negative; this is due to the attractive nature of gravity \cite{Giulini:1994dx} -- it reflects the inherent instability of gravitational systems. Sometimes, this feature is referred to as the ``conformal mode problem'', though it seems more reasonable to consider it a characteristic of gravity rather than a problem.

The path integral was heuristically defined in \eqref{eq:piheuristic} as a sum over all geometries with given boundary conditions. In making this expression precise, one has to face a number of issues: for one, we have not specified contours of integration yet, and moreover we have not dealt with diffeomeorphism invariance. This is important, as in the sum over geometries we would like to include only inequivalent ones, and thus have to eliminate those that are related by changes of coordinates. The way this is done requires techniques that lie somewhat outside of the main thrust of this section, and these techniques are outlined separately in \ref{sec:a2}. The end result is surprisingly simple: the path integral is given by
\begin{align}
G[a_1;a_0] = \int_{\cal C} \mathrm{d}\tilde{N} \int_{a=a_0}^{a=a_1} \mathcal{D} a \, e^{i S(\tilde{N},a)/\hbar}  \,,
\label{eq:gprop}
\end{align} 
where the functional integral over the lapse has been reduced (due to the gauge fixing) to an ordinary integral. We have yet to specify contours of integration, an issue we will return to shortly. The expression above has a simple interpretation: The path integral over the scale factor $\int \mathcal{D}  a e^{iS(\tilde{N},a)/\hbar}$ represents the amplitude for the universe to evolve from scale factor value $a_0$ to the value $a_1,$ in a fixed proper time $\tilde{N}$. Then the integral over the lapse implies that we sum over all possible proper durations that this transition can take. Defined in this way, the path integral corresponds essentially to a gravitational propagator (Green's function), hence the nomenclature $G[a_1;a_0].$ Note that we have implicitly assumed here that the initial wave function corresponds to a ``position'' eigenstate, where the size of the universe is known with infinite precision. This is a convenient idealisation in the present context.

In order to proceed, it turns out to be useful to redefine our variables somewhat, and to write the metric in the following form \cite{Halliwell:1988ik}:
\begin{eqnarray} \label{eq:Metric}
\mathrm{d}s ^2 =  - \frac{N^2}{q(t_q)} \mathrm{d}t_q^2 + q(t_q) \mathrm{d} \Omega _3 ^2 \,.
\end{eqnarray}
That is to say, we call the square of the scale factor $q(t_q),$ and rescale the lapse. This is perhaps a rather unfamiliar writing of the metric, and the new time variable $t_q$ does not have a particular physical meaning, but it has the great advantage that the action becomes quadratic in $q,$ since one obtains
\begin{equation}
S= 2 \pi ^2 \int_0^1 \mathrm{d}t_q \left( -\frac{3}{4 N}\dot{q}^2 + N(3 -  \Lambda q) \right) \,, \label{ActionH}
\end{equation} 
where now $\; \dot{} \equiv d/dt_q$. Here we have again chosen the range of the time coordinate to be $0 \leq t_q \leq 1,$ that is to say $t_q=0$ on the ``initial'' $3-$dimensional boundary, and $t_q=1$ on the ``final'' one, see also Fig. \ref{fig:boundaries}. In deriving the action, the GHY surface terms $\int \mathrm{d}^3y \sqrt{h}K=2\pi^2 \frac{3}{2N}q\dot{q}|_{t_q=0}^{t_q=1}$ have been incorporated, and they have eliminated second derivatives on $q,$ making use also of integration by parts.

Varying the action w.r.t. $q$ and $N$ results in
\begin{align}
    \delta S = 2\pi^2 \int \mathrm{d}t_q \left[\delta q \left(\frac{3}{2N}\ddot{q} -N\Lambda\right) + \delta N \left( \frac{3}{4 N^2} \dot{q}^2 +3 - \Lambda q\right) \right] -\frac{3\pi^2}{N} \dot{q}\delta q|_{t_q=0}^{t_q=1}
\end{align}
This confirms that we can indeed keep $q$ fixed on the boundaries ($\delta q =0$). We set $q(0)=q_0$ and $q(1)=q_1.$ The equation of motion and constraint are then respectively given by
\begin{eqnarray}
\ddot{q}  =  \frac{2\Lambda}{3}N^2; \quad \frac{3}{4 N^2} \dot{q}^2 +3  = \Lambda q \;. \label{qconstraint} 
\end{eqnarray}
The equation of motion can be solved easily, and with the chosen boundary conditions the solution is
\begin{equation}
\label{eq:classicalsolution}
\bar{q}=\frac{\Lambda}{3}N^2 t_q^2 + \left(- \frac{\Lambda}{3}N^2+ q_1- q_0\right) t_q + q_0\,.
\end{equation}
Note that this is just a solution of the equation of motion, and not necessarily of the constraint. 

Now we employ a trick to evaluate the path integral over $q,$ which consists in shifting the variables of integration such that a general history is written as a departure from a classical solution,
\begin{equation}
q(t_q) = \bar{q}(t_q) + Q(t_q)\,.
\end{equation}
For consistency with the boundary conditions, we must set $Q(0)=Q(1)=0,$ but otherwise $Q$ is not restricted (in particular $Q$ is not required to be small). With this shift (which is only a change of variables, and not an approximation), the path integral becomes
\begin{equation}
G[q_1;q_0] = \int_{\cal C} \mathrm{d} N e^{2\pi^2 i S_0/\hbar} \int_{Q[0]=0}^{Q[1]=0} \mathcal{D} Q e^{2\pi^2 i S_2/\hbar}\,,
\end{equation}
with
\begin{eqnarray}
S_0 &=&  \int_0^1 \mathrm{d}t_q \left( -\frac{3}{4 N}\dot{\bar{q}}^2 + 3N - N \Lambda \bar{q} \right) \,, \label{ActionH4}\quad 
S_2 =  -\frac{3}{4 N} \int_0^1 \mathrm{d}t_q\,  \dot{Q}^2 \,. \label{ActionH3}
\end{eqnarray} 
No terms linear in $Q$ have appeared, precisely because $\bar{q}$ solves the equation of motion. But now the integral over $Q$ is a Gaussian, which can be evaluated exactly \cite{Grosche:1998yu} (with a uniquely determined contour of integration), yielding
\begin{equation} \label{Qintegral}
\int_{Q[0]=0}^{Q[1]=0} \mathcal{D} Q e^{2\pi^2 i S_2/\hbar} = \sqrt{\frac{3\pi i}{2N\hbar}} \,.
\end{equation}
The time integral in \eqref{ActionH4} can be evaluated explicitly, so that in the end we are left with an ordinary integral over the lapse function,
\begin{equation}
\label{eq:propagator}
G[q_1;q_0] = \sqrt{\frac{3\pi i}{2\hbar}}\int_{\cal C} \frac{\mathrm{d} N}{N^{1/2}} e^{2\pi^2 i S_0/\hbar}\,,
\end{equation}
with
\begin{equation}
S_0 = N^3 \, \frac{\Lambda^2}{36} + N \left( -\frac{\Lambda}{2}(q_0+q_1) +3 \right) +\frac{1}{N}\left( -\frac{3}{4} (q_1-q_0)^2\right)\,. \label{eq:Naction}
\end{equation}

We will analyse this integral by performing a saddle point approximation. The proper tool for doing this systematically is Picard-Lefschetz theory, which is reviewed in \ref{sec:a3}. First, we must determine the saddle points, defined by $\partial S_0/\partial N = 0.$ There are four of them, located at
\begin{equation}
\label{saddles}
N_\sigma = c_1 \frac{3}{\Lambda} \left[ \left( \frac{\Lambda}{3} q_0-1\right)^{1/2} + c_2 \left( \frac{\Lambda}{3} q_1-1\right)^{1/2}\right]\,,
\end{equation}
with $c_1, c_2\in\{-1,1\}$. Moreover, the action at the saddle points is found to be
\begin{eqnarray}
S_0^{saddle} =  -c_1 \frac{6}{\Lambda}  \left[ \left(\frac{\Lambda}{3} q_0 - 1\right)^{3/2} +c_2 \left(\frac{\Lambda}{3} q_1 - 1\right)^{3/2} \right]\,.
\end{eqnarray}

\begin{figure} 
	\centering
	\includegraphics[width=0.65\textwidth]{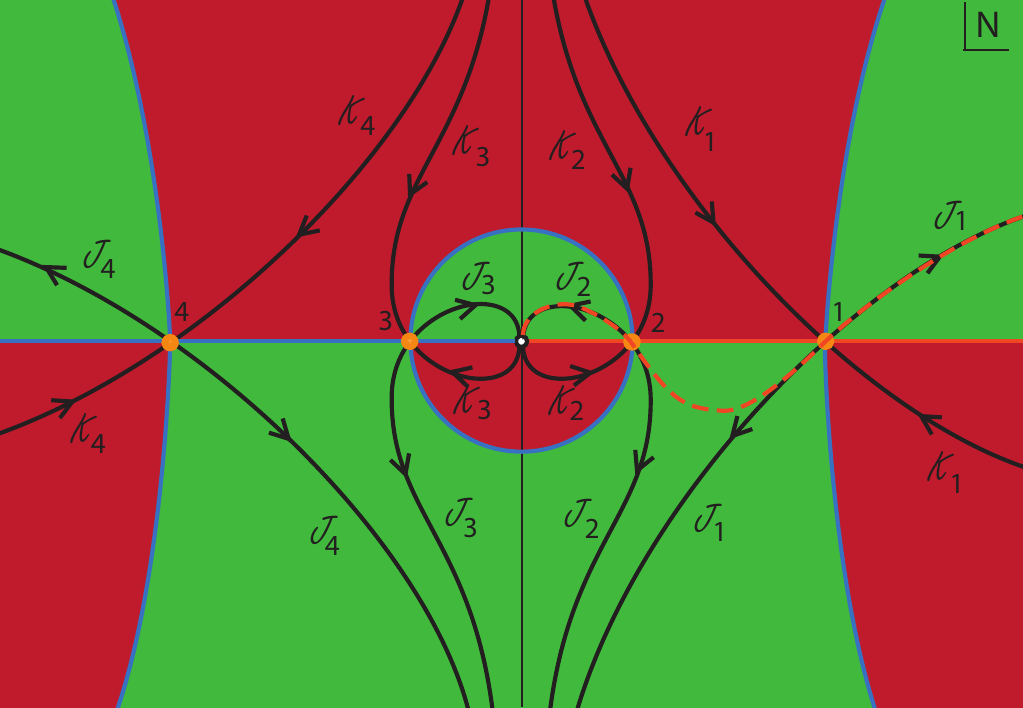}
	\caption{This sketch shows the placement of the steepest descent (${\cal J}$) and ascent (${\cal K}$) contours associated with the four saddle points of the integral \eqref{eq:propagator}, in the complexified plane of the lapse function $N.$ Regions where the integral converges (asymptotically) are shown in green, and regions of asymptotic divergence in red. (The lines delineating red and green regions have constant weighting.) The boundary conditions are such that the initial and final scale factors are larger than the Hubble radius, $q_1>q_0>\frac{3}{\Lambda}.$ The positive real line contour can be deformed into the sum ${\cal J}_2 + {\cal J}_1,$ or equivalently the dashed orange contour. Figure reproduced from \cite{Feldbrugge:2017kzv}. }
	\protect
	\label{fig:classical2}
\end{figure} 

The question we are faced with now is whether all of these saddle points actually contribute to the integral. This turns out to depend both on the boundary conditions, and on the integration contour ${\cal C}.$ \\

\noindent {\it Classical boundary conditions}

We will first look at classical boundary conditions, where $q_0,q_1 > \frac{3}{\Lambda}$ and consequently the saddle points \eqref{saddles} are real. We will also assume that $q_1>q_0.$ As for the contour ${\cal C},$ it makes sense to look back at the definition of the metric \eqref{eq:Metric}: the lapse determines the proper time separation between events. If we want the propagator to correspond to causal evolution, where time moves forward, then we can do this by fixing the sign of $N,$ say to be positive. Moreover, at zero lapse the metric degenerates, and thus zero should be excluded. This suggests that we should take the positive real half-line as integration contour \cite{Teitelboim:1983fh},
\begin{align}
{\cal C} = (0^+,\infty)\,.
\end{align}
Note that \eqref{eq:propagator} contains an essential singularity at $N=0,$ which reflects the physical intuition that a non-trivial transition cannot occur in zero proper time, and supports the exclusion of zero from the contour. 

Along the real $N$ line, the integral \eqref{eq:propagator} is only conditionally convergent. Using Picard-Lefschetz theory, we would like to rewrite it as a sum over absolutely convergent integrals,
\begin{align}
{\cal C} = (0^+,\infty) = \sum_\sigma n_\sigma \cal J_\sigma\,.
\end{align}
where ${\cal J}_\sigma$ are steepest descent contours (thimbles) emanating from the saddle points and the $n_\sigma$ can be $\pm 1$ for contributing thimbles, depending on the chosen orientation of the thimbles. In the present case, the two saddle points on the positive real axis clearly contribute to the integral, and one can indeed rewrite $(0^+,\infty) = {\cal J}_2 + {\cal J}_1,$ where the thimbles are oriented in the increasing $Re[N]$ direction, see Fig.~\ref{fig:classical2}. Thus the propagator can be approximated as a sum of two phases,
\begin{eqnarray} \label{nbwf_classical}
G[q_1;q_0]  
&\approx&  \, e^{-i\frac{\pi}{4}} e^{\frac{i}{\hbar}S(N_{2})} +   \, e^{i\frac{\pi}{4}}e^{\frac{i}{\hbar}S(N_{1})}  \label{eq:wfclassical}\\
  &\approx& \cos\left(\frac{12\pi^2}{ \hbar\Lambda} \left(\frac{\Lambda}{3}q_0 - 1\right)^{3/2} -\frac{\pi}{4}\right)\,  e^{-i \frac{12\pi^2}{\hbar\Lambda}  \left(\frac{\Lambda}{3}q_1 - 1\right)^{3/2}}\,. 
\end{eqnarray} 
In this approximation, we have also included factors of $e^{\pm i\frac{\pi}{4}},$ which arise from the angle of the Lefschetz thimbles at the saddle points (and are straightforwardly determined by looking at the change in $Re(iS_0)$ near the saddle points \cite{Feldbrugge:2017kzv}). 

A number of questions now arise, foremost: What is the meaning of this mathematical expression? We will address this quite generally in sections \ref{sec:formal} and \ref{sec:reconstruct}. For now, we can make progress by recalling the intuition behind the path integral. All possible histories contribute to the transition, but the saddle points represent the dominant ones. Hence it is useful to look at the saddle points in more detail. 

\begin{figure} 
	\centering
	\includegraphics[width=0.35\textwidth]{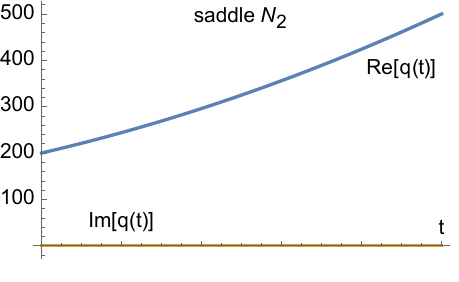} \hspace{1cm}
	\includegraphics[width=0.35\textwidth]{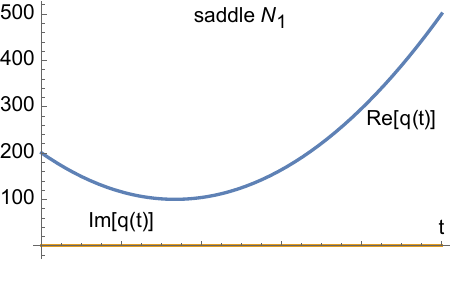}
	\caption{Saddle point geometries for classical boundary conditions. For this example, $\Lambda = \frac{3}{100}, \, q_0=200,\, q_1=500,$ so that $N_2=100, \, N_1 = 300$. Saddle $2$ is purely expanding, while saddle $1$ first shrinks to the de Sitter radius (with $q_{min}=\frac{3}{\Lambda}=100$) and then re-expands.}
	\protect
	\label{fig:saddlesclassical}
\end{figure} 

As a consequence of satisfying $\partial S_0/\partial N = 0,$ the saddle points correspond to geometries that satisfy not only the equation of motion, but also the constraint in \eqref{qconstraint}; they are solutions of the full Einstein equations. We may then find the geometry that they describe by plugging the saddle point values \eqref{saddles} into the expression for the scale factor \eqref{eq:classicalsolution}. The resulting expression is lengthy and not worth writing out in full. However, it is useful to look at the expansion rate at $t=0:$
\begin{equation}
\frac{\mathrm{d}\bar{q}}{\mathrm{d}t_q}(t_q=0) = -\frac{\Lambda}{3}N_{\sigma}^2 + q_1 - q_0 = 2\left(- q_0 + \frac{3}{\Lambda} \right) \pm 2 \sqrt{\left(q_0 - \frac{3}{\Lambda} \right) \left(q_1 - \frac{3}{\Lambda} \right) }\,. \label{eq:initmom}
\end{equation}
Since $q_1>q_0,$ we can infer that saddle point $2$ corresponds to an initially expanding geometry, while saddle $1$ is contracting. Explicit examples are shown in Fig.~\ref{fig:saddlesclassical}. The classical, maximally symmetric solution to the Einstein equations with a positive cosmological constant is de Sitter spacetime. In the closed slicing, where the spatial sections are $3-$spheres, it is well known that de Sitter spacetime can be viewed as a hyperboloid (with minimal radius $\sqrt{\frac{3}{\Lambda}}$) when embedded in a higher-dimensional space. The saddle points correspond to portions of this hyperboloid, with saddle $2$ corresponding to a portion of the hyperboloid on one side of the minimal radius, while saddle $1$ includes the region with minimal radius and thus contains a classical bounce. 

These geometries provide the dominant contribution to the transition from $q_0$ to $q_1.$ Since we fixed only the initial and final sizes, and since two possible classical solutions exist which link $q_0$ to $q_1,$ it is natural that in \eqref{eq:wfclassical} we obtained a sum over the two contributions, in complete analogy with the double slit experiment. In a more realistic model, one would also include matter and fluctuations of the geometry -- such interactions would then cause decoherence and would therefore lead to separate, non-interfering evolutions of the two saddle points.\\

\noindent {\it Non-classical boundary conditions}

\begin{figure} 
	\centering
	\includegraphics[width=0.75\textwidth]{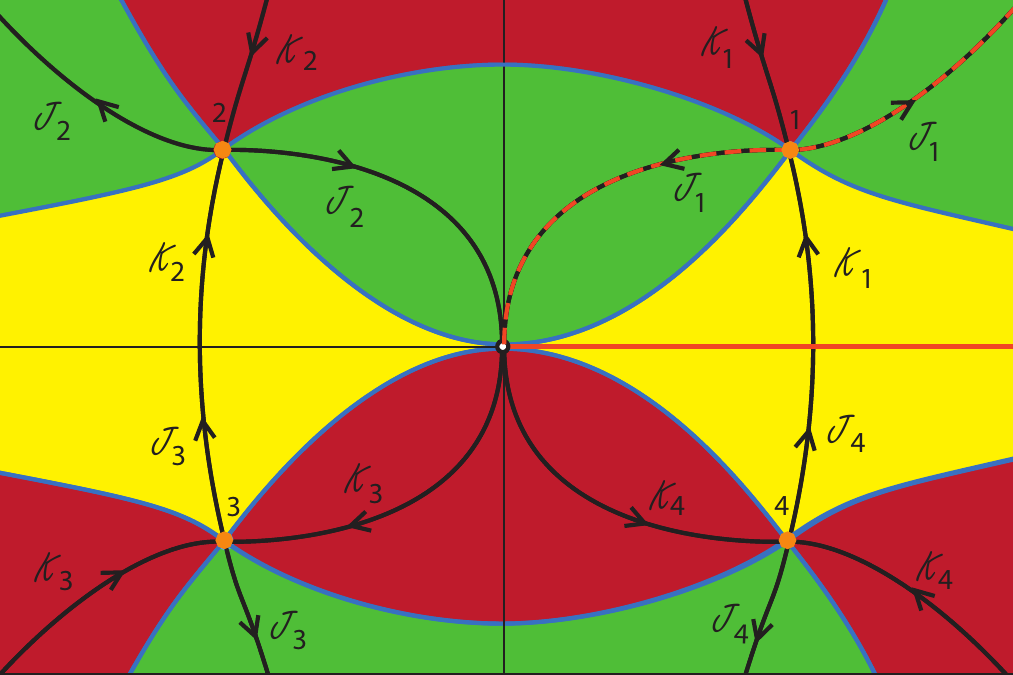}
	\caption{A sketch of the steepest ascent/descent lines for boundary conditions where the first scale factor is smaller than the de Sitter radius ($q_0<\frac{3}{\Lambda}$), and the second larger ($q_1 > \frac{3}{\Lambda}$). In the yellow regions, the weighting is in between the weightings of the adjacent saddle points. The orange integration contour $(0^+,\infty)$ can be deformed into the dashed orange line, which is the thimble ${\cal J}_1$ associated with saddle point $1.$ Thus this is the only saddle point contributing significantly to the transition. Figure reproduced from \cite{Feldbrugge:2017kzv}.}
	\protect
	\label{fig:upper}
\end{figure}

What we discussed above were transitions between large, classically allowed, scale factor values. But it is interesting to see what happens when we set one of the scale factors, say $q_0,$ to a small, classically forbidden value, i.e. $q_0 < \frac{3}{\Lambda}$. Then the saddle points become complex, with
\begin{equation}
\label{saddlessmall}
N_\sigma = c_1 \frac{3}{\Lambda} \left[i \left(1- \frac{\Lambda}{3} q_0\right)^{1/2} + c_2 \left( \frac{\Lambda}{3} q_1-1\right)^{1/2}\right]\,,
\end{equation}
again with $c_1, c_2\in\{-1,1\}$. The action at the saddle points also becomes complex,
\begin{eqnarray}
S_0^{saddle} =  c_1 \frac{6}{\Lambda}  \left[ i \left(1-\frac{\Lambda}{3} q_0 \right)^{3/2} -c_2 \left(\frac{\Lambda}{3} q_1 - 1\right)^{3/2} \right]\,.
\end{eqnarray}
Now the steepest ascent/descent contours look quite different, see Fig.~\ref{fig:upper} for an illustration.

\begin{figure} 
	\centering
	\includegraphics[width=0.55\textwidth]{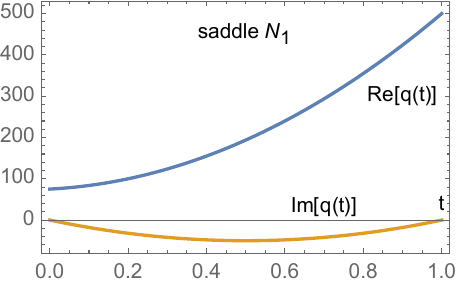} 
	\caption{Saddle point geometry for a transition from non-classical to classical boundary conditions. For this example, $\Lambda = \frac{3}{100}, \, q_0=75,\, q_1=500,$ so that $N_1 = 200 + 50i$. Note that the scale factor starts and ends with real values, but is complex in between.}
	\protect
	\label{fig:saddlenonclassical}
\end{figure}

As the figure shows, the integration contour ${\cal C}=(0^+,\infty)$ can be deformed into the steepest descent contour ${\cal J}_1$ associated with saddle point $1,$ with positive real and imaginary parts. Thus in this case only a single saddle point contributes significantly to the path integral, and the amplitude can be approximated by
\begin{align}
    G[q_1;q_0] &\approx e^{\frac{i}{\hbar}2\pi^2 S_0(N_1)} \\
    &\approx e^{-\frac{12\pi^2}{\hbar \Lambda} \left(1-\frac{\Lambda}{3} q_0 \right)^{3/2}}e^{- i \frac{12\pi^2}{\hbar \Lambda} \left(\frac{\Lambda}{3} q_1 -1 \right)^{3/2}}\,. \label{eq:wfnonclassical}
\end{align}
An example of the associated geometry is shown in Fig.~\ref{fig:saddlenonclassical}. Apart from the end points, the geometry is now complex, reflecting the fact that this transition is classically impossible, and rather represents quantum tunnelling between the two specified scale factor values. In line with this interpretation, note that the amplitude \eqref{eq:wfnonclassical} is suppressed as $e^{-\frac{12\pi^2}{\hbar \Lambda} \left(1-\frac{\Lambda}{3} q_0 \right)^{3/2}}$, indicating that this transition is less likely than classical evolution. Also, the smaller $q_0$ is, the stronger the suppression. Meanwhile, the classically allowed part of the transition, related to $q_1,$ leads to an oscillating factor in the amplitude \eqref{eq:wfnonclassical}. This setting is thus very much in analogy with standard barrier penetration problems in ordinary quantum mechanics.

Let us highlight what may be a surprising feature: even though the saddle points are solutions to the classical equations of motion, they can describe quantum effects. This is because the boundary conditions are classically impossible here, forcing upon us a \emph{complex} solution to the classical field equations. \\

\noindent {\it Wheeler-DeWitt equation}

The transition amplitudes that we derived above satisfy an important equation, called the Wheeler-DeWitt (WdW) equation. Later on, we will derive it formally, but it is already useful to see how it arises for the specific examples we have just studied.

The Lagrangian of our system can be inferred from the action \eqref{ActionH}, and reads
\begin{align}
L = 2\pi^2\left[ - \frac{3}{4N}\dot{q}^2 + 3N - N \Lambda q\right].
\end{align}
Thus, the canonical momentum associated with $q$ is
\begin{align}
p = \frac{\partial L}{\partial \dot{q}} = -   \frac{3 \pi^2}{N}  \dot{q}\,, \label{cancon}
\end{align}
so that the Hamiltonian can be written as
\begin{align}
H =& \dot{q} p -L= -\frac{N}{6 \pi^2} \left[ p^2 + 12 \pi^4 (3 -\Lambda q)\right]=N\hat{H}\,. \label{Ham}
\end{align}
In phase-space the action can thus also be written as
\begin{align}
S = \int \left(\dot{q}p - N\hat{H}\right)\mathrm{d}t_q = \int \left(\dot{q}p + \frac{N}{6 \pi^2} \left[ p^2 + 12 \pi^4 (3 -\Lambda q)\right]\right)\mathrm{d}t_q\,.
\end{align}
Here we can see that the lapse $N$ is a Lagrange multiplier, and implies the classical constraint
\begin{align}
\hat{H}=0\,.
\end{align}
But now we can also quantise the theory directly, in the field representation, by promoting the momentum to its operator equivalent,  $p \mapsto \hat{p} = - i \hbar \frac{\partial}{\partial q}$. This results in the Wheeler-DeWitt equation
\begin{align}
 \hat{H} \Psi = 0 \rightarrow  \,\, & \hbar^2\frac{\partial^2 \Psi}{\partial q^2} + 12 \pi^4 (\Lambda q - 3)\Psi =0\,,
\end{align}
with $\Psi(q)$ the wave-function of the universe. The corresponding propagator $G$ satisfies \cite{Halliwell:1988wc}
\begin{align}
\hat{H}G = - i 6\pi^2 \delta(q_0-q_1)\,,
\end{align}
where the Hamiltonian operator acts either on $q_0$ or on $q_1$ (the factor $6\pi^2$ is also sometimes absorbed into the definition of $\hat{H}$). This last equation is referred to as the inhomogeneous Wheeler-DeWitt equation. 

Now we can go back to the path integral and explicitly show that it satisfies this equation. Starting from \eqref{eq:propagator} and \eqref{eq:Naction}, one can take successive derivatives (momentarily setting $\hbar=1$)
\begin{align}
\frac{\partial G}{\partial q_1}
=& \sqrt{\frac{3\pi i}{2}}\int \frac{\mathrm{d}N}{N^{1/2}} 2\pi^2 i S_{0,q_1} e^{2\pi^2i S_0 }\nonumber\\
=& \sqrt{\frac{3\pi i}{2}}\int \frac{\mathrm{d}N}{N^{1/2}} 2\pi^2 i\left[ -\frac{N}{2}\Lambda - \frac{3}{2N} (q_1-q_0) \right] e^{2\pi^2i S_0}\,,
\end{align}
and analogously
\begin{align}
\frac{\partial^2 G}{\partial q_1^2} 
=& \sqrt{\frac{3\pi i}{2}} \int \frac{\mathrm{d}N}{N^{1/2}} \left[ 2 \pi^2 i S_{0,q_1q_1} - 4 \pi^4 S_{0,q_1}^2 \right] e^{2\pi^2iS_0}\nonumber\\
=& \sqrt{\frac{3\pi i}{2}} \int \frac{\mathrm{d}N}{N^{1/2}} \left[ -4\pi^4 \left(\frac{N}{2}\Lambda + \frac{3}{2N} (q_1-q_0)\right)^2 - \frac{3\pi^2i}{N} \right] e^{2\pi^2iS_0}\,. \label{eq:secondcoming}
\end{align}
Using first the properties of the thimbles, and then integration by parts, one can obtain the relation
\begin{align}
\left[N^{-\frac{1}{2}}e^{2\pi^2 i S_0}\right]_0^\infty =& \int \mathrm{d}N \frac{\mathrm{d}}{\mathrm{d}N}\left[N^{-\frac{1}{2}}e^{2\pi^2 i S_0}\right] \nonumber \\ =&-\frac{1}{2}\int \frac{\mathrm{d}N}{N^{\frac{3}{2}}} e^{2\pi^2 i S_0} + 2 \pi^2 i \int \frac{\mathrm{d}N}{N^{\frac{1}{2}}} S_{0,N}e^{2\pi^2 i S_0}\,.
\end{align}
This is now substituted into \eqref{eq:secondcoming} to obtain
\begin{align}
\frac{\partial^2 G}{\partial q_1^2}
=&
\sqrt{\frac{3\pi i}{2}} \left[\int \frac{\mathrm{d}N}{N^{1/2}} \left[ -12\pi^4 \left(  \Lambda q_1 - 3k 
\right)  \right] e^{2\pi^2iS_0} + 6 \pi^2 i  \left[ N^{-\frac{1}{2}} e^{2\pi^2 i S_0}\right]_0^\infty
\right]\nonumber\\
=& -12\pi^4 \left(  \Lambda q_1 - 3\right) G + 6 \pi^2 i  \sqrt{\frac{3\pi i}{2}}\left[ N^{-\frac{1}{2}} e^{2\pi^2 i S_0}\right]_0^\infty\,,
\end{align}
which is almost the WdW equation already. The Lefschetz thimble we are integrating over is precisely such that the contribution from $N\to \infty$ vanishes. But near the origin, we have to work a little harder: given that the thimble there approaches the origin along the imaginary axis, it helps to write $N=in$, leading to
\begin{align}
\lim_{N\to 0} \frac{e^{2\pi^2 i S_0}}{\sqrt{N}} =& \sqrt{\frac{2\pi}{i}} \lim_{n \to 0} \frac{ e^{-3\pi^2  \frac{(q_1-q_0)^2}{2n}}}{\sqrt{2\pi n}}
=  \sqrt{\frac{2}{3 \pi i}} \delta(q_0-q_1)\,.
\end{align}
Hence, with $\hbar$ reinstated, we indeed recover the WdW (propagator) equation 
\begin{align} 
\hbar^2 \frac{\partial^2 G}{\partial q_1^2} + 12\pi^4 \left(  \Lambda q_1 - 3\right) G 
=&  - 6 \pi^2 i  \delta(q_0-q_1).
\end{align}
For future reference, let us point out that if we had integrated over a contour from $N=-\infty$ to $N=+\infty$ (passing around the singularity at $N=0$), the right hand side would have been zero and we would have obtained the homogeneous WdW equation. 

The WdW equation is the quantum equivalent of the Friedmann equation in cosmology. In some sense, it is the equivalent of the Schr\"{o}dinger equation when gravity is included. (Later on, we will see that it actually contains the Schr\"{o}dinger equation.) Here we see that, by construction, the path integral automatically satisfies this equation. We will discuss it in more detail in the coming sections. \\

\noindent {\it Neumann boundary conditions}

Before proceeding with more formal developments, it is useful to give an example of the path integral where instead of fixing the scale factor on both sides of the transition, we fix its conjugate momentum instead (on one side). This means that we will consider a Neumann, rather than a Dirichlet, condition. For definiteness, we will impose a Neumann condition on the $t=0$ boundary. This can be achieved by not adding a GHY surface term to the action there. That is to say, the minisuperspace action is now given by
\begin{align}
    S=2\pi^2 \int_0^1 \mathrm{d}t_q \left[ \frac{3}{2N}q\ddot{q} +\frac{3}{4N}\dot{q}^2 + N(3-\Lambda q)\right] - \frac{3\pi^2}{N}q\dot{q}|_{t_q=1}
\end{align}
If we now use integration by parts in order to get rid of the second derivative term in the action, then this will cancel the GHY boundary term at $t_q=1$ but instead it will generate a surface term at $t_q=0,$
\begin{align}
     S=2\pi^2\int_0^1 \mathrm{d}t_q \left[-\frac{3}{4N}\dot{q}^2 + N(3-\Lambda q)\right] - \frac{3\pi^2}{N}q\dot{q}|_{t_q=0}
\end{align}
Varying w.r.t. the scale factor $q,$ we obtain the same equation of motion as before, Eq. \eqref{qconstraint}, together with the boundary conditions
\begin{align}
    -\frac{3\pi^2}{N}q \delta(\dot{q})=0|_{t_q=0}\,, \qquad -\frac{3\pi^2}{N}\dot{q} \delta q |_{t_q=1}\,,
\end{align}
confirming that we can specify the momentum $p_0=-\frac{3\pi^2}{N}\dot{q}(t_q=0)$ and the scale factor $q_1=q(t_q=1).$ With these boundary conditions, the solution to the equation of motion is
\begin{align}
    \bar{q} = \frac{\Lambda}{3}N^2 t_q^2 - \frac{p_0 N}{3\pi^2} t_q + q_1 - \frac{\Lambda}{3}N^2 +\frac{p_0 N}{3\pi^2}\,. \label{eq:qbarND}
\end{align}
We can now employ the same trick as before, namely we can shift the integration variable $q=\bar{q} + Q(t_q).$ This time, the fluctuation integral satisfies the boundary conditions $\dot{Q}(t_q=0)=0$ and $Q(t_q=1)=0,$ and with these boundary conditions it simply evaluates to a constant (independent of the lapse) \cite{DiTucci:2020weq},
\begin{align}
    \int {\cal D}Q e^{-\frac{i}{\hbar}\int_0^1 dt_q \frac{3\pi^2\dot{Q}^2}{2N}} = \sqrt{\frac{3\pi i}{2\hbar}}\,.
\end{align}
To leading order, this factor is unimportant and we will ignore it. We are then left again with an ordinary integral over the lapse,
\begin{align}
    G[p_0;q_1] = \int_{\cal C} \mathrm{d}N \, e^{\frac{i}{\hbar}2\pi^2\left[ \frac{\Lambda^2}{9}N^3 - \frac{p_0\Lambda}{6\pi^2}N^2 +(3+\frac{p_0^2}{12\pi^4}-q_1 \Lambda)N +\frac{p_0 q_1}{2\pi^2}\right]}\,, \label{eq:wfND}
\end{align}
where we have not specified the integration contour ${\cal C}$ yet. There are significant differences with the pure Dirichlet case. For one, the integrand is regular at $N=0.$ This makes sense, as we fix the initial momentum and not the initial size. Hence the path integral sums over a range of initial sizes, and this can include a transition where the initial size is already equal to the final size -- precisely the $N=0$ case, cf. also \eqref{eq:qbarND}. Moreover, the integral \eqref{eq:wfND} admits only two saddle points now, located at 
\begin{align}
    N_\pm = \frac{p_0}{2\pi^2 \Lambda} \pm \frac{3}{\Lambda}\sqrt{\frac{\Lambda}{3}q_1 - 1}\,,
\end{align}
and with action
\begin{align}
    S(N_\pm) = \frac{3p_0}{\Lambda} + \frac{p_0^3}{36\pi^4\Lambda} \mp \frac{12\pi^2\left(\frac{\Lambda}{3}q_1 - 1\right)^{3/2}}{\Lambda}\,. \label{NDaction}
\end{align}

\begin{figure} 
	\centering
\begin{minipage}[t]{0.5\textwidth}
\includegraphics[width=\textwidth]{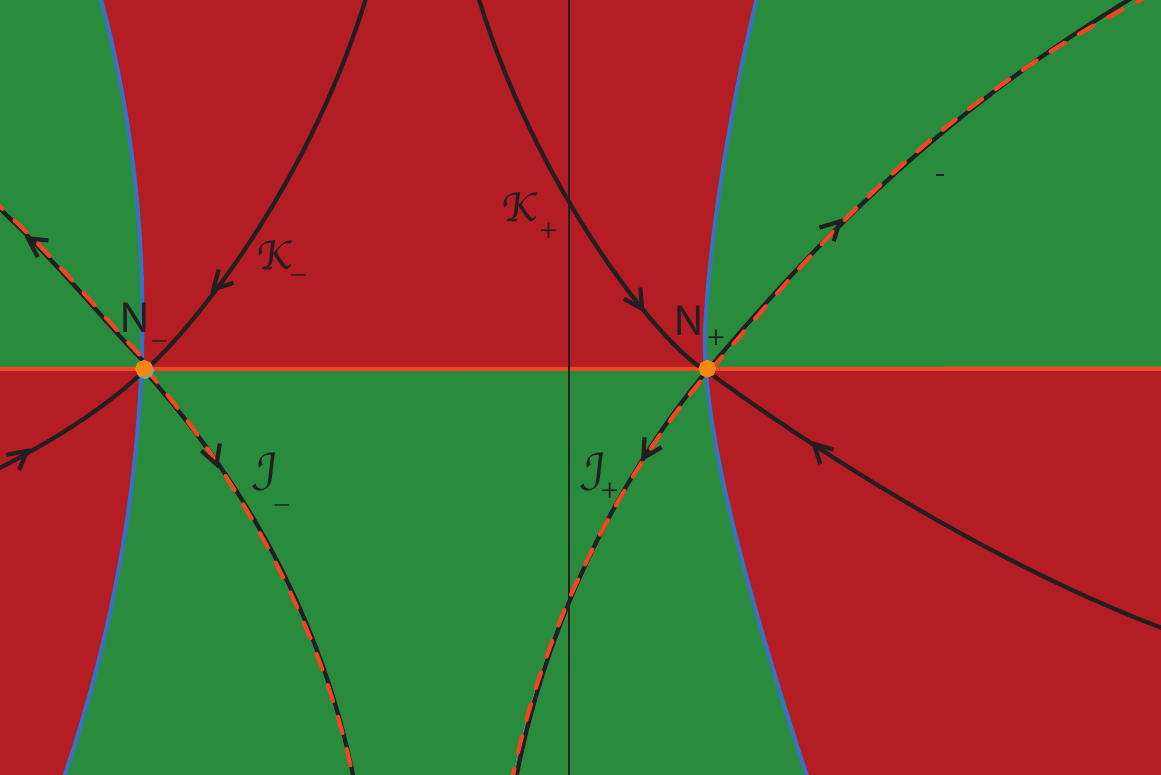}
\end{minipage}
\quad
\begin{minipage}[t]{0.45\textwidth}
\vspace{-4.1cm}
\includegraphics[width=0.65\textwidth]{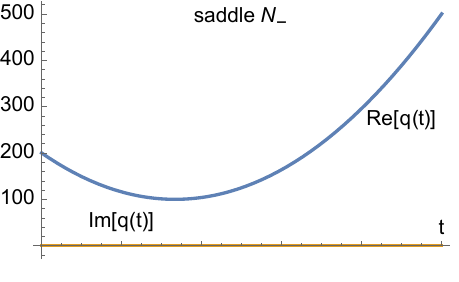}\\
\hspace{1cm}
\includegraphics[width=0.65\textwidth]{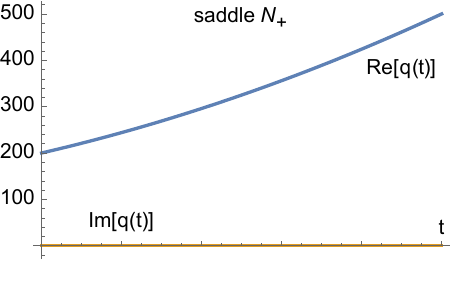} 
\end{minipage}
	\caption{Saddle points and their steepest descent/ascent lines in the complex plane of the lapse function, for the situation where we have a transition with mixed Neumann-Dirichlet boundary conditions. Here $\Lambda = \frac{3}{100}, \, p_0=-6\pi^2,\, q_1=500,$ so that $N_\mp = (-300,+100)$.}
	\protect
	\label{fig:momcl2}
\end{figure} 

To proceed, it is again useful to look at examples. For definiteness, let us use initial momenta $p_0$ that correspond to the momenta for the saddle point geometries encountered above for pure Dirichlet boundary conditions. That is to say, we are choosing initial momenta such that at least one saddle point must be equal to the one from the Dirichlet case. For the example shown in Figs.~\ref{fig:classical2} and \ref{fig:saddlesclassical}, the momenta at $t=0$ are real. For saddle $2,$ which corresponded to an expanding geometry, we have from \eqref{eq:initmom} that $p_0=-3\pi^2\dot{q}/N_2(t=0)=-6\pi^2.$ The corresponding thimbles and saddle point geometries are shown in Fig.~\ref{fig:momcl2}. We can see that indeed the saddle point $N_+$ corresponds to the expanding geometry. But what is more surprising is the saddle $N_-,$ which is again the bouncing geometry that we also encountered in the Dirichlet case. How can this saddle point be present, given that we fixed the initial momentum? The resolution is that it has a negative lapse, so that the combination of an initially contracting geometry with a negative lapse can yield the same momentum as an expanding universe with positive lapse, given that $p=-\frac{3\pi^2}{N}\dot{q}$. The integration contour for the lapse is essentially unique: the real lapse line can be deformed into a sum over the two thimbles associated with $N_\pm,$
\begin{align}
   {\cal C}= (-\infty,\infty) = {\cal J}_- + {\cal J}_+\,.
\end{align}
Note that because of the absence of a singularity at $N=0$ we must integrate over the full lapse line in order to obtain an invariant definition of the path integral. Thus, with this integration contour and with these classical boundary conditions, we recover the analogous semi-classical amplitude as for the pure Dirichlet case, namely a sum over two saddles with phases given by \eqref{NDaction}.

\begin{figure} 
	\centering
	\begin{minipage}[t]{0.5\textwidth}
\includegraphics[width=\textwidth]{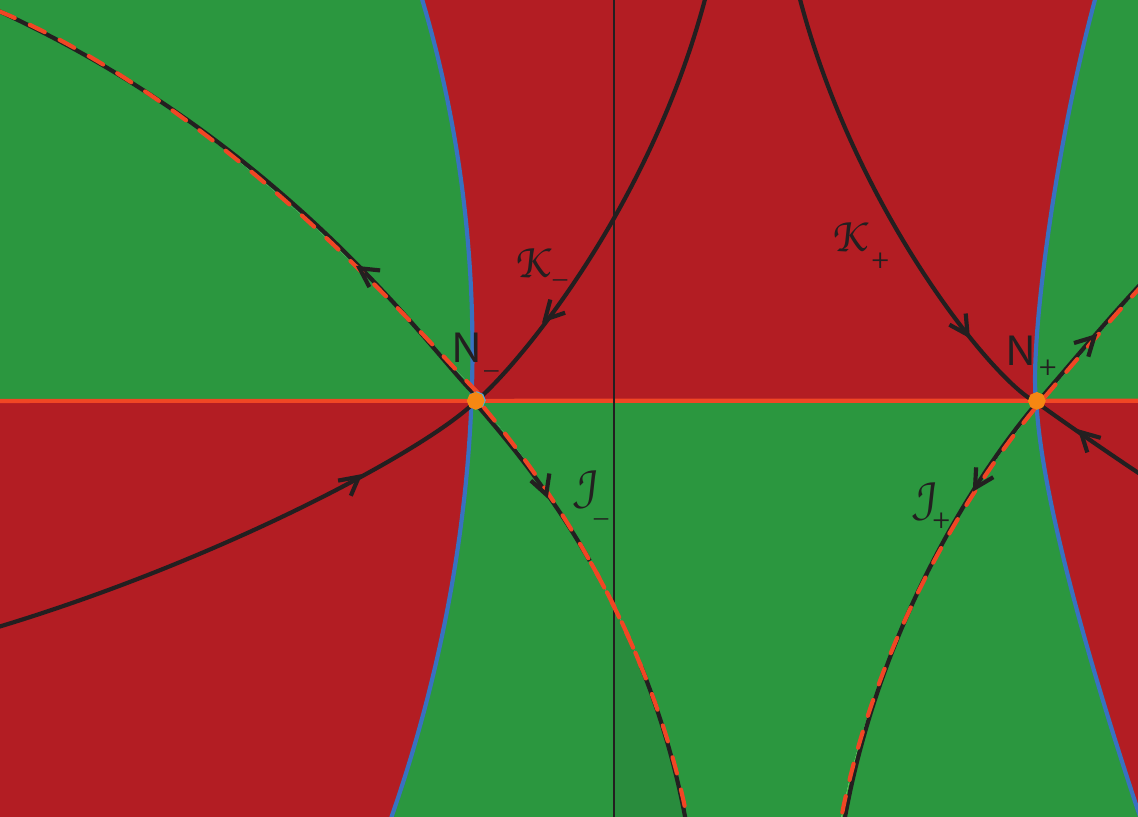}
\end{minipage}
\quad
\begin{minipage}[t]{0.45\textwidth}
\vspace{-4.3cm}
\includegraphics[width=0.65\textwidth]{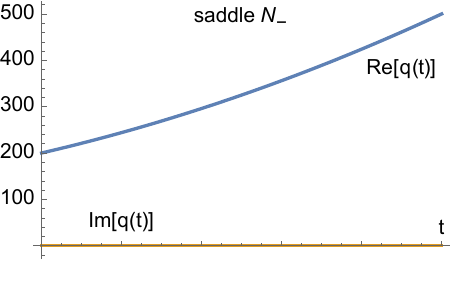}\\
\hspace{1cm}
\includegraphics[width=0.65\textwidth]{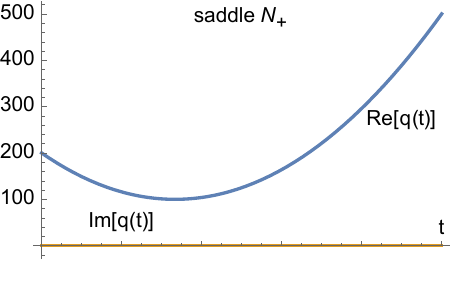} 
\end{minipage}
	\caption{Same as Fig. \ref{fig:momcl2}, but with $\Lambda = \frac{3}{100}, \, p_0=+6\pi^2,\, q_1=500,$ so that $N_\sigma = (-100,+300)$. }
	\protect
	\label{fig:momcl1}
\end{figure} 

When we choose an initial momentum that corresponds to the bouncing saddle (saddle $1$) in Figs.~\ref{fig:classical2} and \ref{fig:saddlesclassical}, we obtain a very similar situation, but this time it is the expanding geometry that arises with negative lapse, see Fig.~\ref{fig:momcl1}. In fact, the saddles in Figs.~\ref{fig:momcl2} and \ref{fig:momcl1} combined reproduce the four saddles of the Dirichlet case.

\begin{figure} 
	\centering
	\begin{minipage}[t]{0.5\textwidth}
\includegraphics[width=\textwidth]{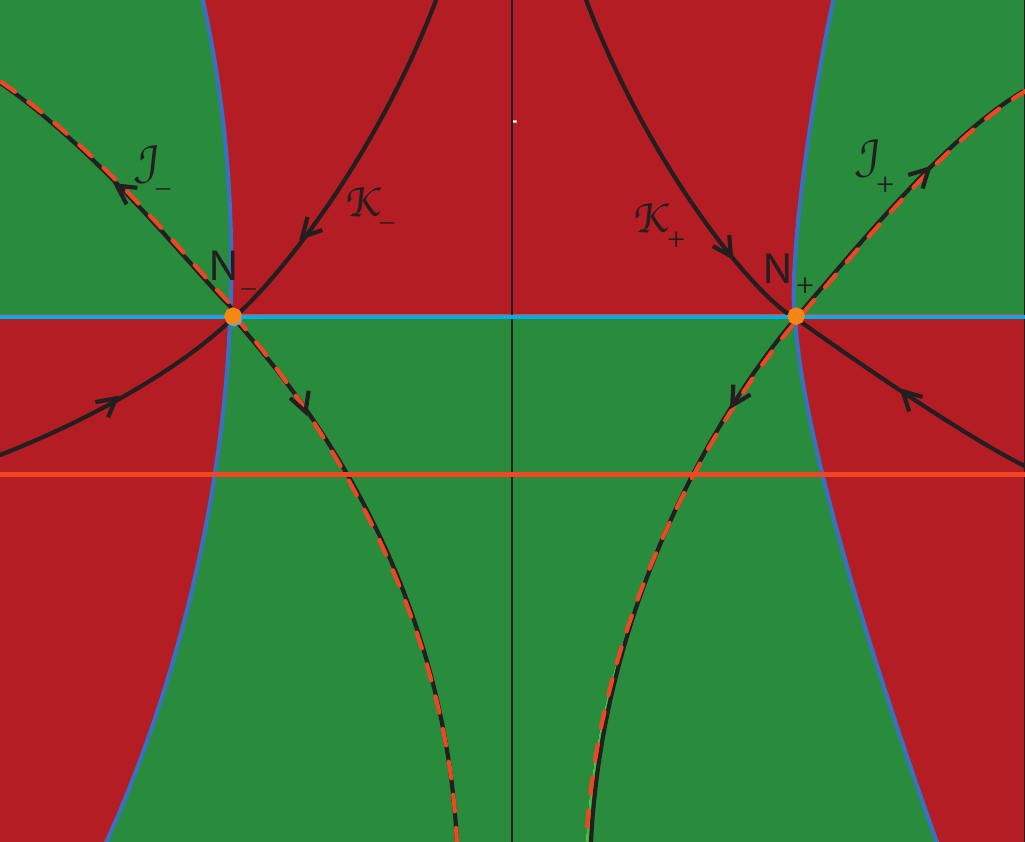}
\end{minipage}
\quad
\begin{minipage}[t]{0.45\textwidth}
\vspace{-4.7cm}
\includegraphics[width=0.65\textwidth]{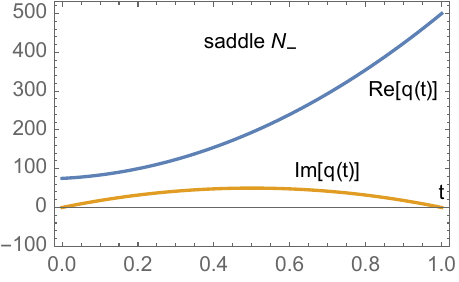}\\
\hspace{1cm}
\includegraphics[width=0.65\textwidth]{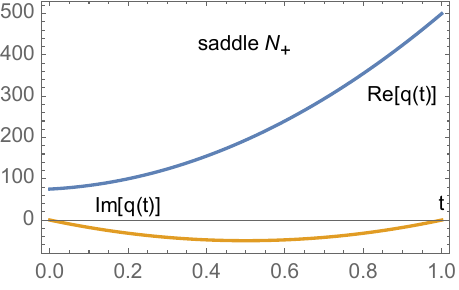} 
\end{minipage}
	\caption{Same as Fig. \ref{fig:momcl2}, but with an imaginary initial momentum. Here $\Lambda = \frac{3}{100}, \, p_0=3\pi^2 i,\, q_1=500,$ so that $N_\pm = \pm 200 + 50 i$. The real lapse line, shown in orange, cannot serve as integration contour in this case, as it runs into the red, asymptotically divergent regions at both large negative and positive values. }
	\protect
	\label{fig:momnoncl}
\end{figure}

The case with non-classical boundary conditions is however a little different. In the pure Dirichlet case (cf. Fig.~\ref{fig:upper}), there was a single saddle point that contributed to the amplitude. Its geometry was complex, though starting and ending with real values. Combining \eqref{eq:initmom} and \eqref{saddlessmall}, we can infer that the initial momentum is then pure imaginary. For an example, see Fig.~\ref{fig:momnoncl}. In the mixed Neumann-Dirichlet case, we recover the same saddle as before, but in addition we get a second saddle with negative real part of the lapse. The two saddle points correspond to geometries that are complex conjugates of each other. In terms of integration contour, a surprise is that we cannot choose the real lapse line as fundamental contour -- the integral simply diverges when integrated over real lapse values. Thus we must define the integration directly on the thimbles. Here we have several choice in principle: we could integrate over ${\cal J}_-,$ ${\cal J}_+$ or a linear combination ${\cal J}_- \pm {\cal J}_+.$ If we sum over both thimbles, we obtain the approximate amplitude 
\begin{align}
    G[p_0,q_1]\approx e^{\frac{i}{\hbar}\left(\frac{3p_0}{\Lambda} + \frac{p_0^3}{36\pi^4\Lambda}\right)} \left( e^{-\frac{i}{\hbar}\frac{12\pi^2\left(\frac{\Lambda}{3}q_1 - 1\right)^{3/2}}{\Lambda}} \pm e^{\frac{i}{\hbar}\frac{12\pi^2\left(\frac{\Lambda}{3}q_1 - 1\right)^{3/2}}{\Lambda}} \right)\,,
\end{align}
which contains a suppression factor (keeping in mind that $p_0$ is imaginary), in line with this describing a sort of tunnelling process from a classically impossible scale factor value to a large value. The no-boundary wave function will turn out to be rather closely related to this amplitude.

The examples provided in this section serve to illustrate the general framework of quantum cosmology. To be more rigorous, one would also have to include perturbations and see if the minisuperspace approximation was actually justified. We will do this once we get to the main topic of interest, namely the no-boundary wave function. However, there are some general properties of quantum gravity amplitudes that are useful to derive first. Plus we will see how amplitudes such as those shown here may best be interpreted. These will be the topics of the next two subsections.


\subsection{Formal Developments: Canonical and Path Integral Quantisations} \label{sec:formal}

The observed homogeneity and isotropy of the early universe imply that for a broad brush analysis, it is useful to start with metrics that already admit these symmetries in their spatial sections. Later on, one can then extend the range of metrics considered, and also add small general perturbations. But such a restriction to symmetric metrics is also technically very useful, as it leads to models that are (at least in part) solvable. This minisuperspace context leads one to consider actions of the following general structure,
\begin{eqnarray}
S = \int dt N \left( \frac{1}{2} G_{AB} \frac{1}{{N}}\frac{\mathrm{d}q^A}{\mathrm{d}t}\frac{1}{N}\frac{\mathrm{d}q^B}{\mathrm{d}t} - U(q^A) \right)\,.\label{minisupact}
\end{eqnarray}
Here $q^A$ are fields that depend solely on time. These could be functions that form a part of the metric, or these could be matter fields. A field that is always present is the scale factor, which determines the size of the universe. This field has the characteristic that it enters with a negative sign kinetic term, cf. \eqref{ActionH}. Thus $G_{AB}$ always contains a negative component. One case that is of particular interest is where one considers the scale factor of the universe, plus a scalar field, so that $q^A = (a,\phi).$ Then with the metric \eqref{eq:LorentzianMetric} the action is given by
\begin{align}
    S = 2\pi^2 \int_0^1 \tilde{N} \mathrm{d}t \left[ -\frac{3}{\tilde{N}^2}a\dot{a}^2 + \frac{1}{2\tilde{N}^2}a^2 \dot\phi^2 +3a - a^3 V(\phi)\right]\,, \label{scalarmodel}
\end{align}
where $V(\phi)$ is the scalar potential, and $G_{aa} = -12\pi^2 a, \, G_{\phi\phi} =  2\pi^2 a^3.$ We will mostly focus on this case when exhibiting examples. 

In the present section we will perform the canonical quantisation directly at the level of the minisuperspace action \eqref{minisupact}. For readers interested in the general case, without symmetry reductions, appendix \ref{sec:a1} provides an overview. To the action \eqref{minisupact} is associated a Hamiltonian
\begin{eqnarray}
{\cal H} = \frac{1}{2} G^{AB} p_A p_B + U \, , \label{Hamiltonian}
\end{eqnarray}
with the canonical momenta $p_a = -12\pi^2a \dot{a}/N, \, p_\phi = 2\pi^2 a^3 \dot\phi/N,$ and where the effective potential is given by
\begin{eqnarray}
U = 2\pi^2\left( -3a + a^3 V\right) \, .
\end{eqnarray}

The Hamiltonian is classically zero and corresponds to the Friedmann equation. If one quantises the theory canonically, by replacing $p_A \rightarrow -i\hbar \frac{\partial}{\partial q^A} \equiv -i\hbar \partial_A,$ one obtains the quantum version of the Hamiltonian constraint, namely the Wheeler-DeWitt (WdW) equation
\begin{eqnarray}
\hat{\cal H} \Psi = \left( - \frac{\hbar^2}{2} \Box + U \right) \Psi = 0\, , \label{WdW}
\end{eqnarray}
where $\Psi = \Psi(a,\phi)$ is the wavefunction of the universe. For the second-derivative operator, there is an ambiguity in terms of the precise placement of the derivatives -- a sensible choice is to fix this ``factor ordering'' issue by writing the operator as $\Box=G^{AB} \nabla_A \nabla_B,$ since then it is covariant under field redefinitions. However, if one is only interested in results to leading order in $\hbar,$ then the ordering is unimportant. As this will be the case for us, we will simply choose a convenient ordering when the situation arises. For the model \eqref{scalarmodel}, the WdW equation explicitly reads
\begin{equation}
\left[ \frac{\hbar^2}{48\pi^4}\left( \frac{1}{a}\frac{\partial^2}{\partial a^2} - \frac{6}{a^3} \frac{\partial^2}{\partial \phi^2}\right) - 3a + a^3 V(\phi) \right] \Psi(a,\phi) = 0 \,. \label{WdWexplicit}
\end{equation}

Mathematically, the WdW equation looks just like the Klein-Gordon equation for a scalar particle. As such, it admits a conserved current, defined as
\begin{align}
    J^A = -\frac{i\hbar}{2}\left(\Psi^\star \nabla^A \Psi - \Psi \nabla^A \Psi^\star \right)\,. \label{current}
\end{align}
One can easily check that $\nabla_A J^A=0$ subject to using the WdW equation. This conserved current will play an important role in the interpretation of the WdW equation, in section \ref{sec:reconstruct}. For now, let us just observe that it is this current, and not $\Psi^\star \Psi,$ that is conserved. Hence the standard assignment of probabilities, familiar from quantum mechanics, cannot automatically be recovered.

\begin{figure}[ht]
	\centering
	\includegraphics[width=0.5\textwidth]{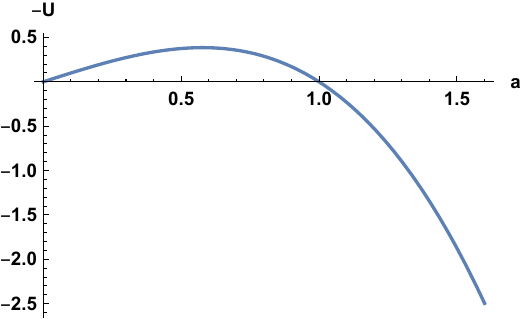}
	\caption{The effective potential $-U$ with constant potential $\Lambda.$ }
	\label{fig:effpot}
\end{figure}

In simple cases the WdW equation can be solved directly. For now, let us just give a qualitative argument as to the nature of solutions, in the context of inflation. A first approximation to inflation is to consider a constant potential $V(\phi)=\Lambda.$ Then one can neglect $\partial_\phi$ in \eqref{WdWexplicit}.  The effective potential looks as shown in Fig. \ref{fig:effpot}. It is reminiscent of a tunnelling problem. And indeed, for large $a,$ we obtain the approximate solutions
\begin{equation}
\Psi \approx e^{\pm i \frac{4\pi^2}{\hbar}\sqrt{\frac{\Lambda}{3}}a^3}\,. \label{largea}
\end{equation}
Note that these solutions oscillate, and roughly corresponds to free propagation (in this case classical expansion of the universe). Meanwhile, at small $a$ the solution is exponentially growing or damped,
\begin{equation}
\Psi \approx e^{\pm  \frac{12\pi^2}{\hbar \Lambda}\left((1- \frac{a^2\Lambda}{3})^{3/2}-1\right)}\,.
\end{equation}
This is the regime where the scale factor takes non-classical values, smaller than the classical minimum radius $a_{min} = \sqrt{\frac{3}{V}}$ of a de Sitter universe. One might try to interpret such a regime as describing the quantum tunnelling of a universe from zero size to $a=a_{min}.$ Later on, we will see that the no-boundary proposal provides a framework to make this much more precise. For now, let us simply point out that a difficulty with solving the WdW equation is that one typically does not know how to fix the integration constants that arise in the solutions. This issue, as we saw, is much clearer in the path integral approach.

In section \ref{sec:qc}, we saw with the help of an explicit example that the path integral satisfies the WdW equation. We can also provide a formal argument showing that this must be the case generally (see e.g. \cite{Leutwyler:1964wn,Halliwell:1988wc}). The argument is surprisingly quick. The propagator/wave function can be decomposed into the lapse integral, plus an amplitude with fixed lapse,
\begin{align}
G[q_1;q_0] = \int_{0}^\infty \mathrm{d}N G[q_1;q_0;N]\,.
\end{align} 
For definiteness, we assumed the pure Dirichlet case, with an integration domain consisting of positive real values of the lapse. The integrand
\begin{align}
G[q_1;q_0;N]= \int_{q=q_0}^{q=q_1}\mathcal{D}q e^{\frac{i}{\hbar}S(N,q)}
\end{align}
is then the amplitude to evolve from $q_0$ to $q_1$ in ``time'' $N$. Just like in ordinary quantum mechanics, this amplitude then satisfies the Schr\"odinger equation
\begin{align}
i\frac{\partial G[q_1;q_0;N]}{\partial N} = \hat{H} G[q_1;q_0;N]\,.
\end{align}
The only subtlety is that we must take into account the coincidence condition that 
\begin{align}
\lim_{N\to 0} G[q_1;q_0;N] = 6\pi^2 \delta(q_0-q_1)\,.
\end{align}
Again, the factor of $6\pi^2$ simply depends on conventions. It then immediately follows that the total propagator of minisuperspace models satisfies the inhomogeneous WdW equation,
\begin{align}
\hat{H}G[q_1;q_0]
=&\int_0^\infty \mathrm{d}N \hat{H}G[q_1;q_0;N]\nonumber\\
=&i\int_0^\infty \mathrm{d}N \frac{\partial G[q_1;q_0;N]}{\partial N}\nonumber\\
=&i G[q_1;q_0;N]\big|_{N=0}^{N=\infty} \nonumber\\
=& - i 6\pi^2 \delta(q_0-q_1)\,.
\end{align}
Had the integration domain for the lapse been the entire lapse line, the right hand side would have been zero above and we would have obtained a solution to the homogeneous WdW equation.

In discussing the WdW equation, we encountered the issue of factor ordering, leading to a potential ambiguity at sub-leading orders in an expansion in $\hbar.$ Can the path integral resolve this? From the point of view of the path integral, the corresponding question is which integration measure one should choose (note that changes in measure are also sub-leading in a saddle point approximation). This could lead to progress, {\it cf.} the gauge fixing procedure outlined in appendix \ref{sec:a2}, where one uses the Liouville measure on the full path integral, including all ghost fields. For further discussion of this question, see also \cite{Halliwell:1988wc}.

\subsection{How to Reconstruct the Universe from the Wave Function} \label{sec:reconstruct}

We just derived the Wheeler-DeWitt equation, and showed that the path integral satisfies it. An obvious and important question is however how the equation should be interpreted. In some sense the WdW equation is the analogue of the Schr\"{o}dinger equation in quantum mechanics, but the presence of gravity changes a few things significantly. Most crucially, time cannot, and does not, appear explicitly. This is because space and time are quantised now, and must emerge from the solutions, rather than being outside constructs. Nevertheless, we will see that the WdW equation contains the Schr\"{o}dinger equation under appropriate circumstances.

\begin{figure}[ht]
	\centering
	\includegraphics[width=0.9\textwidth]{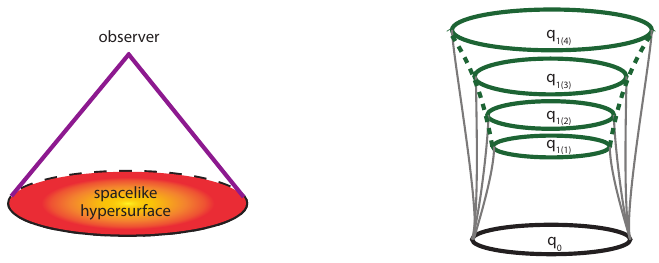}
	\caption{{\it Left panel:} When we observe the universe, we obtain information about our past light cone (in purple). In this sketch time is vertical and space horizontal. It is however equivalent (and mathematically easier) to deal with spacelike hypersurfaces whose causal development would include this light cone. {\it Right panel:} We can imagine a succession of transitions from a given initial condition $q_0$ to a series of final conditions $q_{1(i)}.$ When the wave function is of WKB form, one can infer the physically relevant spacetime from this succession, as indicated by the dotted green line.}
	\label{fig:int}
\end{figure}

But let us start with something more basic. Both the path integral and the WdW equation, as we have formulated them, concern transitions between $3-$dimensional hypersurfaces. But that is not really what we observe. When we look out into the world, we actually receive information about our past light cone. So, strictly speaking, quantum gravity should deal with this past light cone and not with spacelike domains. However, if we take a spacelike boundary at some finite time in our past light cone, then the enclosed spacelike region has the light cone ending at our brain in its causal future, see the left panel of Fig.~\ref{fig:int}. When we look back further and further, ever larger spacelike regions are encapsulated. But since, to a good approximation, we may assume spatial isotropy and homogeneity, this presents no complication and nothing physical is lost by dealing with spacelike hypersurfaces -- mathematically however, this formulation is much more tractable. 

Still, having an amplitude for a transition between two spacelike hypersurfaces is not the same as having a full spacetime manifold. So why do we perceive a continuous evolution in time? In other words, what holds the world together from one instant to the next, and why does this appear seamless?\footnote{The even harder question is: why is our consciousness connected to a particular (small patch of a) hypersurface, but always only one and only for an instant, with a succession of such quasi-instantaneous connections unfolding?} 

It turns out that the answer is already included to a large extent in the saddle point approximation to the path integral. Let us assume that the wave function can be written in the following suggestive form\footnote{In general, the wave function is given by a superposition of such terms, but in applications to quantum cosmology a single term typically dominates.},
\begin{align}
    \Psi = e^{\frac{1}{\hbar}\left({\cal W} + i{\cal S}\right)}\,, \label{wfsuggestive}
\end{align}
where ${\cal W, S}$ are real functions of the fields $q^A$, ${\cal W}$ being the weighting and ${\cal S}$ the phase. We can then expand the WdW equation \eqref{WdW} as a series in $\hbar$ (assuming ${\cal W}$ and ${\cal S}$ to be ${\cal O}(\hbar^0)$), finding to leading and sub-leading orders respectively
\begin{eqnarray}
    -\frac{1}{2} (\nabla {\cal W})^2+\frac{1}{2} (\nabla {\cal S})^2 + U = 0, && \quad \nabla {\cal W} \cdot \nabla {\cal S} = 0, \label{WdW1} \\ \Box {\cal W} =0, && \quad \Box {\cal S} = 0, \label{WdW2}
\end{eqnarray}
where e.g. $\nabla {\cal W}\cdot \nabla {\cal S}\equiv G^{AB} \nabla_A {\cal W} \nabla_B {\cal S}$. There are two equations at each order, one for the real part and one for the imaginary part of the equation. It is also useful to write out the conserved current \eqref{current},
\begin{align}
    J^A =  e^{\frac{2}{\hbar}{\cal W}}\, \nabla^A {\cal S} \,. \label{current2}
\end{align}

Vilenkin then developed the following interpretation \cite{Vilenkin:1988yd}: if we look at the left equation in \eqref{WdW1}, we may notice that if ${\cal W}$ varies slowly compared to ${\cal S},$ then we recover the classical Hamilton-Jacobi equation,
\begin{align} \label{eq:WKBcond}
    (\nabla {\cal W})^2 \ll  (\nabla {\cal S})^2 \quad \rightarrow \quad \frac{1}{2} (\nabla {\cal S})^2 + U \approx 0\,,
\end{align}
with ${\cal S}$ being identified with the classical action, and the canonical momentum assignment
\begin{align}
    p_A = \frac{\partial {\cal S}}{\partial q^A}\,. \label{momassign}
\end{align}
This is in fact the Wentzel-Kramers-Brillouin (WKB) semi-classical approximation often used in standard quantum mechanics, namely that the amplitude of the wave function varies slowly, and the phase fast. The right equation in \eqref{WdW1} then expresses the conservation of the current \eqref{current2} to leading order, while \eqref{WdW2} ensures conservation at sub-leading order as well. 

Several implications immediately follow. The classical action ${\cal S}$ describes a congruence of (classical) trajectories. One can for example think of this congruence as being foliated by constant-${\cal S}$ surfaces. Then one can choose trajectories along a normal vector $n^A$ to these surfaces, such that $\nabla_n {\cal S}>0$ (the locus $\nabla{\cal S}=0$ describes the breakdown of the semi-classical approximation) and define relative probabilities as
\begin{align}
     {\cal P} = e^{\frac{2}{\hbar}{\cal W}} \, \nabla_n {\cal S}\,.
\end{align}
How an overall normalisation can be implemented in a mathematically precise way is an open question, hence the probabilities are only relative. But these relative probabilities are then positive and conserved along classical trajectories, and for small $\hbar$ approximately correspond to the standard probabilities in quantum mechanics, ${\cal P} \approx e^{\frac{2}{\hbar}{\cal W}} = \Psi^\star \Psi$ (up to normalisation). Also, different foliations give the same probabilities, as long as the foliations are chosen so as to intersect each trajectory only once (the foliations must be spacelike with regard to the minisuperspace metric $G_{AB}$). One can improve upon this procedure when the amplitude of the wave function vanishes sufficiently fast, ${\cal W} \to - \infty,$ asymptotically in all directions along the foliation. In this case, a normalisation is possible and one can obtain absolute (approximate) probabilities, which are approximately conserved. We will discuss examples in section~\ref{sec:minisuper}.

When the WKB approximation holds, the wave function \eqref{wfsuggestive} satisfies 
\begin{align}
    p_A \Psi = -i\hbar \, \partial_A \Psi \approx \partial_A {\cal S} \, \Psi \label{pfromS}
\end{align}
to a good approximation, in agreement with the assignment \eqref{momassign}. The wave function is thus peaked on solutions described by the first order relation $p_A = \partial_A {\cal S},$ {\it i.e.} it is peaked on solutions of the classical (Hamiltonian) equations of motion. However, the boundary conditions can force the solutions to be complex valued, when they correspond to conditions that are classically not allowed. This will become clear when we discuss the no-boundary proposal. Note also that a theory of initial conditions, such as the no-boundary proposal, fixes the initial conditions required to calculate the action. Thus, in calculating the action, half of the integration constants are already fixed, and this correspondingly restricts the solutions allowed by Eq. \eqref{pfromS}.

In all the relations above, we are comparing wave functions at different final boundary conditions. This is how one can verify that the amplitude varies much more slowly than the phase, $(\nabla {\cal W})^2 \ll  (\nabla {\cal S})^2.$ Thus one should really think of a family of transitions, from fixed initial conditions to a series of final conditions, {\it cf.} the right panel in Fig.~\ref{fig:int}. When this series of transitions is such that the WKB conditions hold, then one can assign a relative probability ${\cal P} \approx \Psi^\star \Psi$ to the associated evolution. The physical spacetime should then be identified with the collection of such final boundaries, and to leading order in $\hbar$ this sequence of $3-$dimensional hypersurfaces will follow a solution of the Einstein equations (but only for as long as the WKB condition holds). In this vein, it is also sometimes said that quantum cosmology provides probabilities for ``histories''~\cite{Hartle:2008ng}.

There is another interesting consequence that arises for sub-systems \cite{Vilenkin:1988yd}. If we assume that there is a small sub-system characterised by the Hamiltonian $H_2,$ with negligible backreaction on the universe as a whole, then the WdW gets augmented by a term,
\begin{eqnarray}
 \left( - \frac{\hbar^2}{2}\Box + U +H_2\right) \Psi = 0\,.
\end{eqnarray}
If we now expand the WdW again in powers of $\hbar,$ then at sub-leading order we get the additional equation
\begin{equation}
i \hbar \nabla {\cal S} \cdot \nabla \psi_2 = H_2 \psi_2\,,
\end{equation}
where we factorised $\Psi = \psi_{0}\psi_2$ into a product of $\psi_0,$ depending only on the background variables, and $\psi_2,$ depending also on the perturbations. Now comes the crucial observation: If one identifies 
\begin{align}
    \nabla {\cal S} \cdot \nabla \equiv \frac{\partial}{\partial t} \label{tchange}
\end{align}
then one obtains an effective time variable, and one recovers the Schr\"{o}dinger equation
\begin{align}
    i \hbar \frac{\partial}{\partial t} \psi_2 = H_2 \psi_2\,.
\end{align}
This is remarkable. From the space- and timeless WdW equation we thus recover the Schr\"{o}dinger equation for sub-systems, as long as the WKB approximation is applicable. The background wave function itself provides the time. Let us briefly check that this definition of time makes sense, for a simple case: in a de Sitter universe, we saw in \eqref{largea} that at large scale factor the solution is ${\cal S} \approx -4\pi^2\sqrt{\frac{\Lambda}{3}}a^3.$ With $G_{aa}=-12\pi^2 a,$ we obtain
\begin{align}
    \nabla S \cdot \nabla = G^{aa}\partial_a {\cal S}\, \partial_a = \sqrt{\frac{\Lambda}{3}}a\, \partial_a \overset{!}{=}  \frac{\partial}{\partial t}\,,
\end{align}
in agreement with the classical de Sitter solution expressed in physical time $t,$ $a=\sqrt{\frac{3}{\Lambda}}e^{\sqrt{\frac{\Lambda}{3}}t}.$ We should note that in more general cases, when more fields are present, one must make sure that the integrability conditions associated with the change of variables in \eqref{tchange} are everywhere satisfied.

A few more comments are in order. On the mathematical side, let us mention that there exists a body of work that attempts to make the definition of physical states, and their associated Hilbert space, rigorous, at least in minisuperspace models \cite{Marolf:1994wh,Ashtekar:1995zh,Hartle:1997dc,Embacher:1997zi}. This is based on the ``induced'' inner product, and we will briefly sketch the main idea. The norm based on the Klein-Gordon current has the well-known problem that it is not positive definite. For example, for a massive scalar field in flat space we may write out the mode decomposition (with $\omega = \sqrt{k^2+m^2}$)
\begin{align}
   \psi(t,{\bf x}) =  \int \frac{\mathrm{d}^3k}{(2\pi)^{3/2}\omega^{1/2}} \left( a({\bf k})e^{i({\bf k x}-\omega t)} + b({\bf k})e^{i({\bf k x}+\omega t)}\right)\,,
\end{align}
and this leads to the Klein-Gordon norm (where $\Sigma$ is a spacelike hypersurface)
\begin{align}
     \langle\psi_1,\psi_2\rangle_{KG} = \int_\Sigma J_\mu d\Sigma^\mu &\equiv -\frac{i}{2}\int_\Sigma d\Sigma_\mu \left(\psi_1^* \nabla^\mu \psi_2 - \psi_2 \nabla^\mu \psi_1^* \right)  \\ &= \int \mathrm{d}^3k \left(a^*_1({\bf k})a_2({\bf k})-b_1^*({\bf k})b_2({\bf k}) \right)\,.
\end{align}
As we can see, positive $(a)$ and negative $(b)$ frequency modes enter with a relative minus sign. By contrast, the induced inner product starts from the Schr\"{o}dinger product
$\langle\psi_1,\psi_2\rangle = \int \mathrm{d}^4x \sqrt{-g}\psi_1^*\psi_2,$ which is positive definite but can lead to unwanted divergences in the present context. One then uses the WDW operator \eqref{WdW} both to define physical states, and to remove an overcounting of gauge transformations in the product between physical states. This procedure is described in some detail in \cite{Hartle:1997dc,Embacher:1997zi,Halliwell:2009rw}, with the end result that for the massive scalar example above one ends up with
\begin{align}
    \langle\psi_1,\psi_2\rangle_{phys} = 4\pi \int \mathrm{d}^3k \left(a^*_1({\bf k})a_2({\bf k}) + b_1^*({\bf k})b_2({\bf k}) \right)\,.
\end{align}
Effectively, the induced inner product has changed the relative sign between positive and negative frequency parts, and has made the product positive definite. It may thus be used to define and find normalised physical states. As discussed in \cite{Halliwell:2009rw} via application to a number of examples, one then recovers the heuristic interpretation outlined above when the wave function is of WKB form; we will consequently use this formulation in what follows. Given that we will mostly work to leading order in $\hbar$ only, the most important consequence for us will be to deem wave functions normalisable when their weighting falls off asymptotically (${\cal W} \to - \infty$) in all directions in parameter space, while we will discard wave functions as unphysical when the weighting grows asymptotically (${\cal W} \to + \infty$) in at least one direction\footnote{We should alert the reader that this is the best one can do at present -- given that we work in minisuperspace, {\it i.e.} with a restricted number of degrees of freedom, one must keep in mind the possibility that additional degrees of freedom might render an acceptable-looking minisuperspace wave function non-normalisable.}.

An important physical effect is that once one adds matter and perturbations to the universe, then additional classicalisation occurs due to decoherence. This arises from the interactions present in the system \cite{Joos:1986iw,Kiefer:1987ft}, with short wavelength modes acting as an environment that decoheres the long wavelength ``background''. Thus the WKB classicality which we have used above is presumably mostly relevant for the very early universe. Especially if there was an early phase of evolution when no matter was present yet, or if we want to consider the simultaneous emergence of spacetime and matter from a fully quantum state, then WKB classicality is absolutely crucial in guaranteeing a classical background. 

In general, the wave function is approximated not just by a single saddle point, but by a collection thereof,
\begin{align}
    \Psi \approx \sum_i e^{\frac{1}{\hbar}\left({\cal W}_i + i {\cal S}_i \right)}\,.
\end{align}
In such a case, decoherence also plays an important role, as it effectively isolates the different saddle points into separate universes \cite{Halliwell:1989vw}. For each such universe, the above treatment of relative probabilities and sub-systems then separately applies. In fact, it is possible to approach the entire subject of quantum cosmology from the point of view of decoherence, focussing on quasi-classical histories that evolve effectively independently of each other (for a suitably coarse-grained description). This is the \emph{decoherent histories} program, see {\it e.g.} \cite{Hartle:1992as} for a comprehensive introduction.

It is important to realise that the notion of probability derived above is only approximate, valid semi-classically, to leading order in $\hbar$. Thus unitarity is also only an approximate concept \cite{Vilenkin:1988yd}. This is directly related to the fundamental role that time plays in quantum mechanics and quantum field theory. Here time is just part of another quantum field, the metric, and hence our standard intuition cannot be taken for granted. That said, it is clear that it would be desirable to further clarify the definition of probabilities in quantum cosmology -- it seems clear that much more can be found out about this topic \cite{Chataignier:2019kof}, and we will mention a few ideas later on.

Now, given the crucial importance of the WKB approximation, we may ask under what circumstances we can expect the wave function to take this form. In fact, at the moment only two early universe scenarios are known to automatically lead to WKB classicality. They can be determined by noting that, in order for the WKB approximation to be valid, to leading order the wave function must be an oscillating function. Then, if we go back to the WdW equation in the presence of a scalar field, Eq. \eqref{WdWexplicit}, we can identify the following two regimes:

\begin{itemize}
\item {Inflation \cite{Guth:1980zm,Linde:1981mu,Albrecht:1982wi} (for a review see \cite{Baumann:2009ds}):} $V(\phi)>0$ and $\frac{|V_{,\phi}|}{V} < \sqrt{2}.$ \\ A good approximation to inflation is that the scalar field changes little while the universe expands fast, so we can assume  $\frac{1}{a^2}\frac{\partial^2}{\partial \phi^2} \ll \frac{\partial^2}{\partial a^2}$. Then at large scale factor $a$ we are left with the equation
\begin{equation}
\left( \frac{\hbar^2}{48\pi^4 a}\frac{\partial^2}{\partial a^2}  + a^3 V \right) \Psi = 0 \, \end{equation}
which admits an oscillating solution, $\Psi \sim e^{i a^3/\hbar}.$
\item {Ekpyrosis \cite{Khoury:2001wf} (for a review see \cite{Lehners:2008vx}):} $V(\phi)<0$ and $\frac{|V_{,\phi}|}{|V|} > \sqrt{6}.$\\ An ekpyrotic phase is essentially the complement to inflation. In this case the universe is slowly contracting, and the scalar field races down a steep, negative potential. Thus the first approximation is now that $\frac{1}{a^2}\frac{\partial^2}{\partial \phi^2} \gg \frac{\partial^2}{\partial a^2},$ but since the potential is negative we again get an effective equation with oscillating solutions,
\begin{equation}
\left(\frac{\hbar^2}{8\pi^4 a^3} \frac{\partial^2}{\partial \phi^2}  - a^3 V \right) \Psi = 0\,. 
\end{equation}
\end{itemize}
Thus, rather surprisingly, the two dynamical models that can potentially explain the homogeneity, isotropy and flatness of the early universe, while also providing mechanisms for amplifying quantum fluctuations into seeds of structure, have the additional property of rendering the universe classical \cite{Lehners:2015sia}. This can be traced to the fact that both mechanisms act as dynamical attractors, and we will discuss this property in more detail in the next section,  when discussing no-boundary solutions.

The concepts and formalism developed in this section can now be used, and arguably find their most intriguing application, in the no-boundary proposal -- the main topic of this review, whose exploration we now begin.


\section{The No-Boundary Proposal} \label{sec:nb}

\subsection{Heuristic Motivations} \label{sec:nbheurist}

The no-boundary proposal is a theory of initial conditions for the universe. It is a complement to our theories of dynamics, such as quantum field theories or general relativity, which can be seen as describing what kinds of dynamics are possible. Rather, given a theory of dynamics, the no-boundary proposal has as its aim to provide a theory for what kind of dynamics is likely, and what is unlikely. It is a theory for the quantum state of the universe, and it involves gravity in a crucial way -- thus it can only be formulated in quantum gravity. We built up the necessary background in the previous section, but before proceeding to a more technical discussion, it is useful to provide several heuristic motivations that explain rather intuitively what the no-boundary proposal is about.\\

\noindent{\it Avoiding an infinite regression}

Why did the apple fall from the tree? Physics is built on cause and effect. Thus we might have noticed that a bird tried to grab the apple, and inadvertently knocked it out of the tree. But why did the bird try to grab the apple? Because it was hungry. And why was it hungry? This kind of questioning, taken sufficiently far, typically ends up with the question: how did the universe begin? 

\begin{figure}[ht]
	\centering
	\includegraphics[width=0.8\textwidth]{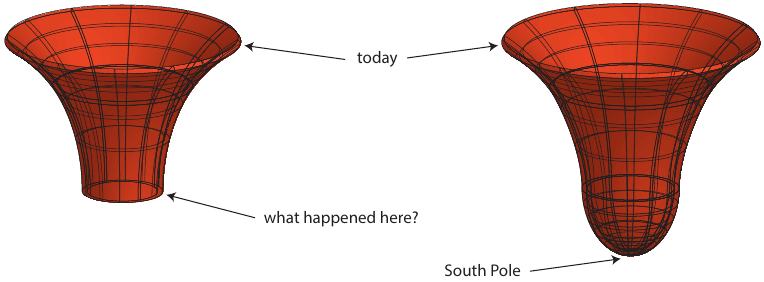}
	\caption{{\it Left panel:} When we investigate the history of our universe, then we have to supply boundary conditions at ever earlier times. {\it Right panel:} The no-boundary proposal cuts off the potentially infinite regression by stipulating that there was no boundary in the past. This in itself then provides initial conditions for the evolution of the universe.}
	\label{fig:regression}
\end{figure}

In cosmology, we know that before the current dark energy era, there was a period of matter domination. And before that one of radiation dominance. Going back in time, at each transition from a prior phase of evolution, we have to supply ``initial'' conditions, which we can think of as being specified on the boundary of that period of evolution, {\it cf.} Fig.~\ref{fig:regression}. This potentially leads to an infinite regression, where we go to earlier and earlier phases, without end. The no-boundary proposal may be seen as a way of cutting the Gordian knot, by stipulating that there was no boundary in the past \cite{Hawking:1981gb}. Then there is no need to specify initial conditions at the ``beginning'' -- the fact that there is no boundary already supplies the required condition. 

An immediate consequence is that the universe is self-contained, both in space and in time. This is because, if there was a boundary, we could meaningfully ask what is on the other side. But if there is no boundary, the universe can be finite and yet entirely self-contained, just like the surface of a ball. The analogy with the surface of a ball is in fact rather accurate: in Lorentzian (pseudo-Riemannian) geometry, it is not possible to round off the geometry of the universe in such a manner. If one tried, one would run into a spacetime singularity, which would again constitute a boundary. But if one allows the geometry of the universe to become Euclidean, then such a smoothing out is indeed possible. Thus we may foresee that the no-boundary proposal will induce us to go beyond Lorentzian geometry. \\

\noindent{\it Ground state of the universe and quantum creation}

The way in which the no-boundary proposal is motivated in the seminal paper \cite{Hartle:1983ai} by Hartle and Hawking is by drawing an analogy with ground states in quantum mechanics. Ground states can be defined by considering a Euclidean path integral, and integrating from configurations of vanishing action in the infinite (Euclidean) past, 
\begin{align}
    \psi_0(x,0) = \int {\cal D}x \, e^{-\frac{1}{\hbar}I_E[x(\tau)]}\,, \label{eq:ground}
\end{align}
where we ignored an overall normalisation factor and where Euclidean time $\tau$ is related to physical time via $t=-i\tau.$ The Euclidean action $I_E$ is related to the Lorentzian one via $I_E=-iS.$ To see that this defines the ground state, consider the amplitude for a particle to propagate from $(x=0, \, t=t^\prime)$ to $(x, t=0),$
\begin{align}
    \langle x,0 \mid 0, t^\prime \rangle & = \sum_n \psi_n(x)\bar\psi_n(0) \, e^{iE_n t^\prime} \\ 
    & = \int {\cal D}x \, e^{iS[x(t)]}\,. \label{eq:qmpi}
\end{align}
In the first line we inserted a complete set of energy eigenstates $\psi_n$ with energy $E_n,$ and the second line is the definition of the amplitude as a path integral. Then if we Wick rotate  $t^\prime = -i \tau^\prime$ and take the limit $\tau^\prime \to -\infty,$ then only the lowest energy eigenstate will be left, and consequently  \eqref{eq:qmpi} becomes \eqref{eq:ground}. Thus an integral from the infinite Euclidean past defines the ground state (vacuum state) of the system. Put differently, the integration over Euclidean time is an alternative manner to implement the ground state as initial state.

We would now like to propose an analogous definition when gravity is included. The question then becomes: what should play the role of the infinite Euclidean past? As discussed by Hartle and Hawking \cite{Hartle:1983ai}, there are two natural choices that come to mind. One is Euclidean flat space, and the other are compact Euclidean metrics. Euclidean flat space would be more appropriate for a scattering amplitude, where particles are modelled to come in from infinity and fly out to infinity again. But in cosmology we are only measuring the universe at late (finite) times, and we are measuring from the inside of the universe. This means that the second option, namely summing over compact metrics, is more appropriate for cosmology, and this is the proposal of Hartle and Hawking. Note that this prescription then obviates the need to insert an initial state explicitly, {\it cf.} Eq. \eqref{eq:piheuristic}, the idea being that the Euclidean integral puts the universe in its ground state.

As we saw in section \ref{sec:qc}, the wave function is a function of three-dimensional spatial slices. The path integral over compact metrics may then be seen as an amplitude from a slice where the $3-$dimensional volume goes to zero, to a final slice with metric $h_{ij},$
\begin{align}
    \Psi_{HH}[h_{ij}] = {\cal N} \int_{\cal C} {\cal D}g_{\mu\nu} \, e^{-I_E[g_{\mu\nu}]}\,, \label{eq:HHpi}
\end{align}
where the integral is over all (inequivalent) compact metrics ${\cal C}$ that contain a surface with metric $h_{ij}.$ ${\cal N}$ here is a normalisation factor. This definition may be given the interpretation of a transition amplitude from zero size to a given final size, {\it i.e.} it can be interpreted as being the amplitude for the universe to tunnel from nothing (for early incarnations of this idea, see \cite{Lemaitre:1931zzb,Tryon:1973xi,Brout:1977ix}). This nothing should be thought of as an \emph{absolute nothing}, {\it i.e.} as the absence of space, time and matter. Incidentally, when matter is included, then the integral is to be performed over compact metrics and regular field configurations on these geometries -- if the matter configurations were not regular, they would induce spacetime singularities, which are explicitly avoided here. 

Note that, comparing with \eqref{eq:ground}, the definition \eqref{eq:HHpi} cannot be interpreted as the lowest energy state of the universe. This is because in a closed universe, energy cannot be defined unambiguously (in general relativity, energy is usually defined asymptotically, with respect to a fixed reference metric \cite{Arnowitt:1962hi}). Rather, the definition \eqref{eq:HHpi} should be seen as defining a state of minimum excitation. Thus we would expect, and we will see that this is in fact true, that spacetimes with fewer wrinkles in them come out as preferred over more crumpled spacetimes. Thus there is the hope that the no-boundary proposal might be able to explain why the early universe was in such a simple state.

The idea of describing the origin of the universe as tunnelling out from nothing was also proposed independently by Vilenkin, and goes by the name of tunnelling proposal \cite{Vilenkin:1982de,Vilenkin:1983xq,Vilenkin:1984wp}. Conceptually, the two proposals are very similar, but there are some technical differences that we will discuss in later sections. Interestingly, the idea that the ground state might be enough to explain the structure of the universe was also mentioned already by Dirac as early as 1939, in a lecture\footnote{Here are Dirac's words, delivered on presentation of the James Scott prize in 1939 \cite{Dirac:1939}: ``Let us now return to dynamical questions. With the new cosmology the universe must have been started off in some very simple way. What, then, becomes of the initial conditions required by dynamical theory? Plainly there cannot be any, or they must be trivial. We are left in a situation which would be untenable with the old mechanics. If the universe were simply the motion which follows from a given scheme of equations of motion with trivial initial conditions, it could not contain the complexity we observe. Quantum mechanics provides an escape from the difficulty. It enables us to ascribe the complexity to the quantum jumps, lying outside the scheme of equations of motion. The quantum jumps now form the uncalculable part of natural phenomena, to replace the initial conditions of the old mechanistic view.''}. Dirac's observation was that the quantum mechanical ground state is not empty, and might ultimately account for all the structure we see in the universe. This is very close to the modern point of view. The ground breaking contribution of Hartle and Hawking was to specify how to include gravity in this scheme. 

We should point out a few immediate consequences of the definition \eqref{eq:HHpi}, which at this point remains rather schematic. The first is that the wave function is real valued, as the integral is, at least formally, over Euclidean compact metrics. We will later discuss how this can nevertheless lead to an operational definition of probabilities, in the vein of section \ref{sec:reconstruct}. The second is that, even though the formal definition describes a sum over Euclidean metrics, somehow our Lorentzian universe must come out. As we will see, this is because the saddle points of the path integral \eqref{eq:HHpi} will turn out to be complex. Third, by definition the big bang singularity is avoided. This is possible because the geometry is not forced to remain Lorentzian in regions where the universe shrinks to zero size -- rather, the origin of the geometry is viewed more like a point on the surface of a ball, and sometimes this point is referred to as the ``South Pole'' of the geometry, {\it cf.} the right panel in Fig.~\ref{fig:regression}. \\

\begin{figure}[ht]
	\centering
	\includegraphics[width=0.8\textwidth]{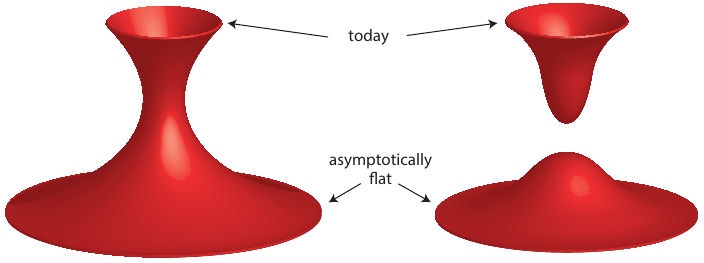}
	\caption{If the initial state were taken to be Euclidean flat space, then in addition to geometries directly connecting our current universe to that state, there would be topologically non-trivial configurations in which the early phase pinches off and a new universe forms from nothing. The conjecture is that such configurations would dominate the path integral, effectively recovering a sum over purely compact metrics without boundary \cite{Hawking:2006ur}.}
	\label{fig:topol}
\end{figure}

\noindent{\it The path integral is effectively of no-boundary type anyway}

What if we had chosen a different initial state in the definition \eqref{eq:HHpi}? In particular, what if we had chosen asymptotically Euclidean flat space as ``in'' state? Then, apart from geometries that would connect this asymptotic region directly to the final hypersurface with metric $h_{ij},$ there would also be geometries that are disconnected, i.e. for which the initial asymptotic Euclidean region would round off and come to an end, combined with a configuration where a new universe is nucleated and this then connects to the final hypersurface. In fact, there would be many more such topologically non-trivial geometries contributing to the path integral, and they would in all likelihood vastly dominate over the configurations directly connected to infinity. Hence, there is a  conjecture that effectively this path integral would again reduce to the one where one sums over compact metrics \cite{Hawking:2006ur}. This idea is illustrated in Fig.~\ref{fig:topol}.\\

\noindent{\it Finite action}

In physics, the action plays a central role, since for just about all theories of physical interest the action provides the definition of the theory. It encodes the classical equations of motion upon variation of the fields and it provides the weighting for quantum amplitudes, in Feynman's path integral approach. Thus a natural requirement on any physically sensible theory might be to ask for the action to be well defined and, in particular, finite (at least at the saddle points) \cite{Barrow1988,Barrow:2019gzc,Jonas:2021xkx}. This is a crucial condition in ensuring that the theory makes sense at the semi-classical level. However, such a requirement is far from trivial. In particular, it immediately rules out the standard hot big bang model, once quantum corrections are taken into account. Let us briefly review this argument \cite{Jonas:2021xkx}. We can focus on the best case scenario, where spatial sections are homogeneous and isotropic, i.e. we can consider a Robertson-Walker (RW) metric
\begin{align}
    \mathrm{d}s^2 = -\tilde{N}^2 \mathrm{d}t^2 + a(t)^2 \left( \frac{\mathrm{d}r^2}{1-k r^2} +r^2( \mathrm{d}\theta^2 + \sin^2\theta \mathrm{d}\phi^2) \right)\,, \label{RW}
\end{align}
where $k \in (-1,0,1)$ parameterises the spatial curvature (respectively open, flat, closed). When quantising gravity, we expect that loop corrections lead to additional terms in the action, proportional to higher powers of the Riemann tensor \cite{Goroff:1985th} (there can also be terms involving derivatives of the Riemann tensor, but for simplicity we will ignore these here). In a RW background, such terms simplify greatly, as there are only two independent non-trivial components of the Riemann tensor, namely (for $i \neq j,$ and with no summation over repeated indices)
\begin{align}
    R^{ij}{}_{ij} = \frac{\dot{a}^2+k\tilde{N}^2}{a^2\tilde{N}^2}\equiv A_1\,,\qquad R^{0i}{_{0i}}=\frac{\ddot{a}}{a\tilde{N}^2}\equiv A_2\,.\label{eq:Riemann}
\end{align}
The action will be a function $f$ of (positive) powers of these terms only, more specifically
\begin{align}
    S=\int \mathrm{d}^4x\sqrt{-g}f(\textrm{Riemann}) = 2\pi^2 \int \mathrm{d}t \tilde{N} a^3 \sum_{p_1,p_2} c_{p_1,p_2} A_1^{p_1} A_2^{p_2}\,, \label{eq:actionRiemann}
\end{align}
where $c_{p_1,p_2}$ are coefficients, and the power of the Riemann terms is given by $p_1+p_2.$ Now the point is that for a big bang type model, we have in physical time $\tilde{N}=1$ that the scale factor goes to zero as some power of time, $a(t) \propto t^s$ with $s>0.$ But then $A_1 = \frac{s^2}{t^2} + \frac{k}{t^{2s}}\,, \, A_2 = \frac{s(s-1)}{t^2}.$ Given that quantum corrections arise at arbitrarily high order, {\it i.e.} $p_1, p_2$ can be arbitrarily large, the integral in \eqref{eq:actionRiemann} will diverge (some terms can be eliminated by using a constraint, but not all \cite{Jonas:2021xkx}) and the action will not be finite. At the semi-classical level, a Lorentzian big bang is thus ruled out. 

There are few ways known that render the action well defined and finite. One possibility is quadratic gravity. In the quantisation of quadratic gravity, no further higher derivative terms appear \cite{Stelle:1976gc}, and hence it is potentially protected from the problem mentioned above. In fact, inflationary solutions, starting from zero size, arise with finite action \cite{Lehners:2019ibe}. However, that theory is potentially plagued by ghosts \cite{Stelle:1977ry}. Another possibility is to have an emergent phase, that is to say a phase of asymptotic flatness out of which the universe arose, see e.g. \cite{Brandenberger:1988aj,Creminelli:2010ba}. However, the previous motivating paragraph has provided a counter argument, by conjecturing that in such a situation the quantum amplitude would in fact be dominated by no-boundary geometries. 

This then leads us to the best understood example that ensures finite action, namely the no-boundary proposal itself. The smooth rounding-off of the geometry, combined with regular matter fields, is the most robust example known to lead to finite action solutions. Technically, this is due to the Euclidean nature of the geometry near the South Pole at $t=0:$ there, as we will derive in below, the solution for the scale factor is of the form $a(t) = \pm it + {\cal O}(t^3)$ with $k=+1,$ leading to $\dot{a}^2+k={\cal O}(t^2)$ and consequently regularity of the Riemann terms $A_1 \sim A_2 \sim {\cal O}(t^0)$ as $t \to 0.$ This will be discussed in detail in section \ref{sec:robust}.

All the above motivations lend an air of inevitability to the no-boundary proposal in semi-classical gravity. That said, a physical theory should not only be intuitive and offer good explanations, but must first of all be in agreement with observations. Thus, we should analyse the no-boundary proposal in detail before passing judgment.

\subsection{Simple Inflationary Examples} \label{sec:inflex}

We just discussed a series of intuitive arguments suggesting that the wave function of the universe might be given by a path integral over compact metrics. Defining such a path integral in general and in detail is rather complicated, and we will discuss progress that was made towards this goal later on. From the discussions in section \ref{sec:qc} and \ref{sec:a3} we know that path integrals can be well approximated by their saddle points, so we may ask a much more tractable question first: do compact and regular saddle point solutions actually exist? 

The most relevant context for answering this question is that of gravity coupled to a scalar field $\phi$ with potential $V(\phi)$, with action
\begin{eqnarray}
S = \int_{\cal M} \mathrm{d} ^4 x  \sqrt{-g} \left( \frac{R}{2} - \frac{1}{2}g^{\mu\nu}\partial_\mu\phi\partial_\nu\phi - V(\phi) \right)  + \int_{\partial \cal M} \mathrm{d}^3 y \sqrt{h}K \,. \label{LorentzianActionwithScalar}
\end{eqnarray}
We will again specialise to FLRW backgrounds \eqref{eq:LorentzianMetric} and a time dependent scalar field only. It is useful to redefine the time coordinate, via $\tilde{N}\mathrm{d}t=-i\mathrm{d}\tau.$ This means that when $\tau$ takes real values, it will correspond to Euclidean time. But it will be useful to consider $\tau$ to be complex valued in general. The metric ansatz is then very simple,
\begin{align}
\mathrm{d}s ^2 = \mathrm{d} \tau ^2 + a ^2( \tau) \mathrm{d} \Omega _3 ^2 \;,
\end{align}
and the Euclidean action $I_E=-iS$ becomes
\begin{equation} \label{eq:complexAction}
I_E = 2\pi^2 \int \mathrm{d} \tau \left( - 3 a a ^{\prime 2} - 3 a + a^3 \left( \frac{1}{2} \phi ^{\prime 2} + V \right) \right) \;,
\end{equation}
where $\; ^{\prime} \equiv \mathrm{d}/\mathrm{d} \tau$. The equations of motion are
\begin{align}
\phi '' + 3 \frac{ a'}{a} \phi' - V_{, \phi} =  0 &\;,\label{eomphi}\\
a'' + \frac{a}{3} \left( \phi ^{\prime 2} + V \right) =  0 &\;, \label{eoma}
\end{align}
while the constraint, arising from time reparameterisation invariance, is
\begin{equation}
a ^{\prime 2}  -1 = \frac{a ^2}{3} \left( \frac{1}{2} \phi ^{\prime 2} - V \right) \;. \label{eq:complexFried}
\end{equation}
In cosmology this equation is usually called the Friedmann equation. Using this equation, we can simplify the action when it is evaluated on a solution of the equations of motion (this is called the ``on-shell'' action)
\begin{equation} \label{eq:onshellAction}
I_E ^{on-shell} =  4 \pi ^2 \int \mathrm{d} \tau\, \left(- 3 a + a ^3 V \right) \;.
\end{equation}

The no-boundary wave function, for now loosely defined via the formal path integral (cf. \eqref{eq:HHpi})
\begin{equation}
\Psi( b, \chi) = \int_{\cal{C}} {\cal D}a {\cal D} \phi \, e^{-I_E(a,\phi)} \sim \sum e^{- I_E(b, \chi)} \;,
\end{equation}
depends on $b$ and $\chi,$ the (late-time) values of the scale factor and scalar field on the final hypersurface. As indicated, we assume that it can be approximated by (a sum of) saddle point contributions. These saddle points must satisfy a number of mathematical and physical requirements \cite{Halliwell:2018ejl}: evidently, they must satisfy the equations of motion and constraints. But moreover, we would like them to be physically meaningful, and for this reason they should yield normalisable wave functions. Moreover, they should lead to physically sensible results, implementing in particular the idea that in the early universe matter fields were in their ground states. Concretely, we must find out if there exist solutions $(a(\tau), \phi( \tau))$, satisfying the following conditions \cite{Hawking:1983hj,Hartle:2008ng}:
\begin{itemize}
\item The solution must be compact, so somewhere we must have $a(0) = 0.$ Here we have shifted the time coordinate such that $\tau=0$ corresponds to the South Pole of the solution, {\it cf.} again Fig.~\ref{fig:regression}. At this point, the solution must also be regular, by which we mean that the fields and their derivatives must take finite values. The Friedmann equation \eqref{eq:complexFried} then implies $a'(0)=\pm 1,$ expressing the fact that the geometry must be Euclidean at the South Pole. The choice of sign for $a'$ is important -- we will show later on that for physical consistency (normalisability) we must choose \begin{align}
    a'(0)=+1\,. \label{regularity}
\end{align} The equation of motion \eqref{eoma} then implies that $a=\tau + {\cal O}(\tau^3).$ Meanwhile, the equation of motion for $\phi,$ Eq. \eqref{eomphi}, shows that regularity implies the condition $\phi'(0)=0.$ This means that no-boundary solutions can be characterised/labelled by the value of the scalar field at the South Pole, $\phi_{SP}=\phi(0).$ This value will be complex in general.
\item There must exist a point $ \tau_f$ in the complex $\tau$ plane where the fields take their specified values $b,\chi \in \mathbb{R},$ that is to say that on the final hypersurface we must have 
\begin{align}
    a(\tau_f)=b \quad \textrm{and} \quad \phi(\tau_f)=\chi, \label{finalcondition}
\end{align} 
with $b,\, \chi$ being the arguments of the wave function. The non-trivial requirement is that the fields must take the specified values simultaneously. If this cannot occur, no solution exists. 
\end{itemize}
Let us add three remarks: first, note that the action $I_E(b, \chi)$ will be evaluated along a path starting at $\tau=0$ and ending at $\tau=\tau_f.$ The choice of path is irrelevant, due to Cauchy's theorem, as long as there are no singularities or branch points present in the complex $\tau$ plane. Second, the complex time path we are talking about here has nothing to do with the complex integration contours (which are contours over the fields) used to define path integrals using the Picard-Lefschetz method discussed in \ref{sec:a3}. Rather, we are talking about a saddle point of such integrals. But this saddle point itself is complex, and can be represented in different ways using different complex time paths. These different representations are physically equivalent, since the action is invariant under changes of path. And third, a note on nomenclature: solutions of the type specified above, satisfying the no-boundary conditions, are often called no-boundary \emph{instantons}. This is a slight abuse of notation, as instantons usually refer to purely Euclidean (finite-action) solutions. Here the instantons are typically complex, and for this reason they have also sometimes been called ``fuzzy instantons''. 

\begin{figure}[ht]
	\centering
	\includegraphics[width=0.8\textwidth]{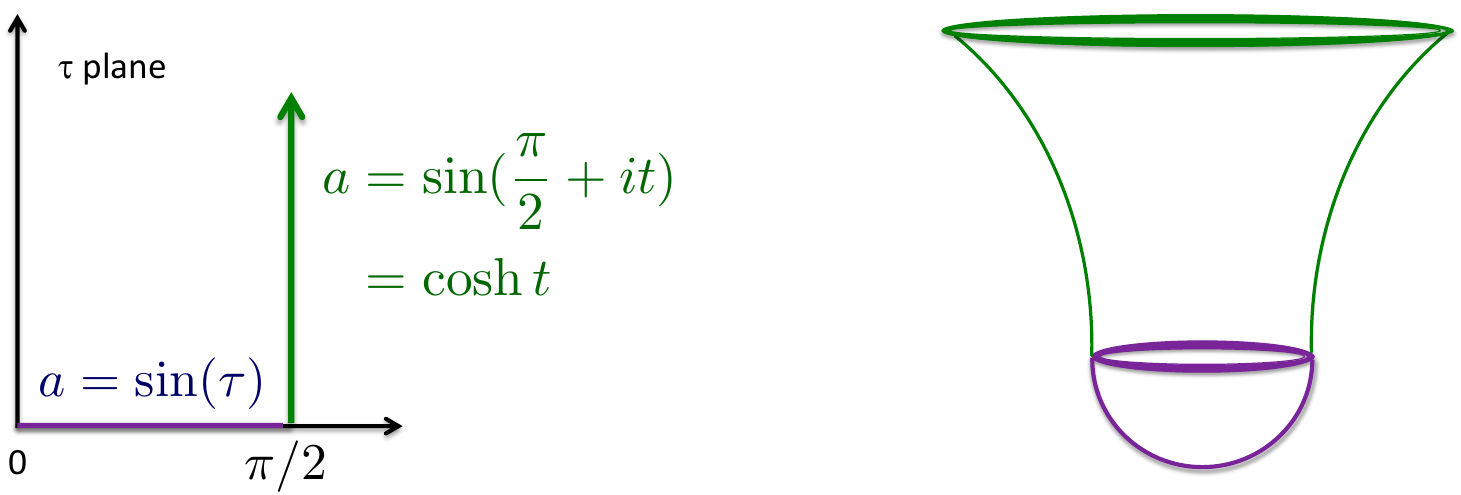}
	\caption{The simplest example of a no-boundary instanton corresponds to a section of complexified de Sitter spacetime, which can be seen as half of a $4-$sphere, in purple, glued at the Hubble radius to half of the Lorentzian de Sitter hyperboloid, in green. The corresponding path in the complex time plane is shown on the left; here we have set the Hubble radius to unity.}
	\label{fig:proto}
\end{figure}

Let us start with the simplest example of a no-boundary solution, which arises when the scalar potential is constant, $V=3H^2.$ In this case the scalar field is constant, and the solution for the scale factor corresponds to the maximally symmetric, constant-curvature de Sitter spacetime, with Hubble radius $1/H.$ In terms of Euclidean time, the solution is given by
\begin{align}
    a(\tau)=\frac{1}{H} \sin(H \tau)\,. \label{Hscale}
\end{align}
Note that this indeed satisfies $a(0)=0$ and $a'(0)=1.$ The $4-$sphere reaches maximum radius at $\tau=\pi/(2H),$ where the scale factor attains the Hubble radius $a=1/H.$ This is the equator of the $4-$sphere, onto which we can glue half of the Lorentzian de Sitter solution by continuing in the Lorentzian time direction,  $\tau \equiv \pi/(2H) + i t,$ with $t\geq 0,$ so that the scale factor evolves as
\begin{align}
    a = \frac{1}{H} \sin(\pi/2 + i H t) = \frac{1}{H} \cosh(H t)\,. \label{Hcont}
\end{align}
The chosen complex time path is illustrated in Fig.~\ref{fig:proto}. Along this path the scale factor takes real values, and thus we can extend the instanton arbitrarily far into the future, until the desired final value $b$ of the scale factor is reached. The corresponding final time is explicitly given by $\tau_f=\pi/(2H) + i\, \textrm{arcosh}(Hb)/H.$ This is the famous Hartle-Hawking instanton, whose ``shuttlecock'' shape is shown in Fig.~\ref{fig:proto}. It is the prototype for all no-boundary solutions. 

We can also calculate the action of this solution, using Eq. \eqref{eq:onshellAction}. The result is straightforwardly found to be\footnote{The easiest way to represent the solution, and to calculate the action, is to use the complex time path which combines a Euclidean and a Lorentzian segment. But if one prefers a smoother representation, without the $90$ degree turn in the complex time plane, then one is in principle free to choose a path that links $\tau=0$ to $\tau_f$ in a smooth, infinitely differentiable, manner.}
\begin{align}
    I_E=\frac{4\pi^2}{H^2}\left[ -1+i\, (H^2b^2 -1 )^{3/2} \right]\,, \label{actiondSsaddle}
\end{align}
so that the no-boundary wave function becomes approximately
\begin{align}
    \Psi \approx e^{\frac{4\pi^2}{\hbar H^2}\left[ 1-i\, (H^2b^2 -1 )^{3/2} \right]}\,.
\end{align}
If we now think about a series of instantons with successively larger scale factor values, we see that the phase of the wave function grows roughly in proportion to the spatial volume $b^3,$ while the amplitude remains constant. Thus, using the results of section \ref{sec:reconstruct}, we can assign a relative probability
\begin{align}
    \Psi^\star\Psi \approx e^{\frac{8\pi^2}{\hbar H^2}} \label{dSprob}
\end{align}
to the corresponding classical history. Of course, in the present case this is a rather trivial statement, as there is only a single history, corresponding to the expanding de Sitter spacetime. However, we could imagine a potential with several approximately flat regions at different potential heights $V=3H^2,$ i.e. we could imagine that $H$ could vary. Wherever the potential is very flat, solutions such as the one above would exist, with their corresponding classical histories. Then we can see an important feature of the no-boundary proposal: smaller potential values come out as preferred. The implications of this will be discussed in section \ref{sec:ob}.

New features enter when we include a non-trivial scalar potential. The simplest potential to consider is that of a mass term, $V(\phi)=\frac{1}{2}m^2\phi^2.$ Current observations \cite{Planck:2018jri} disfavour this model of inflation, but it remains very useful as an example of no-boundary solutions that can be approximated analytically. This was first done in \cite{Lyons:1992ua} (see also \cite{Esposito:1988aw} and the generalisation to generic slow-roll models in \cite{Janssen:2020pii}). To understand the nature of solutions, it is useful to first work out a series expansion near the South Pole, imposing the no-boundary conditions $a(0)=0$ and $a'(0)=1.$ This yields
\begin{align}
a(\tau) &= \tau -\frac{1}{36}m^2\phi_{SP}^2\, \tau^3  + {\cal{O}}(\tau^5)\,, \\
\phi(\tau) & = \phi_{SP} + \frac{1}{8}m^2\phi_{SP}\, \tau^2 + {\cal{O}}(\tau^4) \,, \label{expansionphi}
\end{align}
which shows that the entire solution depends on a single parameter, namely the scalar field value at the South Pole, $\phi_{SP}.$ As we will see momentarily, this must typically be taken to be complex. However, the analytic approximation is most accurate when $\phi_{SP}$ is almost real and large, hence we will assume $|\phi_{SP}^I| \ll 1 \ll |\phi_{SP}^R|,$ where we are denoting real and imaginary parts by the superscripts $R, I.$ 

Near the South Pole, $\phi$ remains approximately constant, and thus Eq. \eqref{eoma} implies that the corresponding solution for $a$ is given by a sinus function, 
\begin{align}
    a \approx \frac{\sqrt{6}}{m \phi_{SP}^R}\, \sin\left( \frac{m\phi_{SP}^R}{\sqrt{6}}\tau\right)\, \qquad \phi \approx \phi_{SP}^R\,. \label{smallasol}
\end{align}
This is again approximately the solution for a $4-$sphere with radius determined by the location of the scalar field on the potential. The equator of the $4-$sphere is reached at time $\tau_{max}^R = \frac{\sqrt{6}}{m \phi_{SP}^R} \frac{\pi}{2}.$ In fact, the series expansion \eqref{expansionphi} shows that $\phi$ remains constant up to this point with fractional error $1/(\phi_{SP}^R)^2.$

We can also find an approximate solution when $a$ is large. This is nothing other than the usual slow-roll approximation for inflationary solutions, but expressed in Euclidean time,
\begin{align}
    a(\tau) & \approx a_0 e^{-i\frac{m\phi_{SP}}{\sqrt{6}}\tau +\frac{m^2}{6}\tau^2}\,, \label{solalarge2}\\
\phi(\tau) & \approx \phi_{SP}+i \sqrt{\frac{2}{3}}m\tau \,. \label{solalarge}
\end{align}
The scalar field slowly rolls down the potential, to good approximation linearly with time, while the scale factor expands exponentially when evolving in the imaginary $\tau$ direction (that is to say, in Lorentzian time). This approximate solution is valid until the second term in the exponent for $a$ overtakes the first term, {\it i.e.} until $\tau^I \sim \phi_{SP}^R/m.$ 

There is an important feature to the scalar field solution \eqref{solalarge}, which is that when we move in the imaginary $\tau$ direction, then the imaginary part of $\phi$ does not change anymore. We can use this property to match onto a desired real value $\chi$ at late times. Also, the late time solution for $a$ will remain approximately real, when matched to the equator of the $4-$sphere solution at small $a.$ Thus, matching at $\tau^R=\tau^R_{max},$ we obtain a (approximately) real solution at late Lorentzian times if we choose ({\it cf.} \eqref{smallasol} and \eqref{solalarge})
\begin{align}
    a_0 \approx \frac{i \sqrt{6}}{m \phi_{SP}^R}\,, \qquad \phi_{SP}^I = - \sqrt{\frac{2}{3}}m\tau_{max}^R = -\frac{\pi}{\phi_{SP}^R}\,, \label{SPcond}
\end{align}
which refines the expression for $\phi$ in \eqref{smallasol}. One obvious lesson of this is that the scalar field must be complex at the South Pole in order for it to reach real values at a late time. This demonstrates explicitly the earlier claim that no-boundary instantons are typically complex. 

We can also approximate the action of these solutions, by using \eqref{eq:onshellAction}. The integral from $\tau=0$ to $\tau_{max}^R$ yields a real value $I_E^R \approx - \frac{24\pi^2}{m^2\chi^2}$ for the Euclidean on-shell action. For the integral along the Lorentzian direction up to $\tau_f,$ one can neglect the term linear in $a$ in \eqref{eq:onshellAction}, since the scale factor becomes exponentially large. Then one may approximate it as follows,
\begin{align}
2\pi^2 \int^{\tau_f}_{\tau_{max}^R} \mathrm{d}\tau \, m^2\phi^2 a^3 = 2\pi^2\int \mathrm{d}\tau\,  m\phi a^2 \sqrt{6} a' & =  2\pi^2 \sqrt{\frac{2}{3}}\int \mathrm{d}\tau \, m\phi \frac{\mathrm{d}}{\mathrm{d}\tau}(a^3) \nonumber \\ & \approx 2\pi^2 \sqrt{\frac{2}{3}}m\chi b^3\,,
\end{align}
where we assumed slow-roll in the last step. The total contribution of this instanton to the wave function is thus
\begin{align}
    \Psi(b,\chi) \, \approx \, e^{\frac{12\pi^2}{\hbar V(\chi)}-i\, 2\pi^2 \sqrt{\frac{2}{3}}m\chi b^3}\,. \label{wfscalarmodel}
\end{align}
We find again that low values of the potential come out as preferred, and that the phase grows in proportion to the spatial volume. 

Note that the wave function is of WKB form, which we recall is a prerequisite for a probabilistic interpretation, as discussed in section \ref{sec:reconstruct}. It is fairly obvious by inspection, as the weighting does not depend on the scale factor $b$ while the phase grows very fast as the universe expands. Nevertheless, we may calculate this more precisely, using $(\nabla I)^2=G^{AB}\partial_A I \partial_B I$ and $G_{bb}=-12\pi^2 b,\, G_{\chi\chi}=2\pi^2 b^3,$
\begin{align}
    \frac{(\nabla I_E^R)^2}{(\nabla I_E^I)^2} \sim \frac{1}{m^6 \chi^8 b^6}\,,
\end{align}
which is indeed driven to tiny values.

The histories implied by the wave function, i.e. the sequences of $(b,\chi)$ values with constant weighting, are in fact those classical solutions that the instantons approach at late times. Explicitly, they can be obtained by eliminating $\tau$ from \eqref{solalarge2} and \eqref{solalarge}, resulting in
\begin{align}
    b \approx \frac{i \sqrt{6}}{m\phi_{SP}^R} e^{\frac{1}{4}(\phi_{SP}^2 - \chi^2)}\,,\label{classsol}
\end{align}
and they correspond to an approximately fixed scalar field value $\phi_{SP} = \phi_{SP}^R -i\frac{\pi}{\phi_{SP}^R}$ at the South Pole. What is important to point out here is that these histories are parameterised by a single real parameter, namely $\phi_{SP}^R,$ whereas general classical solutions would have been parameterised by two real parameters. This shows one aspect of the predictivity of the no-boundary proposal, namely that it restricts the possible solutions to a theory. In other words, not all $(b,\chi)$ combinations may arise, only those that can be related as in \eqref{classsol}. 

Having analytically demonstrated the existence of no-boundary saddle points in a simple inflationary model, we will end this section with a few remarks regarding the consistency of the no-boundary framework. An obvious point is that no-boundary solutions actually exist. They have finite action, as seen in \eqref{wfscalarmodel}, and lead to the prediction of classical expanding universes. These saddle point solutions are everywhere regular, but what is more, the curvature does not become large. In fact the spacetime curvature is roughly constant, as these solutions are close to pure de Sitter spacetime. This means that the curvature radius is given by the Hubble radius, $\sim \sqrt{3/V(\phi_{SP}^R)} = \sqrt{6}/(m\phi_{SP}^R).$ Hence if the potential energy is well below the Planck scale, then the curvature radius $R_H$ of the instantons is correspondingly larger than the Planck length $l_{Pl}$. This means that the expected higher curvature corrections, arising from quantum corrections to the action, will be suppressed by positive powers of $l_{Pl}/R_H \ll 1.$

A final point of importance is the sign choice that we made in \eqref{regularity}. If we had chosen the opposite sign, then we would have found essentially the same solutions, but the Euclidean action would have come out with the opposite sign (this alternative sign choice is the one made in the tunnelling proposal \cite{Vilenkin:1982de}). This would have flipped the relative probabilities, e.g. in \eqref{dSprob} we would have obtained $ \Psi^\star\Psi \approx e^{-\frac{8\pi^2}{\hbar H^2}} $. Then high values of the potential would have been preferred instead of low values. We will discuss this further in section \ref{sec:ob}, and derive the statements that we make here. For now, let us just state the result, which is that with the alternative choice of sign, perturbations around these solutions are unstable, in the sense that solutions with large perturbations are more likely than solutions with small perturbations. Thus, the whole approximation framework of starting with FLRW metrics becomes inconsistent. With the choice of sign that we made here, and which was advocated for these reasons by Hartle and Hawking from the inception of the no-boundary proposal, perturbations are suppressed and the framework is consistent.

\subsection{Numerical Techniques} \label{sec:numtech}

In the previous section we demonstrated the existence of (inflationary) no-boundary saddle points, by using analytic approximations in a simple model containing a massive scalar field. For general potentials, it is however impossible to find explicit no-boundary solutions by analytic methods. In such cases, we have to resort to numerical methods. We will describe these here, and show that they can easily produce large numbers of no-boundary solutions. Moreover, they have the advantage of being able to access parameter regions that were not reachable by the approximations made in the previous subsection.

\begin{figure}[htbp]
\begin{center}
\includegraphics[width=0.9\textwidth]{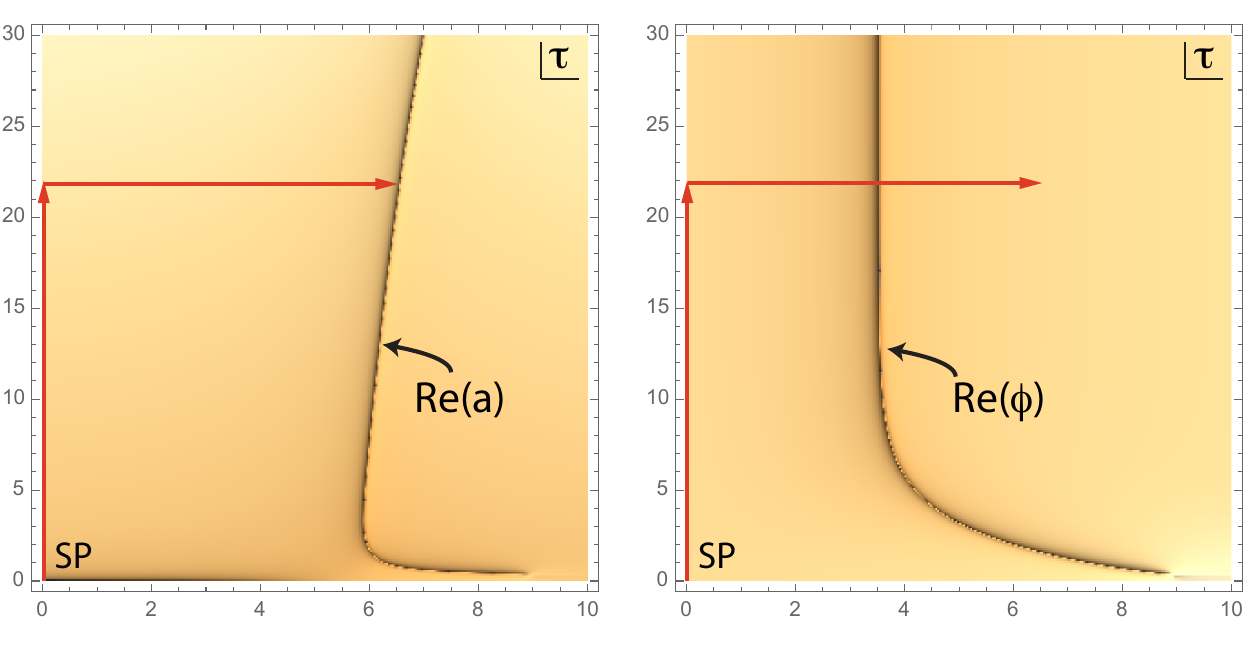}
\end{center}
\caption{ \label{fig:one} These figures show the complexified $\tau$ plane, on the left for the scale factor and on the right for the scalar field. The dark lines show the locus where the fields take real values (more precisely, the figures show a density plot of $ \log{ | \textrm{Im}(a(\tau), \phi(\tau))|}$ for a dense grid of values of $\tau$). The red arrows indicate a useful contour of integration, used in the optimisation procedure needed to find numerical no-boundary solutions. The present figures show what this looks like when the desired solution has \emph{not} been reached yet: the dark lines, where scale factor and scalar field are real, are located in different regions of the complex~$\tau$ plane. For this example, $V(\phi)=\frac{1}{2}m^2\phi^2,$ $m=1/10,$ and we used the trial value $\phi_{SP} = 6.5-0.3i.$ The values reached at $\tau_r \approx 6.58 + 22i$ are $a\approx 376$ and $\phi \approx 4.99+0.25i.$} \mbox{}\vspace{0.2cm} \\
\begin{center}
\includegraphics[width=0.9\textwidth]{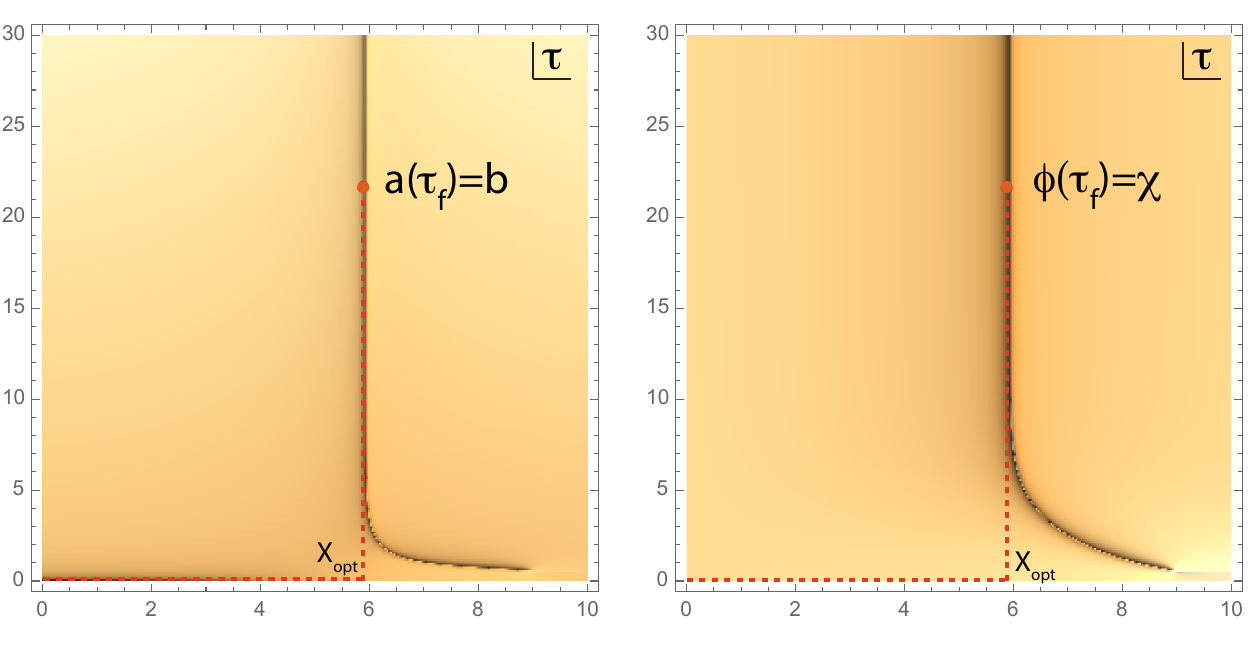}
\end{center}
\caption{ \label{fig:two} Now we show the same model, but with optimised values $\phi_{SP}\approx 6.451-0.5037i.$ The real $ a$ and real $ \phi$ lines now (asymptotically) overlap and a no-boundary solution is obtained. More precisely, the real values $b = 300$ and $\chi=5$ are obtained at $\tau_f \approx 5.943+21.34i.$}
\end{figure}

The strategy for finding numerical no-boundary solutions is as follows \cite{Hartle:2007gi,Hartle:2008ng}. First we can choose a value for the scalar field at the South Pole,
\begin{equation}
\phi_{SP} = \phi_{SP}^R + i \phi_{SP}^I = |\phi_{SP}| e^{i\theta}\,,
\end{equation} 
where we will sometimes use a polar representation with argument $\theta.$ It is useful to start with a value of $\phi_{SP}^R$ or $|\phi_{SP}|$ that lies in a region of the potential where inflationary solutions may be expected to be found, {\it i.e.} in a relatively flat region of the potential -- this will make it easier to find the first solution. Then one should integrate the equations of motion \eqref{eoma} and \eqref{eomphi} from the South Pole, along a chosen contour in the complexified $\tau$ plane. The South Pole is a regular singular point, hence one cannot integrate directly from it, but rather from somewhere close by. This can be done to arbitrary numerical precision, using the series expansion of no-boundary solutions near the South Pole,
\begin{align}
a(\tau) &= \tau -\frac{V}{18}\tau^3 + \frac{8V^2-27V_{,\phi}^2}{8640} \tau^5 + {\cal{O}}(\tau^7)\,, \\
\phi(\tau) & = \phi_{SP} + \frac{V_{,\phi}}{8}\tau^2 + \frac{2VV_{,\phi}+3V_{,\phi}V_{,\phi\phi}}{576} \tau^4 + {\cal{O}}(\tau^6) \,. 
\end{align}
This series expansion is derived by imposing the regularity conditions at $\tau=0$ that were described in section \ref{sec:inflex}. In practice, it is useful to start the integration at $\tau=\epsilon,$ with $|\epsilon|$ of order $10^{-5}$ or even smaller. The integration contour should be chosen for convenience. One choice that works well is to integrate first in the imaginary $\tau$ direction, followed by a segment in the real $\tau$ direction, see Fig.~\ref{fig:one}. A contour of this shape typically allows one to cross the locus where the scale factor takes real values, let us denote this point as $\tau_r.$ However, the scalar field will in general not be real at this location, but will take a complex value. The idea now is to  \emph{tune} the four real parameters $\phi_{SP}, \, \tau_r$ such that one reaches $\textrm{Re}(a)=b,\, \textrm{Im}(a)=0,\, \textrm{Re}(\phi)=\chi,\, \textrm{Im}(\phi)=0$ to the desired accuracy. The tuning can be done by using a Newtonian algorithm, which can straightforwardly be implemented in such a higher-dimensional parameter space. However, some trial and error is typically required at first, in order to already find oneself in the basin of attraction of the solution. Here it is useful to choose target values $b$ and $\chi$ that are close to the values $a(\tau_r)$ and $\textrm{Re}[\chi(\tau_r)].$ The result of a successful optimisation is shown in Fig.~\ref{fig:two}, where the endpoint, at which the fields reach the real values $b, \chi,$ is now denoted as $\tau_f.$ 

\begin{figure}[ht]
	\centering
	\includegraphics[width=0.45\textwidth]{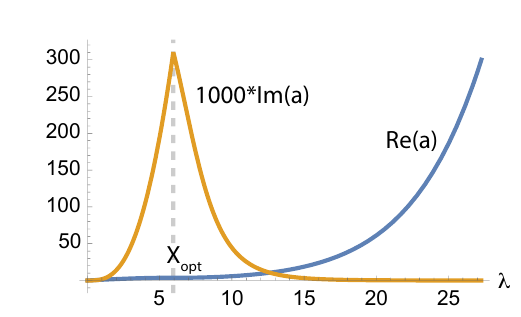}
	\includegraphics[width=0.45\textwidth]{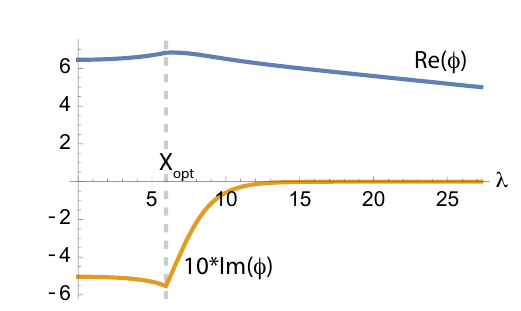}
	\caption{Here we show the field values of the optimised instanton from Fig.~\ref{fig:two}, along the path indicated by the dotted line in that figure. The path first evolves from the South Pole in the Euclidean direction to $X_{opt}\approx 5.943,$ and then in the Lorentzian direction up to $\tau_f.$ $\lambda$ is a parameter along the path. The left panel shows the real and imaginary values of the scale factor, while the right panel is for the scalar field. The imaginary parts have been enhanced for better viewing. Note that the fields become real at the final time, and reach the desired values $b=300, \chi=5.$}
	\label{fig:exfields}
\end{figure}

In Fig.~\ref{fig:exfields} we also plot the field values for the optimised solution, but now following the more standard contour where we evolve in the Euclidean direction first, up to $X_{opt} \equiv \textrm{Re}(\tau_f),$ and then in the Lorentzian direction up to $\tau_f.$ Note that the imaginary parts of the fields quickly decay along the Lorentzian part of the contour. We can also compare the optimised values with the analytic estimates of the previous section. There we had found that $\phi_{SP}^I=-\pi/\phi_{SP}^R.$ Here we would thus expect $\phi_{SP}=5-\frac{\pi}{5}i \approx 5-0.63 i,$ whereas the optimised value we obtained was $\phi_{SP} \approx 6.451-0.5037i.$ The discrepancy can be attributed to the fact that the scalar field value is not very large here, whereas an assumption in section \ref{sec:inflex} was that $\phi_{SP}^R \gg 1.$ But this is the advantage of the numerical method, that one can find solutions in regions of the potential where analytic estimates don't work well.

Optimised no-boundary solutions typically exhibit another feature, clearly visible in Fig.~\ref{fig:two}: there exist lines at constant $X$ along which the scale factor and the scalar field remain approximately real when evolving in the imaginary (Lorentzian) time direction. This is due to the inflationary attractor, which has pulled one close to a classical solution of the equations of motion. However, strictly speaking, if the scalar field is not residing at an extremum of the potential (where the instanton is pure de Sitter spacetime), the scale factor and the scalar field only take the real values $b$ and $\chi$ simultaneously at one location $\tau_f,$ and not along a line segment. Still, due to the inflationary attractor, values of $b,\, \chi$ that are related by classical evolution will correspond to instantons with essentially the same optimised $\phi_{SP}, \, X=\textrm{Re}(\tau_f)$ values, and will only differ in the Lorentzian time location $\textrm{Im}(\tau_f)$ where the final field values are reached. Also, quite generally, once one has found the first such no-boundary solution, one can usually easily find nearby solutions with slightly different $b,\, \chi$ values, even when they belong to different classical histories. This allows us to gain a more global understanding of the existence of solutions.

Before embarking on this task, let us note a simplification: in order to perform the numerical calculations, we can scale the potential to an arbitrary overall normalisation. Indeed, starting from the Euclidean action
\begin{equation}
S_E = - \int \mathrm{d} ^4 x  \sqrt{g} \left( \frac{R}{2} - \frac{1}{2} g ^{\mu \nu} \partial _{\mu} \phi\, \partial _{\nu} \phi - V( \phi) \right) \;,
\end{equation}
we can perform the scalings (with constant $V_0$)
\begin{equation} \label{eq:scaling}
\phi  \equiv \bar{ \phi} \;, \quad
V  \equiv  V_0 \bar{V} \;,\quad
g_{\mu\nu} \equiv   V_0 ^{-1} \bar{ g}_{\mu\nu} \;,
\end{equation}
which result in a new action
\begin{equation}
S_E = -\frac{1}{V_0} \int \mathrm{d} ^4x \sqrt{ \bar{ g}} \left( \frac{ \bar{R}}{2} - \frac{1}{2} \bar{g} ^{\mu \nu} \partial _{\mu} \bar{ \phi} \partial _{\nu} \bar{ \phi} - \bar{V} \right) \;.
\end{equation}
Now the overall scale of the potential appears out front. In fact, this scaling is not only useful for the numerics, but it also explains the functional dependence of the wave function on the potential, {\it cf.} Eqs. \eqref{dSprob} and \eqref{wfscalarmodel}. We will make use of this scaling freedom below.

\begin{figure}[ht]
\begin{minipage}{0.47\textwidth} \flushleft
\includegraphics[width=\textwidth]{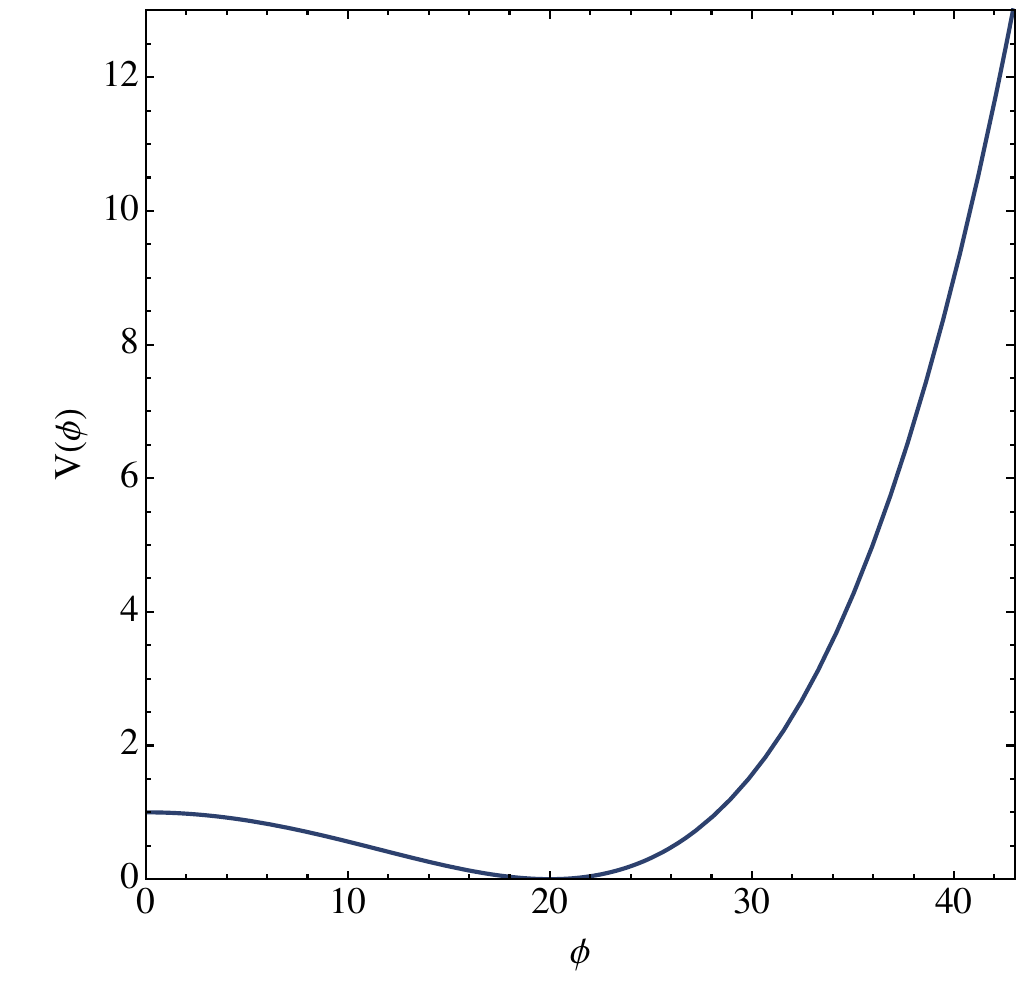}
\end{minipage}%
\begin{minipage}{0.47\textwidth} \flushleft
\includegraphics[width=\textwidth]{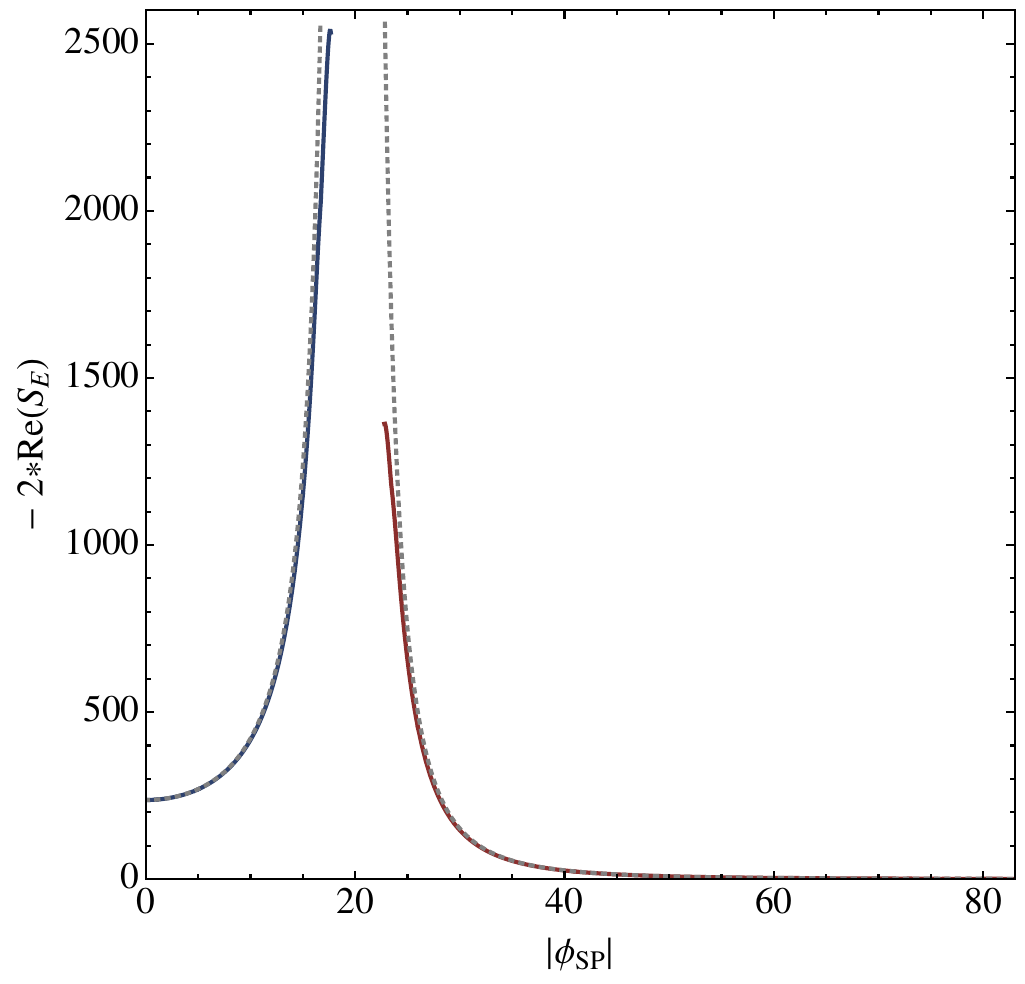}
\end{minipage}%
\caption{\label{fig:compPotential} {\it Left panel:} A toy inflationary landscape potential of the form \eqref{eq:HiggsPotential}. It contains a plateau region at small field values, and a power law region at large field values. It is normalised here such that $V(0)=1$. This potential serves to illustrate general features of (inflationary) no-boundary solutions in a potential landscape. {\it Right panel:} For the no-boundary solutions in Fig.~\ref{Fig:Higgs1}, we show here the weighting of the action, {\it i.e.} the logarithm of the relative probability, again as a function of $|\phi_{SP}|$. The dotted lines represent the approximation of the instanton by a pure de Sitter solution. Low values of the potential come out as preferred. Figures reproduced from \cite{Battarra:2014kga}.}
\end{figure}

\begin{figure}[ht]%
\begin{minipage}{0.48\textwidth} \flushleft
\includegraphics[width=\textwidth]{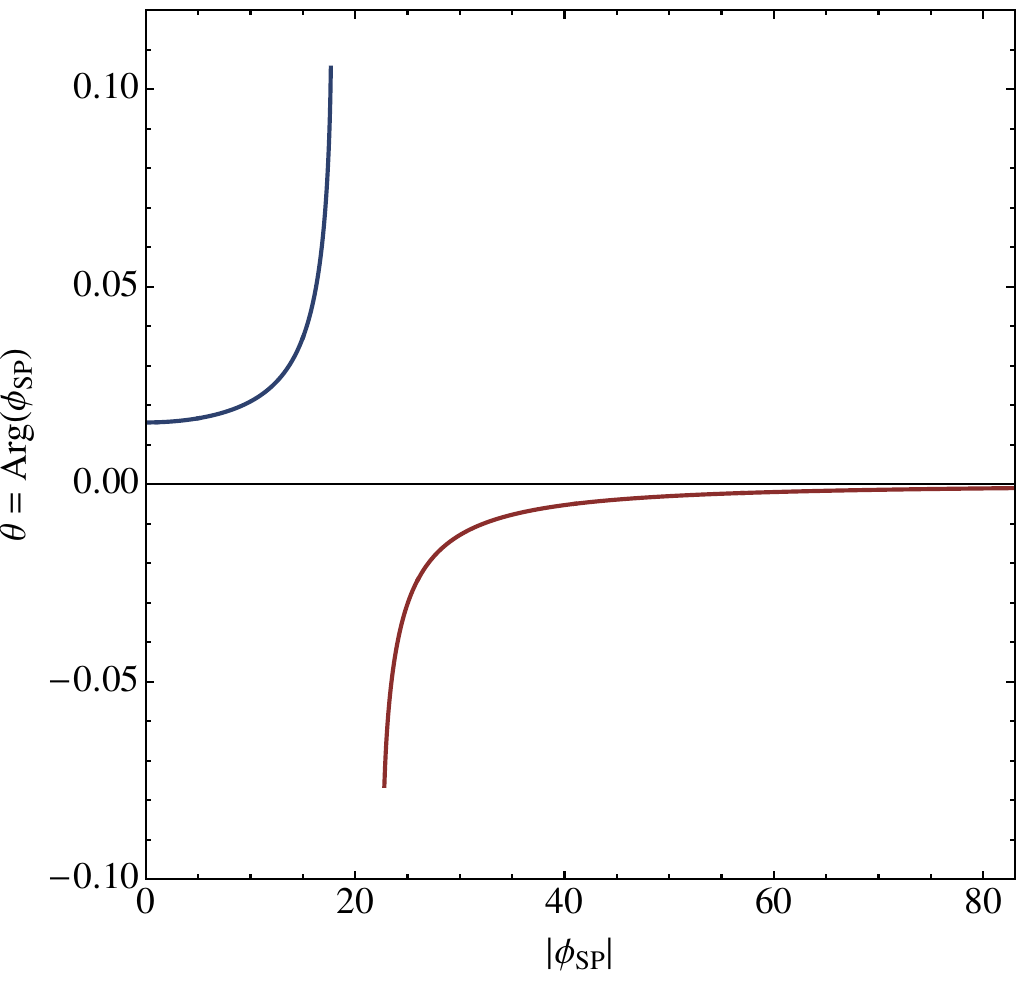}
\end{minipage}%
\begin{minipage}{0.48\textwidth} \flushleft
\includegraphics[width=\textwidth]{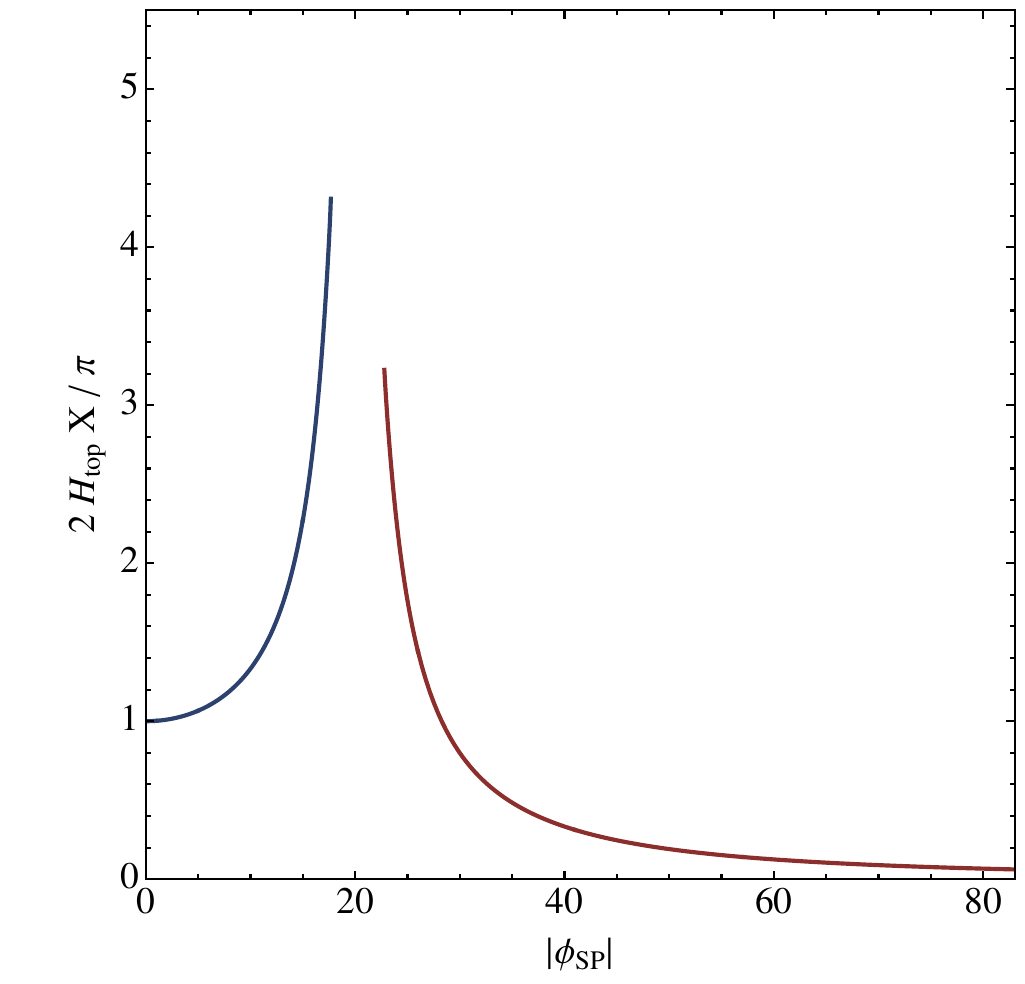}
\end{minipage}%
\caption{\label{Fig:Higgs1} For no-boundary solutions in the potential in Fig.~\ref{fig:compPotential}, we show plots of the optimised phase $\theta$ at the South Pole (left panel) and the location $X$ where real values are reached at late times (right panel), both as a function of $|\phi_{SP}|$. Here we chose $v=20$ while $X$ is expressed in terms of $H_{top} \equiv \sqrt{V(\phi=0)/3} = 1/\sqrt{3}.$ All results can be scaled to any overall scale of the potential using Eqs. \eqref{eq:scaling}. Figures reproduced from \cite{Battarra:2014kga}.}
\end{figure}

A model that serves well to illustrate the properties of inflationary no-boundary solutions is a toy model for a potential energy landscape, with potential \cite{Battarra:2014kga}
\begin{equation} \label{eq:HiggsPotential}
V(\phi) = \frac{1}{v^4} (\phi^2- v^2)^2\;.
\end{equation}
The potential is illustrated in the left panel of Fig.~\ref{fig:compPotential}. It consists of two inequivalent inflationary regions, one of plateau type at small $\phi,$ and an approximately quartic potential at large $\phi,$ with a potential minimum in between. This potential admits no-boundary solutions in both inflationary regions. The optimised scalar field values at the South Pole are shown in Fig.~\ref{Fig:Higgs1}. Perhaps the most important thing to note about this figure is that no-boundary solutions do not exist everywhere on the potential. Rather, when the potential becomes too steep to support a prolonged inflationary phase, no-boundary solutions cease to exist. This occurs near the potential minimum at $\phi=20,$ and arises because there is no dynamical attractor at play there, which could drive the solution towards a classical solution of the equations of motion. In other words, when there is no attractor, it becomes impossible for the scale factor and scalar field to simultaneously become real valued, while also satisfying the regularity conditions at the South Pole (which, we should recall, require the scalar field to start out with a complex value, {\it cf.} \eqref{SPcond}). We will discuss the attractor in more detail in section \ref{sec:class}.

For these solutions, one can also calculate the action. The most interesting part is the weighting, {\it i.e.} the imaginary part of the action (or equivalently minus the real part of the Euclidean action), and this is shown in the right panel of Fig.~\ref{fig:compPotential}. Note that the approximation 
\begin{align}
    \ln \Psi^\star \Psi = -\frac{2}{\hbar} S_E^R = \frac{24\pi^2}{\hbar V(|\phi_{SP}|)}\,,
\end{align}
adapted from the pure de Sitter case in \eqref{dSprob} and indicated by the dotted lines, works very well. As one can see, lower values of the potential come out as preferred. This means that the plateau region of the potential (at small $\phi$) is preferred over the power law region at large $\phi.$ This is in good qualitative agreement with observations of the cosmic microwave background \cite{Hertog:2013mra}. However, we should also point out that within each inflationary region, starting lower on the potential is again preferred, so that short inflationary phases come out as preferred over long ones, which leads to some tension with observations. A detailed comparison with observations will be the subject of section \ref{sec:ob}.

\subsection{Ekpyrotic Examples} \label{sec:ekex}

The numerical techniques that we just described can also be used to find no-boundary instantons of a very different type, namely in ekpyrotic theories containing a steep and negative potential for the scalar field. In such models, the ekpyrotic phase, during which the universe slowly contracts, plays an analogous role to the inflationary phase and replaces it \cite{Khoury:2001wf,Erickson:2003zm}. A contracting phase with high pressure (in fact with pressure larger than the energy density) has the effect of rendering the universe homogeneous, isotropic and flat. Moreover, if one adds a second scalar field, such models can also generate density perturbations in agreement with observations of the cosmic microwave background \cite{Lehners:2007ac,Ijjas:2014fja}. The least understood aspect of these models is how they link the contracting phase with a subsequent hot expanding phase, but for progress on this issue see \cite{Ijjas:2016vtq}.

It may sound surprising to look for no-boundary solutions when the universe is supposed to contract rather than expand -- how can a contracting universe arise from nothing? But one has to keep in mind that the South Pole region of no-boundary solutions does not correspond to a classical spacetime. The idea, roughly speaking, is that Euclidean space is generated from nothing and then turns into a Lorentzian spacetime as the universe contracts \cite{Battarra:2014xoa}. 

\begin{figure}[ht]%
\begin{minipage}{0.47\textwidth} \flushleft
\includegraphics[width=\textwidth]{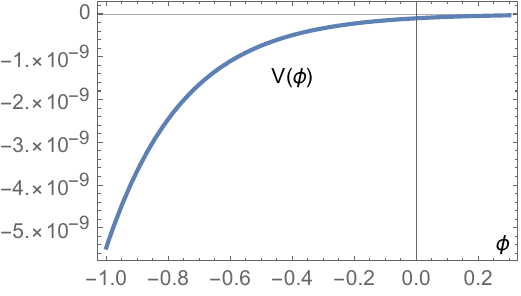}
\end{minipage}%
\begin{minipage}{0.47\textwidth} \flushleft
\includegraphics[width=\textwidth]{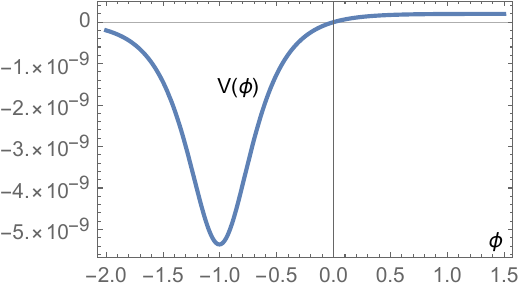}
\end{minipage}%
\caption{\label{Fig:ekpot} Representative shapes of ekpyrotic (left) and cyclic (right) scalar field potentials. In the cyclic case, the ekpyrotic part of the potential is sandwiched between a bounce phase at more negative $\phi$ values, and a dark energy plateau at positive $\phi.$}
\end{figure}

\begin{figure}[ht]%
\begin{minipage}{0.48\textwidth} \flushleft
\includegraphics[width=\textwidth]{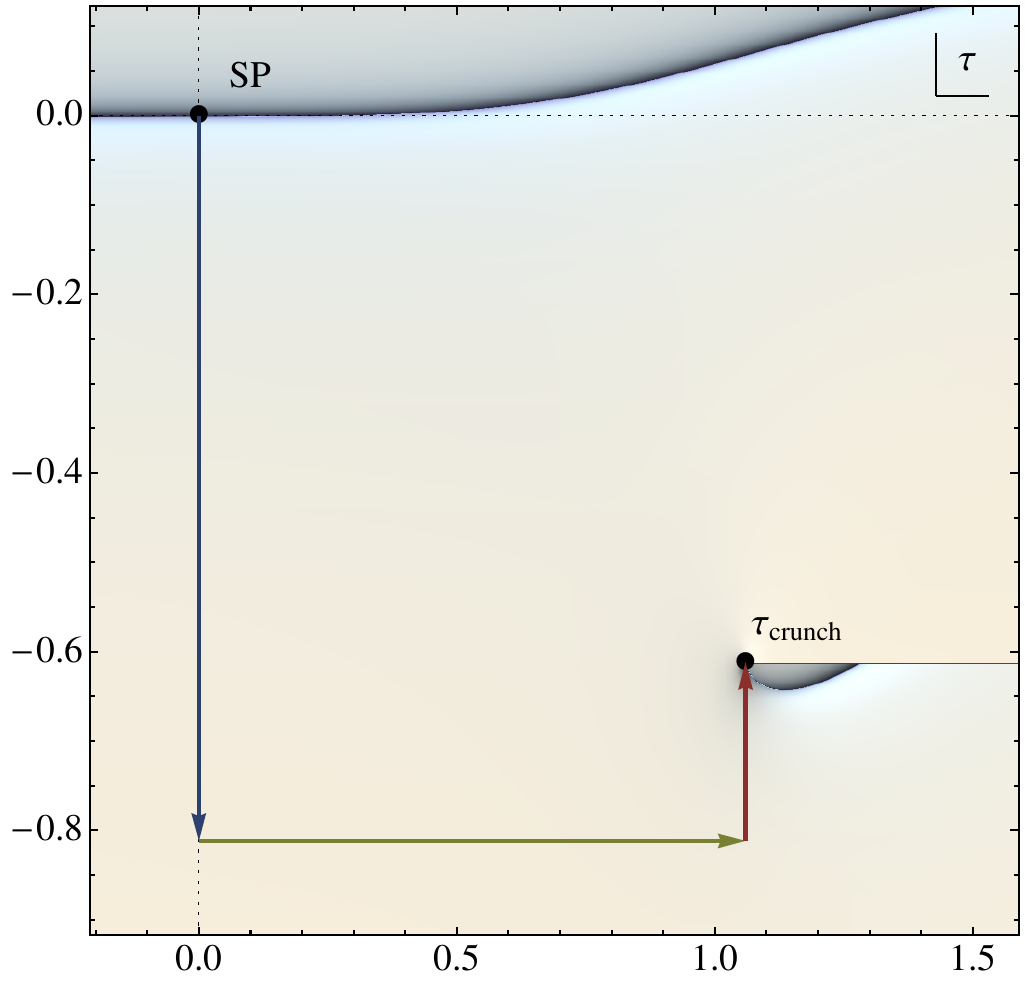}
\end{minipage}%
\begin{minipage}{0.48\textwidth} \flushleft
\includegraphics[width=\textwidth]{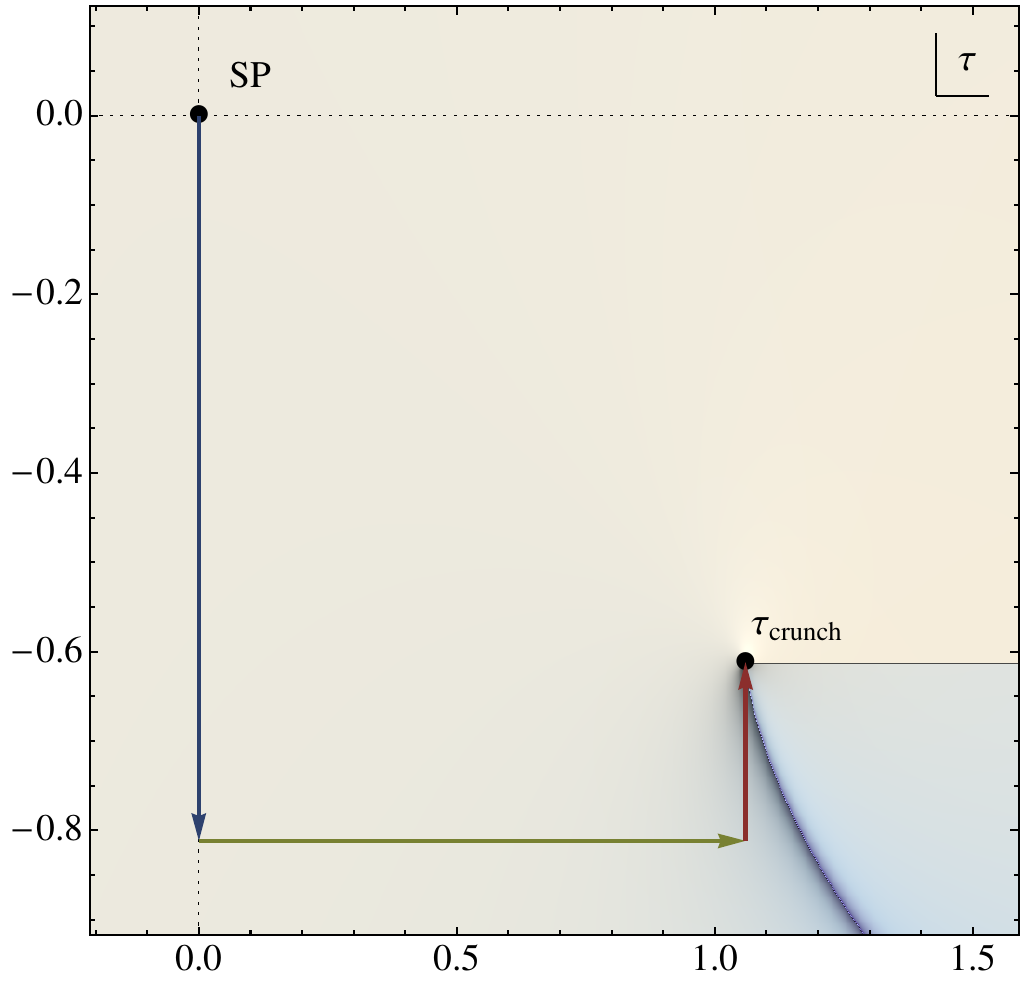}
\end{minipage}%
\caption{\label{Fig:ekex} An example of an ekpyrotic instanton. The coloured arrows indicate a contour of integration that proves useful in finding solutions of this type. Here $c=\sqrt{8}$ and the optimised South Pole value is $\phi_{SP}=0.000-1.481 i.$ Dark lines again show the locus of real field values, {\it cf.} Figs.~\ref{fig:one} and \ref{fig:two}. Figures reproduced from \cite{Battarra:2014kga}.}
\end{figure}

\begin{figure}[ht]%
\begin{minipage}{0.48\textwidth} \flushleft
\includegraphics[width=\textwidth]{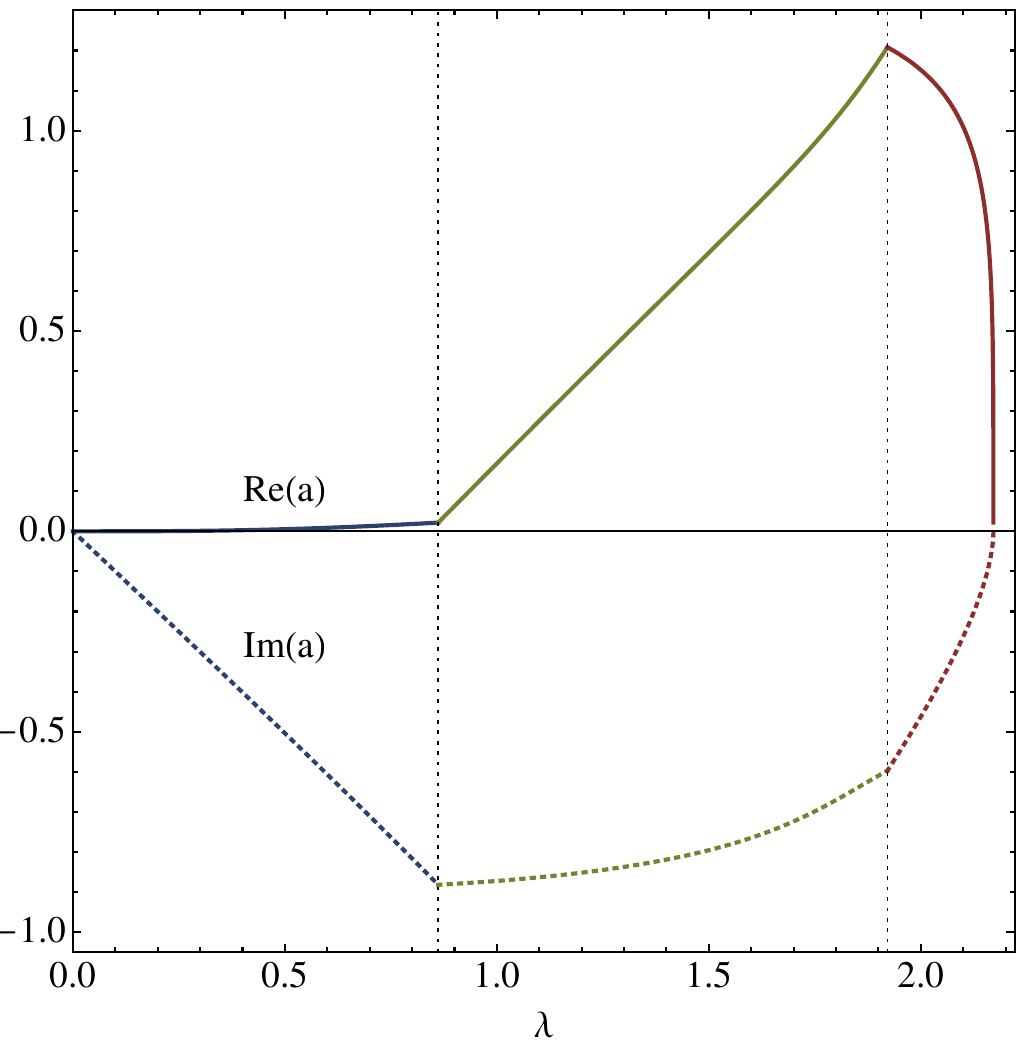}
\end{minipage}%
\begin{minipage}{0.48\textwidth} \flushleft
\includegraphics[width=\textwidth]{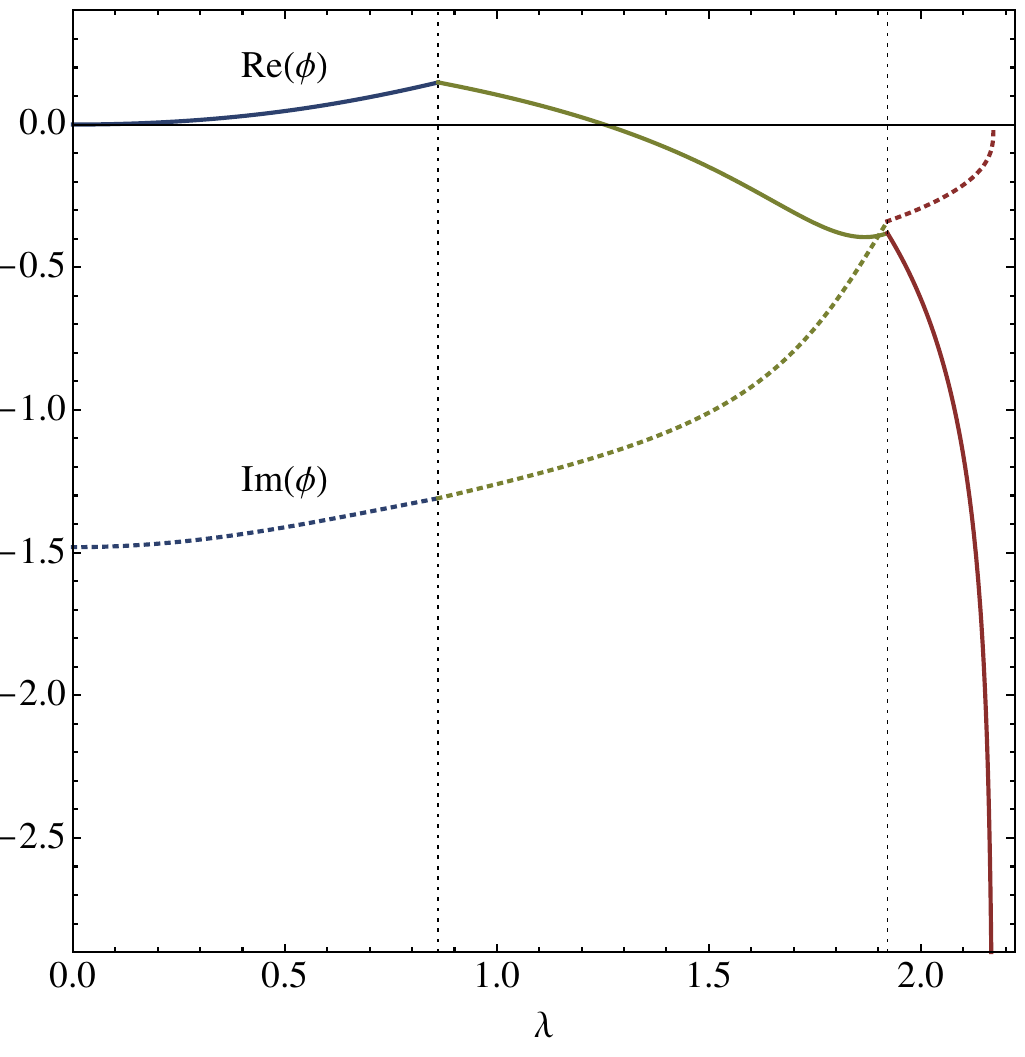}
\end{minipage}%
\caption{\label{Fig:ekfields} The evolution of the scale factor and scalar field for the ekpyrotic instanton shown in Fig.~\ref{Fig:ekex}, along the contour indicated in that figure. Note that along the final segment (in red), the fields become increasingly real as the universe contracts and the scalar rolls down the potential. Figures reproduced from \cite{Battarra:2014kga}.}
\end{figure}

A simple example of an ekpyrotic potential is a negative exponential,
\begin{align}
    V(\phi) = V_0 e^{-c\phi}\,,
\end{align}
where we take $V_0<0$ and $c>\sqrt{6}.$ The latter condition ensures that the contracting solution is an attractor, cf. section \ref{sec:class}. Using analogous techniques to those of the previous section, one may then tune the values of the scalar field at the South Pole in order to obtain simultaneously real $b,\, \chi$ values of the scale factor and scalar field as the universe contracts. An example is shown in Fig.~\ref{Fig:ekex}, taken from \cite{Battarra:2014kga}. Note that one has to use a somewhat different contour in the complex time plane to find such solutions (the field values along the indicated contour are shown in Fig.~\ref{Fig:ekfields}): from the South Pole, the contour first has to run in the negative imaginary direction. Along this part of the contour, the solution corresponds to a portion of a large $4-$sphere, with $a(\lambda) \approx -i \lambda$. Then the contour runs in the Euclidean direction, and the fields are fully complex there. Finally, the contour runs in the Lorentzian time direction, and it is along this segment that the fields become real valued. As one can see from Fig.~\ref{Fig:ekfields}, the scale factor is shrinking there while the scalar field rolls down the potential fast. In this simple model, no bounce is included, and hence the evolution necessarily ends in a crunch. This is also the reason why one cannot see a long vertical time segment in Fig.~\ref{Fig:ekex} along which both the scale factor and scalar field are real to high precision: the crunch occurs shortly after the fields have reached real values. 

The action for these solutions can be calculated numerically, and is found to take a functional form that is very closely related to the one in the inflationary case, in particular one finds for the weighting that
\begin{align}
    \Psi^\star\Psi \approx e^{\frac{s}{\hbar |V(\phi_{SP})|}}\,,
\end{align}
where $s$ is a positive numerical factor that depends on the steepness of the potential \cite{Battarra:2014xoa}. This implies that ekpyrotic instantons receive a very high probability if they start high up on the potential. Thus, in contrast to the inflationary case, a large number of e-folds of ekpyrosis comes out as preferred.

\begin{figure}[ht]%
\begin{center}
\includegraphics[width=0.5\textwidth]{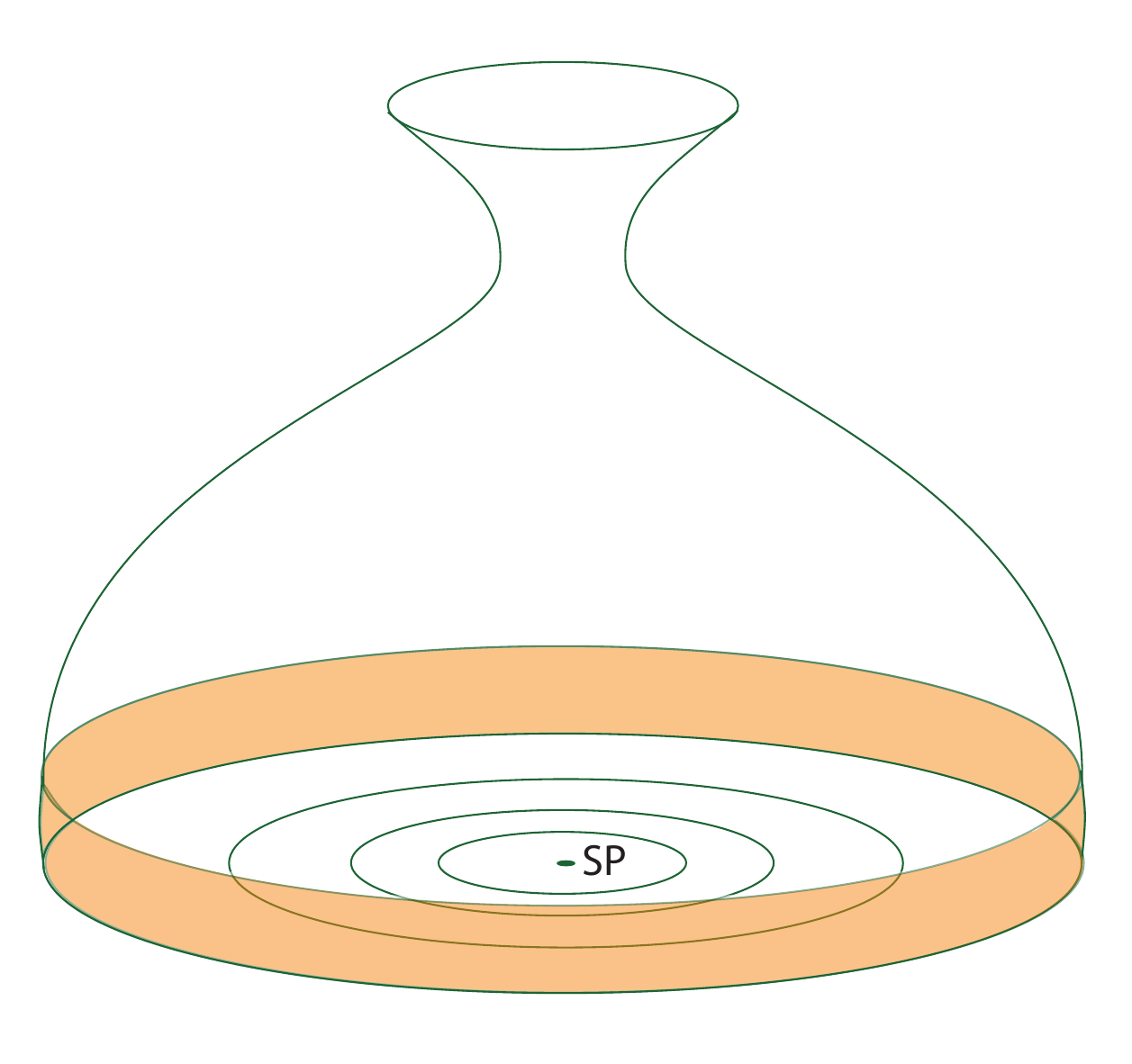}
\end{center}
\caption{\label{fig:carafe} A cartoon of the shape of ekpyrotic instantons. A large Euclidean space is created from nothing, starting from the South Pole. Then, as the universe contracts, the spacetime and scalar field become classical and real valued. In the figure, a bounce into the expanding phase of the universe is indicated. Figure reproduced from \cite{Battarra:2014kga}.}
\end{figure}

The overall shape of ekpyrotic no-boundary solutions is shown in Fig.~\ref{fig:carafe}, and it takes the shape of a decanter. In the figure, it is imagined that a bounce into an expanding phase actually takes place. In \cite{Lehners:2015efa} a ghost condensate model of a bounce was included, because it is a simple example of how to model a classical bounce. In this case, no-boundary instantons with this shape indeed arise. However, the ghost condensate model is known to contain instabilities \cite{Arkani-Hamed:2003pdi}, and thus is not consistent on a quantum level. This is in fact the main problem with classical bounces. In all cases that are known to date, they occur in theories that also allow for unstable solutions (even if the bounce solution itself is stable). It remains unclear whether one may consistently treat such theories as quantum theories, since fluctuations away from the solution of interest might be able to reach unstable solutions. Thus it remains unknown whether quantum gravity allows cosmic bounces or not. Note that this question is also important in assessing whether one of the heuristic motivations given in section \ref{sec:nbheurist} holds up, namely whether the quantum wave function is effectively of (inflationary) no-boundary type even when the universe has an asymptotic flat region to the past, {\it cf.} again Fig.~\ref{fig:topol}.

If bounces do make sense, then there exists another possibility for no-boundary solutions, this time in cyclic extensions of the ekpyrotic model \cite{Steinhardt:2001st}. In those extensions, there exists a positive plateau region in the potential, where the current dark energy phase takes place. In the far future, the dark energy decays and the universe contracts in a renewed ekpyrotic phase. Then the universe bounces again into a hot big bang phase, followed by dark energy, and another cycle starts. For a potential of this type, it is possible to find no-boundary solutions that start in the dark energy phase \cite{Battarra:2014kga}. This has the advantage of leading to vastly higher probabilities, since the potential is very low there, {\it cf.} \eqref{dSprob}. However, this possibility, in the same way as all the other possibilities discussed in this section, hinges on the viability of cosmic bounces in quantum gravity.

The examples of this section highlight an important point, which is that the no-boundary proposal is a theory of initial conditions that is independent of the dynamics, in particular it is logically independent of inflation. Rather, it can be applied to any dynamical model of the universe. However, as we will discuss in more detail now, whether no-boundary saddle points exist depends rather crucially on whether or not the dynamical theory in question exhibits an attractor.

\subsection{Classical Histories from the No-Boundary Wave Function} \label{sec:class}

We have encountered two separate classes of no-boundary solutions so far, those of inflationary and those of ekpyrotic type. In both cases, satisfying both the no-boundary conditions and the late-time reality conditions on the fields depended crucially on the presence of a dynamical attractor. Since this is such a basic requirement, it is useful to make it somewhat more precise.

We can proceed analytically by focusing on models with constant equation of state. Such models arise when the scalar potential is of exponential form
\begin{align}
    V(\phi) = V_0 e^{-c\phi}\,.
\end{align}
Assuming a flat Robertson-Walker background $\mathrm{d}s^2=-\mathrm{d}t^2 + a(t)^2 \mathrm{d}\bf{x}^2,$ the equation of motion and constraint read
\begin{align}
    \ddot\phi +3H\dot\phi -cV =0\,, \qquad 3H^2 = \frac{1}{2}\dot\phi^2 + V\,, \label{expeq}
\end{align}
where $H=\dot{a}/a.$ They admit a scaling solution, given by
\begin{align}
    a(t) = a_0 |t|^{\frac{2}{c^2}}\,,& \quad H=\frac{2}{c^2 t}\,, \label{scal1}\\
    \phi(t) = \frac{1}{c}\ln\left(\frac{c^4 V_0}{12-2c^2} t^2\right)\,,& \quad V=\frac{12-2c^2}{c^4 t^2}\,.\label{scal2}
\end{align}
In the above relations, we have combined two cases: when $V_0>0,$ we assume $t>0$ and this corresponds to an expanding, inflationary, accelerating solution when $c^2<2.$ Meanwhile, when $V_0<0,$ we take $t<0$ and then this corresponds to a contracting, ekpyrotic solution where we will momentarily see that we must take $c^2>6.$ These solutions are called scaling solutions since all the terms in the equation of motion and constraint scale in the same way, as $t^{-2}.$ If we had added a spatial curvature term $\pm 1/a^2$ to the constraint in \eqref{expeq}, then it would have become increasingly subdominant, which justifies our neglect of this term from the start. The equation of state $w$ is given by the ratio of pressure to energy density,
\begin{align}
    w = \frac{p}{\rho} = \frac{\frac{1}{2}\dot\phi^2-V}{\frac{1}{2}\dot\phi^2+V} = \frac{c^2}{3}-1\,, \label{eqeos}
\end{align}
and it is indeed constant, as advertised above. In the inflationary case $w < -\frac{1}{3},$ while for ekpyrosis we have $w>1.$

We can now perturb this solution to assess its stability. In what should be obvious notation, we obtain
\begin{align}
    \ddot{\delta\phi} + 3 \dot\phi\, \delta H + 3H\, \dot{\delta\phi} +c^2V\delta\phi=0\,, \quad 6H\, \delta H = \dot\phi\, \dot{\delta\phi} -cV\delta\phi\,, \label{scalperteq}
\end{align}
which can be combined into
\begin{align}
      \ddot{\delta\phi} + 3H (1+\frac{\dot\phi^2}{6H^2}) \dot{\delta\phi} +\frac{c^2}{2}V\delta\phi = 0\,.
\end{align}
This equation is solved by
\begin{align}
    \delta\phi \propto t^{-1},\, t^{1-\frac{6}{c^2}}\,.
\end{align}
The first solution simply corresponds to a shift in initial conditions, {\it cf.} \eqref{scal2}, so only the second solution is relevant. From \eqref{scalperteq}, it implies that the fractional change in the scale factor also evolves as $\frac{\delta a}{a} \sim t^{1-\frac{6}{c^2}}.$ Hence the scaling solution is an attractor if either the universe is expanding and $c^2<6,$ or if the universe is contracting and $c^2>6$ \cite{Heard:2002dr}. Thus it is an attractor in both the inflationary and ekpyrotic cases. This is the important property that allows one to reach real values for the scale factor and scalar field simultaneously, and thus allows for classicality at late times.

We can even be a little more precise, and derive just how fast the wave function reaches WKB form \cite{Battarra:2014kga,Lehners:2015sia}. For this it is helpful to realise that, when the potential is exponential, the action transforms in a simple manner under shifts of the scalar field. Taking the Euclidean action as the starting point,
\begin{equation}
S_E = - \int \mathrm{d} ^4 x  \sqrt{g} \left( \frac{R}{2} - \frac{1}{2} g ^{\mu \nu} \partial _{\mu} \phi\, \partial _{\nu} \phi - V_0 e^{-c\phi} \right) \;,
\end{equation}
one can shift the scalar and perform a related scaling of the metric, thereby extending the scaling \eqref{eq:scaling},
\begin{equation}
\phi  \equiv  \bar{ \phi} + \Delta \phi \;, \quad
g_{\mu\nu}  \equiv  \frac{e^{c \Delta \phi}}{ |V_0|} \bar{ g}_{\mu\nu} \;,\label{eq:metricscaling}
\end{equation}
to find that the action has transformed into 
\begin{equation} \label{eq:actionrescaled}
S_E = -\frac{e^{c\Delta \phi}}{|V_0|} \int \mathrm{d} ^4x \sqrt{ \bar{ g}} \left( \frac{ \bar{R}}{2} - \frac{1}{2} \bar{g} ^{\mu \nu} \partial _{\mu} \bar{ \phi} \partial _{\nu} \bar{ \phi} \mp e^{-c\bar\phi}\right) \;,
\end{equation}
where the $\mp$ sign in front of the potential corresponds to the inflationary resp. ekpyrotic case. To avoid cluttering due to absolute value signs, we will now continue the calculation for the inflationary case only, but the ekpyrotic case works analogously.  

We will set $V_0=1,$ so that the transformed theory retains the same potential, {\it i.e.} the transformation leaves us in the same theory. Then the relations \eqref{eq:metricscaling} imply that the field equations are invariant under
\begin{eqnarray}
\bar{a} ( \bar{ t}) & = & e^{c \Delta \phi / 2} \, a \left( e^{- c\, \Delta \phi /2} \bar{ t} \right) \;, \label{Rescaling1}\\
\bar{ \phi}( \bar{ t}) & = & \phi\left( e^{- c\, \Delta \phi /2} \bar{ t} \right) + \Delta \phi \;, \label{Rescaling2}
\end{eqnarray}
where overbars denote the transformed quantities. Under this transformation, the scaling solution \eqref{scal1}, \eqref{scal2} morphs into
\begin{align} \label{eq:rescalinga0}
\bar{ a} = \bar{a}_0 \, (\bar{ t}) ^{2/c^2} \;, \quad \bar{a}_0 = \textrm{exp} \left( \frac{ (c^2-2) \, \Delta \phi}{2c}  \right) \, a_0 \;,\quad
V( \bar{ \phi}) =  \frac{12 - 2c^2}{c^4} \frac{1}{ \bar{ t} ^2} \;,
\end{align}
which shows that $a_0$ is a constant of motion,
\begin{equation} \label{eq:labela0}
a_0 = a\,\left( \frac{c^4}{12-2c^2}V \right)^{1/c^2} \;.
\end{equation}
This means that we can label different solutions by their value of $a_0.$

As argued above, at late Lorentzian times, the attractor pulls the solution close to a real classical solution, and thus the imaginary part of the Euclidean action scales as (with $\mathrm{d}\tau = i \mathrm{d}t$) 
\begin{equation}
S_E^I \sim i \, \int \mathrm{d} t\, a ^3\, V \sim i \,a_0 ^3\, (t) ^{- 1 + \frac{6}{c^2}} \sim i \, a_0 ^3\, V ^{ \frac{1}{2} - \frac{3}{c^2}} \;.
\end{equation}
Using the constant of motion \eqref{eq:labela0}, one can thus determine the dependence on the final values $b,\, \chi$ to be
\begin{equation}
S_E^I \sim  i \, b ^3\, V( \chi) ^{1/2} \;.
\end{equation}
Meanwhile, the scaling of the real part of the Euclidean action is governed by the scaling/shift symmetry, and from \eqref{eq:actionrescaled} we obtain
\begin{equation}
\bar{S}_E ^{R} = e^{ c\, \Delta \phi} S_R = \left( \frac{ \bar{a}_0}{a_0} \right) ^{\frac{2c^2}{ c^2-2}} S_E ^{R} \;,
\end{equation}
so that
\begin{equation}
S_E ^{R} \sim a_0 ^{\frac{2c^2}{ c^2-2}} \sim b ^{\frac{2c^2}{ c^2-2}} V( \chi) ^{\frac{2}{ c^2-2}} \;.
\end{equation}

Now it becomes straightforward to work out the WKB condition \eqref{eq:WKBcond}, which says that the amplitude of the wave function should vary slowly compared to its phase. Recalling that $(\nabla I)^2=G^{AB}\partial_A I \partial_B I$ with $G_{bb}=-12\pi^2 b,\, G_{\chi\chi}=2\pi^2 b^3,$ we obtain
\begin{align}
    \frac{(\nabla S_E^R)^2}{(\nabla S_E^I)^2} \sim \frac{b ^{\frac{4c^2}{ c^2-2}-3} V ^{\frac{4}{ c^2-2}}}{b^3 V}\,\sim b^{c^2-6}\,.
\end{align}
Thus we see a confirmation that classicality is reached under the same conditions under which we had found a dynamical attractor, namely either for an expanding universe with $c^2<6,$ or for a contracting one with $c^2>6.$ Given that during inflation the scale factor expands exponentially fast, we may also infer that classicality is obtained exponentially quickly. 

What is interesting is that this is a purely dynamical way of obtaining classicalisation. The usual procedure in quantum mechanics is to invoke decoherence \cite{Joos:1984uk,Zurek:2003zz}, {\it i.e.} the loss of quantum coherence due to interactions of a system with its environment. Such a process is highly relevant on Earth, where interactions are extremely common. But it is not available at the creation of the universe, for two reasons: on the one hand, there is no environment as the universe contains everything by definition, and is still empty. And on the other hand decoherence is a process that happens over time, while here we must first address the classicalisation of space and time. In other words: once time has become classical, decoherence can take place. Here we have just seen how time (and space) can become classical in the first place, purely due to cosmological dynamics. In this way, the no-boundary proposal can explain the classicality of the early universe.

\subsection{Implementations in Minisuperspace} \label{sec:minisuper}

So far we looked only at saddle points, {\it i.e.} at (usually complex) solutions of the classical equations of motion satisfying the no-boundary conditions \eqref{regularity} and \eqref{finalcondition}. It is reassuring that such saddle points exist, but can we do better? What exactly are the saddle points approximating? Can we refine the heuristic definition \eqref{eq:HHpi} and define the no-boundary path integral more precisely? In doing so, there are a number of issues that must be faced:
\begin{itemize}
    \item Gravity is not renormalisable, which means that we expect an infinite number of correction terms involving ever higher powers of the Riemann tensor and ever higher numbers of derivatives. This might however still be fine, as long as the curvature remains well below the Planck scale. At least for the saddle points we looked at so far, this happened to be true: they all had curvatures on the order of the Hubble scale in the early universe, which for inflationary examples is constrained by observations to be $H/M_{Pl} \lesssim 10^{-5}$ \cite{Planck:2018jri}. Under such circumstances, we may expect the path integral to yield reliable semi-classical results (see e.g. \cite{Woodard:2014jba} and also section \ref{sec:robust}).
    \item The path integral involves the action, which is integrated over ranges of coordinates. Hence, even though the idea of the no-boundary proposal consists of the notion that there should be no boundary to the past, we must still integrate from somewhere, {\it i.e.} we must still impose some form of boundary conditions. The (somewhat Zen-flavoured) question then is: which boundary conditions correspond best to no-boundary conditions?
    \item The no-boundary condition that we imposed for saddle points cannot be imposed as a condition on the full path integral, as it is in conflict with the uncertainty principle. This is because it is a condition on both the field value $a(\tau=0)=0,$ ensuring compactness, and on the velocity/expansion rate $a'(\tau=0)=1,$ ensuring regularity. Thus we cannot define a path integral that sums over metrics that are both compact and regular. It has to be one or the other, or perhaps a condition on a linear combination of field value and momentum. Note that imposing compactness will not guarantee that the saddle points turn out to also be regular, and vice versa. We will have to check this at the end of the calculation. 
    \item In the action for gravity coupled to matter, the kinetic term for the scale factor of the universe enters with a different sign than all of the other kinetic terms, both those of anisotropic components of the metric and those of matter fields,
    \begin{align}
        S \sim  \int dt\left[ -\, 3a\dot{a}^2 + \frac{1}{2}a^3\dot\phi^2 + \cdots \right]
    \end{align}
    This is known as the \emph{conformal mode problem}, so-called because the scale factor can be seen as the conformal mode of spatial sections. If the path integral is defined as a Euclidean integral, then the integrand will thus be unbounded above and below, regardless of the overall choice of sign. Hence it is doubtful that the Euclidean path integral might make sense. A Lorentzian path integral seems more promising as it would not have this problem (by being a sum over phases), but there we have the issue that the integral is only conditionally convergent and it has to be defined carefully to make its meaning unambiguous.
    \item In situations where several no-boundary saddle points exist, is the definition of the path integral unique \cite{Halliwell:1989dy}? Does it uniquely specify which saddle points contribute to the wave function, and which do not?
    \item Finally, it should be expected that the general sum over 4-manifolds is very difficult to make precise. One has the freedom to sum over metrics that can differ at all spacetime points. Moreover, in analogy with the paths integrated over in quantum mechanics (see {\it e.g.} \cite{feynman2010quantum}), one might expect the required manifolds in general not to be differentiable anywhere. What is more, even the topology of $4-$manifolds remains ill understood. Hence it seems difficult at present to properly define a sum over $4-$manifolds. We will take a more pragmatic approach, and restrict to metrics that have certain symmetries, in particular cosmologically relevant symmetries. This is the framework of minisuperspace, where the metric is parameterised by a finite number of functions of time. One may object to this on the basis that we are neglecting infinitely many degrees of freedom, and what is worse, we are setting both these degrees of freedom and their conjugate momenta to zero simultaneously. Nevertheless, we know from observations that the early universe was highly symmetric. Hence, we should expect a realistic theory of initial conditions to predict high probability for precisely these minisuperspace kinds of metrics. What is crucial then is to check at the end of our calculations whether or not metrics that are perturbed around the minisuperspace representation come out as suppressed. If so, then we may have confidence in our calculations, and justifiably refer to minisuperspace as an ``approximation''.
\end{itemize}

\noindent{\it Dirichlet boundary condition}

In their original paper \cite{Hartle:1983ai}, Hartle and Hawking envisioned the no-boundary proposal as corresponding to a Euclidean path integral over compact metrics. This then implicitly provides no-boundary conditions: we will take them to mean that we keep the scale factor fixed at zero size on the initial hypersurface. We will try to see to what extent such an integral can be realised in a simple setting, namely gravity with a cosmological constant, with action
\begin{eqnarray}
S = \int_{\cal M} \mathrm{d} ^4 x  \sqrt{-g} \left( \frac{R}{2} - \Lambda \right)  \pm \int_{\partial {\cal M}_{0,1}} \mathrm{d}^3 y \sqrt{h}K \,. \label{LorentzianActionrepeat}
\end{eqnarray}
The GHY surface terms are essential as we intend to keep the field values fixed both on the initial ($\partial {\cal M}_{0}$) and on the final ($\partial {\cal M}_{1}$) hypersurface, {\it i.e.} we are imposing Dirichlet conditions at both ends. This is the setting first studied by Halliwell and Louko in \cite{Halliwell:1988ik}. For closed Robertson-Walker (RW) metrics, parameterised again with an especially useful definition of the time coordinate $t_q$,
\begin{eqnarray} \label{eq:Metricrepeat}
\mathrm{d}s ^2 =  - \frac{N^2}{q(t_q)} \mathrm{d}t_q^2 + q(t_q) \mathrm{d} \Omega _3 ^2 \,,
\end{eqnarray}
the action simplifies and becomes quadratic in the scale factor squared $q,$
\begin{equation}
S_q= 2 \pi ^2 \int_0^1 \mathrm{d}t_q \left( -\frac{3}{4 N}\dot{q}^2 + N(3 -  \Lambda q) \right) \,. \label{ActionHrepeat}
\end{equation} 
Details of this calculation were already presented in section \ref{sec:ex}, so we will not repeat all of them here. However, it may be useful to recall that we can choose the time coordinate to run over the values $0 \leq t_q \leq 1,$ with $t_q=0$ being the initial hypersurface on which we fix the scale factor to zero, $q(t_q=0)=0.$ On the final hypersurface we will fix $q(t_q=1)=q_1,$ with the scale factor being larger than the Hubble radius, $q_1 > \frac{3}{\Lambda}.$ The total time elapsed between initial and final hypersurfaces is determined by the integral $\int_0^1 \mathrm{d}t_q \frac{N}{\sqrt{q}},$ and will thus depend on the lapse $N.$ 

The no-boundary wave function (with Dirichlet conditions at both ends) is thus given by the path integral
\begin{align}
    \Psi_{DD}(q_1) = \int_0^{q_1} {\cal D}q \int_{\cal C} \mathrm{d}N \, e^{\frac{i}{\hbar}S_q}\,,
\end{align}
where the contour ${\cal C}$ for the lapse integral remains to be determined. With the above boundary conditions, the equation of motion $\ddot{q}  =  \frac{2\Lambda}{3}N^2$ is solved by
\begin{equation}
\label{eq:classicalsolutionnb}
\bar{q}(t_q)=\frac{\Lambda}{3}N^2 t_q^2 + \left(- \frac{\Lambda}{3}N^2+ q_1\right) t_q \,.
\end{equation}
As described in section \ref{sec:ex}, the path integral over $q$ can then be performed by shifting variables $q(t_q) = \bar{q}(t_q) + Q(t_q),$ with the result that 
\begin{equation}
\label{eq:propagatorDD}
\Psi_{DD}(q_1) = \sqrt{\frac{3\pi i}{2\hbar}}\int_{\cal C} \frac{\mathrm{d} N}{N^{1/2}} e^{\frac{i}{\hbar}2\pi^2  S_0}\,,
\end{equation}
with
\begin{equation}
S_0 = N^3 \, \frac{\Lambda^2}{36} + N \left( 3-\frac{\Lambda}{2}q_1  \right) -\frac{3}{4N}q_1^2\,. \label{eq:Nactionnb}
\end{equation}
Note that the lapse integral contains an essential singularity at $N=0.$ This can be understood physically from the impossibility of evolving from size zero to size $q_1 \neq 0$ in vanishing time.

\begin{figure}%
\begin{center}
\includegraphics[width=0.8\textwidth]{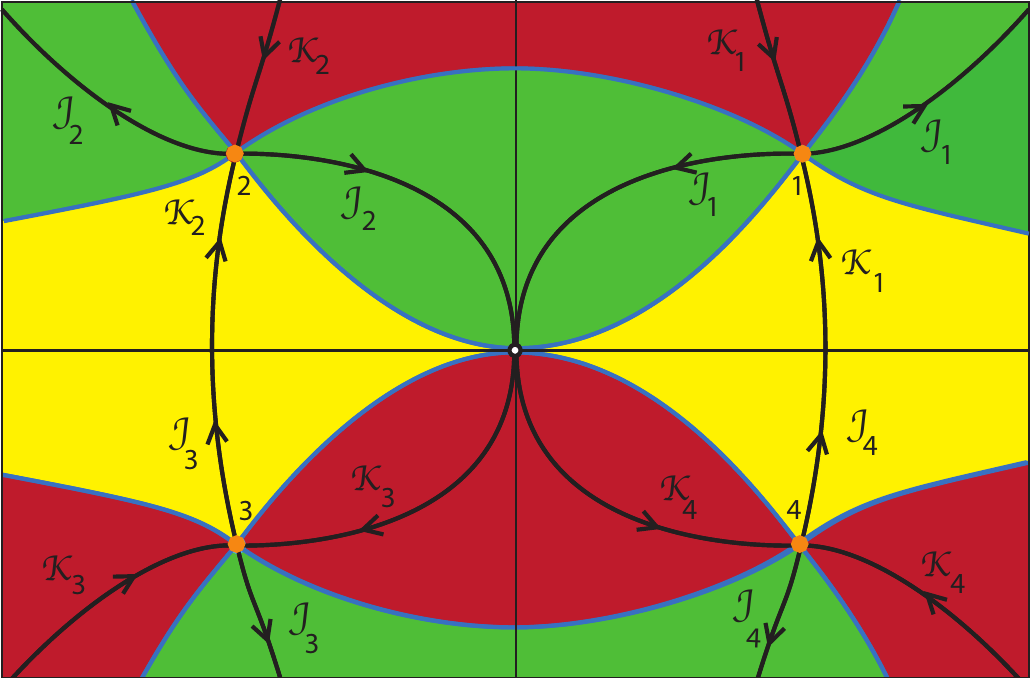} \\ \vspace{0.2cm}
\includegraphics[width=0.8\textwidth]{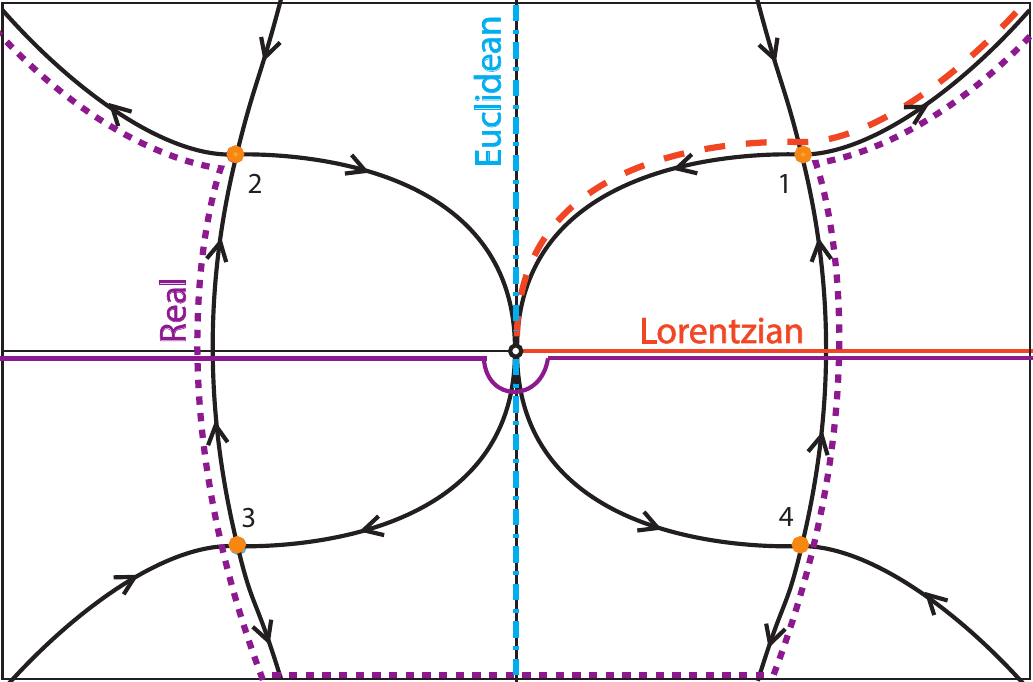}
\end{center}
\caption{\label{fig:DD} Saddle points and their associated steepest ascent (${\cal K}_i$) and descent (${\cal J}_i$) paths, in black, shown in the complexified plane of the lapse function $N.$ Arrows indicate the direction of descent. Blue lines are lines of equal weighting. In the upper panel, green regions have lower weighting than the saddles, red regions have higher weighting and yellow regions have a weighting that is in between those of the adjacent saddle points. Asymptotically, {\it i.e.} either at infinity or near $N=0,$ the integral converges in green regions and diverges in red regions. In the lower panel, we exhibit the integration contours described in the main text.}
\end{figure}

We will analyse this integral by performing a saddle point approximation, using the tools made available by Picard-Lefschetz theory and reviewed in \ref{sec:a3}. There are four saddle points, determined by $\partial S_0/\partial N = 0,$ located at
\begin{equation}
\label{saddlesnb}
N_\sigma = c_1 \frac{3}{\Lambda} \left[ i + c_2 \left( \frac{\Lambda}{3} q_1-1\right)^{1/2}\right]\,,
\end{equation}
with $c_1, c_2\in\{-1,1\}$. The action at the saddle points is given by
\begin{eqnarray}
S_0^{saddle} =  c_1 \frac{6}{\Lambda}  \left[ i - c_2 \left(\frac{\Lambda}{3} q_1 - 1\right)^{3/2} \right]\,. \label{sadact}
\end{eqnarray}
The first thing to note is that the saddle points, as well as the corresponding actions, are complex. This was to be expected, as they describe a combination of quantum nucleation and classical evolution. The presence of the term containing $i$ in \eqref{sadact} implies that two saddle points will have a suppressed weighting $e^{-12\pi^2/(\hbar\Lambda)}$ (those with $c_1=+1$) and two will have an enhanced weighting $e^{+12\pi^2/(\hbar\Lambda)}$ (those with $c_1=-1$). The four saddle points and their associated steepest descent (${\cal J}_i$) and ascent (${\cal K}_i$) lines in the plane of the lapse function are shown in Fig.~\ref{fig:DD}. The suppressed saddle points are located in the upper half plane, and the enhanced ones in the lower half plane. Here we must recall an important point mentioned at the end of section \ref{sec:inflex} and derived in detail in section \ref{sec:perts}: the saddle points with suppressed weighting are associated with enhanced fluctuations, and vice versa. More precisely, if we add a (linear) gravitational wave perturbation with amplitude $h_1$ and wave number (frequency) $k,$ then the total weighting becomes   $e^{-12\pi^2/(\hbar\Lambda) + 3\pi^2 h_1^2/(\hbar\Lambda)}.$ This means that larger fluctuations, {\it i.e.} more lumpy universes, come out as preferred. In other words, these saddle points are unstable when perturbations are included, and thus do not warrant the use of minisuperspace. By contrast, the saddle points in the lower half plane admit a Gaussian distribution of perturbations $e^{+12\pi^2/(\hbar\Lambda) - 3\pi^2 h_1^2/(\hbar\Lambda)},$ and are stable. It is these that we must hope the path integral will pick up.

This brings us to the crucial question of contours of integration for the lapse $N,$ see also Fig.~\ref{fig:DD}. Let us first discuss the Euclidean contour, {\it i.e.} a contour along imaginary values of the lapse function. This contour was originally proposed by Hartle and Hawking. However, as we can see immediately from Fig.~\ref{fig:DD}, there are regions of asymptotic divergence both at large positive imaginary lapse values, and at small negative imaginary lapse. This means that the Euclidean contour, whether defined in the upper or lower half planes, leads to a divergent, mathematically meaningless, integral. Thus, this simple example shows that a Euclidean contour is not actually viable. The root of the obstruction is the conformal mode problem discussed at the beginning of this section.  

If we cannot use a Euclidean contour, then could we use a Lorentzian one? After all, physics as we know it takes place in Lorentzian spacetime, and thus a Lorentzian contour would seem to be a physically sensible choice\footnote{Let us recall however that the saddle points we are interested in are fully complex, yet they describe the evolution of a Lorentzian universe, as described in section \ref{sec:class}. From this point of view, the Lorentzian contour is not {\it a priori} distinguished -- it is sufficient that the contour should pick up the saddle point(s) of interest.} \cite{Feldbrugge:2017mbc}. Given the singularity of the integrand at $N=0,$ we can either define it over positive or negative real values of the lapse, but these two choices are just a matter of convention. If we choose the positive real line, then we can see from Fig.~\ref{fig:DD} that it is crossed by a single steepest ascent contour, namely ${\cal K}_1.$ Thus the Lorentzian integration contour can be deformed to the thimble ${\cal J}_1$ passing through saddle point $1.$ The arc at infinity linking ${\cal J}_1$ to the real line gives a vanishing contribution to the integral, as ensured by Picard-Lefschetz theory and as one can verify explicitly \cite{Feldbrugge:2017kzv}. Thus the Lorentzian integral picks up a single saddle point, though unfortunately it is one of the unstable ones,
\begin{align}
    \Psi_{DD,Lorentzian} \approx e^{-\frac{12\pi^2}{\hbar \Lambda} - i \frac{12\pi^2}{\hbar\Lambda}\left(\frac{\Lambda}{3} q_1 - 1\right)^{3/2}}\,.
\end{align}
As argued above, the inclusion of perturbations enhances the weighting, and signals the breakdown of the minisuperspace ``approximation''. This signals an inconsistent calculation -- certainly, this choice of integration contour does not lead to a ground state wave function, as was intended. It is in fact rather surprising that what appears to be the most sensible contour on physical grounds ends up giving entirely non-physical answers.

To circumvent this problem, and based on early considerations in \cite{Halliwell:1989dy}, it was suggested in \cite{DiazDorronsoro:2017hti} that one should take the contour labeled ``real'' in Fig.~\ref{fig:DD}. This contour may be seen as an integral over the entire real lapse line, but passing below the singularity at $N=0.$ The contour is crossed by all four steepest ascent lines, and thus all four saddle points contribute to the resulting path integral,
\begin{align}
    \Psi_{DD,real} \approx e^{-\frac{12\pi^2}{\hbar \Lambda} - i \frac{12\pi^2}{\hbar\Lambda}\left(\frac{\Lambda}{3} q_1 - 1\right)^{3/2}} + e^{+\frac{12\pi^2}{\hbar \Lambda} - i \frac{12\pi^2}{\hbar\Lambda}\left(\frac{\Lambda}{3} q_1 - 1\right)^{3/2}} + \, \textrm{c.c.}\,, \label{DDreal}
\end{align}
where the sum over all saddles renders the wave function real and thus explains the name given to the contour. This contour was chosen explicitly so that the desired saddle points in the lower half plane are picked up. It should be emphasised that this contour, even though it superficially appears to sum mainly over real lapse values, actually obtains its largest contribution from the large-weighting region just below $N=0.$ This explains how the wave function can obtain the enhanced weighting manifest in \eqref{DDreal}, even though one must always flow down from the original integration contour to saddle points that are relevant ({\it cf.} appendix \ref{sec:a3}). The trouble with the real contour is that it also picks up the unstable saddle points in the upper half plane. Thus it leads to a competition of weightings as the fluctuations become large, of the form
\begin{align}
    e^{+\frac{12\pi^2}{\hbar\Lambda} - \frac{3\pi^2}{\hbar\Lambda} h_1^2} + e^{-\frac{12\pi^2}{\hbar\Lambda} + \frac{3\pi^2}{\hbar\Lambda} h_1^2}\,,
\end{align}
and it is simply unknown what happens to the integral when $h_1$ becomes large (for discussions of this issue see \cite{Feldbrugge:2017mbc,Janssen:2019sex}). At some point backreaction on the geometry can no longer be neglected, yet a full understanding would be required to see what happens for large perturbations. This would be necessary in order to assess whether this contour gives physically sensible results. In the absence of a reliable non-perturbative calculation, we cannot trust this contour to provide us with the definition of the sought-after no-boundary wave function.

At this point, one may realise that it is pointless to explore different contours in this setting. This is because the steepest descent lines ${\cal J}_{3,4}$ emanating from the desirable saddle points $3$ and $4$ directly lead to the undesirable saddle points $1$ and $2.$ This arises because the action \eqref{eq:Nactionnb} is a real function of $N$, and thus complex saddle points arise in complex conjugate pairs which lead to the same phase (and inverse weightings) in the wave function. Since steepest ascent/descent lines are defined by having constant phase, the stable and unstable saddle points are necessarily linked. Thus any contours picking up the stable saddle points will also include the unstable saddles, leading to the identical problem with large fluctuations discussed above \cite{Feldbrugge:2017mbc}. Thus in the end we must conclude that the sum over compact metrics, at least in this simple setting, does not lead to a trustworthy no-boundary (ground state) wave function.\\

\noindent{\it Neumann boundary condition}

If a sum over compact metrics is problematic, then how can we define the no-boundary wave function? Heuristically, a possible solution was suggested above \cite{DiTucci:2019dji}: the uncertainty principle says that we can either put a condition on the initial size, or on the initial expansion rate. But the initial expansion rate is Euclidean, and in fact requires a choice of sign, {\it cf.} \eqref{regularity}. This choice of sign is precisely what distinguishes the stable from the unstable saddle points that we just discussed. One may also think of this choice of sign as the choice of Wick rotation, {\it i.e.} do we define $t=+i\tau$ or $t=-i\tau$? Only one sign assignment leads to stable, Gaussian distributed fluctuations, and this is the choice $a'(0)=+1.$ So can we define a path integral with this boundary condition?

A related issue is that when we defined the path integral with fixed initial size, for consistency we had to include the GHY surface term in \eqref{LorentzianActionrepeat}. But the spirit of the no-boundary proposal is that there should be no boundary. Hence why should one include a boundary term? And where should it be placed, if the intention is that there should be no boundary? From this point of view it seems much more natural not to include a boundary term. If we do this, it changes the variational problem, and in fact leads to a Neumann problem, as shown in section \ref{sec:ex}, allowing us precisely to fix the initial momentum.

\begin{figure} 
	\centering
	\begin{minipage}[t]{0.45\textwidth}
\vspace{-5.2cm}
\includegraphics[width=0.75\textwidth]{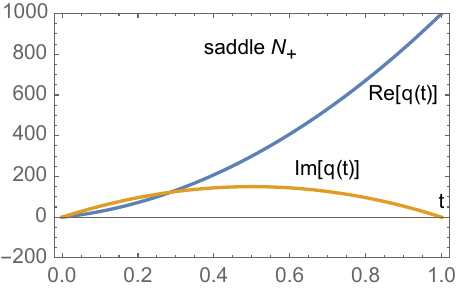}\\
\hspace{1cm}
\includegraphics[width=0.75\textwidth]{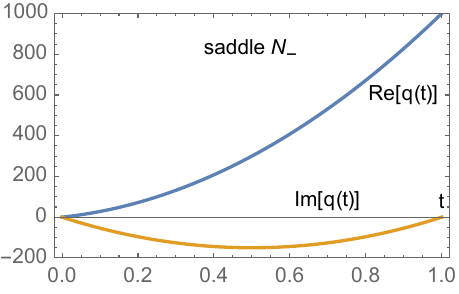} 
\end{minipage}
\quad
	\begin{minipage}[t]{0.5\textwidth}
\includegraphics[width=\textwidth]{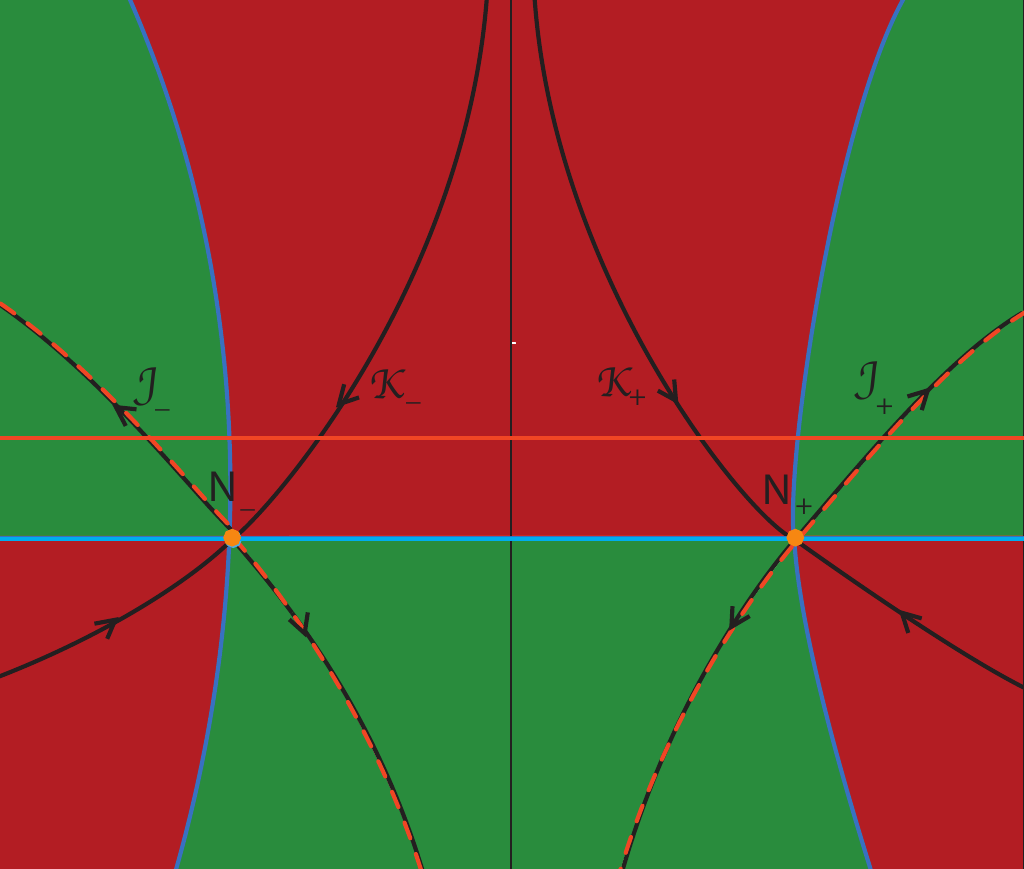}
\end{minipage}
	\caption{{\it Left panel:} an example of the evolution of the scale factor squared for the saddle points of the no-boundary wave function defined with an initial Neumann/momentum condition, with $\Lambda=3/100, \, q_1=1000,$ so that the saddle points are at $N_\pm = \pm 300 -100i$. One can see that the saddle points are compact, as they start at zero size. {\it Right panel:} The saddle points and their associated steepest ascent/descent lines in the complex plane of the lapse. The Lorentzian integration contour may be deformed into a sum of the 2 thimbles ${\cal J}_{\pm}.$}
	\protect
	\label{fig:nbnd}
\end{figure}

We will now see how this works for the simple setting of gravity plus a cosmological constant, with closed RW metrics considered above \cite{DiTucci:2019bui}. We will \emph{not} add any surface term on the initial hypersurface, yet we will again include a GHY term on the final boundary, as we wish to keep the metric fixed there,
\begin{eqnarray}
S = \int_{\cal M} \mathrm{d} ^4 x  \sqrt{-g} \left( \frac{R}{2} - \Lambda \right)  + \int_{\partial {\cal M}_{1}} \mathrm{d}^3 y \sqrt{h}K \,. \label{LorentzianActionND}
\end{eqnarray}
We will specialise to the metric \eqref{eq:Metricrepeat}. Using integration by parts in the action, the surface term at $t_q=1$ is eliminated by the GHY term, while a surface term is then generated at $t_q=0,$
\begin{align}
     S=2\pi^2\int_0^1 \mathrm{d}t_q \left[-\frac{3}{4N}\dot{q}^2 + N(3-\Lambda q)\right] - \frac{3\pi^2}{N}q\dot{q}|_{t_q=0}\,.
\end{align}
As shown in section \ref{sec:ex}, variation of this action leads to the equation of motion $\ddot{q}  =  \frac{2\Lambda}{3}N^2$ and the boundary condition that we can fix $\dot{q}/N$ at $t_q=0.$ Recall that $N\,\mathrm{d}t_q/q^{1/2}=\mathrm{d}t,$ so that consequently $\dot{q}/N = 2 \mathrm{d}a/\mathrm{d}t$ implying that we should fix
\begin{align}
    \frac{\dot{q}}{2N} = +i\,. \label{regularityq}
\end{align}
This is the no-boundary regularity condition \eqref{regularity} expressed in our currently used variables. On the final hypersurface we will again set $q(t_q=1)=q_1>3/\Lambda.$ With these boundary conditions, the solution to the equation of motion reads
\begin{align}
    \bar{q}(t_q) = \frac{\Lambda}{3}N^2 t_q^2 +2Ni t_q + q_1 - \frac{\Lambda}{3}N^2 -2Ni\,. \label{eq:qbarND2}
\end{align}
Evaluating the path integral over $q$ in the same manner as before, we then obtain the following expression for the Neumann-Dirichlet wave function,
\begin{align}
    \Psi_{ND}(q_1) = \int_{\cal C} \mathrm{d}N \, e^{\frac{i}{\hbar}2\pi^2\left[ \frac{\Lambda^2}{9}N^3 +i\Lambda N^2 -q_1 \Lambda N -3iq_1 \right]}\,, \label{eq:wfND2}
\end{align}
where once again the contour of integration ${\cal C}$ for the lapse integral remains to be specified. But before discussing contours, we may already notice some important differences with the Dirichlet action \eqref{eq:Nactionnb}. There is no singularity at $N=0,$ since it is now possible (even if unlikely) that the initial geometry, satisfying $\dot{q}=2Ni,$ already coincides with the final geometry, satisfying $q=q_1.$  Also, the action now contains explicit factors of $i,$ due to the boundary condition \eqref{regularityq}. Thus we do not expect saddle points to come in complex conjugate pairs anymore. In fact, there are only two saddle points this time, and they are located at
\begin{align}
    N_\pm = \frac{3}{\Lambda} \left[-i \pm \frac{3}{\Lambda}\sqrt{\frac{\Lambda}{3}q_1 - 1}\right]\,. \label{saddlesneumann}
\end{align}
The corresponding field evolutions are shown for an example in Fig.~\ref{fig:nbnd}. Note that the saddle points are not only regular, which they are by design, but also compact: at the saddle points, one may see and verify that $q(t_q=0)=0.$ Also note that the fields are only becoming real right at the end of the time evolution, as this now does not correspond to a Euclidean-plus-Lorentzian contour in the complex time plane, but rather to a solution with fixed (complex) lapse. Comparing to \eqref{saddlesnb}, we see that in fact only the two saddle points in the lower half plane are left. That is to say, the unstable saddle points have been eliminated, and we are left purely with the stable saddle points!

We still have to figure out which integration contour to take. For this, see Fig.~\ref{fig:nbnd}, in which the saddle points and their steepest descent paths are shown. We can now proceed again with an examination of suitable contours. We may notice right away that the Euclidean contour once again does not work. Since there is no singularity at $N=0,$ it would have to be defined over the entire imaginary lapse line in order to yield an invariant result. However, we can see that there is again a divergence at large positive imaginary values, and again the Euclidean definition does not make sense.

By contrast, the Lorentzian contour works \cite{DiTucci:2019bui}. The integral converges both at negative and positive infinity and since there is no singularity at $N=0,$ we must integrate over the entire real $N$ line in order to obtain an invariant definition. Note that the Lorentzian contour is crossed by both steepest ascent paths, and hence both saddle points contribute to the integral. In fact, the real lapse line can be deformed into the sum of both thimbles ${\cal C} = {\cal J}_+ + {\cal J}_-$, with orientations chosen in the direction of increasing real parts of the lapse. The wave function then becomes approximated by
\begin{align}
    \Psi_{ND}(q_1) \approx e^{+\frac{12\pi^2}{\hbar \Lambda} - i \frac{12\pi^2}{\hbar\Lambda}\left(\frac{\Lambda}{3} q_1 - 1\right)^{3/2}} + \, e^{+\frac{12\pi^2}{\hbar \Lambda} + i \frac{12\pi^2}{\hbar\Lambda}\left(\frac{\Lambda}{3} q_1 - 1\right)^{3/2}}\,, \label{wfND}
\end{align}
Note that it is real, as it is the sum of two complex conjugate contributions, even though the integral is not defined over a Euclidean contour. This is due to the symmetry between negative and positive real parts of the lapse. One further remark: Picard-Lefschetz theory implies that relevant saddle points must have a lower weighting than that of the defining contour. One may thus wonder how it is possible to obtain an enhanced weighting from a Lorentzian integral. This is because even though the starting action \eqref{LorentzianActionND} is real, the boundary condition \eqref{regularityq} is complex, and this results in a positive weighting in \eqref{eq:wfND2}, even at real values of the lapse.

In fact, there are only a few other contours we could contemplate. One possibility is to sum over the two thimbles, but with opposite orientations ${\cal C} = {\cal J}_+ - {\cal J}_-$, {\it i.e.} to sum from negative imaginary infinity up to negative real infinity, plus an integral from negative imaginary infinity to positive real infinity. This choice would give a pure imaginary wave function -- however, since we are ignoring the prefactor, which could be imaginary too, this must be seen as equivalent to a real wave function. At the semi-classical level, the implications are however largely unaffected by this choice of orientation of the thimbles. This is because the two saddle points behave effectively independently, as soon as perturbations and the resulting decoherence is taken into account. Indeed, as shown in \cite{Halliwell:1989vw}, perturbations quickly decohere the two saddles as the universe grows, already separating their evolutions when the universe is only a little larger than the Hubble radius.

The only remaining contours we could envision are to single out one of the thimbles ${\cal J}_\pm$, by summing from negative imaginary infinity either to positive or negative real infinity. In this case, we would have no need to talk about decoherence, as the integral would be approximated by a single saddle point. However, the wave function would not be real. Still, one attraction of such contours is that they would constitute a kind of compromise between a Euclidean and a Lorentzian one.

\begin{figure}[ht]%
\begin{center}
\includegraphics[width=0.7\textwidth]{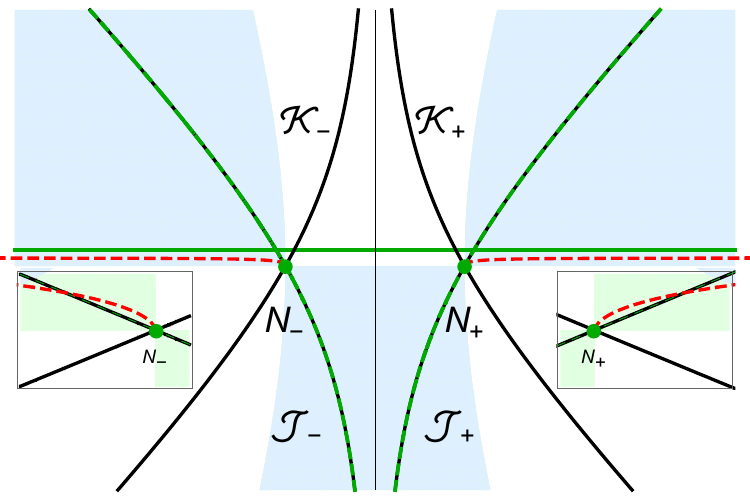} 
\end{center}
\caption{\label{fig:NDzeros} For the no-boundary wave function with a Neumann initial condition, {\it cf.} Fig.~\ref{fig:nbnd}, this graph shows the location of singular geometries in which the scale factor passes through zero (indicated by the red dashed lines) in the plane of the lapse function. The insets show a zoom of the regions near the saddle points, and demonstrate that the thimbles intersect the lines of singular geometries. Figure reproduced from \cite{DiTucci:2019bui}.}
\end{figure}

Let us now focus our attention a little more on the thimbles. Since the path integral is redefined by integrals over thimbles, we may wonder what kind of geometries are actually summed over. In general, there is no particular distinguishing feature to these complex geometries. However, a few special locations in the lapse plane may be singled out. For instance, at large negative imaginary values of the lapse, the geometries may straightforwardly be seen to be essentially very large Euclidean $4-$spheres with their North Pole cap removed at radius squared $q_1$. The asymptotic regions at large $|\textrm{Re}(N)|$ are similar, but the radius of the sphere is complex there. We may also wonder whether singular geometries are included. For this, we can ask whether $\bar{q}(t_q)$ passes through zero at some $t_q$ (with $0 \leq t_q \leq 1$). The locus of such geometries is shown in Fig.~\ref{fig:NDzeros}. As seen there, and calculated in \cite{DiTucci:2019bui}, the thimbles actually pass through this locus, {\it i.e.} the thimbles also contain a singular geometry. It is not clear that this should be considered pathological. In general, one may expect gravitational path integrals to contain geometries that are not differentiable anywhere, so this may not be problematic. Also, such singular geometries most likely obtain a divergent action once perturbations are included, and may thus eliminate themselves automatically by virtue of having zero weighting. These considerations also raise the question of whether it is problematic when $q$ changes sign, signalling a signature reversal in the metric. This is a question of ongoing research, and we will return to this issue in section \ref{sec:allowability}. The quick answer is that this is not known yet.

Having found a minisuperspace path integral implementation of the no-boundary wave function, we may also study its relation to the WdW equation \cite{Lehners:2021jmv}. For this purpose, recall from \eqref{cancon} and \eqref{Ham} that the Hamiltonian is given by
\begin{align}
H = -\frac{N}{6 \pi^2} \left[ p^2 + 12 \pi^4 (3 -\Lambda q)\right]=N\hat{H}\,,
\end{align}
with the canonical momentum $p=-\frac{3 \pi^2}{N}\dot{q}.$ The WdW equation then corresponds to the operator version of this equation,
\begin{align}
\hat{H} \Psi = 0\,,
\end{align}
with $\Psi$ being the wave function of the universe. Now we have to pay attention to the boundary conditions we imposed, namely Dirichlet on the final hypersurface and Neumann on the initial one. The canonical commutation relation $[q,p]=i$ must be implemented correspondingly. On the final hypersurface, where we work in field space, so we replace the momentum by a derivative operator $p \mapsto \hat{p} = - i  \frac{\partial}{\partial q}$, leading to  
\begin{align}
\hat{H}_{(q)} \Psi = 0 \rightarrow  \,\, & \frac{\partial^2 \Psi}{\partial q^2} + 12 \pi^4 (\Lambda q - 3)\Psi =0\,. \label{WdWq}
\end{align}
By contrast, on the initial hypersurface we impose a momentum condition, so as to obtain the WdW equation in momentum space we substitute $q \mapsto \hat{q} =  i  \frac{\partial}{\partial p}$, leading to 
\begin{align}
\hat{H}_{(p)} \Psi = 0 \rightarrow  \,\, & (p^2 + 36 \pi^4) \Psi + 12 \pi^4 \Lambda i \frac{\partial \Psi}{\partial p} =0\,. \label{WdWp}
\end{align}
There is one subtlety here: because we are imposing this equation on the initial boundary, we had to flip the sign $\frac{\partial}{\partial p}\to - \frac{\partial}{\partial p}$ \cite{Lehners:2021jmv}.

\begin{figure}[ht]
	\centering
	\includegraphics[width=0.6\textwidth]{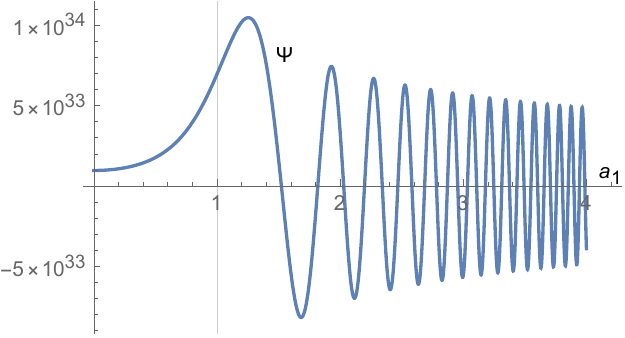}
	\caption{The no-boundary wave function as an Airy function, see \eqref{WdWAirysol}. Here the wave function $\Psi$, which is real, is plotted as a function of the final radius $a_1=\sqrt{q_1}.$ The cosmological constant is set to $\Lambda=1.$ Figure reproduced from \cite{Lehners:2021jmv}.}
	\label{fig:Airy}
\end{figure}

The momentum space equation \eqref{WdWp} is of first order and yields an essentially unique solution, the exponential of a cubic in $p_0$. Meanwhile, the position space equation \eqref{WdWq} can be identified as an Airy equation, with two linearly independent solutions, the $Ai$ and $Bi$ functions. Choosing a particular linear combination is directly related to the choice of contour in the path integral approach \cite{DiTucci:2020weq}, for instance the Lorentzian contour yields the $Ai$ function and the contour summing both thimbles from negative imaginary infinity yields the $Bi$ function. Explicitly, if we stick to the Lorentzian integration contour, then the equivalent, exact, solution to the WdW equation is
\begin{align}
\Psi(p_0,q_1) = e^{\frac{3}{\hbar\Lambda}ip_0+\frac{1}{36\pi^4\hbar \Lambda}ip_0^3}\, Ai\left[ \left(\frac{18\pi^2}{\hbar\Lambda}\right)^{2/3}\left(1-\frac{\Lambda}{3}q_1\right)\right]\,. \label{WdWAirysol}
\end{align}
with initial momentum given by \eqref{regularityq}, that is to say
\begin{align}
p_0 =  -   \frac{3 \pi^2}{N}  \dot{q}(0) = -6\pi^2 i\,. \label{momcond}
\end{align} 
The wave function is shown in Fig.~\ref{fig:Airy} as a function of the final size $q_1=a_1^2.$ As one can see there, the wave function rises exponentially (from a non-zero value) at $q_1=0$ and then starts to oscillate once the universe has become larger than the Hubble radius. These two regimes correspond to quantum tunnelling from nothing, and subsequent classical evolution, respectively. Note that it remains unclear how to interpret the fact that the wave function is non-zero at zero size. In \cite{Hartle:1983ai} it was suggested that this could be due to the contribution of non-trivial topologies, but the present calculation did not contain additional topologies. This question thus deserves further study.

For the particular value of the initial momentum \eqref{momcond}, which was chosen to ensure regularity, we find from \eqref{WdWp} that at the no-boundary point, the wave function satisfies the additional relation \cite{Lehners:2021jmv}
\begin{align}
i \frac{\partial}{\partial p}\Psi = \hat{q} \, \Psi = 0 \qquad \qquad \textrm{no-boundary condition}\,.  \label{newnbc}
\end{align}
This is particularly suggestive: it says that the no-boundary wave function satisfies the momentum space equivalent of the zero size condition, that is to say the regularity condition we imposed on the wave function turns out to be equivalent to the operator expression for the zero size condition! This certainly conforms well with the spirit of the no-boundary proposal. It also means that at the nucleation of the universe, there is no momentum transfer into the universe. In other words, this condition also expresses the notion that the universe is self-contained.

A final comment: {\it a priori}, one might think that field space and momentum space definitions might be equivalent, as they can be Fourier transformed into each other. However, the Fourier transform would sum over all possible initial momenta, and would thus also include momenta that correspond to unstable Wick rotations. This argument also implies that the results obtained so far are mutually consistent. It might however be interesting to see if a viable field space wave function could be obtained from a partial Fourier transform of the momentum space wave function, where only initial momenta of the appropriate sign are included. This does not seem to have been explored so far.\\

\noindent {\it A model with anisotropies -- biaxial Bianchi IX spacetimes}

From the simplest dynamical model above, we learned that a path integral from zero size gives us saddle points that come in two kinds, with stable (Gaussian distributed) perturbations, and those with unstable (inversely Gaussian) perturbations. This is a general feature, and it makes the definition of the no-boundary wave function as a path integral from zero size questionable. However, we also saw that a definition in which we impose a regularity condition instead works rather well. The regularity condition is a condition on the momentum conjugate to the spatial metric, and is obtained when we do not add any surface terms to the action. It is instructive to see how this prescription may be implemented in more complicated models, with anisotropic metrics and, later, with a scalar field added.

In fact, only a handful of minisuperspace models are known, which are tractable in the sense that all integrals except for that over the lapse can be done easily (and in some cases fully analytically) \cite{Halliwell:1989vu,Halliwell:1990tu,Garay:1990re,DiazDorronsoro:2018wro,Feldbrugge:2018gin,Janssen:2019sex,Jonas:2021ucu,Fanaras:2021awm}. Here we will focus on two representative examples, the first with biaxial Bianchi IX metrics, and the second with an inflationary scalar field. The techniques used to analyse these models are analogous to the techniques exhibited above, so we will be much briefer, and leave some details to the original references.

First we will stick to the action consisting of gravity with a positive cosmological constant $\Lambda,$ but now consider metrics of the form
\begin{align} \label{metricbb9}
\mathrm{d}s^2 = - \frac{N^2}{q}\mathrm{d}t^2 + \frac{p}{4} (\sigma_1^2+\sigma_2^2) + \frac{q}{4} \sigma_3^2\,,
\end{align}
where $p(t),q(t)$ are time dependent scale factors and $\sigma_1 = \sin\psi \mathrm{d}\theta - \cos \psi \sin \theta \mathrm{d}\varphi$, $\sigma_2 = \cos \psi \mathrm{d}\theta + \sin \psi \sin \theta \mathrm{d} \varphi$, and $\sigma_3 = \mathrm{d}\psi + \cos\theta \mathrm{d}\varphi$ are differential forms on the three sphere with coordinate ranges $0 \leq \psi \leq 4 \pi$, $0 \leq \theta \leq \pi$, and $0 \leq \varphi \leq 2 \pi.$ This metric describes Bianchi IX spacetimes on the axes of symmetry (in the notation of Misner \cite{Misner:1969hg} this would correspond to pure $\beta_+$ perturbations with $\beta_-=0$). This is also known as the Taub spacetime. Locally it can be seen as a product of a $2-$sphere with radius squared~$p$ and a circle with radius squared~$q.$ Globally, the metric describes a fibration of $S^1$ over $S^2,$ and when $p=q$ we recover the round $3-$sphere. When $p\neq q,$ we may think of it as a squashed, anisotropic $3-$sphere (see \cite{Daughton:1998aa}).

We have to be slightly more specific with regards to the action that we are considering. We would like to sum over manifolds with the metric held fixed on the final boundary, hence we must add the GHY surface term there. But on the initial ``no-boundary'' surface, we will not add any surface term. Thus the action reads
\begin{align}
S & = \frac{1}{2}\int \mathrm{d}^4x \sqrt{-g} \left(R - 2 \Lambda\right) \, + \, \left(p\Pi_p + q \Pi_q\right)\mid_{t=1}\,,
\label{act1}
\end{align}
where we expressed the GHY term in terms of the scale factors and their conjugate momenta
\begin{align}
    \frac{1}{2\pi^2} \Pi_p = -\frac{1}{2N}\left(\dot{q} + \frac{q}{p}\dot{p} \right)\,, \quad \frac{1}{2\pi^2}\Pi_q = -\frac{1}{2N} \dot{p}\,.
\end{align}
When the action is varied (see e.g. \cite{DiazDorronsoro:2018wro,Feldbrugge:2018gin,Janssen:2019sex}), one obtains the following boundary terms at $t=0$
\begin{align}
    p \delta \Pi_p=0\,, \quad q \delta \Pi_q =0\,.
\end{align}
These must be set to zero in order to obtain a consistent variational problem. Note that we cannot set both a variable and its conjugate momentum to zero simultaneously, as this would be in conflict with the commutation relations. But we see that we also cannot fix both momenta. Thus the geometrical Neumann condition that we imposed translates into a combination of Dirichlet and Neumann conditions on the variables used. There are two appropriate combinations of interest \cite{Janssen:2019sex}: we can either choose  $p_0=0, \Pi_q(t=0) = -2\pi^2 i$ (NUT, $S^2$ shrinks to zero) or $q_0=0, \Pi_p(t=0) = -2\pi^2 i$ (Bolt, $S^1$ shrinks to zero), where in both cases the signs of $\Pi_{q,p}(t=0)$ are again fixed so as to correspond to the usual, stable sign for the implied Wick rotation.

\begin{figure}[ht]%
\begin{center}
\includegraphics[width=0.5\textwidth]{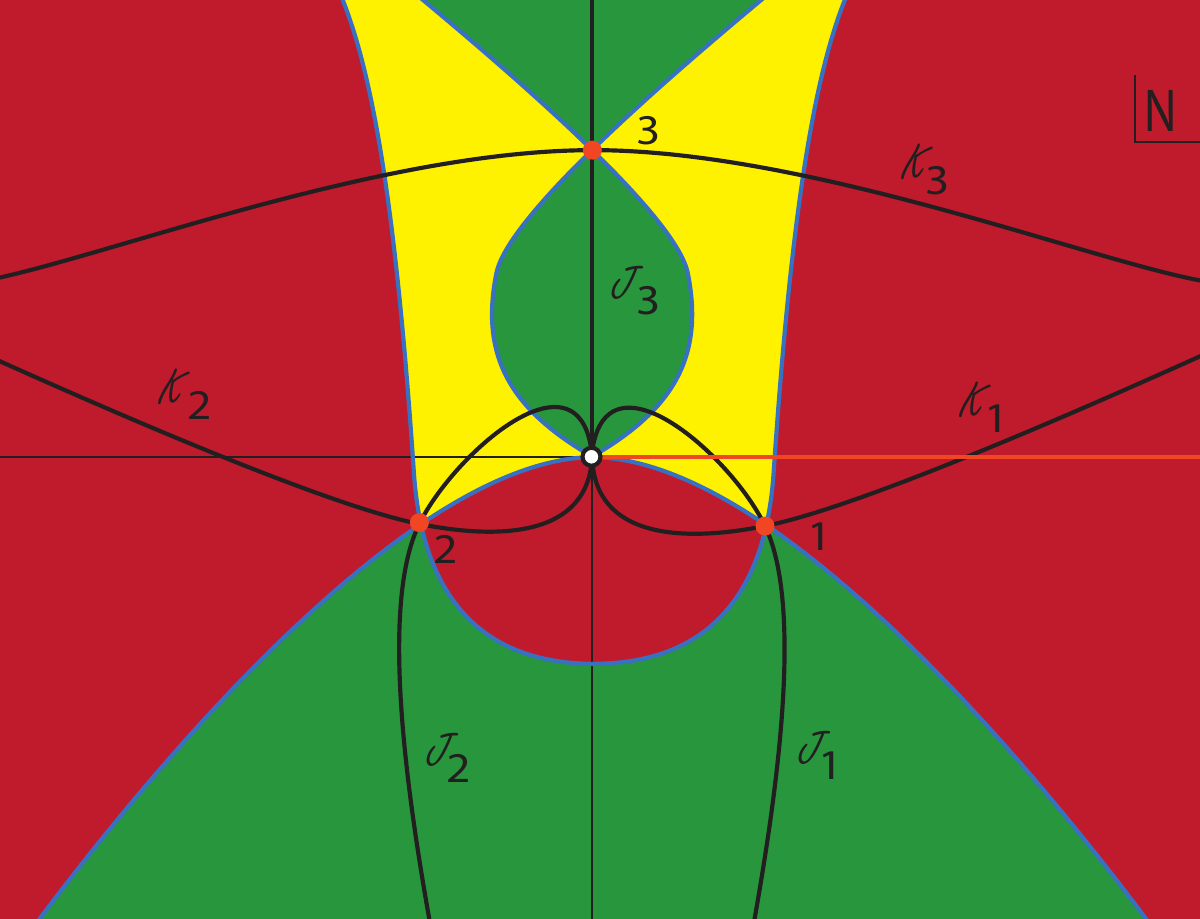} 
\end{center}
\caption{\label{fig:bb9} Saddle points (in orange), steepest ascent (${\cal K}$) and descent (${\cal J}$) contours for Taub-NUT metrics, with lapse action \eqref{actionNUT}. The saddles numbered $1$ and $2$ are physically relevant. Figure reproduced from \cite{Feldbrugge:2018gin}.}
\end{figure}

The first choice leads to a Taub-NUT-dS spacetime, in which the $2-$sphere is shrunk to zero size initially. When the equations of motion for $p$ and $q$ are solved with these boundary conditions, one is left with an action that depends purely on the lapse
\begin{align} 
\frac{1}{2\pi^2}S_{NUT}(N) = -\frac{p_1 q_1}{N} + i q_1  + N \left(4 - \frac{\Lambda}{3}p_1\right)  - i  \frac{\Lambda}{3} N^2 \,. \label{actionNUT}
\end{align} 
This action admits three saddle points, see Fig.~\ref{fig:bb9}. One of these is purely Euclidean, and does not lead to a classical spacetime. The other two saddle points are of physical interest (they yield a complex action, with a real part that grows rapidly with increasing $p_1,q_1$), and are simply related by a reflection in the real part of $N.$ They are picked up by the path integral if one chooses a contour of integration for the lapse that contains their thimbles, {\it i.e.} we must sum over ${\cal J}_1 \pm {\cal J}_2.$ Note that a Lorentzian integration contour, as we had in the isotropic case, is not possible this time as the integral simply diverges along such a contour. If one chooses the contour ${\cal J}_1 + {\cal J}_2,$ then note that it can be deformed into a closed, circular contour around the singularity at $N=0.$ Whether a closed contour makes sense is debatable \cite{Feldbrugge:2018gin}: if we imagine performing the lapse integral first, before doing the integrals over $p$ and $q,$ then there is no singularity at $N=0$ and the contour can be shrunk to zero, leading to a vanishing result. This argument indicates that there can be no fundamental meaning to a closed contour, and thus it might be preferable to sum from negative Euclidean infinity to the origin along both thimbles. However, in that case we also rely on the fact that there is a singularity at $N=0$ on which the thimbles end. Hence we see from this example that one cannot claim at present that it is understood how to define integration contours from first principles. We will take a pragmatic approach here, and assume that the saddles $1$ and $2$ are picked up, with the expectation that a better justification for the required integration contour will be found in future work.

\begin{figure}[ht]%
\begin{center}
\includegraphics[width=0.31\textwidth]{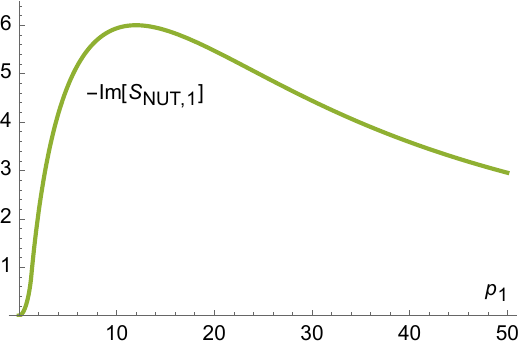} 
\includegraphics[width=0.31\textwidth]{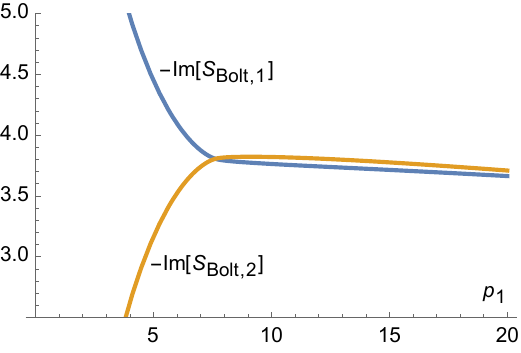}
\includegraphics[width=0.31\textwidth]{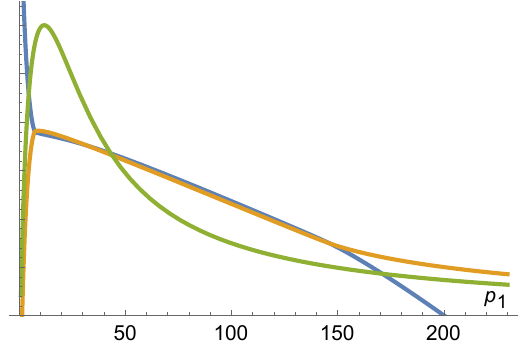}
\end{center}
\caption{\label{fig:bb9weight} The weightings of different saddle points of interest for Taub metrics. The NUT case is shown in the left panel, the Bolt case in the middle, and a superposition of these two cases (with the same colouring) in the right panel. In all cases, we set $\Lambda=1$ and $q_1=12.$ For a detailed description, see the main text.}
\end{figure}

When these saddle points are relevant, then we may look at their weighting, which is the same for both -- an example is shown in the left panel of Fig.~\ref{fig:bb9weight}, with $q_1$ fixed and as a function of $p_1$. Here we can see that the isotropic case $p_1=q_1$ comes out as favoured, while more anisotropic configurations are suppressed. This supports the notion that the no-boundary wave function describes a state of minimum excitation. It also confirms the choice of sign in the initial momentum $\Pi_{q}(t=0).$ A comment regarding normalisability: the weighting approaches zero for $p_1 \to \infty.$ Thus at this level of approximation, we cannot yet say whether an integral over all $p_1$ values (or perhaps an integral over all configurations with fixed volume) would yield a convergent result, as this will depend on a prefactor. But the prefactor depends on the exact measure in the path integral, and this measure is not known (the related ambiguity in the WdW equation is the question of factor ordering). This is an issue that deserves further work - for more discussion of this point see  \cite{Janssen:2019sex}.

\begin{figure}[ht]%
\begin{center}
\includegraphics[width=0.38\textwidth]{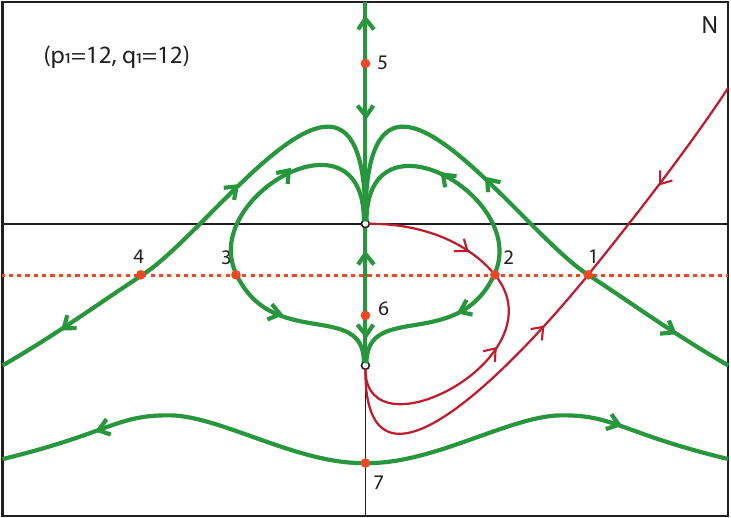} 
\includegraphics[width=0.19\textwidth]{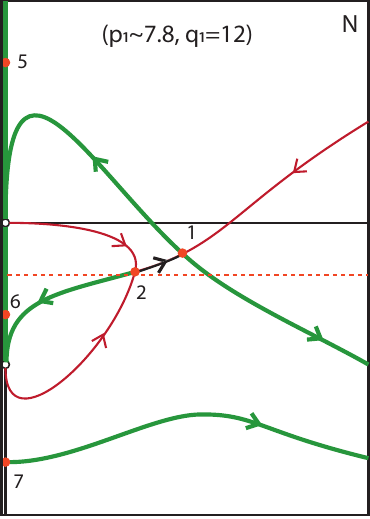} 
\includegraphics[width=0.38\textwidth]{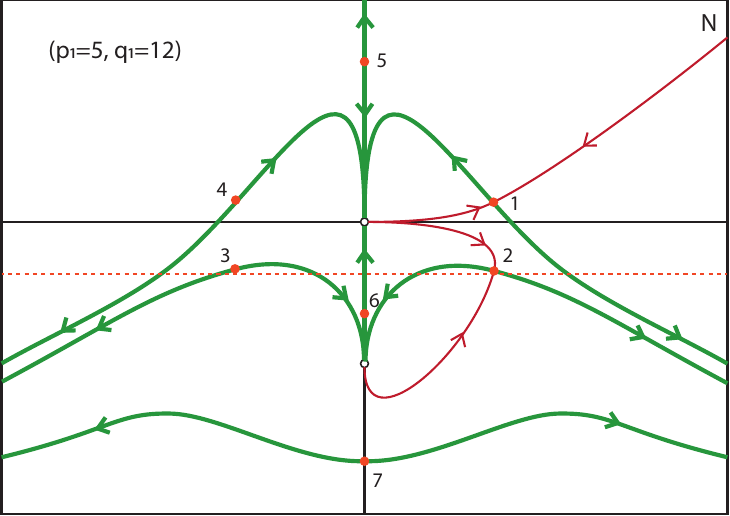} 
\end{center}
\caption{\label{fig:bb9Bolt} Saddle points and associated thimbles (steepest descent contours, in green) for Bolt boundary conditions, for the action \eqref{actionBolt}. Also shown are the steepest ascent contours associated with saddles $1$ and $2$ (in red). Arrows indicate downwards flow. The left panel shows an isotropic example ($p_1=q_1=12$), and the right panel one with much smaller $p_1=5.$ The middle panel shows the cross-over between these two regimes. This is described by a Stokes phenomenon, where the downwards flows from saddles $2$ and $3$ link up with the upwards flow from saddles $1$ and $4$ (this is shown by the black line). The dashed orange line is located parallel to the real $N$ line in the lower half plane, and represents the preferred integration contour. }
\end{figure}

But as we saw above, a second set of boundary conditions is also possible, $q_0=0, \Pi_p(t=0) = -2\pi^2 i$, for which the initial circle is shrunk to zero. This gives rise to so-called Taub-Bolt-dS spacetimes. One can again solve the equations of motion for $p$ and $q$ with these boundary conditions \cite{Janssen:2019sex}, and perform the integrations over the scale factors. This time the resulting lapse action is more complicated, and reads (for $\Lambda=1$)
\begin{align}
    \frac{1}{2\pi^2}S_{Bolt}(N) = & \,\, \frac{p_1 \left(N^4-18 N^2 q_1-36 i N q_1+9 q_1^2\right)}{12 N^2 (N+3 i)} \nonumber \\ &  \, - \frac{N^3+3 i N^2+(p_1-12) p_1 N}{3 p_1}\,. \label{actionBolt}
\end{align}
This action admits $7$ saddle points. Examples of the associated thimbles are shown in Fig.~\ref{fig:bb9Bolt}. Three saddles are Euclidean and not of physical interest. The other four are however of potential physical relevance. They arise again in pairs, with equal weighting. The imaginary parts of the actions of saddles $1$ and $2$ are shown in the middle panel in Fig.~\ref{fig:bb9weight}. We can see that there is a cross-over in likelihoods, with the transition between dominance of the two saddles taking place when $p_1$ is a little smaller than $q_1.$ Also, the weighting diverges at small $p_1,$ for saddle number~$1.$ Thus, the predictions depend rather crucially on which saddles are picked up in the path integral.

The thimbles are shown in Fig.~\ref{fig:bb9Bolt}, both for $q_1$ fixed and various values of $p_1.$  Let us first start with the case where the scale factors are equal, see the left panel in the figure. The obvious contour of interest is the one running parallel to the real $N$ line, in between the two singularities at $N=0$ and $N=-3i$ (in fact, for convergence, the contour must run below $N=-i$). This contour can be deformed into a sum of the four thimbles associated with saddles $1$ to $4,$ and thus all four saddle points are picked up. The ones with the highest weighting will dominate the path integral, in this case saddles $2$ and $3.$

Now comes a crucial point: if it remains true that all $4$ complex saddles contribute to the path integral, then at small $p_1$ (for fixed $q_1$) saddles $1$ and $4$ will dominate, and in fact their weighting grows unboundedly at small $p_1,$ {\it cf.} again Fig.~\ref{fig:bb9weight}. This would render the wave function non-normalisable and we would have to conclude that somehow the no-boundary wave function for biaxial Bianchi IX spacetimes does not exist, {\it cf.} also the discussion in section~\ref{sec:reconstruct}. However, a topological change in the steepest ascent/descent paths occurs as $p_1$ shrinks below a critical value (in our numerical example, for $p_1\approx 7.8$), which is near the point where the dominance of the two saddles switches. This is known as a Stokes phenomenon\footnote{The analysis of this Stokes phenomenon represents original work by the author.}. At the critical $p_1$ value, the steepest descent contour from saddle $2$ coincides with the steepest ascent contour passing through saddle $1,$ {\it i.e.} the actions at both saddles have the same real part (and similarly for saddles $3$ and $4$), though the weightings of saddles $2, 3$ are still higher than those of saddles $1, 4$. Below this critical value, the thimbles of saddles $2$ and $3$ run directly to infinity, and the integration contour for the lapse that we considered before can now simply be rewritten as the sum of the thimbles associated with the saddles $2$ and $3.$ In other words, saddles $1$ and $4$ do not contribute anymore (their steepest ascent paths do not cross the integration contour, see the right panel in Fig.~\ref{fig:bb9Bolt}) after this Stokes phenomenon has occurred. For this to be true, we must however stick to the same defining contour of integration we had for larger $p_1.$ This shows that, despite the misgivings expressed in the analysis of the NUT case, we can see clearly that the integration contour is of paramount importance in figuring out the predictions of gravitational path integrals. For a more in-depth analysis of Stokes phenomena and integration contours in this context see \cite{Lehners:2024kus}.

We can draw two lessons from this analysis: the first is that in general we must expect to sum over ``no-boundary'' boundary conditions, as with the NUT and Bolt cases above. In fact, when both cases are summed, the weighting of the various saddle points is shown in the right panel of Fig.~\ref{fig:bb9weight}. As one can see there, for small $p_1$ and up until $p_1$ comfortably exceeds $q_1,$ the NUT geometry dominates (since, for small $p_1,$ the saddle represented by the blue line is not picked up). Then for larger $p_1$ saddles $1$ and $4$ of the Bolt geometry come to dominate, and at very large $p_1$ it is saddles $2$ and $3$ of the Bolt geometry that give the largest contribution to the path integral. Thus, even in a fairly simple model, interesting phase transitions may occur \cite{Janssen:2019sex}. And the second lesson is that, even though there is no understanding yet of how to define gravitational integration contours from first principles, they are of clear relevance in elucidating the consequences of these integrals. \\

\noindent {\it A model with a scalar field}

The realisation that one must sum over boundary conditions is made even more manifest when considering a model that includes a scalar field. A tractable model is obtained if one chooses \cite{Garay:1990re}
\begin{align}
V(\phi) = \alpha \cosh \sqrt{\frac{2}{3}}  \phi \,. \label{Pot}
\end{align}
This model is of inflationary type, with a de Sitter minimum at $\phi=0.$ The model is not realistic in the sense that the potential is too steep over most of its range to lead to viable primordial perturbations, and it does not include a reheating phase. However, it is useful as a toy model for quantum cosmology. This is because for closed Robertson-Walker metrics of the form
\begin{align}
\mathrm{d}s^2 = -\frac{N^2}{a(t)^2}\mathrm{d}t^2 +a(t)^2 \mathrm{d}\Omega_3^2,
\end{align}
one can perform a redefinition of the fields \cite{Garay:1990re},
\begin{align}
x(t) \equiv a^2(t) \cosh \left( \sqrt{\frac{2}{3}} \phi(t) \right)\,, \quad y(t) \equiv a^2(t) \sinh \left( \sqrt{\frac{2}{3}}  \phi(t) \right)\,, \label{redef2}
\end{align}
which renders the action quadratic
\begin{align}
S = 2\pi^2 \int_0^1 \mathrm{d}t N\left[ \frac{3}{4 N^2}  \left(\dot{y}^2 - \dot{x}^2 \right) +3 - \alpha x  \right] \,\, +  \frac{3\pi^2}{N}(y\dot{y}-x\dot{x})\mid_{t=0}\,. \label{eq:action}
\end{align}
Here, in line with prior discussions, we did not include a GHY surface term at $t=0,$ but did include one at $t=1.$ Integrations by parts then actually remove the boundary term at $t=1$ and generate one at $t=0,$ resulting in the action written above. The original fields can be recovered from the inverse transformations
\begin{align}
a(t)=\left[x^2(t)-y^2(t) \right]^{1/4}\,, \qquad \phi(t) = \sqrt{\frac{3}{2}} \textrm{artanh} \left( \frac{y(t)}{x(t)}\right)\label{invtrfm}\,.
\end{align}
The momenta conjugate to the new variables $x, y$ are given by
\begin{align}
\Pi_x=  -\frac{3\pi^2}{N} \dot{x}\,,  \quad \Pi_y= \frac{3\pi^2}{N} \dot{y}\,.
\end{align}
Our boundary conditions are such that we will impose conditions on these momenta at $t=0,$ while we will fix the final values $x(t=1)=x_1,\, y(t=1)=y_1$ on the final hypersurface. To determine which conditions should be imposed on the initial momenta, we should consider the constraint that will be satisfied by the saddle points. For the action \eqref{eq:action}, it is given by
\begin{align}
\frac{1}{12\pi^4}\left(\Pi_x^2-\Pi_y^2\right) +3 = a^2 V(\phi)\,. \label{constraintwithscalar}
\end{align}
We would like to obtain saddle points that are not just regular, but also closed, {\it i.e.} we would like the saddle points to satisfy $a(t=0)=0.$ Thus we should impose the following regularity condition \cite{Jonas:2021ucu}
\begin{align}
\Pi_x^2-\Pi_y^2 = -36\pi^4 \quad \textrm{at } t=0 \qquad \textrm{(regularity)} \label{regcon}
\end{align}
with in addition $\textrm{Im}(\Pi_x)<0$ for stability.

\begin{figure}[ht]%
\begin{center}
		\includegraphics[width=0.39\textwidth]{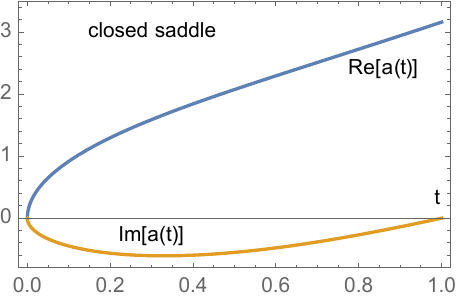}\qquad \qquad
		\includegraphics[width=0.39\textwidth]{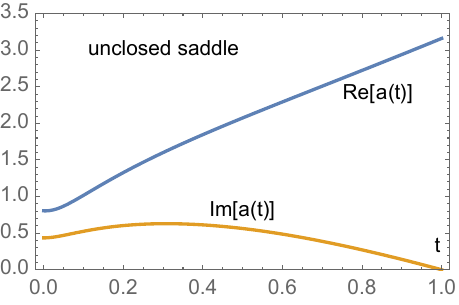}
		\includegraphics[width=0.39\textwidth]{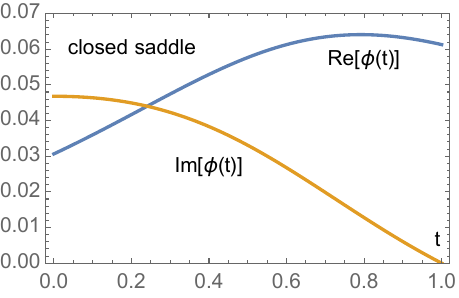}\qquad \qquad
		\includegraphics[width=0.39\textwidth]{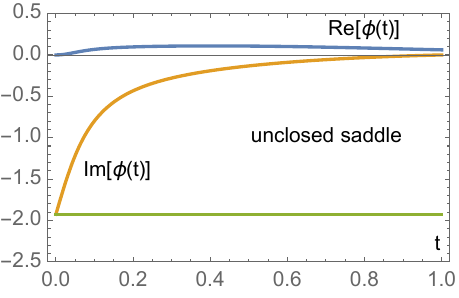} 
\end{center}
\caption{\label{fig:closedunclosed} Examples of closed and unclosed saddle points for fixed initial momenta, in a scalar field model with an inflationary potential. The numerical values used here are $\Pi_x \approx 0.0566 - 59.2i,\, \Pi_y \approx -2.26 + 1.48i,\, x_1=10,\, y_1=0.5,\, \alpha=1$ implying $a_1 \approx 3.16,\, \phi_1\approx 0.0613$. Note that the unclosed saddle has $\phi_{SP}=-\sqrt{\frac{3}{2}}\frac{\pi}{2}i,$ a value indicated by the green line.}
\end{figure}

Perhaps surprisingly, what is found is that two types of saddle points exist. The first type is the expected compact and regular no-boundary geometry, see the left graphs in Fig.~\ref{fig:closedunclosed} for an example. In this case the scale factor starts at zero size and reaches the desired final value, while the scalar field starts at a complex value and also reaches the desired final, real value. This saddle point is completely analogous to the numerical examples we discussed in section \ref{sec:inflex}. However, we should note that it only exists for precisely tuned values of the initial momenta $\Pi_{x,y}(t=0).$ But it is found that, for the same initial values of the momenta, there can be saddles of a second type, which are in fact not closed at $t=0,$ {\it i.e.} they start with a non-zero, complex scale factor, see the right graphs in the figure for an example. This is made possible not by $a$ becoming zero at $t=0,$ but rather by the potential $V(\phi_{SP})$ becoming zero, {\it cf.} the constraint equation \eqref{constraintwithscalar} \cite{Jonas:2021ucu}. That such saddles can exist is implied by Picard's little theorem, which roughly speaking implies that a non-constant entire function assumes all possible values, hence somewhere (or in fact at multiple locations) the potential vanishes when analytically continued. For our potential \eqref{Pot} this occurs when $\sqrt{\frac{2}{3}}  \phi_{SP} = (2n+1) \frac{\pi}{2}i$ with $n \in \mathbb{Z}.$

It was found in \cite{Jonas:2021ucu} that the weighting of the unclosed saddle is subdominant to that of the closed, no-boundary, saddle. Hence the unclosed saddle is unimportant in this example -- however, it is not known whether this occurs for other potentials too. 

This brings us to a conceptually important point. The South Pole values of $\Pi_{x,y}$ must be tuned in order for a saddle point to exist, which is compact at $t=0$ and reaches the desired final values $x_1, y_1.$ This is analogous to the tuning of $\phi_{SP}$ performed in section \eqref{sec:inflex}. But how does the late universe know which values the initial momenta had to have at nucleation? Simply fixing this value by hand is highly non-local, and does not seem plausible. It would be much more natural, and in the spirit of the Feynman sum over histories, to sum over all possible initial values of $\Pi_{x,y},$ subject to the regularity condition \eqref{regcon}. In some sense, locality implies that we really must sum over initial conditions. Such a sum would then automatically include the closed and regular no-boundary saddle. But, as we saw above, it would also include other saddle points, that are regular but not closed. There is no guarantee that these will all be subdominant. Hence it is not at all clear that the wave function, so defined, will be of no-boundary type. 

At this point, let us look ahead somewhat to see how this issue might be resolved. There are two arguments that suggest that the unclosed saddles ultimately do not contribute. The first is that when one performs the same calculation, but with a negative potential, {\it i.e.} with $\alpha<0$ in \eqref{Pot} (where it is arguably better defined in light of AdS/CFT \cite{Maldacena:1997re}), then the unclosed saddles are found to be singular in the sense that the scale factor passes through zero \cite{Jonas:2021ucu}. This means that when perturbations are added, the action will become infinite and such ``saddles'' remove themselves from the path integral. Hence, if the wave function can be analytically continued in $\alpha,$ then one might also expect the unclosed saddles to be spurious when $\alpha>0.$ A second argument is that the scalar field has to become highly complex, in the sense that the imaginary part must become very large, to reach $V(\phi)=0.$ This might simply not be allowed at a fundamental level, as such large imaginary parts can lead to divergences in the path integral for the full theory (when including other matter contributions) \cite{Witten:2021nzp,Lehners:2022xds}. Thus it seems possible, if not plausible, that a proper definition of the full path integral will remove such unclosed saddle points. We will return to this issue in section \ref{sec:allowability}.

\subsection{No-Boundary Saddles with Anisotropies and Black Holes} \label{sec:blackholes}

In the previous section, we discussed definitions of the path integral in the context of minisuperspace models. This was made possible by the simplifications that arose from using restricted classes of metrics, and special parameterisations of the fields. For more general models, such simplifications are not known to occur, and the best we can do is analyse potential saddle points of the path integral, {\it i.e.} try to figure out the properties of compact and regular (typically complex) solutions of the equations of motion and constraints. This is however a good starting point in elucidating the consequences of the no-boundary wave function, and in many situations it seems plausible that the saddle points thus found are also in fact the dominant ones. \\

\noindent {\it Anisotropies -- full Bianchi IX}

Our universe is highly isotropic on the largest scales, but less and less so as we probe smaller scales. It is thus of interest to study anisotropic models, as they provide a more realistic description of the universe. The Bianchi IX metric stands out for several reasons \cite{Misner:1969hg}. Locally, it provides a generic description of spacetime geometry. This is used in the proof that in the approach to a spacelike (big bang-like) singularity the metric (locally) becomes ever closer to Bianchi IX form \cite{Belinsky:1970ew}. Hence it is of interest to see what the no-boundary proposal implies for such metrics, and in particular whether the no-boundary wave function favours more or less isotropic universes. This topic was explored in a number of papers over the years, see in particular \cite{Hawking:1984wn,Wright:1984wm,Amsterdamski:1985qu,Duncan:1988zq,delCampo:1989hy,Fujio:2009my,Bramberger:2017rbv}.

For our analysis, we will again consider a gravity-scalar model with an exponential potential $V(\phi) = V_0 e^{c \phi},$ with $c<\sqrt{2}$ in order to obtain inflationary dynamics. The Bianchi IX metric can be written as \cite{Misner:1969hg}
\begin{align}
\mathrm{d}s_{IX}^2 = - N^2 \mathrm{d}t^2 + \frac{a(t)^2}{4} \left( e^{\beta_+(t) + \sqrt{3}\beta_-(t)} \sigma_1^2 +  e^{\beta_+(t) - \sqrt{3}\beta_-(t)} \sigma_2^2 +  e^{-2 \beta_+(t)} \sigma_3^2\right)\,,
\end{align}
where the 1-forms $\sigma_i$ were defined below \eqref{metricbb9}. The scale factor $a$ thus determines the spatial volume, while the $\beta_\pm$ parameterise shape change. When $\beta_\pm = 0$ one recovers the isotropic case, while $\beta_-=0$ is the biaxial case discussed in section \ref{sec:minisuper}. The Lorentzian action then reduces to
\begin{align}
S = 2\pi^2 \int \mathrm{d}t N a & \left[  \frac{1}{N^2}\left( -3\dot{a}^2 +a^2 \left(\frac{1}{2}\dot{\phi}^2 + \frac{3}{4}\dot{\beta}^2_+ + \frac{3}{4}\dot{\beta}^2_- \right) \right) \right. \nonumber \\ & \left. - \left( a^2  V(\phi) + U(\beta_+, \beta_-)\right)\right]\,,
\end{align}
where a potential arises for the anisotropy parameters,
\begin{align} \label{anisotropypotential}
U(\beta_+, \beta_-)  = & - 2 \left( e^{ 2 \beta_+ } + e^{-\beta_+ - \sqrt{3}\beta_-} + e^{-\beta_+ + \sqrt{3}\beta_-} \right) \nonumber \\ & + \left( e^{ -4 \beta_+ } + e^{2\beta_+ - 2\sqrt{3}\beta_-} + e^{2\beta_+ + 2\sqrt{3}\beta_-} \right)\,.
\end{align}
This potential has a minimum at $U(0,0)=-3$ and at large $\beta_\pm$ exhibits a triangular symmetry in $\beta_\pm-$space.

\begin{figure}[ht] 
\begin{center}
\includegraphics[width=0.75\textwidth]{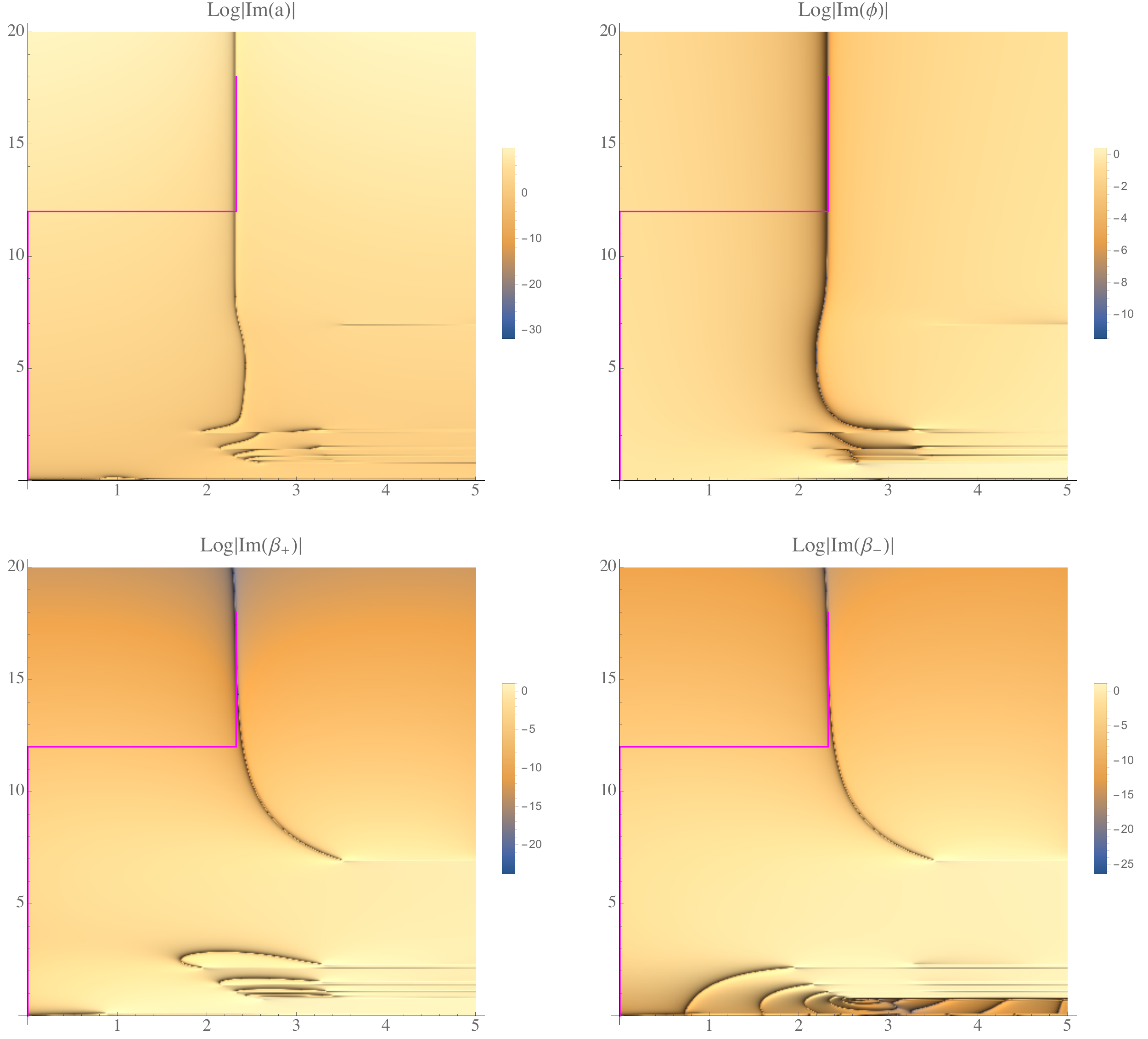}
\caption{An anisotropic instanton, optimised to reach the real values $b = 10000, \chi= -2, b_+=1, b_-=1$ on the final boundary, for $V_0=1, c=1/3$. These values are reached at $\tau_f = 2.33 + 18.0\,i$, with the South Pole values $\phi_{SP}=0.942 - 0.554\,i, \beta_{SP+}'' = -0.926 + 0.173\, i, \beta_{SP-}'' = -0.00373 + 0.000697\, i.$ Note the presence of branch points and associated cuts in the lower right parts of the figures. The style of the figures coincides with that of Figs.~\ref{fig:one} and \ref{fig:two}. Figures reproduced from \cite{Bramberger:2017rbv}.}
\label{fig:afterdip}
\end{center}
\end{figure}

Varying with respect to the lapse $N$ yields the Friedman equation
\begin{align} \label{Friedman}
3\dot{a}^2 =a^2 \left(\frac{1}{2}\dot{\phi}^2 + \frac{3}{4}\dot{\beta}^2_+ + \frac{3}{4}\dot{\beta}^2_- \right)  +  N^2 \left( a^2 V(\phi) + U(\beta_+, \beta_-)\right)\,,
\end{align}
while the equations of motion for $\beta_\pm$ and $\phi$ are given (in constant $N$ gauge) by 
\begin{align}
& \ddot{\beta}_\pm + 3\frac{\dot{a}}{a}\dot{\beta}_\pm  + \frac{2}{3}\frac{N^2}{a^2} U_{,\beta_\pm} = 0\,,  \\
& \ddot{\phi} + 3\frac{\dot{a}}{a}\dot{\phi} + N^2 V_{,\phi} = 0 \,.
\end{align}

No-boundary solutions must be compact and regular. We may again solve the equations of motion and constraint perturbatively around the South Pole, imposing compactness and regularity. The result, in terms of Euclidean time ($N=i$), is \cite{Bramberger:2017rbv}
\begin{align}
a &= \tau -\frac{1}{18} V_0 e^{c\phi_{SP}} \tau^3 \nonumber \\ & \quad +\frac{1}{8640}((-216 ( \beta_{SP+}'' )^2-216 ( \beta_{SP-}'' )^2+(8-27 c^2) V_0^2 e^{2 c \phi_{SP}}) \tau^5 +\cdots \label{seriesa} \\
\phi &= \phi_{SP}+\frac{c}{8}  V_0 e^{c \phi_{SP}} \tau^2+\frac{c (2+3 c^2)}{576}  V_0^2 e^{2 c \phi_{SP}} \tau^4+\cdots \\
\beta_+ &= \frac{1}{2} \beta_{SP+}'' \tau^2 + \frac{1}{144} (45 ( \beta_{SP-}'' )^2+ \beta_{SP+}'' (-45\beta_{SP-}'' +7 V_0 e^{c \phi_{SP}})) \tau^4+\cdots \\
\beta_- &= \frac{1}{2}  \beta_{SP-}'' \tau^2 + \frac{1}{144}  \beta_{SP-}'' (90  \beta_{SP+}''+7 V_0 e^{c \phi_{SP}}) \tau^4 +\cdots\,, \label{seriesbn}
\end{align}
where we used a prime to denote derivatives w.r.t. Euclidean time $\tau=it.$ These expansions are needed to form a well-defined numerical problem, {\it cf.} the discussion in section \ref{sec:inflex}. This time there are three free parameters, which characterise solutions. They are 
\begin{equation} \label{SPvalues}
\phi_{SP}, \quad \beta_{SP+}'', \quad \beta_{SP-}''\,,
\end{equation}
which can all three assume complex values. Note that the regularity condition forces the anisotropy parameters, as well as their first derivatives, to vanish at the South Pole. Thus all instantons are created isotropically, but subsequently anisotropies can grow, as the second derivative can be non-zero.

\begin{figure}[ht] 
\begin{center}
\includegraphics[width=0.65\textwidth]{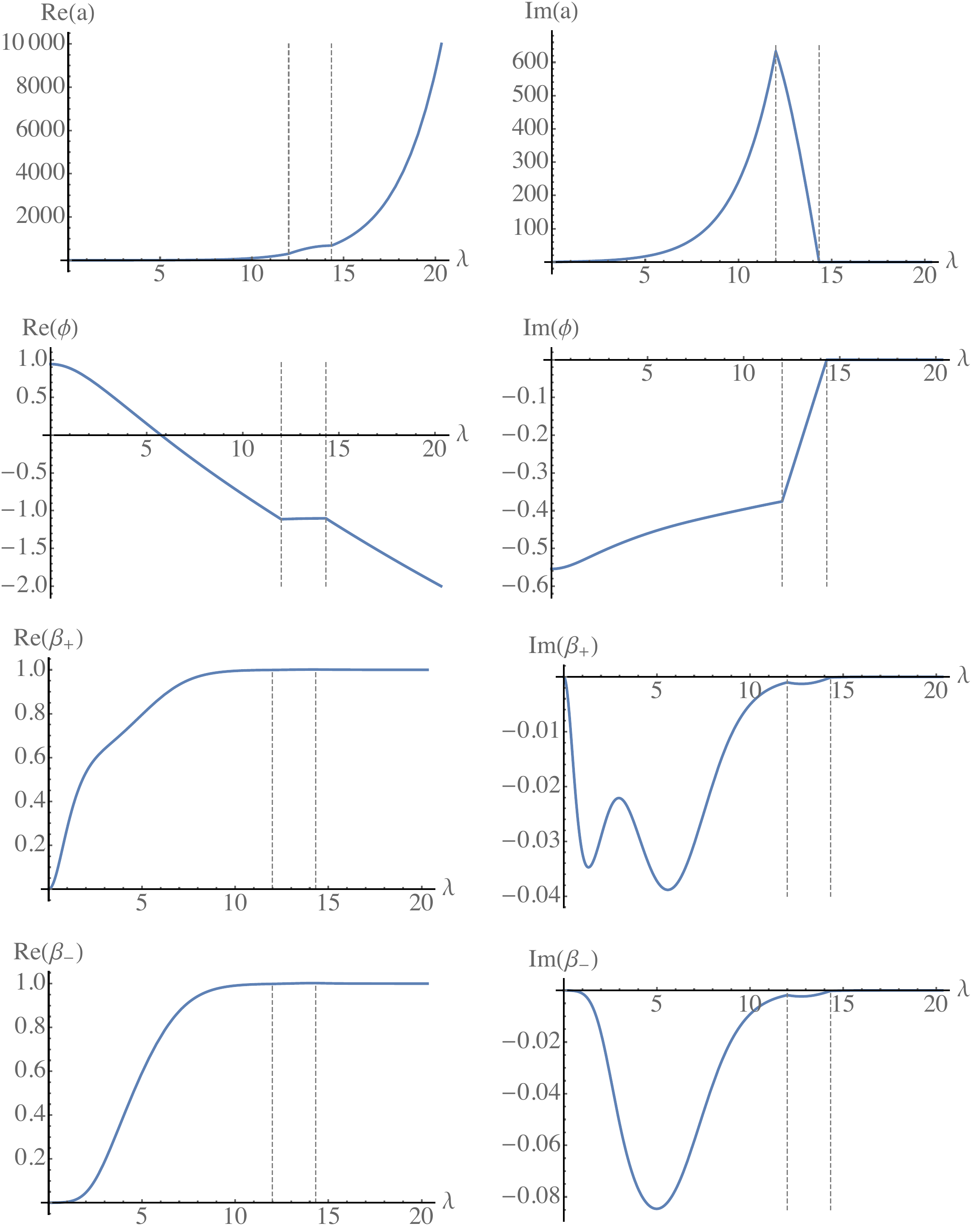}
\caption{The evolution of the fields $a, \phi, \beta_\pm$ along the magenta contour drawn in Fig.~\ref{fig:afterdip}. The contour is parameterised by $\lambda,$ and the dashed lines indicate the changes of direction of the contour. All fields become real on the final hypersurface. Note that the anisotropies start out at zero and then grow to reach the desired final values. Figures reproduced from \cite{Bramberger:2017rbv}.}
\label{fig:fieldevolution}
\end{center}
\end{figure}

\begin{figure}[ht] 
\begin{minipage}{0.5\textwidth}
		\includegraphics[width=0.7\textwidth]{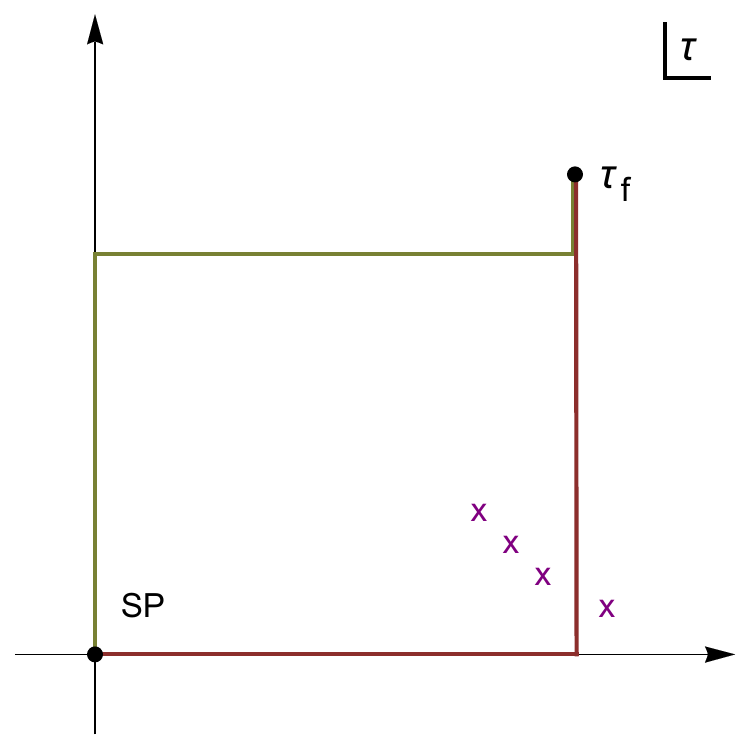}
	\end{minipage}%
	\begin{minipage}{0.5\textwidth}
		\includegraphics[width=0.9\textwidth]{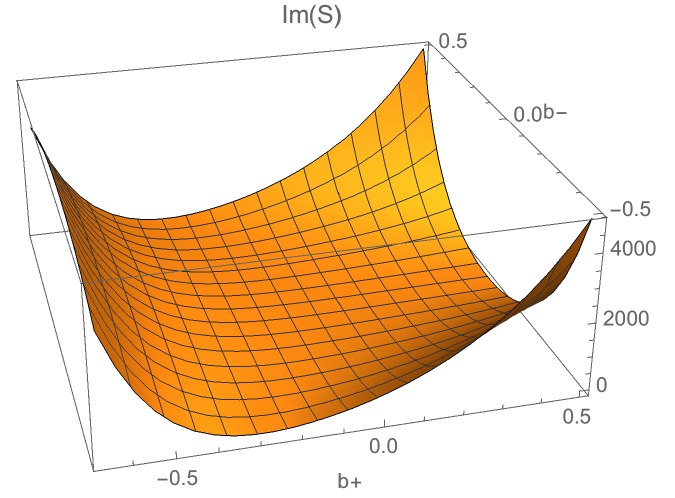}
	\end{minipage}%
	\caption{{\it Left panel:} The presence of singularities (marked by purple crosses) prohibits the use of a standard ``Hawking'' contour (in red). Rather, one must use a contour that stays to the left of the singularities, such as the contour marked in green. {\it Right panel:} The imaginary part of the Lorentzian action, as a function of final anisotropy values $b_\pm,$ for $b=100,\, \chi=-1/2.$ Since the weighting scales as $\textrm{Exp}[-\textrm{Im}(S)/\hbar],$ isotropic solutions have higher weighting.  Figures reproduced from \cite{Bramberger:2017rbv}.}
	\label{fig:action}
\end{figure}

A numerical example of an anisotorpic no-boundary instanton is presented in Figs.~\ref{fig:afterdip} and \ref{fig:fieldevolution}. Obtaining such solutions requires again an optimisation procedure to tune the South Pole values of the scalar field and of the anisotropy parameters. This can be done using a higher-dimensional Newtonian algorithm. Compared to the numerical solutions discussed in section \ref{sec:inflex}, a new feature arises here, namely that singularities appear in the complex time plane, see Fig.~\ref{fig:afterdip}. These imply that one cannot use a Euclidean-plus-Lorentzian contour to reach the final hypersurface. If one tried, one would in fact not reach coincident real field values at all, but one would end up on the wrong sheet of the solution (sticking to such a contour would put a limit on how large anisotropies can be \cite{Fujio:2009my}). It proves useful to use a contour that is Lorentzian first, then Euclidean, and followed by a further Lorentzian segment. The field evolutions along such a contour are shown in Fig.~\ref{fig:fieldevolution}. A sketch of this situation is presented in the left panel of Fig.~\ref{fig:action}. When using this alternative contour, arbitrarily large anisotropies can be obtained.

One can also calculate the action of these no-boundary solutions numerically. The right panel in Fig.~\ref{fig:action} shows the imaginary part of the action as a function of the final anisotropy parameters, for fixed values of the final scale factor $b$ and final scalar field $\chi.$ This graph shows that isotropic solutions receive a higher weighting $e^{-\textrm{Im}(S)/\hbar},$ and are hence preferred. In a sense, this result confirm the minisuperspace approach pursued previously, as anisotropic deviations are seen to be suppressed. It also means that the no-boundary proposal assigns exponentially higher probability to an isotropic than to an anisotropic beginning of the universe. \\

\noindent {\it Black holes}
 
The question of primordial black holes has recently come into renewed focus -- due to the observations of black hole collisions of various masses, as dark matter candidates, and because of the existence of supermassive black holes in the centers of very early galaxies. Primordial black holes could play a central role in explaining these phenomena \cite{Carr:2020gox}. For such an explanation to be viable, a reliable formation mechanism has to be discovered. Many current works focus on specific dynamics during inflation, such as a phase of ultra-slow roll (for an overview, see {\it e.g.} \cite{Ozsoy:2023ryl}). Such scenarios operate when the universe as a whole is already very classical, and the predictions of such scenarios will be largely unchanged by the no-boundary framework. However, another possibility might be that black holes have been formed directly by quantum nucleation at the creation of the universe. This is the question we will be concerned with here. In other words, does the no-boundary wave function predict a significant quantum creation of primordial black holes?

This question was considered in a series of works by Bousso and Hawking \cite{Bousso:1995cc,Bousso:1996au,Bousso:1998na} and extended in \cite{Chao:1997osu,Gregory:2013hja,Draper:2022xzl,Morvan:2022ybp}. To start, it is helpful to look at the relevant solution describing black holes in the presence of a cosmological constant, {\it i.e.} the Schwarzschild-de Sitter (Kottler) solution
\begin{align}
 \mathrm{d}s^2 = -f(r) \mathrm{d}t^2 + \frac{\mathrm{d}r^2}{f(r)} + r^2 \mathrm{d}\Omega_2^2\,,
\end{align}
with $f(r) = 1-\frac{2M}{r}-\frac{\Lambda}{3}r^2.$ There are two horizons, sitting at the positive solutions to the equation $f(r)=0.$ The smaller horizon is the black hole horizon, and the larger one the cosmological horizon. When $M=0,$ the black hole horizon disappears and the solution describes a portion of de Sitter spacetime. Its Euclidean version, as we have seen, is just the $4-$sphere -- in the Euclidean version, the cosmological horizon becomes a regular point, as long as the Euclidean time coordinate is made periodic with a suitable period \cite{Gibbons:1977mu}. The $4-$sphere is rounded off, and thus naturally satisfies the no-boundary requirements of compactness and regularity at the South Pole. 

What if $M \neq 0$? In that case there are two horizons, and in trying to form a regular Euclidean solution, one is faced with the problem of specifying the periodicity of the Euclidean time coordinate. At best one can render one of the horizons smooth, but then at the location of the second horizon there will be a conical deficit. If we insist on having an entirely smooth solution, then, apart from pure de Sitter spacetime, we are only left with the possibility of making the two horizons equal in size, by increasing the black hole parameter $M$ to its maximal value $M_N=1/(3\sqrt{\Lambda})$. This is known as the Nariai spacetime \cite{Nariai}. Interestingly, even though the two horizons are equal in size, the space between them does not vanish, as shown in detail in \cite{Ginsparg:1982rs}. The resulting Euclidean metric can be written as
\begin{align}
    \mathrm{d}s^2 = \mathrm{d}\tau^2 + \frac{1}{\Lambda}\sin(\sqrt{\Lambda}\tau)^2 \mathrm{d}x^2 + \frac{1}{\Lambda}\mathrm{d}\Omega_2^2\,.
\end{align}
This is simply the product of two $2-$spheres, each with radius $1/\sqrt{\Lambda}.$ Again this spacetime is appropriately compact and rounded off, and hence may serve as a no-boundary instanton, with the no-boundary South Pole identified with the South Pole of one of the two spheres. 

The reason for writing the metric in the above form is that one can see that if $\tau$ is analytically continued back to Lorentzian time, the first $2-$sphere turns into a two-dimensional de Sitter line element. In analogy with the simplest no-boundary solutions, {\it cf.} Eqs.~\eqref{Hscale} and \eqref{Hcont}, one performs this analytic continuation at the equator of the sphere containing the no-boundary South Pole. Thus the Lorentzian spacetime that results from this instanton is a product of two-dimensional de Sitter spacetime with a constant $2-$sphere. The spatial sections have topology $S^1 \times S^2,$ which can be regarded as a $3-$sphere with holes punched through the North and South Poles, describing a pair of black holes~\cite{Bousso:1995cc}.

Since we are interested in the probability for creating such black holes, we must calculate the Euclidean action, which will determine the rate of creation. It arises from one $2-$sphere, plus one hemisphere of the second $2-$sphere. A general result for spacetimes with multiple horizons is that it is given by a quarter of the sum of horizon areas (reinstating $8\pi G$ momentarily) \cite{Bousso:1998na,Chao:1997osu,Gregory:2013hja},
\begin{align}
    I_E = - \frac{1}{4G}\left({\cal A}_{\textrm{b.h. horizon}} + {\cal A}_{\textrm{cosm. horizon}}\right)\,. \label{bhaction}
\end{align}
For a product of two $2-$spheres, one obtains, with $8\pi G=1,$ $I_E=-2\pi \cdot 2\cdot {\cal A}=-\frac{16\pi^2}{\Lambda}.$ Since for Nariai black hole pair creation we only include one hemisphere of one of the spheres, we must halve this result to obtain 
\begin{align}
    I_{E,N} = -\frac{8\pi^2}{\Lambda}\,.
\end{align}
Hence the rate $\Gamma_N$ of nucleating regions of spacetime containing Nariai black holes, compared to creating the universe as a purely de Sitter spacetime, is
\begin{align}
    \Gamma_N = \frac{\Psi^\star_N\Psi_N}{\Psi^\star_{dS}\Psi_{dS}} = \frac{e^{-2I_{E,N}}}{e^{-2I_{E,dS}}} = e^{\frac{16\pi^2}{\hbar\Lambda}-\frac{24\pi^2}{\hbar\Lambda}} = e^{-\frac{8\pi^2}{\hbar\Lambda}}\,.
\end{align}
When the vacuum energy scale $\Lambda$ is a few orders of magnitude below the Planck scale, then this rate is heavily suppressed. 

We may also ask what happens if we allow for the creation of smaller mass black holes. As discussed above, these have conical deficits at least at the location of one horizon. But they have actions that are perfectly regular, and again given by the general formula \eqref{bhaction}; the actions interpolate between the Nariai and pure de Sitter cases. These solutions have been argued to arise as constrained instantons, {\it i.e.} as saddle points not of an ordinary path integral, but of one in which an additional constraint has been put on the mass \cite{Draper:2022xzl,Morvan:2022ybp}. One may then integrate over all masses, to find that the total rate of nucleation of black holes in de Sitter spacetime can be approximated as (see \cite{Morvan:2022ybp} for the details of this calculation)
\begin{align}
    \Gamma \approx \frac{M_N}{I_{E,N}-I_{E,dS}}\left(1-e^{2(I_{E,dS}-I_{E,N})} \right) = \frac{\sqrt{\Lambda}}{24\pi^2}\left(1-e^{-\frac{8\pi^2}{\Lambda}} \right)\,.
\end{align}
Interestingly, this rate contains a perturbative contribution, which significantly enhances the result. Still, for inflationary models in agreement with data, we expect the vacuum energy to have been many orders of magnitude below the Planck scale, and thus the rate of quantum formation of black holes from nothing remains small. In a way, this may be seen as further confirmation that the no-boundary wave function prefers the nucleation of homogeneous and isotropic universes. \\


\section{Link to Observations} \label{sec:ob}

The simple models that we considered up to now were particularly suitable to a detailed, and in many parts exact, treatment. However, when we make contact with observations, we must include perturbations describing the actual distribution of matter in the early universe. The most obvious point of contact of the no-boundary wave function with observations occurs for the cosmic microwave background radiation (CMB), which provides us with the earliest electromagnetic picture of the universe. A specific and crucial question is whether the no-boundary wave function, in combination with a suitable dynamical model, can explain not only the homogeneous, isotropic and flat background spacetime, but also the distribution of temperature fluctuations in the CMB. In order to discuss this question, we will first review how perturbations are included in the no-boundary framework. Then we will analyse the implications for observations, and confront no-boundary probabilities with what we know about the early universe.  

\subsection{Perturbations} \label{sec:perts}

We wish to extend our analysis to include cosmological perturbations. The most important ones are scalar and tensor perturbations, leading respectively to density perturbations and gravitational waves. We will treat perturbations at leading order, which means that we must consider their action up to quadratic order in perturbations. For definiteness, we will write out the analysis for tensor perturbations below -- these are always present, as they form a part of the metric. The analysis of scalar perturbations proceeds in close analogy, and we will simply quote the results at the end. The standard theory of cosmological perturbations is discussed in numerous references, see {\it e.g.} \cite{Mukhanov:1990me}.

Tensor perturbations arise as transverse, traceless perturbations of the spatial metric,
\begin{align}
    \delta g_{ij}  \qquad \textrm{with} \qquad {\delta g^i}_i=0,\,\,\, {\cal D}^i \delta g_{ij}=0\,, 
\end{align}
where ${\cal D}_i$ denotes a covariant derivative formed from the spatial background metric, which is the metric on a $3-$sphere here. Tensor perturbations arise in two polarisation states $+,\times$, each of which may be decomposed in terms of harmonics on the $3-$sphere, $\delta g_{jk}^{+,\times}=h(t) G_{jk}^{(l)}=h(t) \sum_{n=2}^l \sum_{m=-n}^n c^l_{nm} (G_{jk})^l_{nm}$, satisfying the eigenvalue equation
\begin{align}
    {\cal D}^i {\cal D}_i G_{jk}^{(l)}= -[l(l+2)-2] G_{jk}^{(l)}\,, \qquad l \geq 2\,, \label{eigen}
\end{align}
where $l$ is the principal quantum number on the sphere and $c^l_{nm}$ are Fourier coefficients \cite{Gerlach:1978gy}. We will consider a single such mode, with amplitude $h(t),$ as at leading order these modes evolve independently. The treatment can be extended straightforwardly to include a collection of modes, with the total wave function becoming a product of the individual wave functions for all modes.  

Thus our aim is to calculate the no-boundary wave function including a single tensor harmonic
\begin{align}
\Psi(a_1,h_1) = \int_{\cal C} \mathrm{d}N \int ^{a_1}\mathcal{D}a\int^{h_1} \mathcal{D}h\,\,\, e^{\frac{i}{\hbar} S}\,,
\end{align} 
where $h_1$ denotes the (real valued) amplitude of the tensor perturbation on the final hypersurface. This type of calculation was first performed by Halliwell and Hawking \cite{Halliwell:1984eu}. The total action $S=S^{(0)}[a,N]+S^{(2)}[a,N,h]$ now consists of the background part we had before, plus a term for the tensor perturbation
\begin{equation}
S^{(2)}[a,N,h]=\frac{1}{2} \int N\mathrm{d}t \left( a^3 \left(\frac{\dot{h}}{N}\right)^2 - a \, l(l+2) \, h^2\right)\,, \label{pertaction}
\end{equation}
where we assumed the background metric $\mathrm{d}s^2=-N^2\mathrm{d}t^2+a(t)^2d\Omega_3^2$ and where the spatial term arises from the eigenvalue equation \eqref{eigen} combined with a curvature term \cite{Garriga:1997wz}.  The equation of motion that follows from this action reads
\begin{equation}
\ddot{h} + 3 \frac{\dot{a}}{a} \dot{h} + \frac{N^2}{a^2}l(l+2) h = 0\,. \label{heom}
\end{equation}

We will assume that the backreaction of the perturbations on the background is negligible. This means that the background part of the path integral can be performed independently, and moreover will be approximated by a collection of saddle points. The path integral over tensor perturbations is then performed with the background fields taking their saddle point values. That is to say, the path integral for tensor modes is a quadratic integral in perturbations, and can thus be performed exactly. All that is required is the saddle point solution. In other words, we must solve the perturbation equation \eqref{heom} on no-boundary saddle point geometries.

\begin{figure}[ht] 
\begin{center}
		\includegraphics[width=0.6\textwidth]{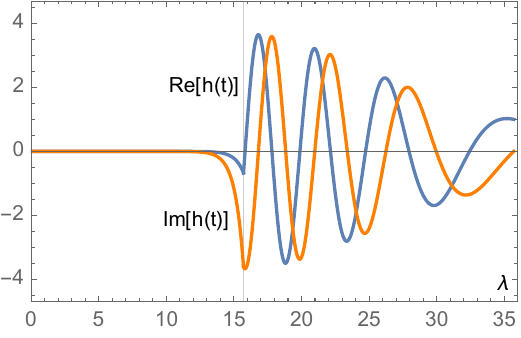}
		\end{center}
	\caption{An example of a tensor perturbation mode, with $H=1/10,\, l=15, h_1=1$. The solution is plotted along the standard Euclidean-plus-Lorentzian contour, parameterised by a parameter $\lambda.$ The transition from Euclidean to Lorentzian is indicated by the thin vertical line at $\lambda=5\pi.$}
	\label{fig:sadperts}
\end{figure}

With a cosmological constant $\Lambda \equiv 3H^2,$ the no-boundary saddle point is given by $a(\tau) = \frac{1}{H}\sin(H\tau),$ cf. section \ref{sec:inflex}. Here $\tau$ denotes Euclidean time, {\it i.e.} $N=i$ and $\tau=it.$ The equation for the tensor perturbations then becomes
\begin{equation}
h_{,\tau\tau}+ 3H\cot(H\tau) h_{,\tau} - \frac{H^2 l(l+2)}{(\sin(H\tau))^2} h = 0\,.
\end{equation}
This equation possesses two solutions,
\begin{align}
    F_1(\tau) &=\frac{(1-\cos(H\tau))^{l/2}(\cos(H\tau)+l+1)}{(1+\cos(H\tau))^{(l+1)/2}}\,, \\ F_2(\tau) &=\frac{(1+\cos(H\tau))^{l/2}(\cos(H\tau)-l-1)}{(1-\cos(H\tau))^{(l+1)/2}}\,,
\end{align}
so that the general solution is a linear combination of these. Note that these solutions behave very differently near the South Pole $\tau=0.$ Most importantly, $F_2(\tau)$ diverges as $\tau \to 0,$ and regularity at the South Pole forces us to eliminate this solution and retain only $F_1(\tau).$ In fact, near $\tau=0,$ we have $F_1(\tau) \propto \tau^l,$ {\it i.e.} this mode vanishes at the South Pole and grows away from there. Moreover, the shorter the mode, the more suppressed it is near the South Pole. Thus the correct solution, satisfying regularity at the South Pole and reaching the value $h(\tau=\tau_f)=h_1$ is given by
\begin{align}
    h(\tau) = \frac{F_1(\tau)}{F_1(\tau_f)} h_1\,. \label{pert1}
\end{align}
For a numerical example see Fig.~\ref{fig:sadperts}. In the example, the mode function is plotted along the usual Hawking contour, {\it i.e.} following a Euclidean time path until the background $4-$sphere has grown to its equator, followed by Lorentzian evolution along the de Sitter hyperboloid. In the Euclidean regime, the mode function is growing monotonically, while in the Lorentzian region it is oscillating. For completeness, let us also write out the relevant solution in physical time,
\begin{equation}
F_1(t) = \left( 1 + \frac{i}{\sinh(Ht)}\right)^{\frac{l}{2}} \left( 1 - \frac{i}{\sinh(Ht)}\right)^{-\frac{l+2}{2}} \left( 1 - \frac{i(l+1)}{\sinh(Ht)}\right)\,. \label{RealFluct}
\end{equation}
This way of writing the solution makes it clear that at late times $F_1 \to 1.$ This solution represents the Bunch-Davies vacuum \cite{Bunch:1978yq}, for a closed spatial slicing. 

We are also interested in the action of the perturbations. On-shell, we may use the equation of motion \eqref{heom} in the action \eqref{pertaction}, which turns it into a surface term
\begin{align}
    S^{(2)}_{\textrm{on-shell}}[a,N,h] & =\frac{1}{2N} \int \mathrm{d}t \left( a^3 \dot{h}^2 + a^3 h \ddot{h} + 3 a^2 \dot{a} h \dot{h}\right) \nonumber \\ & = \frac{1}{2N} a^3 h \dot{h} \mid_{t=t_f} \nonumber \\ & = \frac{ia_1^2}{2}\frac{l(l+2)}{l+1-i\sqrt{H^2 a_1^2 -1}}h_1^2 \\ & = -\frac{l(l+2)a_1}{H}h_1^2 + i \frac{l(l+1)(l+2)}{2H^2}h_1^2 + {\cal O}(a_1^{-1})\,, \label{pertactionlargea}
\end{align}
where we used the fact that the perturbation vanishes on the initial hypersurface, and where we expanded at large final scale factor $a_1$ to obtain the last line. This implies that, for $a_1 \gg l/H,$ the wave function receives the following factors at the saddle points, for each wave number and polarisation,
\begin{align}
    \Psi^{(2)} \approx e^{-\frac{l(l+1)(l+2)}{2\hbar H^2}h_1^2 -i\frac{l(l+2)a_1}{\hbar H}h_1^2}\,.
\end{align}
Note that the weighting is a Gaussian, and independent of $a_1.$ The dependence on $l^3/H^2$ implies that the spectrum is scale invariant (because the background is exact de Sitter space here), and the amplitude of perturbations scales as the inverse of the square of the Hubble rate. The phase grows with the scale factor $a_1,$ as a reflection of the approximate classical behaviour of perturbations on super-Hubble scales. Thus we recover standard results in inflationary cosmology. But there is a crucial difference: here we did not have to assume the initial Bunch-Davies vacuum state \eqref{pert1} or \eqref{RealFluct} -- rather, regularity at the South Pole forced us to choose this solution. Thus the no-boundary wave function automatically implies the Bunch-Davies state for perturbations.

We are now in a position to explain a result that we have used repeatedly in this review, and which concerns the distinction between stable and unstable saddle points. Above, we calculated the wave function for tensor perturbations on a Hartle-Hawking saddle point geometry. This geometry may be specified by a combination of Euclidean and Lorentzian parts, as detailed in Eqs. \eqref{Hscale} and \eqref{Hcont}. The geometry we focused on corresponds to one of the saddle points in the lower half plane of the lapse function in section \ref{sec:minisuper}. However, because the Einstein equations are real, the complex conjugates of these geometries are also solutions to the equations of motion. Moreover, since the final scale factor is real, they also obey the imposed final boundary condition (and in fact, these saddle points are picked out in the tunneling proposal \cite{Vilenkin:1982de,Vilenkin:1983xq,Vilenkin:1984wp}). The only thing that changes is the regularity condition at the South Pole, which changes from $\dot{a}=+i$ to $\dot{a}=-i$ (in physical time). This seemingly innocent change has profound consequences \cite{Feldbrugge:2017fcc}: it changes the mode function \eqref{RealFluct} and the action \eqref{pertactionlargea} to their complex conjugate values (note that the South Pole then resides at $t=-i\frac{\pi}{2H}$ rather than $t=i\frac{\pi}{2H}$). Then, for perturbations, we would obtain factors of the form  
\begin{align}
    \Psi^{(2)}_{\textrm{unstable}} \approx e^{+\frac{l(l+1)(l+2)}{2\hbar H^2}h_1^2 -i\frac{l(l+2)a_1}{\hbar H}h_1^2}\,.
\end{align}
Thus, the distribution of perturbations would be given by an inverse Gaussian. It might appear that this is not problematic {\it per se}, as the path integral converges and leads to a finite result. However, extrapolating this result to large amplitudes one concludes that such a wave function is not normalisable and thus does not represent a physically acceptable state. Even if one were to imagine that at large amplitudes the weighting would turn around and render the wave function normalisable, the physical implications would point to an inconsistency: larger perturbations would be more likely than smaller perturbations. The most likely outcome would then be the largest fluctuations for which we still trust our calculation, for all perturbation modes. This is completely at odds with observations: fields would be in the opposite of their ground state -- all fields would be predicted to fluctuate wildly, which is simply not seen. An isotropic universe would be a highly unlikely outcome.  One might speculate that the problem arises by the choices made near the South Pole, and that different UV-physics ({\it e.g.} in the guise of modified dispersion relations) might cure the problem, but this option has also been shown to be unlikely to work \cite{Matsui:2022lfj}. For these reasons, we term such saddle points unstable, and they must be avoided if we are to trust our calculations.

The last point can also be turned around: in the no-boundary wave function perturbations are predicted to be in their ground state, and as such perturbations are likely to be small. This provides a physical justification for the minisuperspace simplification we used before, since it implies that highly symmetric spacetimes are in fact favoured and this renders the results of minisuperspace calculations believable. 

Analogous results can be obtained for other perturbations, in particular scalar perturbations. These give rise to density perturbations in the early universe. They can be described rigorously in terms of a gauge-invariant curvature perturbation $\zeta$ (see in particular \cite{Garriga:1997wz,Garriga:1998he,Gratton:1999ya})  with action
\begin{align}
    S^{\zeta} = \int \mathrm{d}t \epsilon \left( a^3 \dot{\zeta}^2 - a (n^2-1) \zeta^2 \right)\,,
\end{align}
where $\epsilon = -\dot{H}/H^2 = \dot\phi^2/(2H^2)$ is the slow-roll parameter and with mode numbers $n \in \mathbb{N}^\star$. The action for scalar perturbations is suppressed by the slow-roll parameter $\epsilon$ \cite{Maldacena:2002vr,Baumann:2009ds} (in the exact de Sitter limit, $\epsilon=0,$ the scalar perturbation is pure gauge). The analysis proceeds entirely in parallel with that of tensor fluctuations, and leads to the following approximate weighting factors in the wave function (at large scale factor values)
\begin{align}
    |\Psi^{\zeta}| \approx e^{-\frac{\epsilon n^3}{\hbar H^2}\zeta_1^2}\,.
\end{align}
Compared to the tensor fluctuations, the main difference is that the amplitude of scalar fluctuations is enhanced when the potential is suitably flat, {\it i.e.} when $\epsilon$ is small. But apart from that difference, we obtain the same consequences, namely fluctuations that start with zero amplitude at the South Pole, in the Bunch-Davies vacuum state, and that subsequently grow to reach real values $\zeta_1$ (with a Gaussian probability distribution) at late times.

The analysis of perturbations can be generalised to situations in which the background itself is not isotropic, see for example \cite{Janssen:2019sex}; in a sense one is then dealing with small perturbations superimposed on larger perturbations. 

Moreover, fermions can be included \cite{DEath:1986lxx} (see also the briefer description in \cite{Hertog:2019uhy}, and for extensions including supersymmetry see {\it e.g.} \cite{Moniz:1996pd,Moniz:1997zz,VargasMoniz:2010zz}), and it is again found that they start out being zero at the South Pole. This is a very general feature of the no-boundary proposal, namely that it implies that quantum fields start out in their vacua.

Up to now, we analysed perturbations evaluated at the saddle points of the background. Can we also say something about the perturbations off-shell? For this, we take another look at the simplest minisuperspace model, which is that of section \ref{sec:minisuper} containing a cosmological constant and considering RW metrics. We will specialise to the case where we impose an initial momentum condition\footnote{The case with an initial Dirichlet condition was discussed in \cite{Feldbrugge:2017mbc,Janssen:2019sex}. The results discussed here are original work.}. Off-shell in the lapse, the scale factor is given by ({\it cf.} Eq. \eqref{eq:qbarND2})
\begin{align}
    \bar{q}(t_q) = H^2 N^2 (t_q^2-1) +2Ni (t_q-1) + q_1 \,, \label{eq:qbarND3}
\end{align}
and in these variables, the perturbation equation \eqref{heom} reads
\begin{align}
    \ddot{h} + 2 \frac{\dot{\bar{q}}}{\bar{q}}\dot{h} + \frac{N^2}{\bar{q}^2}l(l+2)h=0\,,
\end{align}
with a dot denoting a derivative w.r.t. $t_q$ here. The solution to this equation is given in terms of Legendre functions,
\begin{align}
    & h(t_q) = \bar{q}^{-1/2} \left( c_1 \, \textrm{LegendreP}[1,\gamma,x] + c_2 \, \textrm{LegendreQ}[1,\gamma,x] \right) \\ & \gamma = \sqrt{1-\frac{l(l+2)}{(H^2N+i)^2-H^2q_1}}\,, \qquad x=\frac{H^2 N t+i}{\sqrt{(H^2N+i)^2-H^2q_1}}\,, \nonumber
\end{align}
where $c_{1,2}$ are integration constants that need to be fixed to satisfy the boundary conditions. We are only concerned here with the question of whether such perturbations are well behaved, or whether they can be singular. Clearly, the solutions above blow up when $\bar{q}$ passes through zero, unless the Legendre functions vanish suitably fast at those events. But this happens only at the saddle points, where from the results above we know that perturbation modes vanish as $h_{\textrm{saddle}} \sim t_q^{l/2}.$ 

Further, it is known that the Legendre functions have branch points at $x=\pm 1.$ In fact, it can be seen straightforwardly that the condition for having branch points is equivalent to the condition that $\bar{q}$ passes through zero. Thus, off-shell, the perturbation modes are well behaved unless the background passes through zero. The locus of such points was already discussed in section \ref{sec:minisuper}, with the results shown in Fig.~\ref{fig:NDzeros}. Thus, the red dashed lines in this figure correspond to locations where the perturbations blow up, and where we cannot trust our analysis. In defining integration contours, it would therefore be prudent to circumvent those singular curves. In the present example, this can be done without changing the asymptotic regions of the contours of integration, {\it i.e.} without changing the results obtained at background level. Note however that the saddle points themselves sit right on the boundary of the singular curves. We will encounter a closely related phenomenon in section \ref{sec:allowability}, where we will discuss further criteria that geometries might have to satisfy to be considered reliable. We should also remark that studies of off-shell perturbations have been rather few in number so far, and that this topic provides opportunities for further research.

\subsection{Probabilities} \label{sec:probs}

In section \ref{sec:reconstruct} we saw how one may define relative probabilities in quantum cosmology. A requirement is that the wave function becomes of WKB form, which we saw occurs for the no-boundary wave function under certain conditions. The two examples we encountered are an inflationary phase and an ekpyrotic phase, in both cases lasting at least a few e-folds, {\it cf.} also section \ref{sec:class}. No-boundary saddle point solutions then receive a significant weighting, which is conserved for a series of instantons with final boundaries following a classical history. These weightings lead to relative probabilities which, in the presence of a scalar field with an appropriate potential, we recall are given by
\begin{align}
    {\cal P}_{\textrm{inf}} \approx e^{\frac{24\pi^2}{\hbar V(\phi_{SP})}}\,, \qquad {\cal P}_{\textrm{ekp}} \approx e^{\frac{s}{\hbar |V(\phi_{SP})|}}\,, \label{probbies}
\end{align}
where $s>0$ is a numerical factor depending on the slope of the potential. The probabilities are relative since the overall normalisation of the wave function is unknown, and presumably depends on the UV completion of the theory under consideration. If several such saddle points contribute to the wave function, then once perturbations are amplified these have the characteristic that they decohere the saddle points \cite{Kiefer:1987ft}, which from then on evolve as essentially separate universes \cite{Halliwell:1989vw}. We may thus focus on individual saddle points at this stage. We will discuss the implications for inflation and ekpyrosis in turn, and at the end compare them.

\begin{figure}[ht] 
\begin{center}
		\includegraphics[width=0.49\textwidth]{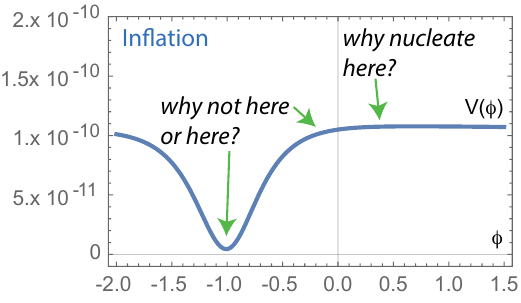}
		\includegraphics[width=0.49\textwidth]{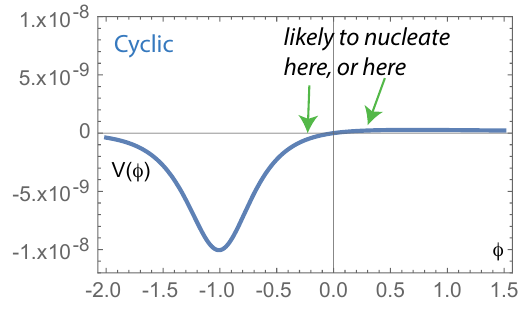}
		\end{center}
	\caption{{\it Left panel:} An inflationary potential. Why should the universe start high up on the potential, when, at least naively, the probability to nucleate at lower values appears to be higher? {\it Right panel:} A cyclic potential. The highest likelihood is to nucleate at low absolute values of the potential, meaning either at the onset of the ekpyrotic phase or during the dark energy phase.}
	\label{fig:probs}
\end{figure}

The most striking feature of the expressions in \eqref{probbies} is that they favour low values of the potential. For inflationary models, this leads to two immediate puzzles. The first is that it would be vastly preferable for the universe to nucleate in the current dark energy phase, rather than at a much higher, inflationary value of the potential \cite{Page:2006hr}, see also Fig.~\ref{fig:probs}. The current dark energy phase is equally suitable in bringing about a classical universe, as it entails the same kind of attractor behaviour as an inflationary phase, only with a vastly lower expansion rate. This would however result in an essentially empty universe. One might object that the thermal fluctuations \cite{Gibbons:1977mu} implied by a quasi-de Sitter phase would eventually produce conscious observers (so-called Boltzmann brains), and that the combined probability of nucleating a vast empty universe and then forming Boltzmann brains would still be higher than nucleating a small, inflationary universe \cite{Page:2006hr}. However, serious doubts as to the long-term stability of de Sitter spacetime have arisen in recent years, in particular the issues codified in the corresponding swampland conjectures \cite{Obied:2018sgi,Garg:2018reu}. From these it seems plausible that the current dark energy phase will simply not last long enough to make the presence of Boltzmann brains a realistic prospect. This is in fact something that can be tested, in the sense that this hypothesis implies that the dark energy equation of state cannot be constant. Future dark energy surveys will eventually be able to tell. 

Given that we are not directly interested in the bare probability for nucleating a universe, but rather in the conditional probabilities for nucleating universes that may include observers (this could be weakened to the requirement of containing large, durable structures such as galaxies) \cite{Hawking:2002af,Hartle:2004qv,Hawking:2006ur}, we should thus focus on no-boundary instantons that start out on the inflationary part of the potential. Here we encounter the second puzzle: the preference for low values of the potential means that, amongst inflationary histories, the favoured ones appear to be those that last only a few e-folds. The most likely histories in fact would be those lasting the bare minimum of e-folds required to satisfy the conditions for {\it e.g.} the occurrence of galaxies. Yet our observations are not compatible with such a short inflationary phase, they indicate that a minimum of about $60$ e-folds is required to explain the observations of the CMB \cite{Baumann:2009ds}. There currently exists no consensus on how this second puzzle may be resolved, or whether it in fact invalidates the no-boundary proposal. We will discuss a few possible resolutions.

The first is to resort to strong anthropic reasoning \cite{Hartle:2004qv}: if it happens to be the case that conscious observers require an old universe, that is roughly spatially flat and contains many galaxies, then a long inflationary phase is essentially required {\it by fiat}. In such a case, the main purpose of the no-boundary proposal would be to explain other features of the universe, such as correlations in the fluctuations of the CMB. Such strong anthropic reasoning can never be logically ruled out, yet (and to some extent because of this) it closes the road to finding deeper, and more useful, explanations. There is a clear risk that the road block is simply due to our lack of imagination. Also, if the anthropic argument really were true, we should expect to be living right at the edge of possible existence. This does not appear to be the case: the universe could have developed more galaxies, or fewer, and the spatial curvature could have been more positive, or more negative, yet this would hardly have affected the emergence of life on Earth.  

It has also been suggested that, because we are interested in the conditional probability of us observing the universe to be in a particular state, we should allow for the fact that we could be in any possible location in the universe. Put differently, if there is a certain likelihood for observers to come into existence, then if the universe is larger then the total probability ought to be higher, precisely in proportion to the spatial volume. This is known as volume weighting \cite{Hartle:2007gi,Hartle:2008ng}. Let us see what the consequences would be for inflationary predictions. If there are $N_e$ e-folds of inflation, the probability measure should thus be adjusted according to
\begin{equation}
{\cal P}_{\textrm{volume weighted}} = e^{3 N_e} \Psi^\star \Psi = e^{3 N_e + \frac{24\pi^2}{\hbar V(\phi_{SP})}} \;.
\end{equation}
For small $N_e$ the probability decreases with increasing potential, but we may hope that for large $N_e$ it will increase again. Thus we may ask where on the potential the minimum probability occurs. This can be estimated as follows, 
\begin{align}
0 = {\cal P}_{,\phi} &\approx  \left( \frac{24\pi^2}{V} + 3 \int\frac{V}{V_{,\phi}} d\phi \right)_{,\phi} \nonumber \\ & = -\frac{24\pi^2 V_{,\phi}}{V^2} + 3 \frac{V}{V_{,\phi}} \nonumber \\  & \rightarrow \quad\frac{V_{,\phi}^2}{V^3} = \frac{1}{8\pi^2}\;.
\end{align}
In terms of the slow-roll parameter the last condition can also be written as $V \sim \epsilon,$ and in fact it corresponds to the onset of eternal inflation. Thus we obtain the result that it is only in the regime of eternal inflation where the volume weighted probability starts going up again. But in eternal inflation the spatial volume in fact becomes infinite. If one took this seriously, after normalization there would be zero probability for non-eternal inflation, and all the probability would be concentrated in the eternal regime, without a clear preference for different parts of the potential. In fact assigning probabilities in the eternal inflation regime is a notoriously intractable problem. The underlying reason for this might simply be that eternal inflation is not physical. It leads to ill-defined semi-classical amplitudes \cite{Jonas:2021xkx}, and in fact is also conjectured to conflict with quantum gravity consistency conditions \cite{Brahma:2019iyy,Rudelius:2019cfh}.

Though volume weighting did not produce the expected turn-around in probabilities, it could still be that we do not understand the total probability measure well enough yet. This might be due to oversimplifying the model. Inflation is a dynamical attractor, and many originally slightly inhomogeneous and anisotropic universes would lead to essentially the same final outcome. Hence, it might be that many more histories are effectively indistinguishable from the Robertson-Walker background we assumed in calculations. These would enhance the total probability for a long inflationary phase. Another effect is that a higher potential also leads to more structure formation, {\it i.e.} more galaxies. It is conceivable, though this is plain conjecture at the moment, that when all these effects are taken into account, the probability for a long inflationary phase could become very significant, especially near a broad maximum in the potential (a broader maximum would allow for more histories with quasi-indistinguishable outcomes). Performing this calculation will however require us to go beyond the minisuperspace framework, and this is currently an open problem.

Let us mention one by-product: if there exist several regions in the scalar potential that could all lead to long inflationary phases and viable universes, then the lowest such region would seem to come out as preferred, due to the basic preference for low potential values \cite{Hertog:2019uhy}. This would imply that plateau models of inflation would be preferred over power-law potentials, and in fact one would then also expect the tensor-to-scalar ratio to be small. From this point of view, the present non-detection of primordial scale-invariant gravitational waves is not surprising.

A further possibility is that the early universe history is more complicated than a purely inflationary expansion phase. In particular, it has been suggested in \cite{Matsui:2019ygj} that after a modest initial expansion, the universe could recollapse, then bounce due to the spatial curvature and subsequently enter a much longer inflationary phase. This would have the advantage that the scalar field could nucleate much lower on the potential (and thus with higher probability) and then run up the scalar potential during the collapse and bounce phases. Multiple bounces are also conceivable, though with each additional bounce the required initial conditions become more finely tuned. In such a case the wave function would contain various histories in competition with each other \cite{Lehners:2024qaw}, but determining which one truly dominates the wave function currently remains out of reach, as the measure on the full set of geometries is unknown.  

All this being said, let us mention that inflationary models are quite generally in tension with string theory, in the sense that no unambiguous embedding of inflation into string theory has been found to date. Simple string compactifications, and their swampland codifications \cite{Obied:2018sgi,Garg:2018reu}, certainly seem to be in conflict with the inflationary no-boundary wave function \cite{Matsui:2020tyd}. Most likely, one will therefore have to go beyond the simplest framework, and include non-perturbative effects as well as typical string theory objects such as orientifolds and various branes, to obtain realistic inflationary vacua \cite{Kachru:2003aw}. This is an active research area.

Turning our attention to ekpyrotic models, we can see from \eqref{probbies} that the situation is reversed. The preference for a low initial absolute value of the potential translates into a preference for a long ekpyrotic phase \cite{Battarra:2014xoa}. If there is a potential landscape with several ekpyrotic regions, this preference for low values would however not easily distinguish between different regions, as the predictions of ekpyrotic models are determined by what occurs further down the potential. In cyclic models, there is also the possibility of nucleating a universe in the dark energy phase \cite{Battarra:2014kga}, see Fig.~\ref{fig:probs}. Though this universe would again be empty at first, just as in inflationary models, it would not remain so, as the universe would collapse and at the bounce matter and radiation would be produced for the subsequent expanding phase. Let us mention that with simple effective models for the bounce, one can extend the quantum cosmology framework to include the bounce phase, and obtain instantons that contain the phases of nucleation, contraction, bounce and expansion \cite{Lehners:2015efa}. However, as discussed previously, it remains an open question whether bounce models can be consistently quantised in general, {\it i.e.} whether they are also consistent and instability-free away from RW backgrounds. Unfortunately, this attaches a question mark to all bounce models. Moreover, ekpyrotic potentials may be just as hard to obtain in string theory as inflationary potentials, due to the extreme steepness they require \cite{Lehners:2018vgi}. But if both ekpyrotic and inflationary models turn out to be possible, then the basic preference for low potential values expressed in \eqref{probbies} would give the edge to ekpyrotic, bouncing cosmologies.

\subsection{Taking Stock: What Does the No-Boundary Proposal Explain?} \label{sec:stock}

It seems like a good point now to take stock, and crystallise the main explanations that the no-boundary wave function may offer. Here we will focus on the achievements of the no-boundary proposal in a $4-$dimensional effective approach to cosmology. A more detailed discussion of links to string theory and open questions will follow in sections \ref{sec:st} and \ref{sec:di}. The list is rather impressive:
\begin{itemize}
    \item The no-boundary proposal specifies a single quantum state for the entire universe, and as such provides a theory of (probabilistic) initial conditions.
    \item In fact, it provides a \emph{unification of quantum gravity dynamics and initial conditions}, expressed in the path integral approach to quantum theory, \cite{Hartle:2002nm}
    \begin{align}
        \Psi(\phi_f) = \int_{\cal C}^{\phi_f} \, e^{\frac{i}{\hbar}S}\,.
    \end{align}
    The dynamics is encoded in the action $S$, and the state in the boundary conditions and integration ranges ${\cal C}$ of the integral. Thus a single mathematical object combines the dynamics and the specification of the state. This is why it is so important to understand the proper definition of gravitational path integrals. Moreover, the wave function depends on the final field values $\phi_f,$ and this dependence of the wave function on spatial hypersurfaces in some sense makes the integral also an intrinsically holographic object. This property will be discussed further in section \ref{sec:ads}. 
    \item The no-boundary wave function explains how space and time became classical at the early stages of evolution of the universe. This classicalisation moreover requires the presence of a dynamical attractor, such as inflation or ekpyrosis. The attractor drives spacetime to classicality, and allows for the definition of probabilities for different classical histories of the universe.
    \item Once spacetime has become classical, one naturally obtains the framework of quantum field theory in curved spacetime.
    \item The big bang singularity is resolved. No-boundary instantons are finite and regular, and simply do not contain a big bang singularity. We can only say things with confidence about the universe in the regime where spacetime has already become effectively classical. Going back in time, there is no operational way of saying something about the phase when the radius of the universe was smaller than the primordial Hubble scale. The latter phase is better thought of as a quantum tunneling/nucleation event, which was required to produce space and time in the first place.
    \item Two immediate implications of the no-boundary idea are that the universe is finite in spatial extent, and that the overall average spatial curvature is positive. However, the spatial curvature might very well have been diluted to such an extent as not to be measurable today.
    \item Quantum fields are predicted to initially have been in their vacua. The dynamics of the universe may then excite certain fields. 
    \item Homogeneous and isotropic initial conditions are favoured over less symmetric ones. The initial state of the universe is thus one of relative simplicity. Combined with the previous item stating that quantum fields start out in their vacua, this provides appropriate low entropy initial conditions for the universe. This also provides an appropriate starting point for the 2nd Law of thermodynamics. 
    \item At the nucleation of the universe, only scalar fields may have a non-zero energy density. Other matter fields are not permitted. This follows from the equation of continuity $\dot\rho + 3 \frac{\dot{a}}{a} (\rho + p)=0.$ At zero scale factor this equation remains regular only if the sum of energy density and pressure $\rho+p$ vanishes, which in turn only a scalar field can (momentarily) achieve, when its kinetic energy is zero, {\it cf.} \eqref{eqeos}. Thus no initial radiation or ordinary matter are allowed -- these must be created later, for example during reheating. This setting fits well with cosmological models that use scalar fields as their main dynamical matter ingredient.
\end{itemize}
Some potential additional explanations that the no-boundary proposal may offer should still be considered to be works in progress. This concerns the predictions regarding inflation in particular. As we saw, it is not yet clear whether the no-boundary wave function assigns high probabilities to long inflationary phases or not, or to which particular type of inflationary background. Tied to this is also the question of whether or not the observed near-flatness of the current universe is explained, {\it i.e.} whether the positive spatial curvature of no-boundary solutions is sufficiently diluted. These important questions will require more work.


\section{Link to String Theory} \label{sec:st}

The no-boundary proposal is formulated in $4$ dimensions, with the theory of gravity being general relativity. However, it is widely expected that quantum gravity will require further, or different, fundamental ingredients. In the present section, we will ask whether the no-boundary proposal is compatible with quantum gravity, and with string theory in particular. Is it perhaps even required under some circumstances? And what do we learn by applying no-boundary ideas to string theory?

Before embarking on this, let us make a comment of principle. In string theory, one starts with the quantisation of a string. Classically, the world sheet admits a conformal/Weyl symmetry, which must be preserved under quantisation. This leads to non-trivial requirements, specifically that the spacetime in which the string moves must satisfy the equations of motion of certain supergravity theories (plus corrections thereof at higher orders in the string length $\alpha^\prime$) \cite{Fradkin:1984pq}. One may now wonder whether it is consistent to consider a path integral of this effective higher-dimensional supergravity theory, since one is then in some sense quantising the theory again. In well-known examples (meaning examples with significant amounts of preserved supersymmetry) this procedure is known to be consistent, for a striking example see \cite{Green:1997as}. However, as quite generally in string theory, not much is known in non-supersymmetric situations. We will proceed with optimism.

\subsection{Robustness} \label{sec:robust}

When gravity is quantised, effective terms with higher derivatives are generated from graviton loops at higher orders in $\hbar$ \cite{Goroff:1985th}. These terms typically arrange themselves as higher powers of the Riemann tensor, and derivatives thereof. A pertinent question, even when sticking to an effective $4-$dimensional description of gravity, is thus whether no-boundary solutions are still good (regular, finite action) solutions of the corrected theory. In other words, are no-boundary solutions robust to the inclusion of quantum corrections? In string theory, such higher order terms also arise, now accompanied by coefficients that contain powers of the string tension $\alpha^\prime.$ Thus, the analogous question arises there too. 

The most crucial aspect of this question is to see whether the rounded-off region near the South Pole of no-boundary instantons remains an acceptable solution. We analysed this question briefly in section \ref{sec:nbheurist}, and in fact used it to motivate the no-boundary proposal. We will fill in a few more details here. 

As we discussed in section \ref{sec:perts}, perturbations with wave number $l$ decay near the South Pole as $t^l.$ Hence we may actually focus on homogeneous and isotropic backgrounds, knowing that perturbations may arise away from the South Pole, but that they play no crucial part in the question of whether such solutions exist at all. As stated in section \ref{sec:nbheurist}, see Eqs. \eqref{eq:Riemann} and \eqref{eq:actionRiemann}, once we restrict to closed RW universes \eqref{RW}, with lapse $N$, the only non-zero components of the Riemann tensor are given in terms of $A_1 =  \frac{\dot{a}^2+N^2}{a^2N^2}$ and $A_2 = \frac{\ddot{a}}{aN^2}.$ An action which is a function of the Riemann tensor may be expanded in terms of these combinations as
\begin{align}
    S=\int \mathrm{d}^4x\sqrt{-g}f(\textrm{Riemann}) = 2\pi^2 \int \mathrm{d}t N a^3 \sum_{p_1,p_2} c_{p_1,p_2} A_1^{p_1} A_2^{p_2}\,, \label{eq:actionRiemann2}
\end{align}
where $c_{p_1,p_2}$ are coefficients, and the power of the Riemann terms is given by $P=p_1+p_2.$ The constraint equation can be found by taking a derivative of the action with respect to the lapse function, with the result \cite{Jonas:2020pos}
\begin{align}
0=\frac{\delta S}{\delta N}=2\pi^2&\sum_{p_1,p_2}c_{p_1,p_2}\bigg[2p_1(p_2-1)\frac{a\dot{a}^2}{N^2}A_2^{p_2}A_1^{p_1-1}+(1-p_2)a^3A_2^{p_2}A_1^{p_1}\nonumber\\
&+p_2(p_2-1)\frac{a\dot{a}a^{(3)}}{N^4}A_2^{p_2-2}A_1^{p_1}-p_2(2p_1+p_2-3)\frac{a\dot{a}^2}{N^2}A_2^{p_2-1}A_1^{p_1}\bigg]\,.\label{Friedmannconstraintnotsimplified}
\end{align}
The equation of motion for the scale factor follows from taking a time derivative of the constraint, and hence we do not need to consider it separately. The constraint is in fact invariant under the following transformation,
\begin{equation}
\left\lbrace
\begin{aligned}
&t\to -t,\\
&a \to -a,
\end{aligned}
\right.
\quad\Rightarrow\quad a(-t)=-a(t)\,,\label{aisodd}
\end{equation}
which implies that the scale factor is odd in $t$. The no-boundary ansatz, with its Euclidean rounding off at $a=0$ then corresponds to a series expansion
\begin{equation}
\left\lbrace
\begin{aligned}
&a(t)=a_1t+\frac{a_3}{6}t^3+\frac{a_5}{120}t^5+O(t^7)\,;\\
&a_1^2=-N^2\,.
\end{aligned}
\right.\label{ansatz}
\end{equation}
In physical time, we have $a_1^2=-1$ {\it i.e.} $a_1 = \pm i.$ For any solution, there exists a time-reversed solution obtained by sending $t\to-t,$ and moreover the solutions come in complex conjugate pairs. Hence there are always four related solutions (two of which can be eliminated by fixing the initial expansion rate). With this ansatz, the components of the Riemann tensor read
\begin{equation}
\left\lbrace
\begin{aligned}
&A_1=-\frac{a_3}{a_1^3}+\frac{\big(a_3^2-a_1a_5\big)}{12a_1^4}\,t^2+\frac{\big(a_3a_5-a_1a_7\big)}{360a_1^4}\,t^4+O(t^6)\,;\\
&A_2=-\frac{a_3}{a_1^3}+\frac{\big(a_3^2-a_1a_5\big)}{6a_1^4}\,t^2-\frac{\big(10a_3^3-13a_1a_3a_5+3a_1^2a_7\big)}{360a_1^5}\,t^4+O(t^6)\,.
\end{aligned}
\right.\label{A1A2expansion}
\end{equation}
Importantly, the series expansions start at order $t^{0},$ and not $t^{-2}$ as naively expected, and this is the main reason why no-boundary solutions can have finite action. Solving the constraint equation leads to the following expressions at the leading orders,
\begin{align}
&\mbox{Order $t$ :}\quad\sum_{p_1,p_2}\frac{c_{p_1,p_2}}{N^{2P}}a_1^{4-P}a_3^{P-1}\big(p_2-p_1\big)=0\,;\label{LOcondition}\\
&\mbox{Order $t^3$ :}\quad\sum_{p_1,p_2}\frac{c_{p_1,p_2}}{N^{2P}}\,a_1^{3-P}a_3^{P-2}\,\Big(a_3^2\cdot G_3[p_1,p_2]+a_1a_5\cdot G_{5}[p_1,p_2]\Big)=0\,,\label{NNLOcondition}
\end{align}
with \begin{align}
& G_3[p_1,p_2]=\frac{1}{6}\left(p_1^2-15p_1+6-4p_2^2+12p_2\right)\;, \\
& G_5[p_1,p_2]=\frac{p_1(1-p_1)}{6}-\frac{2p_2(1-p_2)}{3}\,.
\end{align}
Again it is a consequence of the no-boundary ansatz that there are no non-trivial conditions at negative powers of $t.$ The order $t$ condition is most easily solved by $c_{p_1,p_2}=c_{p_2,p_1},$ which turns out to be satisfied quite generally, in particular for all terms that are powers of the Ricci scalar $R=6(A_1+A_2)$ and all terms involving the quadratic combinations  $R_{\mu\nu\rho\sigma}R^{\mu\nu\rho\sigma}=6\left(A_1^2+A_2^2\right)$ and $R_{\mu\nu}R^{\mu\nu}=12\left(A_1^2+A_1A_2+A_2^2\right)$ \cite{Hawking:1984ph,Cano:2020oaa}. The condition at order $t^3$ then fixes $a_5$ in terms of $a_3,$ \cite{Jonas:2020pos}
\begin{equation}
a_5\cdot\sum_{p_1,p_2}\frac{c_{p_1,p_2}}{N^{2P}}\,a_1^{3-P}a_3^{P-2}G_5[p_1,p_2]=-\frac{a_3^2}{a_1}\cdot\sum_{p_1,p_2}\frac{c_{p_1,p_2}}{N^{2P}}\,a_1^{3-P}a_3^{P-2}G_3[p_1,p_2]\,.\label{a5expr}
\end{equation}
At higher orders, the next coefficients $a_7, a_9,$ {\it etc.} are then fixed in turn. Thus, a full solution, with manifestly finite action, is obtained, with free parameter~$a_3.$ This parameter specifies the early expansion rate and, in the case where a scalar field is added, is linked to the initial value of the scalar field. An explicit example, considering the inclusion of a Gauss-Bonnet term, appeared in \cite{Narain:2022msz}. 

In string theory, additional terms appear, in particular with derivatives acting on Riemann tensors. It is not possible to analyse such terms in full generality, but it was shown in \cite{Jonas:2020pos} that the first few correction terms, both in heterotic and type IIB string theory, still admit no-boundary solutions with finite action. This indicates a basic compatibility of string theory with the no-boundary proposal.

\subsection{Link to AdS/CFT and Holographic Definition} \label{sec:ads}

The best understood example of quantum gravity is the AdS/CFT correspondence \cite{Maldacena:1997re}. In this correspondence, quantum gravity with fields that asymptotically approach Anti-de Sitter (AdS) spacetime is related to a conformal field theory (CFT) that lives on a spacetime that is given by the conformal geometry of the asymptotic boundary. The manifold that the conformal field theory lives on has one dimension fewer than the bulk gravitational theory, which is why this setting is holographic. The great advantage is that conformal field theories are very well understood, so that one can use this knowledge to learn about quantum gravity. The drawback is that this only works for asymptotically AdS spacetimes, which do not correspond to our universe. Nevertheless, it seems plausible that quantum gravity contains universal features which one can uncover in this setting.

There are two approaches that have been pursued in connection with quantum cosmology: the first is to study AdS/CFT in order to gain insight into gravitational path integrals. One can then try to translate this knowledge into a cosmological context. And the second is to try to directly define the wave function holographically, by linking it to AdS/CFT. We will describe both approaches in turn.\\

\noindent {\it Gravitational path integrals in AdS}

Gravity in asymptotically AdS spacetimes is generally better understood than gravity in dS-like spacetimes. To some extent this is due to the fact that in asymptotically AdS spacetimes there exists a clear (timelike) boundary, and it is straightforward to fix the asymptotic geometry\footnote{In dS this is conceptually more complicated due to the presence of a (observer dependent) horizon, which shields observers from far-away regions.}. Connected to this is the conjecture that one may describe the physics of such spacetimes by a dual quantum field theory that lives on the (fixed) asymptotic geometry, and thus does not involve gravity at all. In this way arose the realisation that a quantum field theory can encode the same physics as a theory containing gravity and describing dynamical spacetime of one dimension higher \cite{Maldacena:1997re,Aharony:1999ti}. This can be used in some situations to check gravitational calculations by comparing with known quantum field theory results. These considerations motivate us to look at path integrals in the presence of a negative cosmological constant. In this way we may hope to learn general lessons for quantum cosmology \cite{DiTucci:2020weq}.

Specifically, we will analyse two setups: one in which we perform the analogue of the Robertson-Walker calculation of the wave function in section \ref{sec:minisuper}, and one in which we will add black holes. These calculations turn out to support the implementation of the no-boundary wave function as a path integral with a momentum condition. We will be brief here -- full details were presented in \cite{DiTucci:2020weq,Lehners:2021jmv}. As in \cite{DiTucci:2020weq}, we will keep Newton's constant explicit here and write the cosmological constant $\Lambda \equiv -\frac{3}{l^2}$ in terms of the radius of curvature $l$ (not to be confused with the wave number of perturbations in section \ref{sec:perts}).

We will once again look at the path integral \eqref{eq:HHpi} in minisuperspace, with RW metrics of the form \eqref{eq:Metricrepeat}. We will impose a final Dirichlet condition $q(t=1)=R_3^2$ ({\it i.e.} the ``final'' radius of the $3-$sphere is fixed to be $R_3$) and an initial momentum condition which we will parameterise by
\begin{equation}
p_0  =  -  \frac{3\pi}{4G}\alpha\,, \label{momentumcondition}
\end{equation}
intentionally leaving $\alpha$ arbitrary at first. Then, in complete analogy with the calculations surrounding Eq. \eqref{eq:wfND2} in section \ref{sec:minisuper}, one may derive that the saddle points are located at
\begin{subequations}
\begin{align} 
    N_\pm & = \alpha l^2 \pm i l \sqrt{R_3^2 + l^2}\,, \label{S3saddles} \\
    \frac{4 G}{\pi} S_0(N_\pm) &= \alpha (3+\alpha^2) l^2 \pm \frac{2i}{l}\left(R_3^2 + l^2 \right)^{3/2}\,, \label{saddleactionN}
\end{align}
\end{subequations}
where we also wrote out the action at the saddle points. It is instructive to look at the saddle point geometries, which are given by
\begin{align} \label{saddlemetric}
    \bar{q}(t_q)\mid_{N_\pm} = - \left( \alpha l \pm i \sqrt{R_3^2 + l^2}\right)^2 t_q^2 + 2\alpha \left( \alpha l^2 \pm i l \sqrt{R_3^2 + l^2}\right) t_q - l^2 (1+\alpha^2)\,.
\end{align}
We can immediately see that if we would like to have saddle point geometries that close at $t=0$, then we should choose $\alpha = \pm i.$ This is necessary, as otherwise we do not include the entire bulk spacetime. With $\alpha$ being imaginary, note that the saddle points reside on the Euclidean lapse axis, and as a consequence $t$ has become a spatial coordinate. The saddle point geometries are then purely Euclidean geometries.

\begin{figure}[ht]
	\centering
	\includegraphics[width=0.59\textwidth]{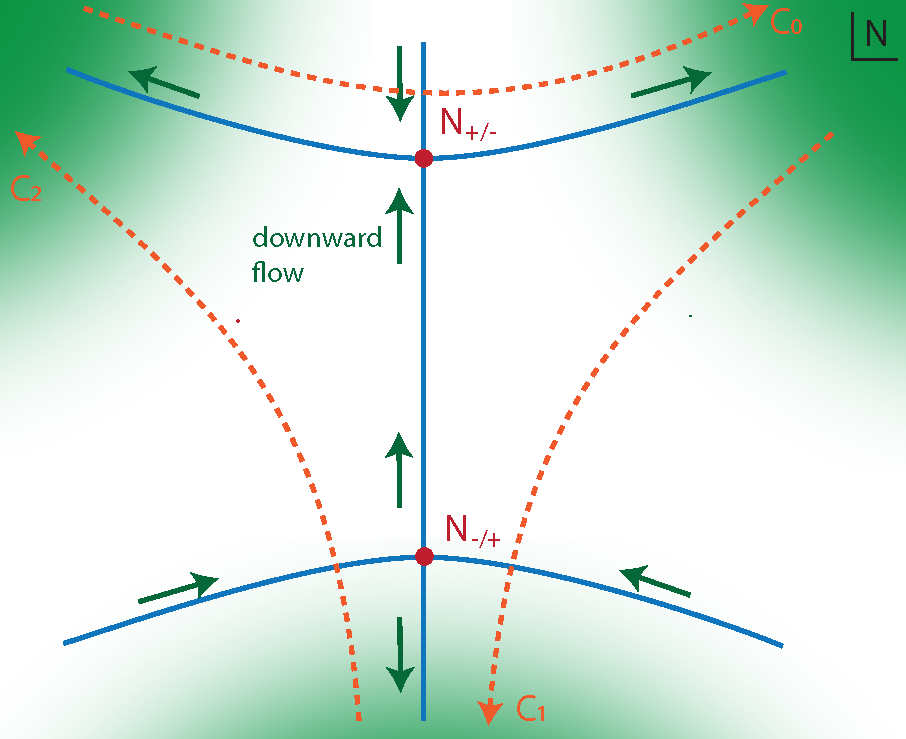}
	\includegraphics[width=0.4\textwidth]{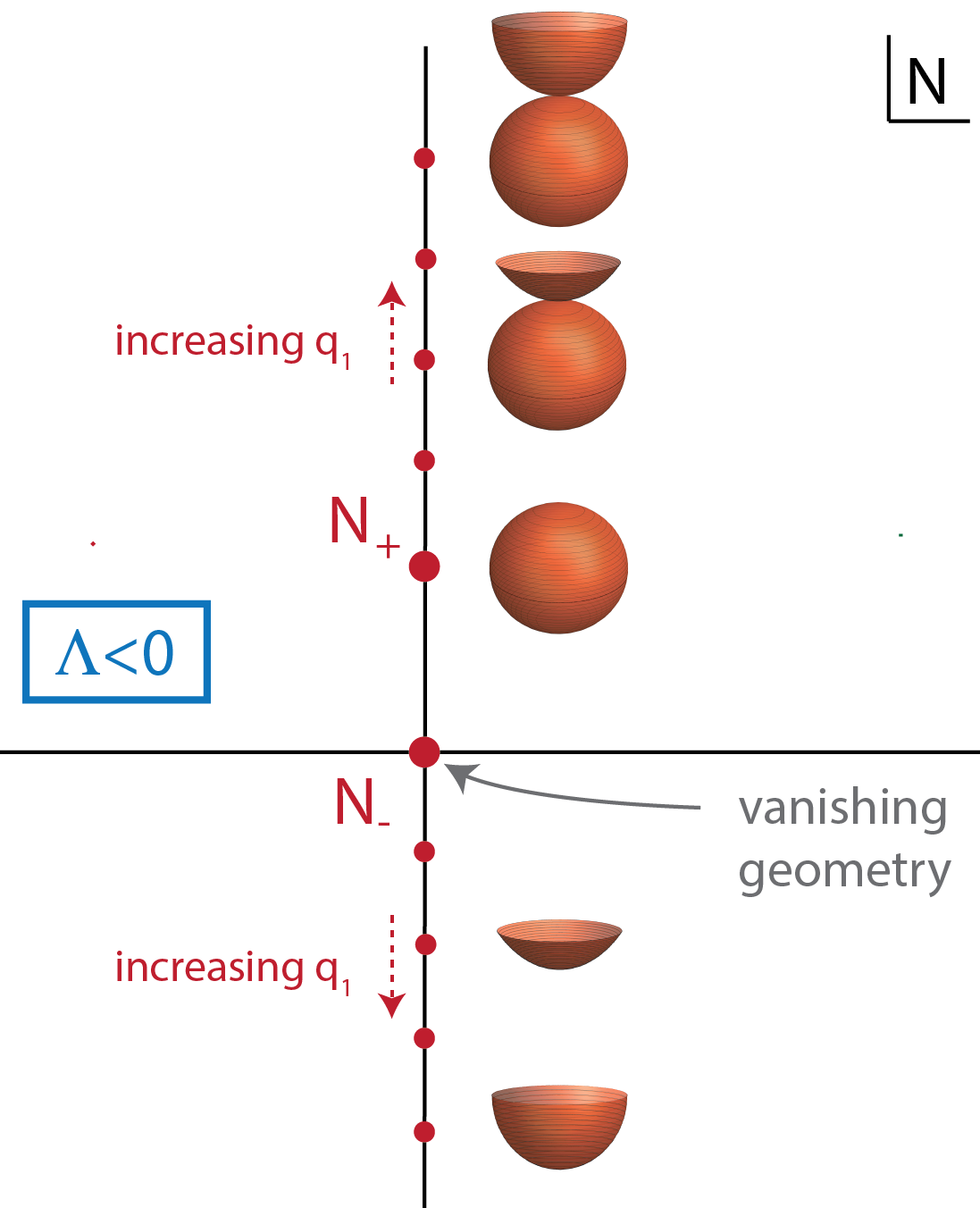}
	\caption{{\it Left panel:} Saddle points and steepest descent lines, for the case of AdS spacetimes (all $q_1=R_3^2$) or for dS (small universes with $q_1<\frac{3}{\Lambda}$), in the complexified plane of the lapse $N$. Green regions indicate asymptotic convergence at angles $0<\theta<\frac{\pi}{3},$ $\frac{2\pi}{3}<\theta<\pi$ and $\frac{4\pi}{3}<\theta < \frac{5\pi}{3}$. When $\Lambda<0,$ the upper saddle point is $N_+$ and the lower $N_-,$ and vice versa for $\Lambda>0.$ {\it Right panel:} Geometry of the saddle points for $\Lambda <0$. As the final scale factor $q_1=R_3^2$ is increased, the saddle points move apart. Figures reproduced from \cite{Lehners:2021jmv}.}
	\label{PLneg}
\end{figure}

We can restrict $\alpha$ further. When $\alpha=+i,$ the saddle point $N_-$ corresponds to Euclidean AdS space. The second saddle point, $N_+,$ also describes a section of complexified AdS space, but this time it includes a piece of reversed-signature Euclidean AdS glued onto a part of Euclidean AdS. In other words, this geometry contains a point where the scale factor passes through zero. Once perturbations are included, we expect them to blow up at this point. Therefore, this saddle point can actually be treated as being singular. Meanwhile, when $\alpha=-i$ the situation is reversed and the two geometries are exchanged. The flow lines in Fig.~\ref{PLneg} indicate that $N_-$ is always the dominant saddle point. Thus we find that we must choose $\alpha=+i$ in order for the dominant geometry to be regular and closed. This is in fact the same condition that we imposed in the no-boundary case, with positive cosmological constant, in \eqref{momcond} (and it is the choice that leads to stable perturbations). Thus we see that the momentum condition is vindicated by the AdS calculation.

But there is more. The only integration contour for the lapse function that completely projects out the singular saddle point is the combination of contours ${\mathcal C}_2 - {\mathcal C}_1,$ see Fig.~\ref{PLneg}. This combination in fact results in an Airy Bi function, of the form \cite{DiTucci:2020weq}
\begin{align}
    \Psi(R_3) \propto  \, Bi\left[ \left(\frac{3\pi}{4 G \hbar l} \right)^{\frac{2}{3}} \left( R_3^2+l^2\right)\right] \,. \label{Z3result}
\end{align}
In AdS one has to add counter terms to regulate the volume divergence (since the volume contributes to the weighting of the wave function rather than to the phase, as in the dS case). When these terms are added, one finds complete agreement with the CFT result, which is known \cite{Fuji:2011km,Marino:2011eh}, and is also given by an Airy function. Since this takes us away from the cosmological context of interest, we refer to \cite{Caputa:2018asc,DiTucci:2020weq} for detailed discussions. 

For us, two things are worth noting: the first is that the required contour of integration is neither Lorentzian nor Euclidean, but fully complex. The ``average'' of the two paths ${\mathcal C}_2,\, - {\mathcal C}_1$ is the Euclidean contour, but the individual contributions are fully complex. Thus, even though the wave function is real, we are forced to sum over complex metrics. The second point is that we can use the AdS result and analytically continue it to positive values of the cosmological constant. {\it A priori,} it is not clear that this is justified, as there is no understanding what the intermediate complex values of the cosmological constant might mean. However, mathematically we can simply perform the continuation to see what we obtain. Reinstating $\Lambda,$ we can use the following formula
\begin{align}
Bi\left[ \left(\frac{18\pi^2}{-\hbar\Lambda}\right)^{2/3}\left(1-\frac{\Lambda}{3}q_1\right)\right] =  \sqrt{3} \, Ai\left[ \left(\frac{18\pi^2}{\hbar\Lambda}\right)^{2/3}\left(1-\frac{\Lambda}{3}q_1\right)\right]\,,
\end{align}
which shows that for positive $\Lambda$ the wave function is better thought of as being proportional to an $Ai$ rather than a $Bi$ function. When this calculation is done in full \cite{Lehners:2021jmv}, one recovers precisely the result \eqref{WdWAirysol}.

\begin{figure}[ht]
	\centering
	\includegraphics[width=0.7\textwidth]{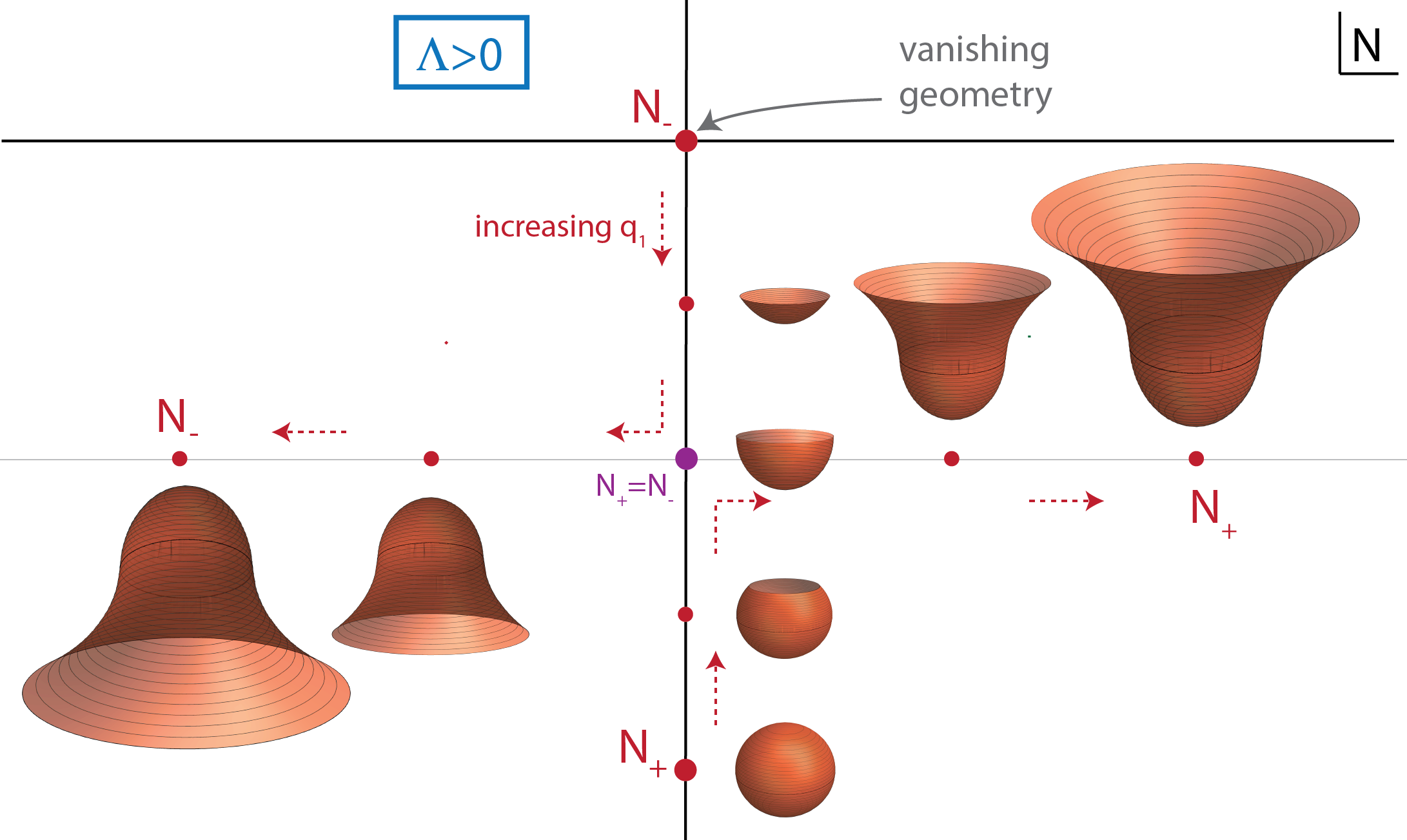}
	\caption{This graph shows the geometry of the saddle points when $\Lambda >0.$ At small scale factor $q_1 \leq \frac{3}{\Lambda}$ they are Euclidean, but then turn complex for $q_1 > \frac{3}{\Lambda}$, with increasingly large Lorentzian dS sections. Figure reproduced from \cite{Lehners:2021jmv}.}
	\label{saddlespos}
\end{figure}

Let us mention a few more properties of the Airy function, which may clarify some aspects of the calculation performed in section \ref{sec:minisuper}. When the final radius of the universe is small, $q_1 < \frac{3}{\Lambda},$ then the saddle points are Euclidean (we are still in the nucleation phase), and the flow lines are essentially identical to the AdS case in Fig.~\ref{PLneg}. But to obtain the $Ai$ result rather than $Bi,$ the contour must be different, and this time the required contour is  ${\mathcal C}_0,$ which picks up the upper saddle point only. In the dS case, this is the geometry that is regular and does not have the scale factor passing through zero, see also Fig.~\ref{saddlespos}. Note that this means that at the nucleation of the universe, when $q_1=0,$ only the vanishing geometry contributes and not the full sphere. In a sense this indicates that non-trivial topologies do not contribute to the no-boundary wave function, even though the wave function is non-zero when $q_1=0$ -- the latter property was originally interpreted as suggesting that non-trivial topologies must be the reason why $\Psi(0)\neq 0$ \cite{Hartle:1983ai}. Here we see that an explicit calculation does not support this interpretation, and a better explanation should be searched for.

\begin{figure}[ht]
	\centering
	\includegraphics[width=0.49\textwidth]{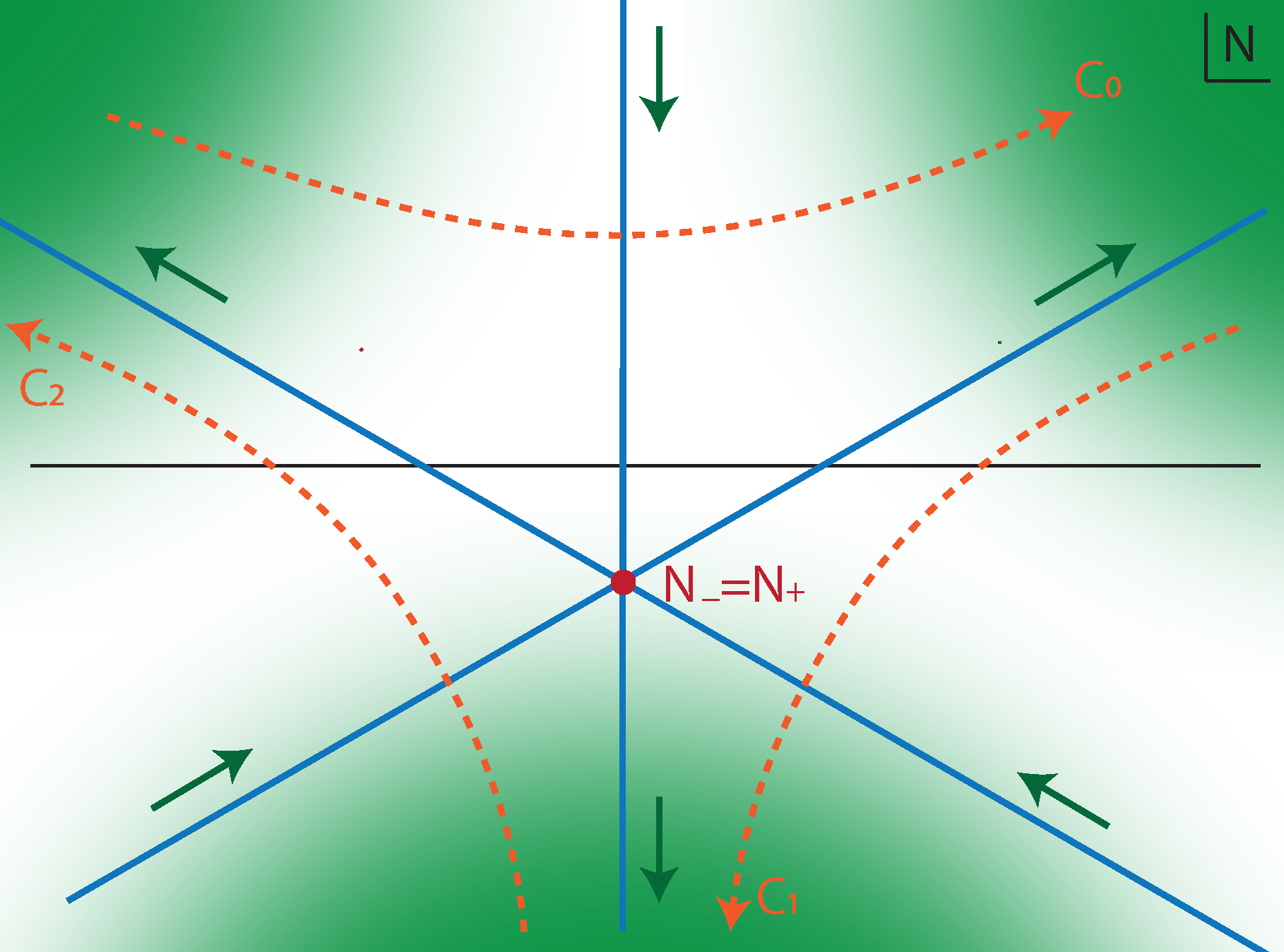}
	\includegraphics[width=0.49\textwidth]{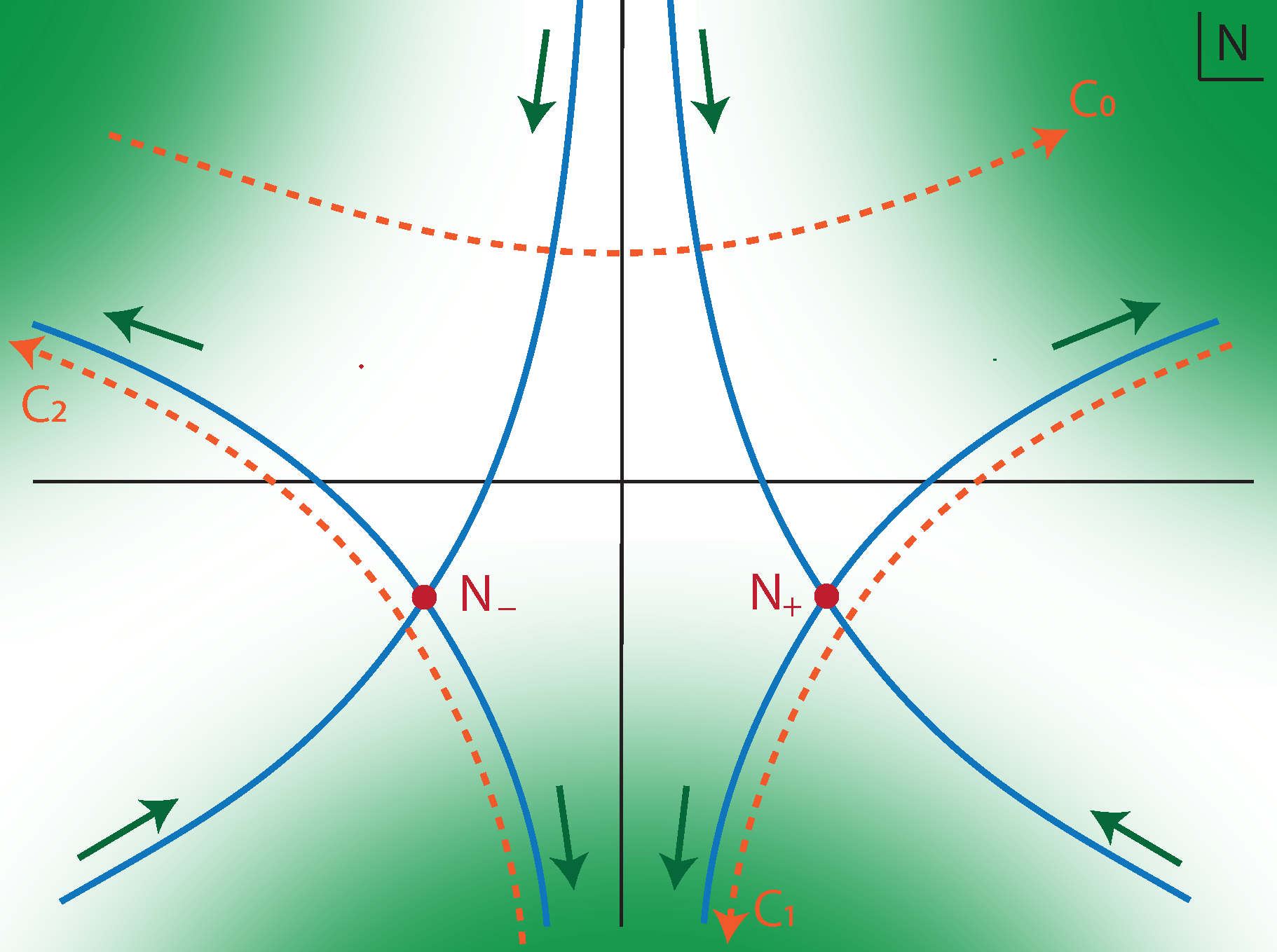}
	\caption{Saddle points and steepest descent lines when $q_1=\frac{3}{\Lambda}$ (left panel) and when $q_1 > \frac{3}{\Lambda}$ (right panel). A Stokes phenomenon occurs, and for large scale factors there are two saddle points contributing to the path integral. This Stokes phenomenon is related to the appearance of time. Figure reproduced from \cite{Lehners:2021jmv}.}
	\label{PLdegenerate}
\end{figure}

As the universe grows, the saddle points approach each other until they coalesce at $q_1=\frac{3}{\Lambda},$ see Fig.~\ref{PLdegenerate}. At larger values of the final scale factor, $q_1>\frac{3}{\Lambda},$ this degenerate saddle point splits into two saddles, which are now both relevant to the path integral. Thus a Stokes phenomenon has occurred, and from having a single Euclidean relevant saddle point we have gone to having two complex geometries that are equally dominant. These complex saddle points contain a Lorentzian section near the final boundary, that is to say time has emerged. The wave function still remains real, as it is a sum over two complex conjugate contributions, one from each complex saddle point. Thus, the overall wave function remains timeless, yet the two complex saddles each contain time -- in a sense time flows in opposite directions in both saddles. It remains an open question whether there could have been any significant interference effects between these two saddle points in the very early stages of the universe.

More insights can be learned by including black holes in the AdS calculation. The calculation is relatively lengthy and we will simply mention the results here -- for details see \cite{DiTucci:2020weq}. Euclidean Schwarzschild-AdS black holes are described by the line element
\begin{align}
\mathrm{d}s^2 = \frac{\mathrm{d}\rho^2}{\left( \frac{\rho^2}{l^2} + 1 - \frac{2M}{\rho}\right)} + \left( \frac{\rho^2}{l^2} + 1 - \frac{2M}{\rho}\right) \mathrm{d}\tau^2 + \rho^2 \mathrm{d}\Omega_2^2\,, \label{AdSbhmetric}
\end{align}
where $M$ denotes the mass of the black hole. The horizon radius $r_+$ is given by the real root of $\frac{\rho^3}{l^2} + \rho -  2M=0.$ One can then invert this relation to obtain the mass as a function of the horizon size
\begin{align}
M=\frac{1}{2} r_+ \left( 1 + \frac{r_+^2}{l^2} \right)\,. \label{mass}
\end{align}
The Euclidean manifold contains a conical singularity at $\rho=r_+$ unless one periodically identifies the $\tau$ coordinate, with period \cite{Hawking:1982dh}
\begin{align}
\beta = \frac{4\pi l^2 r_+}{3r_+^2+l^2}\,. \label{horizon}
\end{align}
AdS/CFT then relates the mass $M$ in \eqref{mass} to the expectation value of the Hamiltonian of a conformal field theory in a thermal state at an inverse temperature~$\beta$~\cite{Balasubramanian:1999re}.

The path integral can be performed by using a suitable choice of metric. The appropriate variables were found in \cite{Halliwell:1990tu}, and read
\begin{align}
\label{eq.bhmetric}
\mathrm{d}s^2 = \frac{c(r)}{b(r)} \mathrm{d} \tau^2 - \frac{b(r)}{c(r)}N^2 \mathrm{d}r^2 + b^2(r) \mathrm{d}\Omega_2^2\,.
\end{align} 
There is a radial direction $r$ with $0 \leq r \leq 1,$ and the spatial slices have the topology $S^1 \times S^2.$ As ``outer'' boundary conditions, we must fix the size of the spatial slices to be
\begin{align}
b(r=1) \equiv R_2\,, \qquad \sqrt{\frac{c(r=1)}{b(r=1)}}\Delta \tau \equiv R_1\,. \label{Dr1}
\end{align}
On the inner boundary one can impose a regularity condition $\omega$ (which is a condition on $\dot{c}/(Nb)(r=0)$) that implements the periodicity \eqref{horizon}. 

\begin{figure}
\begin{center}
\includegraphics[width=0.55\textwidth]{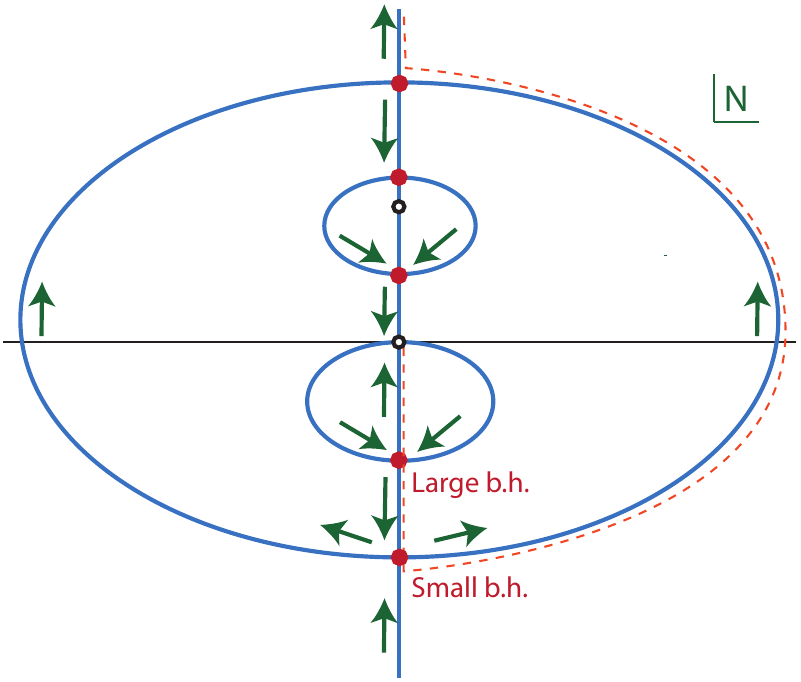}
\end{center}
\caption{Saddle points and steepest descent contours for $R_1=12, R_2 = 10, l=1.$ Arrows indicate directions of descent. There are two singularities indicated by the small circles, one at the origin and one on the positive imaginary axis. The dashed orange line is a possible contour of integration, capturing the large black hole saddle, but also including the small black hole and a further subdominant saddle point. Figure reproduced from \cite{DiTucci:2020weq}.} \label{fig:bh2}
\end{figure} 

The path integral/partition function then becomes
\begin{align}
  Z(R_1,R_2) =  \int_\omega \int \mathrm{d}N e^{\frac{i}{\hbar}\left(S_{ND}(N) - S_{EAdS}\right)}\,,
\end{align}
where the background Euclidean AdS action is subtracted to regulate the volume divergence. The path integral admits five saddle points, whose nature and locations depend on the final boundary conditions \cite{DiTucci:2020weq}. An illustrative example is given in Fig.~\ref{fig:bh2}. Two of the saddle points describe black holes, one large and one small. There are three additional saddle points that describe Euclidean geometries -- these are subdominant, or irrelevant, depending on the contour of integration. Their CFT counterpart is currently unknown. A possible integration contour for the lapse is shown in the figure. As one can see, it is again necessarily complex, and to obtain a real wave function (as expected in the CFT) one should sum this contour with its reflection across the imaginary lapse axis. The large black hole is always found to dominate over the small black hole, which agrees with thermodynamic expectations \cite{Hawking:1982dh}.

The partition function is interpreted as representing the canonical ensemble at fixed temperature $1/\beta,$ given that the size of the outer boundary is kept fixed. One obtains
\begin{align}
\ln Z  = \frac{R_2}{T \emph{l}^2_P}\left(\sqrt{1+ \frac{R_2^2}{l^2}-\frac{2M}{R_2}} - \sqrt{1+ \frac{R_2^2}{l^2}} \right) + \frac{\pi r_+^2}{\emph{l}^2_P}\,.
\end{align}
where $\emph{l}_P = \sqrt{\frac{G \hbar}{c^3}}$ is the Planck length. Using standard thermodynamic relations \cite{Brown:1994gs} one can calculate the expectation value of the energy 
\label{energy}
\begin{align}
\langle E \rangle =k_B T^2 \frac{\partial \ln Z}{\partial T} = \frac{k_B R_2}{\emph{l}^2_P} \left(\sqrt{1+ \frac{R_2^2}{l^2}} - \sqrt{1+ \frac{R_2^2}{l^2}-\frac{2M}{R_2}} \right) 
\end{align}
as well as the entropy
\begin{align}
{\cal S} & =k_B \ln Z + \frac{\langle E \rangle}{T} =  \frac{k_B}{\emph{l}^2_P} \, \pi r_+^2 = \frac{k_B}{\emph{l}^2_P} \, \frac{Area}{4} \label{entropy}\,.
\end{align}
These relations satisfy the Quantum Statistical Relation \cite{Gibbons:1976ue}
\begin{align}
- k_B T \ln Z = \langle E \rangle - T{\cal S}\,.
\end{align}
Now comes a crucial point: if we had imposed a Dirichlet condition at $r=0,$ instead of the regularity condition $\omega,$ then we would have obtained an additional surface term of magnitude $\pi r_+^2$ at the horizon $r_+,$ and the partition function would have come out as
\begin{align}
    -k_B \ln Z \approx \frac{\langle E \rangle}{T} - {\cal S} + {\cal S} = \frac{\langle E \rangle}{T}\,.
\end{align}
This would have corresponded more closely to the microcanonical ensemble, in which one considers states at fixed energy. However, this does not agree with the fact that we kept the size of the outer boundary fixed, which corresponds to fixing the temperature. Hence, a proper physical interpretation of the AdS black hole calculation requires the absence of a surface term on the inner boundary. By direct analogy, this provides further support for the implementation of the no-boundary wave function with a momentum condition. 

To summarise, we have found that AdS path integrals share many common features with the no-boundary proposal, in particular the absence of a surface term and the ensuing regularity/momentum condition at the ``no-boundary'' point; the sign of the momentum condition corresponding to the standard Wick rotation of quantum field theory; a sum over all field values implementing the momentum condition; and a sum over complex (that is to say neither Lorentzian nor Euclidean) metrics.\\

\noindent {\it A Holographic Definition}

The previous discussion was rather conservative in that it tried to compare no-boundary path integrals with similar integrals performed in the context of AdS/CFT, and this showed that surprisingly similar features arise in the two settings. However, there has also been a program, initiated by Hertog and Hartle in \cite{Hertog:2011ky} and developed in \cite{Hartle:2012qb,Hartle:2012tv,Hartle:2013vta,Hertog:2015nia,Conti:2017pqc}, to \emph{define} the no-boundary wave function more directly in terms of AdS/CFT (earlier ideas in this direction include \cite{Hawking:2000kj,Horowitz:2003he,Ooguri:2005vr}). We will review this proposal here.

\begin{figure}[ht] 
\begin{center}
		\includegraphics[width=0.6\textwidth]{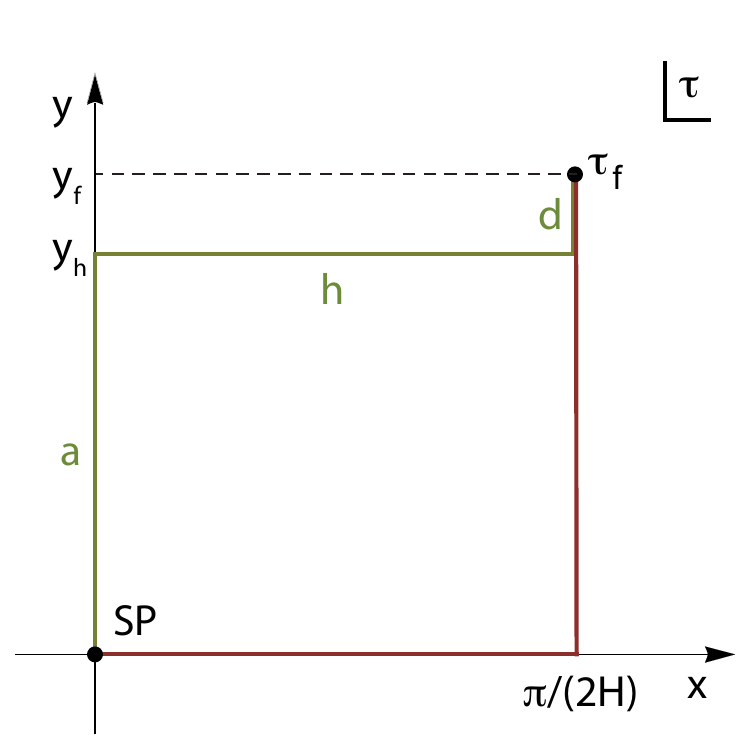}
		\end{center}
	\caption{Time contours used to represent the de Sitter instanton: the standard Euclidean-plus-Lorentzian contour in red, and the ``holographic'' contour in green.}
	\label{fig:contHH}
\end{figure} 

To understand the proposal, we must return to the de Sitter saddle point geometry, as described at the beginning of section \ref{sec:inflex}. In terms of Euclidean time $\tau$ and with $\Lambda=3H^2,$ the dS solution is given by
\begin{align}
    \mathrm{d}s^2 = \mathrm{d}\tau^2 + \frac{1}{H^2}\sin^2(H\tau) \mathrm{d}\Omega_3^2\,.
\end{align}
The usual contour used to represent this solution runs from the South Pole in the Euclidean direction to the equator of the $4-$sphere at $\tau=\pi/(2H),$ followed by a segment in the Lorentzian direction defined by $\tau=\pi/(2H) + i y$ with $0 \leq y \leq y_f$ along which the metric is that of Lorentzian dS
\begin{align}
    \mathrm{d}s^2 = -\mathrm{d}y^2 + \frac{1}{H^2}\cosh^2(Hy) \mathrm{d}\Omega_3^2\,. \label{dmetric}
\end{align}
The scale factor then reaches the final value $b=\frac{1}{H}\cosh(Hy_f)$ at time $\tau_f=\pi/(2H) + i y_f,$ {\it cf.} also Fig.~\ref{fig:contHH}.

But the physical consequences are unchanged if we deform the time contour, as long as no singularities are present. In particular, the value of the action of the saddle point (which is the quantity that enters the semi-classical wave function) does not change. Consider then the contour marked in green in Fig.~\ref{fig:contHH}, and which we will refer to as the holographic contour. The first segment, labelled ``$a$'', starts at the South Pole in a Lorentzian direction, for $\tau=iy$ with $0 \leq y \leq y_h.$ Along this contour the metric reads
\begin{align}
    \mathrm{d}s^2 = -\mathrm{d}y^2 - \frac{1}{H^2}\sinh^2(Hy) \mathrm{d}\Omega_3^2\,.
\end{align}
This is the metric of Euclidean AdS (EAdS) spacetime, with cosmological constant $-\Lambda,$ except that there is an overall minus sign in the metric, {\it i.e.} the signature has been reversed. The scale factor reaches a final value that is imaginary, with magnitude $b_h = \frac{1}{H}\sinh(Hy_h).$ The second segment, labelled ``$h$'', then interpolates horizontally between the EAdS region and the Lorentzian dS region along $\tau = x+ i y_h$ with $0 \leq x \leq \frac{\pi}{2H}.$ The metric is fully complex along this segment. Finally, on the last segment marked ``$d$'', the metric is that of Lorentzian dS spacetime, see Eq. \eqref{dmetric}. The final value of the scale factor is $b=\frac{1}{H}\cosh(Hy_f).$ It is larger than the magnitude of the scale factor $b_h$ reached along the EAdS part of the contour.

It is straightforward to evaluate the action along the various segments. For this we may use the action \eqref{eq:miniact} with $\tilde{N}$ appropriately chosen along the different segments, to find
\begin{align}
    S_a &= -i \frac{4\pi^2}{H^2} \left[1-\cosh^3(Hy_h) \right]\,, \\
    S_h &=  -i \frac{4\pi^2}{H^2} \left[\cosh^3(Hy_h) -i \sinh^3(Hy_h)\right]\,, \\
    S_d &= \frac{4\pi^2}{H^2} \left[\sinh^3(Hy_h) - \sinh^3(Hy_f)\right]\,.
\end{align}
The sum of these actions of course recovers the standard result for dS saddle points, {\it cf.} Eq. \eqref{actiondSsaddle},
\begin{align}
    S_{total} = S_a + S_h + S_d = \frac{4\pi^2}{H^2} \left[-i - \sinh^3(Hy_f)\right]\,.
\end{align}
But the holographic contour suggests a new interpretation based on the actions along the different segments. 

First consider the action along the first segment. It diverges in the large $y_h$ limit, but since the geometry along this part of the contour is EAdS, this is in fact expected. In AdS/CFT one adds counter terms to cancel the volume divergence. These counter terms are constructed from geometrical quantities involving solely the boundary metric. In $4$ dimensions, these are given by \cite{Balasubramanian:1999re,deHaro:2000vlm}
\begin{align}
    S_{ct}(y_h) = i \int \mathrm{d}^3y \sqrt{-h} \left( 2H + \frac{1}{2H}R^{(3)}\right) = i \frac{4\pi^2}{H^2} \left[ \sinh^3(Hy_h) + \frac{3}{2}\sinh(Hy_h)\right]
\end{align}
These cancel the volume divergence and keep the action finite in the $y_h \to \infty$ limit. In fact, if we define the regulated action via $I_{AdS}^{reg}(b_h) \equiv \frac{4\pi^2}{H^2},$ then the result along the $a$ contour can be rewritten as
\begin{align}
    S_a = -i I_{AdS}^{reg} + S_{ct}(y_h) + {\cal O}(e^{-Hy_h})\,.
\end{align}
Next note that the total weighting is given precisely by the regulated EAdS weighting
\begin{align}
    \textrm{Im}(S_{total}) = -I_{AdS}^{reg}\,.
\end{align}
Thus one can say that the requirement to reach the classical part $d$ of the contour regulates the divergence automatically, as the horizontal connecting part precisely implements the counter terms, up to terms that vanish in the large volume limit, 
\begin{align}
    i\textrm{Im}(S_h) = - S_{ct}(y_h) + {\cal O}(e^{-Hy_h})\,.
\end{align}
In the end, one can write the total action as
\begin{align}
    S_{total} = -iI_{AdS}^{reg}(b_h) +iS_{ct}(y_f) + {\cal O}(e^{-Hy_f})\,, \label{actionrewrite}
\end{align}
where the counter terms are now evaluated at the final point $\tau_f = \frac{\pi}{2H} + iy_f.$ As seen from the final dS part of the contour, the counter terms in fact provide the phase of the wave function, which is responsible for the classical evolution. By contrast, the weighting, which remains constant as $y_f$ evolves to ever later times, is determined by the EAdS part of the contour alone.

This rewriting of the action now suggests a new definition of the no-boundary wave function. The idea is to use the EAdS part of the contour to relate the wave function to the partition function of a dual quantum field theory. (It is not a conformal field theory as the boundary resides at a finite value). From the Euclidean version of AdS/CFT one expects that in the supergravity limit
\begin{align}
    e^{-I_{AdS}^{reg}(b_h,\chi_h)} = Z_{QFT}(b_h,\chi_h)\,, \label{EAdSQFT}
\end{align}
where the correspondence can be extended to include matter fields, here indicated in the form of a scalar field with boundary value $\chi_h.$ (The precise definition of the dual quantum field theory depends on the matter content of the gravitational theory, see \cite{Conti:2017pqc} for an example.) Putting \eqref{actionrewrite} and \eqref{EAdSQFT} together, we arrive at the proposal for a holographic no-boundary wave function \cite{Hertog:2011ky},
\begin{align}
    \Psi(b,b_h,\chi,\chi_h,\zeta)=\frac{1}{Z_{QFT}(b_h,\chi_h,\zeta)}e^{\frac{i}{\hbar}S_{ct}(b,\chi)}\,. \label{holonbwf}
\end{align}
It is implicitly understood that the final EAdS scale factor $b_h$ and the final dS scale factor $b$ are related via the asymptotic equations of motion, at the saddle point. When a scalar field is added, the contour is slightly shifted to smaller Euclidean times, in analogy with the examples described in section \ref{sec:inflex}. However, the general analysis proceeds in complete analogy with that of pure dS \cite{Hertog:2011ky}, and thus we will not spell it out here.

The definition \eqref{holonbwf} is semi-classical in nature, as it builds on the supergravity limit of the gravitational side of the theory, and moreover implicitly assumes the validity of the saddle point equations of motion. Related to this is the fact that the definition naturally includes a cut-off labelled $\zeta$ above. Consider for instance linearised perturbations around the cosmological backgrounds, as described in section \ref{sec:perts}. There we saw that these only behave classically when they have been stretched to super-Hubble scales, the requirement on the wave number is roughly  $l \lessapprox Hb.$ This suggests that there is a length scale $\zeta \sim 1/(Hb)$ below which we simply do not treat the spacetime as classical. Put differently, the spacetime is coarse-grained on those scales. When the universe is still small, the coarse-graining is significant. As it grows, $\zeta$ shrinks and more and more modes have become classical. In the infinite $b$ limit we obtain the most fine-grained description. In the dual QFT, $\zeta$ becomes a high energy cut-off, specifying the energy scale above which modes are to be integrated out. In this way, the cosmological evolution becomes related to (inverse) renormalisation group flow of the QFT, realising an idea suggested in \cite{Strominger:2001pn,Strominger:2001gp}.

The main advantage of the holographic definition is that it provides a direct implementation of the no-boundary proposal in string theory. It relies on the expectation that the no-boundary condition is ``universal'', applying generally in gravitational theories, and being implied by dual quantum field theories. Several nice properties emerge, in particular that the counter terms need not be put in by hand, but rather arise from the requirement for classicality. Also, as we have just described, cosmological evolution and renormalisation group flow become linked.  

Many open questions remain, which offer multiple avenues for further analysis. A general question is whether the proposal is not only conceptually attractive, but whether the dual quantum field theory leads to new observational predictions for early universe cosmology. Another general question is whether signature reversal of metrics is admissible in quantum gravity. We will discuss this issue in a little more detail in section \ref{sec:allowability}. Somewhat related to this is the question of how to include generic matter fields. For vector fields, for example, it seems that the sources on the QFT side must be complex \cite{Mithani:2013ed,Hartle:2013vta}. Is this consistent from the QFT point of view? Also, the EAdS part of the gravitational theory has the inverse potential to the potential in the dS part. Thus, when the AdS potential contains additional fields with positive, stable potentials, these turn into unstable directions in the cosmological part. Even if such unstable modes are set to be zero at background level, do their fluctuations and couplings to other fields cause instabilities \cite{Hartle:2013vta}? Another question concerns the fact that the EAdS and dS parts of the contour are separated (in the example above, these were the $a$ and $d$ segments, respectively). Is this separation guaranteed, and unique? Should the wave function not depend solely on measurable quantities, and thus not depend on the (fiducial) EAdS part of the contour? This concern would be eliminated if one could show how to uniquely determine the EAdS part ($b_h,\chi_h$) from the arguments ($b,\chi$), in general. And how does this proposal work for non-inflationary potentials, for example for ekpyrotic ones for which, as we saw in section \ref{sec:ekex}, no-boundary solutions also exist, yet are markedly different? Opportunities for further research abound.

\subsection{A Filter on the Landscape} \label{sec:filter}

String theory predicts the existence of additional spatial dimensions. In order to be compatible with observations, these extra dimensions should either be sufficiently small in volume \cite{Candelas:1985en}, or be highly curved \cite{Randall:1999vf,Crampton:2014hia}, so that gravity appears $4-$dimensional at observationally accessible scales. The volume and shape of the additional dimensions determine the features of the observable universe, in particular the nature of fundamental forces and their coupling constants. Since coupling constants have not been measured to vary over the currently probed history of the universe \cite{Uzan:2002vq}, a further requirement is that the additional spatial dimensions must be stable and essentially non-evolving over the last $13.8$ billion years. This leads to immediate questions of cosmological relevance: what determines the size and shape of extra dimensions? How are compactified spacetimes created in the first place? Can the compactification change over time?

Added to these questions is the obvious question of whether suitable compactifications, giving rise both to realistic particle physics and cosmological evolution, exist. This turns out to be a far harder question than initially thought. Even though myriads of solutions of string theory were conjectured \cite{Taylor:2015xtz}, it turned out that constructing concrete examples is severely hampered by general quantum gravitational consistency conditions, known as swampland constraints \cite{Palti:2019pca}. In particular, it is thought to be impossible to find de Sitter solutions in perturbative string theory \cite{Obied:2018sgi}. A more realistic goal is to search for inflationary potentials, but even those are hard to find \cite{Garg:2018reu,Agrawal:2018own} (and likewise, ekpyrotic potentials may be just as difficult to construct \cite{Lehners:2018vgi}). The most promising examples to date require a careful balancing of perturbative and non-perturbative effects \cite{Kachru:2003aw}, with many approximations that remain debated (for recent criticism, see {\it e.g.} \cite{Lust:2022lfc}).  

Still, it seems plausible that non-perturbative string theory contains solutions that undergo accelerated expansion (in fact, it is necessary if string theory is to be compatible with the current era of dark energy domination that we find ourselves in). We will proceed on the assumption that this is the case. Then, in these solutions, the shape and volume of the extra dimensions will be described by many so-called moduli fields, which can be thought of as the parameters of the solutions. These must be stabilised, which can be achieved for example with the inclusion of non-trivial flux fields in the internal dimensions \cite{Sethi:1996es,Giddings:2001yu}. 

If we then ask which compactifications are preferred, we are really asking which values of the moduli fields are preferred. In turn, this corresponds to searching for a probability distribution over the ingredients of the compactifications, {\it i.e.} over fluxes, branes, orientifolds {\it etc.} In this section, we would like to analyse whether the no-boundary wave function could provide precisely such a probability distribution. In other words, we are asking whether the no-boundary wave function can act as a vacuum selection principle, determining which kinds of universes are likely to be created from nothing, and which are unlikely \cite{Hawking:2006ur,Hartle:2010dq}. In addition, we must keep in mind the possibility of transitions within an existing universe, to another vacuum/compactification, {\it e.g.} via nucleation of membranes \cite{Brown:1988kg}. This certainly seems possible, but is always suppressed since it is a non-perturbative process. We will restrict our attention here purely to the creation phase of space, time and matter.

In fact, since not much detailed knowledge exists about the landscape of realistic string theory solutions, all we can do is study a toy model \cite{Lehners:2022mbd}. Still, this serves to illustrate how a probability distribution over compactifications might ultimately arise.

\begin{figure}[ht]
	\centering
	\includegraphics[width=0.45\textwidth]{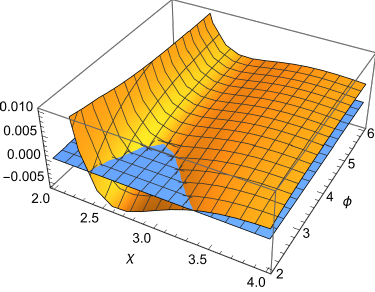}
	\includegraphics[width=0.45\textwidth]{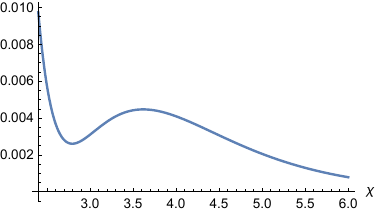}
	\caption{The potential \eqref{twofpot} for $\tilde\alpha=1$ and $n_4=13.$ {\it Left panel:} Two-field potential, with $V=0$ indicated in blue for reference. {\it Right panel:} A slice of the potential at $\phi=6.$ One can see that there is the possibility for $\chi$ to be stabilised at $\chi_{min} \approx 2.8.$ Figures reproduced from \cite{Lehners:2022mbd}.}
	\label{fig:8dpot}
\end{figure}

We will use a toy model that is defined in $8$ dimensions, and includes a non-perturbative $R^4$ correction term \cite{Ketov:2017aau,Otero:2017thw}. This term leads to an inflationary potential, very much in analogy to the Starobinsky model in $4$ dimensions \cite{Starobinsky:1980te}. Moreover, the model includes $4-$form flux. Both features are known to arise in $11-$dimensional supergravity \cite{Cremmer:1978km,Green:1997as}. The action is given by
\begin{align}
    S &= \frac{1}{2}\int \mathrm{d}^{8} x \sqrt{-\hat{g}}\left(\hat{R}+ \alpha \hat{R}^4 - \frac{1}{2\cdot 4!}e^2 F_{(4)}^2\right) \,,
\end{align}
where $e$ is a coupling constant. We then perform two steps: the first is to redefine the metric via a conformal transformation in order to go to Einstein frame. Specifically, we define $ \hat{g}_{\mu\nu} \equiv e^{\frac{2}{\sqrt{42}}\phi} g_{\mu\nu}$ with $e^{\sqrt{\frac{6}{7}}\phi} = 1+4\alpha \hat{R}^3.$ The second is to dimensionally reduce on a $4-$sphere, with $4-$form flux wrapping the sphere, in order to land in $4$ spacetime dimensions,
\begin{align}
    \mathrm{d}s_8^2 = e^{-\frac{2}{\sqrt{3}}\chi}\mathrm{d}s_4^2 + e^{\frac{1}{\sqrt{3}}\chi}\mathrm{d}\Omega^{2}_4\,, \quad F_{(4)} = 2 n_4 \textrm{vol}(S^4)\,.
\end{align}
The resulting theory in $4$ dimensions contains gravity with $2$ scalars and a potential,
\begin{align}
 S   &=  \frac{16\pi^4}{3} \int \mathrm{d}t  \left(-3\frac{a \dot{a}^2}{N}+\frac{a^3}{2N}\left( \dot\phi^2 + \dot{\chi}^2 \right) +3Na - N a^3 V(\phi,\chi)\right) \nonumber \\ & \quad + \left[3\frac{a^2\dot{a}}{N} -\frac{a^3\dot{\chi}}{\sqrt{6}N} \right]_{surface}\,, \\
 V(\phi,\chi) &= \tilde\alpha\left(1 - e^{-\sqrt{\frac{6}{7}}\phi} \right)^{\frac{4}{3}} e^{- \frac{2}{\sqrt{3}}\chi}  + n_4^2 e^{-2\sqrt{3}\chi} - 6 e^{-\sqrt{3}\chi}\,. \label{twofpot}
\end{align}
where $\tilde\alpha$ is a constant. The surface term on the final boundary is removed by the inclusion of a GHY boundary term. On the initial boundary, we do not add a surface term, as discussed several times in this review. However, it vanishes in any case when the saddle point geometry is compact, $a(t=0)=0.$ In the above, it is important that the flux on the $4-$sphere is quantised \cite{Henneaux:1986ht}, and thus $n_4$ is proportional to an integer, 
\begin{align}
    n_4 = \frac{2\pi}{2 e \, \textrm{vol}(S^4)}z = \frac{3}{8\pi e} z\,, \qquad z \in \mathbb{Z}\,.
\end{align}
The shape of the potential is shown in Fig.~\ref{fig:8dpot} for an interesting example of parameter values. One can see that it contains a valley at $\chi \approx 2.8,$ where $\chi$ and thus the size of the $4-$sphere can be stabilised. In the orthogonal $\phi$ direction, inflation can occur. The model is not fully realistic, as the inflationary phase eventually ends when the potential drops to negative values. However it is realistic enough to address the nucleation of universes.

\begin{figure}[ht]
	\centering
	\includegraphics[width=0.7\textwidth]{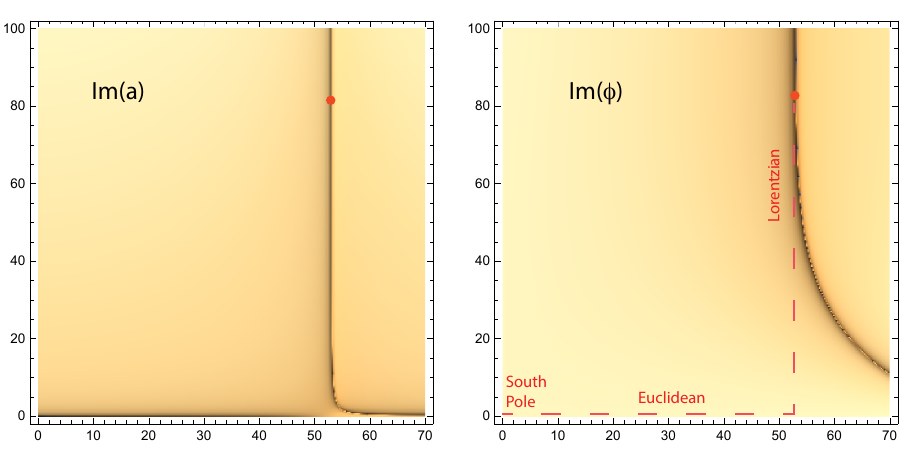}
	\caption{An example of a no-boundary instanton in the potential \eqref{twofpot}, with $\chi$ stabilised on the valley floor at $\chi \approx 2.8.$ The dark lines show the locus of real $a$ and $\phi$ values, and the red dot indicates the final time $\tau=53.185+83.538i,$ at which the fields reach the designated real values $a_1=200,\phi_1=6$. For this, the scalar field value has been tuned at the South Pole to the value $\phi_{SP}=6.1104-0.09991i.$ Figures reproduced from \cite{Lehners:2022mbd}.}
	\label{fig:ex8d2_1}
\end{figure}

\begin{figure}[ht]
	\centering
	\includegraphics[width=0.35\textwidth]{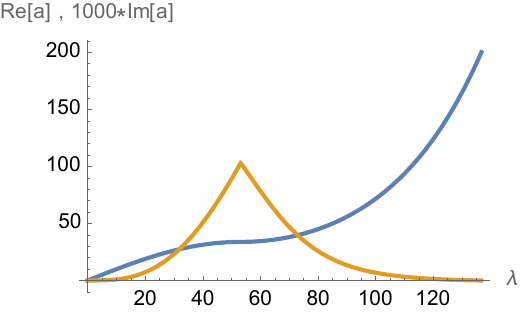}
	\includegraphics[width=0.35\textwidth]{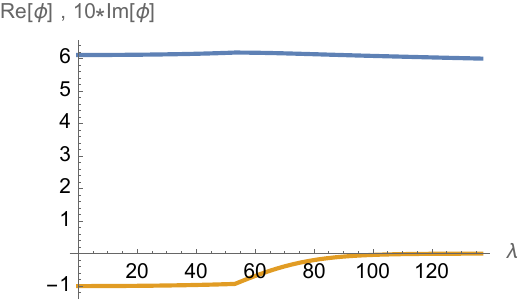}
	\caption{Field evolutions along the dashed line path indicated in Fig.~\ref{fig:ex8d2_1}. The real parts are shown in blue, and the imaginary parts (magnified for better visibility) in orange. Figures reproduced from \cite{Lehners:2022mbd}.}
	\label{fig:ex8d2_2}
\end{figure}

As we saw in section \ref{sec:inflex}, it is imperative that a dynamical attractor exists in order for no-boundary solutions to exist. Here the inflationary valley can play precisely this role. Thus no-boundary solutions are expected to exist, with $\chi$ stabilised and $\phi$ slowly rolling down the valley floor. Using the numerical techniques described in section \ref{sec:numtech}, this expectation is borne out -- an example is shown in Fig.~\ref{fig:ex8d2_1}, with the field values in Fig.~\ref{fig:ex8d2_2}. It has all the usual no-boundary characteristics, with the fields reaching a quasi-Lorentzian evolution at late times. Closely related solutions, with inflation lasting more or less long, can then be constructed with the same techniques \cite{Lehners:2022mbd}. In this potential, there also exist no-boundary solutions at large $\chi,$ {\it cf.} the inflationary slope shown in the right panel of Fig.~\ref{fig:8dpot}. However, for these solutions the sphere size modulus $\chi$ rolls to large values, and consequently the solutions decompactify and are not of phenomenological relevance. 

\begin{figure}[ht]
	\centering
	\includegraphics[width=0.6\textwidth]{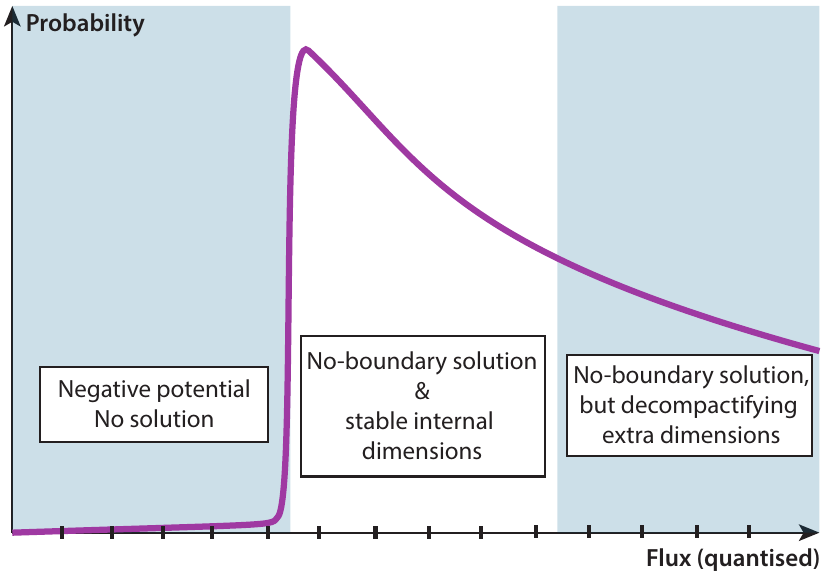}
	\caption{The no-boundary wave function implies a probability distribution over compactifications, linked here to a probability distribution over fluxes. In this example \cite{Lehners:2022mbd}, no-boundary instantons with stable internal dimensions are only possible for a narrow range of fluxes. Figure reproduced from \cite{Lehners:2022mbd}.}
	\label{fig:summary}
\end{figure}

What is interesting about this model is not just that inflationary no-boundary instantons, with stable internal dimensions, exist at all. The point is that they only exist for a range of values of the flux parameter $n_4.$ When $n_4$ is larger, the inflationary valley floor rises until the valley actually disappears, and only decompactifying solutions are left. And when $n_4$ is too small, the valley floor sinks to negative values of the potential, eliminating the possibility of an inflationary solution. In the present example, the viable range is found to be \cite{Lehners:2022mbd}
\begin{align}
    11.7 \lessapprox n_4 \lessapprox 15.1 \quad (\tilde\alpha=1)\,.
\end{align}
Note that the no-boundary solutions that exist in this range have the property that the $4-$sphere is present from the outset, {\it i.e.} the universe nucleates as a product of a fixed $4-$sphere with a geometry that starts at zero size and then grows into a Lorentzian quasi-dS spacetime. (This is a higher-dimensional analogue of the Nariai instanton described in section \ref{sec:blackholes}.) As usual, no-boundary solutions that exist for lower potential values come out as preferred. This means that in the present toy model, the no-boundary wave function implies a probability distribution of initial conditions, where the initial conditions include the parameters of the $8-$dimensional compactification \cite{Lehners:2022mbd}. This translates into a probability distribution over the fluxes, as sketched in Fig.~\ref{fig:summary}.

In more realistic compactifications, we expect many more parameters to be present. Once a better understanding of realistic compactifications becomes available, including both particle physics and cosmological aspects, it will be interesting to see how these different facets of the solutions influence each other. A crucial question will be whether universes like ours come out as being rather likely or unlikely. In particular, one will be able to investigate which kinds of effective lower-dimensional laws of physics are likely, and which not. Only those tied to a universe that can actually come into existence by virtue of its cosmological dynamics stand a chance of being assigned a high probability. In this way, as one can already see in embryonic form in the example described above, the micro and macro properties of the universe become intimately linked, and the no-boundary proposal can act as a filter on the possible higher-dimensional worlds.

\subsection{Allowable Metrics} \label{sec:allowability}

Throughout this review, we have seen that the no-boundary proposal is intimately connected with complex metrics. The usefulness of complex metrics became appreciated through the study of complex black hole metrics, which provide the quickest way of deriving (and to some extent, understanding) the thermodynamic properties of black holes \cite{Gibbons:1976ue}. However, not all complexified metrics make sense. Witten gives the example of zero-action wormholes, which would render tunnelling via wormholes just as likely as classical evolution, if they were permitted \cite{Witten:2021nzp}\footnote{These are easy to construct. Take flat space in polar coordinates $\mathrm{d}s^2=\mathrm{d}R^2 + R^2 d\Omega_3^2$ and promote $R \to R(u)$ for a real parameter $u.$ Now if $R(u)$ interpolates between asymptotic regions $R\to \pm \infty$ while avoiding $R=0$ by passing around the origin in the complexified $R$ plane, then this solution interpolates between two asymptotically flat regions of spacetime and describes a complex wormhole in between. But since the metric is simply obtained by a coordinate change from flat space, the Ricci curvature remains zero and so does the action.}. Hence, there must be some kind of criterion that tells us which complex metrics should be included, and which not. This is important, as in our examples in sections \ref{sec:inflex}, \ref{sec:ekex} and \ref{sec:minisuper} in particular, we saw that it is not only the saddle points that are complex, but the integration contours are typically also over complex metrics. Thus the definition of gravitational path integrals is potentially sensitive to any such allowability criterion.

Louko and Sorkin analysed this question in a simplified two-dimensional context in \cite{Louko:1995jw}, allowing complex metrics only when they admit a well-defined (that is to say, convergent) scalar field theory on them. In a similar vein, though independently, Kontsevich and Segal proposed to define quantum field theories on fixed complex backgrounds under the condition that the complex backgrounds allow for well-defined theories of arbitrary $p-$form matter fields \cite{Kontsevich:2021dmb}. The reason for highlighting $p-$forms of arbitrary rank is that these lead to local covariant stress-energy tensors \cite{Weinberg:1980kq}, and as such provide a rather general description of matter suitable for local quantum field theories. Let us briefly review the criterion here: the idea is to require a path integral over real valued $p-$form matter fields, with field strengths $F_{j_1 j_2 \cdots j_{p+1}},$ to converge. It is enough to focus on the kinetic terms, which already provide all of the necessary conditions. Thus, we require
\begin{align}
& |e^{\frac{i}{\hbar}S}|<1 \,\,\, \textrm{or} \,\,\, |e^{-\frac{1}{\hbar}I_E}|<1 \,\,\, \textrm{implying} \\  & Re\left[ \sqrt{g} g^{j_1 k_1} \cdots g^{j_{p+1} k_{p+1}} F_{j_1 \cdots j_{p+1}} F_{k_1 \cdots k_{p+1}}\right] > 0\,. \label{KS}
\end{align}
Pointwise, one can always write the metric in diagonal form
\begin{align}
g_{jk} = \delta_{jk} \lambda_j
\end{align}
where the $\lambda_j$ are now complex numbers. As an example, for $p=0$ and in $4$ dimensions, the condition \eqref{KS} becomes 
\begin{align}
-\pi < Arg(\lambda_1) +Arg(\lambda_2) + Arg(\lambda_3) + Arg(\lambda_4) <\pi\,.
\end{align}
For higher $p,$ some of the signs are flipped. Writing these conditions out for all $p,$ {\it i.e.} for all possible sign combinations, leads to the concise condition \cite{Kontsevich:2021dmb}
\begin{align}
\Sigma \equiv \sum_{j} |Arg(\lambda_{j})| < \pi\,, \label{bound}
\end{align} 
which must hold everywhere in spacetime. 

When dynamical gravity is included, it is not clear that this is the correct condition. However, Witten observed that it indeed eliminates known pathological metrics, while allowing many useful ones \cite{Witten:2021nzp}. Thus, it makes sense to investigate the consequences that this criterion would have on no-boundary path integrals. This question was studied in some detail in \cite{Witten:2021nzp,Lehners:2021mah,Jonas:2022uqb,Lehners:2022xds} and we will review the main results below.

First note that standard Lorentzian metrics saturate the bound \eqref{bound} (they have $\Sigma=\pi$), and thus reside right on the edge of the allowed domain of metrics. This is reasonable, as Lorentzian path integrals are only conditionally convergent, and the bound expresses the condition for absolute convergence. Any Lorentzian metric can be easily regulated to satisfy the bound, {\it e.g.} for RW metrics we may write
\begin{align}
    \mathrm{d}s^2 = - (1 \mp i \epsilon) \mathrm{d}t^2 + a(t)^2 \mathrm{d}\Omega^2\,,
\end{align}
for a small real number $\epsilon.$ One can then imagine taking the limit $\epsilon \to 0$ at the end of calculations. However, it is crucial that $\epsilon$ must not change sign. Thus, in a sense, the bound \eqref{bound} already divides the space of metrics into two.

\begin{figure}[ht]
	\centering
	\includegraphics[width=0.3\textwidth]{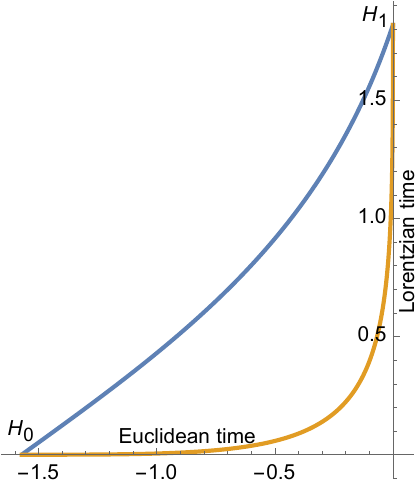} \hspace{1cm}
	\includegraphics[width=0.45\textwidth]{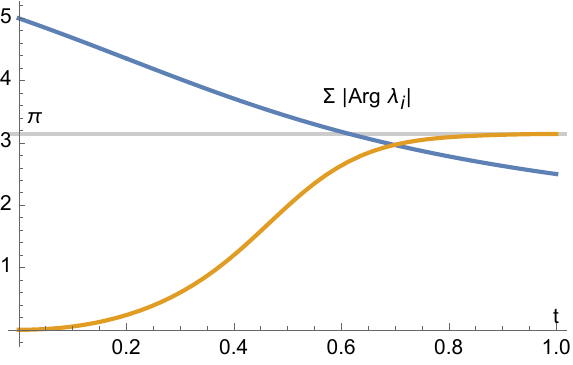}
	\caption{{\it Left Panel:} A no-boundary saddle point solution, interpolating between an initial hypersurface $H_0$ (South Pole) and a final one $H_1,$ via two different paths in the complex time plane. The paths are described in the main text. {\it Right panel:} the sum of arguments $\Sigma,$ as defined in \eqref{bound}, along the two time contours. Figures reproduced from \cite{Lehners:2021mah}.}
	\label{timecontours}
\end{figure}

What can we say about no-boundary geometries? The saddle points in section \ref{sec:minisuper} were obtained in a gauge where the lapse is constant, {\it cf.} Eq. \eqref{saddlesneumann}. These may be transformed to physical time by defining $\frac{N}{\sqrt{q}}\mathrm{d}t \equiv \mathrm{d}T,$ which leads to 
\begin{align}
\sqrt{\frac{\Lambda}{3}}\, T(t) = 2i \operatorname{arsinh}\left( \sqrt{\frac{\Lambda Nt}{6i}}\right) - \frac{\pi}{2} \,.
\end{align} 
In Fig. \ref{timecontours} the resulting path in the complex $T$ plane is plotted in blue in the left panel. In the right panel, we plot the sum $\Sigma$ of absolute values of arguments of the metric components, as defined in \eqref{bound}. What may come as a surprise is that the bound is seen to be violated. Thus, in the constant lapse gauge, the saddle points appear to violate the allowability bound \eqref{bound}. However, as we discussed previously, the action, and thus also the physical consequences, are unchanged when the path is deformed. In fact, the original Euclidean-plus-Lorentzian contour gives $\Sigma=0$ along the Euclidean segment, and $\Sigma=\pi$ along the Lorentzian one, and thus implies that no-boundary saddle points actually saturate the allowability bound. But one may worry that this contour is non-smooth. Let us therefore consider a smooth family of paths, obtained by specifying $T=\theta(t)$ with  
\begin{align}
\theta(t)=-\frac{\pi}{2}(1-t)^n + T(1)\, t^n\,, \quad 0 \leq t \leq 1\,. \label{modtime}
\end{align}
In Fig. \ref{timecontours} an example with $n=3$ is plotted in orange. There we can see that now the bound is indeed satisfied, and saturated only at the end point.

\begin{figure}[ht]
	\centering
	\includegraphics[width=0.42\textwidth]{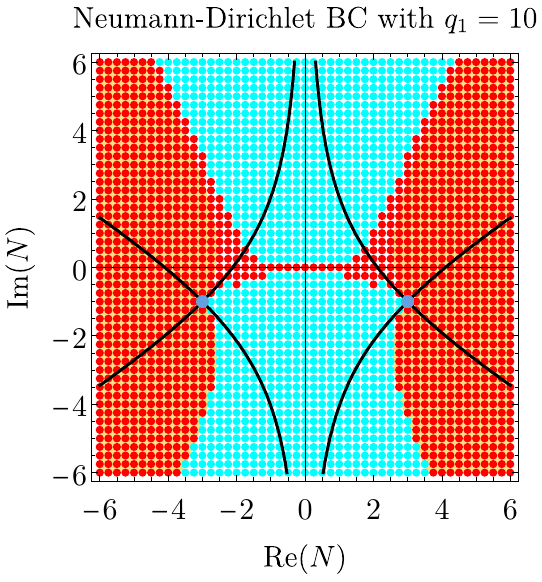} \hspace{1cm}
	\includegraphics[width=0.45\textwidth]{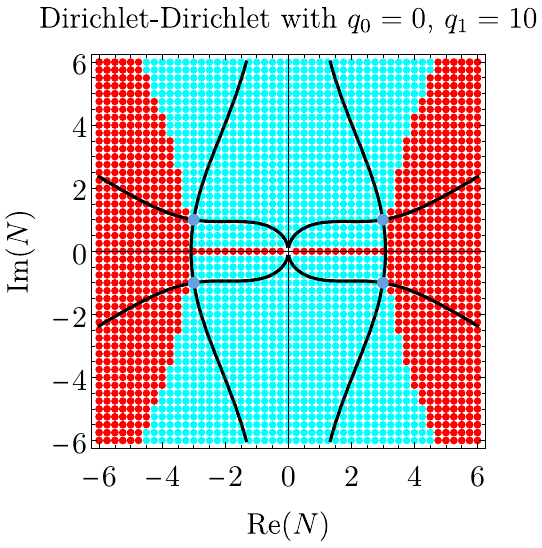}
	\caption{Allowable metrics (in light blue) in the plane of the complex lapse function; disallowed metrics are shown in red. {\it Left panel:} with an initial Neumann condition. {\it Right panel:} with an initial Dirichlet condition. The saddle points and steepest descent contours are also shown. Here $\Lambda=3, q_1 = 10.$ Figures reproduced from \cite{Jonas:2022uqb}.}
	\label{fig:allowable}
\end{figure}

The previous example should make it clear that it is in general difficult to assess whether a metric is allowable or not, if we permit such changes of time path. Techniques were developed in \cite{Jonas:2022uqb} to deal with this situation, and we refer to this paper for details. The results for no-boundary integrals, both with a Neumann initial condition and for a Dirichlet initial condition, are shown in Fig.~\ref{fig:allowable}. Both cases share the characteristic that the real lapse line constitutes a boundary that cannot be traversed. And in both cases, the steepest descent contours run into regions that are not allowed. This means that, if this allowability criterion is strictly enforced, then we can no longer define the sums over metrics along thimbles. It is unclear at present what this implies. 

One interesting feature in the Neumann case is that, as seen in the plane of the lapse function, the saddle points reside right at the edge of the allowable domain. This can be understood analytically: for the path integral with Neumann initial conditions, the initial size of the geometries is not fixed. In fact, this initial size can become complex off-shell, and by itself it can cause the allowability bound to be violated. Indeed, near the saddle point let us write $N=N_\pm + \Delta,$ and work to linear order in $\Delta.$ Then from \eqref{eq:qbarND2} we obtain
\begin{align}
q(0) \approx  \pm 2 \Delta (\frac{\Lambda}{3}q_1-1)^{1/2}\,,
\end{align} 
where we assume that $q_1>\frac{3}{\Lambda}.$ This implies that for $N_-$ we get the condition $3|Arg(\Delta)|<\pi.$ Consequently, starting from the saddle point, the allowed directions are limited to  $-\frac{\pi}{3} < Arg(\Delta) < \frac{\pi}{3}.$ For $N_+$ one analogously finds $\frac{2\pi}{3} < Arg(\Delta) < \frac{4\pi}{3}.$ Hence the saddle points, even though they are fully complex, are at the edge of the allowed domain, and thus the steepest descent contours are cut in ``half'' by the allowability criterion, with only the lower portion remaining.

For the case with Dirichlet initial conditions, the fact that the upper half plane gets separated from the lower half plane of the lapse function may be interesting. As we saw in section \ref{sec:perts}, the unstable saddle points are in the upper half plane, while the stable ones are in the lower half plane. Therefore, if one is not allowed to cross the real lapse line, then a definition of the integral in terms of exclusively stable metrics becomes a possibility. This is an idea worth pursuing.

The bound \eqref{bound} was derived with the assumption that the matter fields take real values. However, we saw in sections \ref{sec:inflex} and \ref{sec:ekex} that when a scalar field is added, it is typically required to take complex values at the South Pole. Might this lead to a conflict? It has been observed in \cite{Lehners:2022xds} that there are situations  in which it may be natural to allow for complex scalars. In particular, when there are additional dimensions that are compactified, then the lower-dimensional theory contains scalar fields (moduli) that arise from the higher-dimensional metric. We saw an example in section \ref{sec:filter}. In this case, if the bound \eqref{bound} is imposed in the higher-dimensional parent theory, then it is clear that it will allow for the lower-dimensional scalars to be complex, to a certain extent. What one finds is that the imaginary part of the scalar is bounded. The precise bound depends on the situation; we will just give one example, focusing on the scalar field $\phi$ that determines the volume of the internal manifold. The compactification ansatz reads \cite{Duff:1986hr}
\begin{align}
g_{MN} dx^M dx^N = e^{2a\phi} g_{\mu\nu}dx^\mu dx^\nu + e^{2b\phi} g_{ij} dx^i dx^j\,, \label{KKansatz}
\end{align}
with Latin indices running over the $(D-d)-$dimensional internal manifold, and Greek indices over the $d$-dimensional external manifold. To obtain a canonically normalised scalar, we must choose 
\begin{align}
a=-\frac{D-d}{d-2}b\,, \quad b= \sqrt{\frac{d-2}{(D-d)(D-2)}}\,.
\end{align}
We can immediately see from \eqref{KKansatz} that it will be the imaginary part of $\phi,$ rather than its argument, that will contribute to $\Sigma.$

Now, as an example, if we look near the South Pole of putative no-boundary solutions, and assume that $g_{\mu\nu},\, g_{ij}$ are Euclidean there, then we get
\begin{align}
\Sigma = \left[d\sqrt{\frac{D-d}{(D-2)(d-2)}} + \sqrt{\frac{(D-d)(d-2)}{D-2}}\right] 2 |\textrm{Im}\,(\phi)|\,.
\end{align}
For instance, if we compactify from $D$ down to $4$ dimensions, we get the bound
\begin{align}
6\sqrt{2}\sqrt{\frac{D-4}{D-2}}|\textrm{Im}\,(\phi)| < \pi\,, \quad \textrm{or} \quad |\textrm{Im}\,(\phi)| < \sqrt{\frac{D-2}{D-4}}\frac{\pi}{6\sqrt{2}} \approx \frac{\pi}{10}\,. \label{Euclibound}
\end{align}
Meanwhile, no-boundary solutions require (see section \ref{sec:inflex} and \cite{Janssen:2020pii})
\begin{align}
\textrm{Im}(\phi_{SP}) \approx  - \frac{V_{,\phi}}{V} \frac{\pi}{2}\,. \label{SPestimate}
\end{align}
which then translates into  $\frac{|V_{,\phi}|}{V} \lessapprox \frac{1}{5}.$ Hence only sufficiently flat potentials would allow for no-boundary solutions in this example \cite{Lehners:2022xds}. 

Note that we have only analysed what happens near the South Pole in this example, the full no-boundary geometry might very well lead to a stronger condition on the potential. This has started being explored very recently and leads to promising results concerning observational predictions, in particular regarding the tensor-to-scalar ratio~\cite{Hertog:2023vot} and the overall size of the universe~\cite{Lehners:2023pcn}. This direction of research appears highly promising at present.


\section{Discussion and Open Questions} \label{sec:di}

According to our current understanding, the fundamental principles for physical laws are the principles of quantum theory. When they are applied to the universe as a whole, it follows that the universe must admit a quantum state. If we knew this state, we could infer probabilities for different initial conditions and for subsequent evolutions of the universe. The no-boundary proposal provides a prescription for calculating this quantum state. 

The prescription combines quantum theory, gravity, and what one may term a containment principle, namely the idea that the universe is entirely self-contained in space and time. When formulated in terms of gravitational path integrals, the idea is that the dominant geometries, {\it i.e.} the saddle points of the path integral, consist of closed and regular spacetimes, with regular matter configurations on them, admitting as their only boundary the present-day configuration of the universe. In other words, there is no boundary to the past at which ``outside'' conditions might come into relevance. This may be interpreted as describing the emergence of the universe out of nothing -- or, perhaps, one should rather say that it describes the existence of the universe in a self-consistent manner.

The no-boundary condition appears as a very natural, almost inevitable, condition to put on path integrals. This is also confirmed by studies of analogous integrals in a non-cosmological, asymptotically AdS, setting. That said, the precise mathematical implementation of the no-boundary condition can be tricky and has so far been studied only on a case by case basis, and only in simple minisuperspace models. These studies have suggested that instead of summing over closed metrics, as initially advocated by Hartle and Hawking, it might be more appropriate to impose a regularity condition on the geometries that are summed over. However, a general prescription is still lacking, and this is one of the outstanding open questions related to the no-boundary wave function. 

But before discussing open issues, let us briefly recap what the no-boundary wave function already manages to explain (a fuller discussion was already presented in section \ref{sec:stock}). The most important feature of the no-boundary wave function is that it can explain the emergence of space and time, and how spacetime becomes classical. In doing so, the big bang singularity is automatically avoided, as only initially regular geometries enter the path integral. A further consequence is that matter fields are predicted to have been in their ground states at the nucleation of the universe. All these features explain aspects of our universe that, without clear justification, were simply assumed to hold. But according to the no-boundary wave function, not every type of universe can emerge from nothing. In fact, a dynamical attractor is required in order for no-boundary solutions to exist. Two such attractors are currently known, inflation and ekpyrosis, and we discussed the corresponding no-boundary solutions in sections \ref{sec:inflex} and \ref{sec:ekex}. The requirement of an attractor shows that the no-boundary proposal also leads to vacuum selection, by strongly restricting the possible early universe dynamics.

What is currently less well understood are the precise predictions for cosmological observables. Naively, the no-boundary measure favors short inflationary phases at low values of the potential, and long ekpyrotic phases. However, as discussed in section \ref{sec:probs}, these probabilities have so far only been inferred from highly simplified models, and may still be subject to revision. Also, in the ekpyrotic case, the transition from contraction to expansion remains ill understood at the quantum gravitational level. These are certainly important topics for future work.

The remarks above lead us to the many opportunities for further research that the no-boundary framework brings into the open. Let us begin with more mathematical questions. A basic one is whether a general prescription may be found that would determine the integration contours over fields, in particular the integration over the lapse function. As we saw in section \ref{sec:allowability}, if there exist physical restrictions on the complex metrics that are summed over, then this will have important consequences for the possible contours of integration. Let us highlight here that it appears so far that neither of the two ``obvious'' physical choices, that is to say neither Euclidean nor Lorentzian contours of integration, have been found to work in general. This fact encourages us to study complex metrics much more seriously in the future. A related open question is which kinds of singularities ought to be allowed in the off-shell geometries, and whether or not such singularities play a role. An evident long-term goal is to extend the treatment of path integrals beyond minisuperspace. This is a topic that unfortunately has not seen much progress over several decades.

And then there is a host of questions that are more conceptual in nature. For example, what is the meaning of the fact that the no-boundary wave function does not vanish at zero size? The simplest non-trivial topologies do not seem to contribute to the wave function (see section \ref{sec:ads}), thus reinforcing the puzzle. Also, saddle points typically come in pairs, which may be thought of as time reverses of each other. Can there be interesting interference effects between these saddles when the universe is still very small? And can the definition of probabilities be refined? So far, it is based on mathematical/WKB properties of the wave function. But can it be made more physical, highlighting the importance of interactions between sub-systems in the universe? Very generally, what is the meaning of probability when we only get to observe a single universe? Another very general question is whether there can be any other dynamical attractors, besides inflation and ekpyrosis, that allow for no-boundary solutions. If so, then this might yet again significantly affect our thinking about the early universe.

A final set of questions concerns the interplay of the no-boundary proposal and string theory. For instance, can the effects of winding modes be included? How does one describe no-boundary solutions containing branes and orientifolds? Can one construct models with at least semi-realistic particle physics and cosmological dynamics, to see how these features affect each other -- in particular, it would be interesting to see which vacua receive high probability and which are excluded. Can a bounce be included in a consistent manner? And two very general questions to end: first, is the holographic definition, described in section \ref{sec:ads}, correct? And second, if the cobordism conjecture (which states that all possible asymptotic field configurations can be related by interpolating spacetimes \cite{McNamara:2019rup}) is correct, then how does it affect the no-boundary framework? At least naively, it would seem to imply the existence of no-boundary saddles with arbitrarily large numbers of transitions between various cosmological epochs. Will the simple examples that have been studied so far end up being good approximations to the preferred route to our present day conditions?

One can look forward to the insights that will be gained from pursuing these questions.

\section*{Acknowledgments}

I have learned a great many things (and not just about quantum cosmology) from discussions with Andr\'{e}s Anabal\'{o}n, Lorenzo Battarra, Sebastian Bramberger,  Alice Di Tucci, Job Feldbrugge, Shane Farnsworth, Jonathan Halliwell, Jim Hartle, Arthur Hebecker, Michal Heller, Thomas Hertog, Oliver Janssen, Caroline Jonas, Claus Kiefer, Axel Kleinschmidt, Michael K\"{o}hn, George Lavrelashvili, Rahim Leung, Vincent Meyer, Hermann Nicolai, Burt Ovrut, J\'{e}r\^{o}me Quintin, Laura Sberna, Paul Steinhardt, Kelly Stelle, Stefan Theisen, Neil Turok, and Alex Vilenkin. Thank you!

I gratefully acknowledge the support of the European Research Council via the ERC Consolidator Grant CoG 772295 ``Qosmology''.

\appendix

\section{Canonical Quantisation} \label{sec:a1}

In the main text, we concentrated on minisuperspace models where spatial isotropy and homogeneity is assumed from the outset. This was the most relevant example, but some readers may be interested in seeing the more general framework, where such a symmetry reduction is not made. We will focus on the gravitational part of the action, with the matter content being left implicit. Thus the action is taken to be (setting $8\pi G = 1$)
\begin{align}
S = \frac{1}{2} \left[ \int _M \mathrm{d}^4 x \sqrt{-g}
(R-2\Lambda) + S_{boundary} \right] +
S_{matter} 
\end{align}
The choice of boundary term determines which boundary conditions can be consistently imposed. It is then useful to write the metric in $(1+3)$ decomposition \cite{Arnowitt:1962hi}, 
\begin{align}
\mathrm{d}s^2=   -N^2 \mathrm{d}t^2+ h_{ij} \left( \mathrm{d}x^i + N^i \mathrm{d}t\right) \left(\mathrm{d}x^j + N^j \mathrm{d}t\right)\,, 
\end{align}
where $N$ is the lapse and $N_i$ the shift. A useful quantity is the extrinsic curvature
\begin{align}
K_{ij} = \frac{1}{2N} \left[ - \frac{\partial h_{ij}}{\partial t}
+ 2 D_{(i} N_{j)} \right] \,, \label{eq:extrinsic}
\end{align}
where $D_i$ is the covariant derivative on the three-surface. The aim is to rewrite the action in terms of $N, N^i$, $h_{ij}$ and $K_{ij}.$ This can be done using the Gauss-Codazzi relation between 4-curvature and 3-curvature, yielding
\begin{align}
S= \frac{1}{2} \int \mathrm{d}^3x \ \mathrm{d}t \ N \sqrt{h} \left[ K_{ij}
K^{ij} -K^2 + {^3 R} - 2 \Lambda \right] + S_{matter} \,.
\end{align}

The Hamiltonian form
of the action is given by \cite{Arnowitt:1962hi}
\begin{align}
S=\int \mathrm{d}^3x \ \mathrm{d}t\left[ \dot h_{ij} \pi^{ij}
-N{\cal H}-N^i{\cal H}_i\right] 
\end{align}
where $\pi^{ij}=\frac{\delta {\cal L}}{\delta \dot{h}^{ij}}=-\frac{\sqrt{h}}{2} \left( K^{ij} - h^{ij}K\right)$  are the momenta conjugate to
$h_{ij}$.  The Hamiltonian is a sum of
constraints, with the lapse $ N $ and shift $ N^i$ being Lagrange multipliers. There is the momentum constraint,
\begin{align}
{\cal H}^i=-2D_j \pi^{ij} +{\cal H}_{matter}^i=0 \,,
\end{align}
and the Hamiltonian constraint
\begin{align}
{\cal H}=  2 G_{ijkl}\pi^{ij} \pi^{kl}- \frac{1}{2}   \sqrt{h} ({^3}R-2 \Lambda) +{\cal H}_{matter}=0\,,
\end{align}
where $ G_{ijkl} $ is the DeWitt metric \cite{DeWitt:1967yk}
\begin{align}
G_{ijkl}=\frac{1}{2\sqrt{h}}  \left(h_{ik} h_{jl}+h_{il}
h_{jk}-h_{ij} h_{kl}\right) \,.
\end{align}
These constraints are essentially equivalent to the $0i$
and $00$ components of the classical Einstein equations. The
constraints play a central role in the canonical quantisation
procedure.

Canonical quantisation amounts to imposing the constraints as operator equations, in the field representation with the substitution
\begin{align}
\pi^{ij}\to -i\frac{\delta}{\delta h_{ij}}
\end{align}
and similarly for the matter momenta. This results in four equations: the
momentum constraint
\begin{align}
{\cal H}^i\Psi= 2i D_j \frac{\delta \Psi}{\delta h_{ij} } + {\cal
H}_{matter}^i \Psi = 0 \,,
\end{align}
and the Wheeler-DeWitt equation \cite{DeWitt:1967yk,Dewitt:1968lxx}
\begin{align}
{\cal H}\Psi (h_{ij},\Phi_{matter}) = \left[ - G_{ijkl} \frac{\delta}{\delta h_{ij} }
\frac{\delta}{\delta h_{kl} } - \sqrt{h} ({^3}R-2 \Lambda) +{\cal
H}_{matter} \right] \Psi = 0 \,. \label{eq:wdw}
\end{align}
We should point out that there is an ambiguity in factor ordering, as the precise placement of the functional derivatives is not fixed. In explicit examples, sensible choices can often be found, {\it e.g.} by requiring invariance under field redefinitions \cite{Halliwell:1988wc}.

Since the constraints are so central, it is worthwhile investigating their meaning. To understand the momentum constraint better \cite{Higgs:1958mh},
consider a change of coordinates on the three-surface, $x^i \rightarrow x^i - \xi^i
$. Then
\begin{align}
\Psi[h_{ij}+D_{(i}\xi_{j)}] = \Psi[h_{ij}] + \int \mathrm{d}^3 {\bf x} \
D_{(i}\xi_{j)} \frac{\delta \Psi}{\delta h_{ij} } 
\end{align}
Integrating by parts in the last term, and dropping the boundary
term (assuming the three-manifold is compact), one finds that the
change in $ \Psi $ is given by
\begin{align}
\delta \Psi = - \int \mathrm{d}^3 {\bf x} \ \xi_j D_i \left( \frac{\delta \Psi}{\delta h_{ij}} \right) = -\frac{1}{2i} \int \mathrm{d}^3 {\bf x} \
\xi_i {\cal H}^i \Psi
\end{align}
This will be zero when the momentum constraint is imposed. Hence it expresses spatial diffeomorphism invariance.

The Wheeler-DeWitt equation \eqref{eq:wdw}, due to its association with the lapse function, is similarly related to time reparameterisation invariance. It combines spacetime geometry and matter into a single quantum equation, as expected in quantum gravity. One variable stands out: the scale factor (or size) of the universe, given by the appropriate power of $det(h_{ij})$: it enters with a negative sign in the DeWitt metric, as we also saw in explicit examples, see {\it e.g.} \eqref{eq:miniact}. Other metric deformations, as well as matter fields, enter with a positive sign.
Note that the Wheeler-DeWitt equation does not contain any explicit dependence on coordinates, in particular on time. This is different from ordinary quantum mechanics, where time plays a privileged role. It makes sense here, since we do not measure time directly, but rather correlations between field configurations ({\it e.g.} the arrow on the watch points towards the $12$ and the sun is high in the sky).

The Wheeler-DeWitt equation should be solved at every point in spacetime. For general metrics this is technically impossible. The general configuration space consists of spatial metrics and matter configurations, up to diffeomorphisms (this is called superspace). One then typically restricts to homogeneous spatial metrics, containing just a few time-dependent functions and fixed spatial dependence. This is known as minisuperspace. One drawback is that one has set all other metric deformations to zero, including their momenta (which is in conflict with the uncertainty principle). However, we know that our universe is rather homogeneous, hence one may hope to obtain a self-consistent ``approximation''. In practice one has to see if perturbations around the minisuperspace geometries are suppressed, and above we found that for no-boundary saddle points this was always the case.

\section{Gravitational Path Integrals} \label{sec:a2}

When gravity is included, the path integral includes a sum over geometries. Defining this requires some care, because of diffeomorphism invariance. One has to make sure not to overcount, as geometries might be related to each other via changes of coordinates. Hence it is important to properly fix the gauge. The general procedure was first worked out by Teitelboim in \cite{Teitelboim:1981ua,Teitelboim:1983fk}, and applied to minisuperspace models by Halliwell in \cite{Halliwell:1988wc}. Here we will outline how this is done; additional details can be found in the original papers.

For simplicity, we will consider a minisuperspace action with a single degree of freedom, namely the scale factor $a$ of the universe (and its conjugate momentum $p$). Models with more degrees of freedom can be dealt with analogously. The action is thus given by
\begin{align}
    S=\int_0^1 \mathrm{d}t \left( p\dot{a} - N H\right)
\end{align}
with the Hamiltonian $H=\frac{1}{2}p^2 + U(a).$ Classically, the Hamiltonian would vanish. The metric is of the form $\mathrm{d}s^2 = - N(t)^2 \mathrm{d}t^2 + a(t)^2 \mathrm{d}\Omega_3^2,$ and overcounting may arise due to time reparameterisation invariance. Hence we must fix a gauge. This gauge must be chosen such that any history can be deformed into one that satisfies the gauge condition, and the gauge must be fixed completely. An appropriate choice is  \cite{Teitelboim:1981ua}
\begin{align}
    \dot{N}=f(a,p,N)\,,
\end{align}
where $f$ is an arbitrary function of the fields, but not of their time derivatives. Such a gauge choice can be implemented with a Lagrange multiplier $\Pi(t)$, by adding the following term to the action
\begin{align}
    S_{gf} = \int_0^1 \mathrm{d}t \Pi \left( \dot{N} - f \right)\,.
\end{align}
This term fixes the gauge, but this is not enough yet. We also have to make sure that the path integral is actually independent of the choice of gauge fixing function $f.$ This at first seems rather tricky to establish, but can be achieved using the formalism developed by Batalin, Fradkin and Vilkovisky \cite{Fradkin:1975cq,Batalin:1977pb} (based on \cite{Faddeev:1973zb}). The idea is first to add ghost fields. More specifically, one adds the anticommuting fields $C,\bar{C}$ and their conjugate momenta $P,\bar{P}.$ Then one uses the extended action to define a new symmetry, a so-called BRS symmetry, which will help establish invariance under changes of gauge. Some educated guesses are required at this step, see {\it e.g.} \cite{Halliwell:1988wc}. It turns out that if the action for the ghosts is taken to be
\begin{align}
    S_{gh}=\int_0^1 \mathrm{d}t \left( \bar{P} \dot{C} + \bar{C} \dot{P} - \bar{P} P + C \{ f,H\} \bar{C} + P \frac{\partial f}{\partial N} \bar{C}\right)\,, \label{eq:ghost}
\end{align}
where $\{ f,H\}=\frac{\partial f}{\partial a}\frac{\partial H}{\partial p} - \frac{\partial f}{\partial p}\frac{\partial H}{\partial a}$ is the Poisson bracket, then the total action is invariant under a BRS transformation with anticommuting parameter $\lambda,$
\begin{align}
   & \delta a = \lambda C \frac{\partial H}{\partial p}\,, \,  \delta p = -\lambda C \frac{\partial H}{\partial a}\,, \, \delta N = \lambda P\,, \, \delta \Pi =0 \,, \nonumber \\ & \delta C = 0 \,, \, \delta P = 0 \,, \, \delta \bar{C} = -\lambda \Pi \,, \, \delta \bar{P} = -\lambda H\,, \label{eq:brs}
\end{align}
subject to the boundary conditions that $\Pi, C$ and $\bar{C}$ vanish at the end points $t=0,1.$ Now the path integral is defined by using the Liouville measure and integrating over the total action, including the gauge fixing and ghost terms,
\begin{align}
    \Psi = \int {\cal D}a{\cal D}p{\cal D}N{\cal D}\Pi{\cal D}C{\cal D}P{\cal D}\bar{C}{\cal D}\bar{P} \, e^{\frac{i}{\hbar}\left( S + S_{gf} + S_{gh}\right)}\,.
\end{align}
The punch line now is that this path integral is indeed independent of the choice of gauge fixing function $f.$ This can be shown explicitly by performing a BRS transformation with the special choice of parameter
\begin{align}
    \lambda = \frac{i}{\hbar} \int_0^1 \mathrm{d}t (f-\tilde{f})\,.
\end{align}
The total action is evidently invariant, since the transformations \eqref{eq:brs} were chosen specifically for this to be the case for any $\lambda$. And when calculating the (super-)Jacobian $J$ of this transformation, one obtains a factor \cite{Henneaux:1985kr}
\begin{align}
J=e^{\frac{i}{\hbar}\int \mathrm{d}t \left( C \{ \tilde{f}-f,H\} \bar{C} + P \frac{\partial (\tilde{f}-f)}{\partial N} \bar{C}\right)}\,,
\end{align}
which has the effect of replacing $f$ by $\tilde{f}$ in the ghost action \eqref{eq:ghost}. Thus the path integral is indeed independent of the choice of $f.$ Note that this greatly simplifies the analysis. Not only can one perform the path integral with $N$ fixed for metrics of the form \eqref{eq:LorentzianMetric}, but also for more involved choices such as the useful version in \eqref{eq:Metric}. 

For now, we may make use of this freedom and choose $f=0$ to further evaluate the integrals. With this choice, the ghost integrals factorise out, and one can see immediately that they must yield a purely numerical factor, as they have become independent of the other fields. The integrals over anti-commuting variables can be evaluated straightforwardly \cite{Halliwell:1988wc}, resulting in a factor of unity
\begin{align}
    \int {\cal D}C{\cal D}P{\cal D}\bar{C}{\cal D}\bar{P}e^{\frac{i}{\hbar}\int_0^1 \mathrm{d}t \left( \bar{P} \dot{C} + \bar{C} \dot{P} - \bar{P} P \right)} = e^{[\bar{C}(1)-\bar{C}(0))(C(1)-C(0)]} = 1\,,
\end{align} 
where one has to make use of the boundary conditions that $C(0)=C(1)=\bar{C}(0)=\bar{C}(1)=0$. Meanwhile, the integral over the lapse and its conjugate momentum also simplifies drastically,
\begin{align}
    \int {\cal D}N {\cal D}\Pi \, e^{\frac{i}{\hbar} \int_0^1 \mathrm{d}t \dot{N}\Pi} = \int {\cal D}N \delta(\dot{N}) = \int \mathrm{d}N\,,
\end{align}
that is to say, the path integral over the lapse reduces to an ordinary integral over $N.$ This is a huge simplification, which is at the root of the tractability of minisuperspace models. In the end, we are therefore left with the expression
\begin{align}
    \Psi = \int \mathrm{d}N {\cal D}a {\cal D}p \, e^{\frac{i}{\hbar}\int_0^1 \mathrm{d}t \left(p\dot{a} - NH \right)}\,.
\end{align}
One can of course also switch from this phase space integral to one purely in field space, as is used in the main text. For further details, for example on how to extend the analysis to first loop order, see {\it e.g.} \cite{Esposito:1992xz}.


\section{Picard-Lefschetz theory} \label{sec:a3}

At various instances in this review, we have to evaluate an oscillating integral (usually for the lapse), of the form
\begin{align}
\label{eq:integral}
I = \int_{\cal C} \mathrm{d}x\, e^{\frac{i}{\hbar}S[x]}\,.
\end{align}
We take $S[x]$ to be a real function, $\hbar$ is a real parameter and ${\cal C}$ is the domain of integration, typically the positive real line, or the full real line. Such an integral is only conditionally convergent, as the integrand has modulus $1$ everywhere, in particular there are no regions where the modulus drops off and where convergence could be guaranteed. Still, intuitively one might guess that the oscillations lead to cancellations, and that many integrals of this type might in fact lead to sensible results. 

The problem with such integrals is best illustrated by an example using a discrete version of an oscillating integral, {\it e.g.} the series $\sum_{n=0}^\infty (-1)^n.$ The value of such a sum depends on how we define the order of summation, e.g. $[1 + (-1)] + [1 + (-1)] + \cdots = 0,$ or $1+[-1+1]+[-1+1]+\cdots =1.$ And some sums and integrals of this sort do not converge at all. Picard-Lefschetz theory is a useful tool in dealing with such integrals \cite{Lefschetz:1975ta}. Its main idea is to rewrite the conditionally convergent integral as a sum of absolutely convergent integrals. It also shows when a rewriting of this kind is possible, and when not. When it is, this procedure defines the integral unambiguously. We will give an elementary overview here (based on \cite{Feldbrugge:2017kzv}), sufficient for our purposes. For a more in depth treatment, see \cite{Witten:2010cx}. A few additional aspects are discussed in \cite{Feldbrugge:2017kzv,Feldbrugge:2017mbc}.

The first step is to let the variable $x$ become complex, $x  \in \mathbb{C},$ and to view $S[x]$ as a holomorphic function of $x.$ We will also write $x=u^1+i u^2$ for real $u^1, u^2.$ Then one can try to deform the integration domain ${\cal C},$ using Cauchy's theorem, into a complex contour such that the integral becomes manifestly convergent. Ideal integration contours are steepest descent contours, along which the modulus of the integrand falls off as fast as possible. Such contours are also called ``Lefschetz thimbles''. They are associated with critical points (in fact saddle points\footnote{For a complex function $f(x),$ critical points are necessarily saddle points: expanding near a critical point $x_c$ one finds $f(x) \approx f(x_c) + \frac{1}{2}f''(x_c) [(u^1)^2 - (u^2)^2 +2iu^1u^2],$ {\it i.e.} there is always at least one direction along which the function increases, and at least one along which it decreases.}) of $S[x],$ {\it i.e.} they fall off from saddle points. (If a neighbourhood of a critical point is like a horse's saddle, then the thimble is made of the legs of the rider.) 

As a simple example, consider $S[x] =x^2$, which has a critical point at $x=0$. Then  Re$[iS[x]]=-2 u^1 u^2$. The magnitude of the integrand decreases most rapidly along the contour $u^1=u^2,$ and this is the Lefschetz thimble. Conversely, the modulus of the integrand increases most rapidly along $u^1=-u^2$, and this is the steepest ascent contour. We denote thimbles by ${\cal J}$ and steepest ascent contours by ${\cal K}.$

\begin{figure}[ht] 
\begin{minipage}{0.5\textwidth}
		\includegraphics[width=0.7\textwidth]{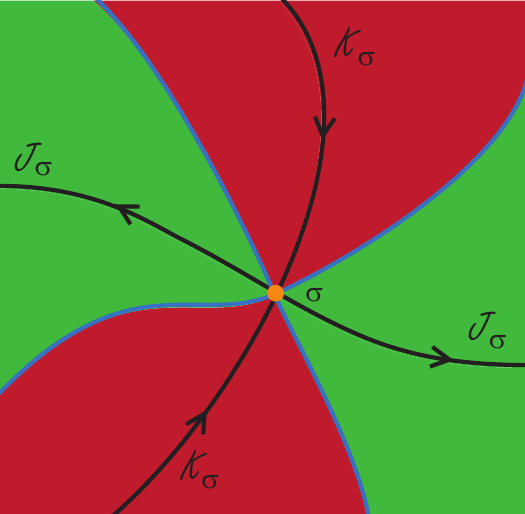}
	\end{minipage}%
	\begin{minipage}{0.5\textwidth}
		\includegraphics[width=0.95\textwidth]{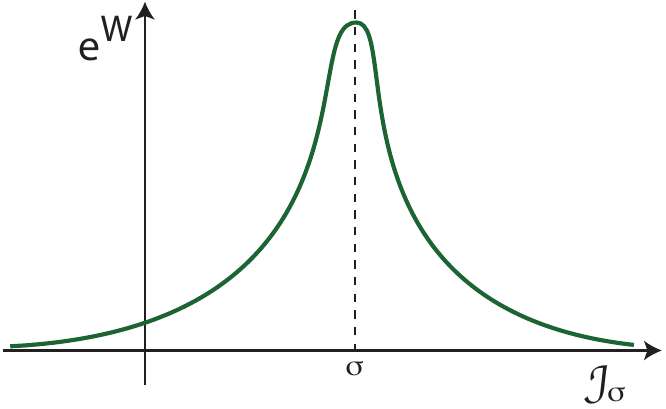}
	\end{minipage}%
	\caption{{\it Left panel:} A sketch of the complex $x$ plane. A saddle point $\sigma$ is shown, with its steepest descent (${\cal J}_\sigma$) and ascent ${\cal K}_\sigma$) contours. Arrows indicate the direction of descent.  Regions in which the weighting is smaller than at the saddle point are shown in green, and regions with higher weighting in red.  These regions are separated by blue lines, along which the weighting equals that of the saddle point.  {\it Right panel:} Sketch of the weighting along a Lefschetz thimble (steepest descent contour).}
	\protect
	\label{fig:thimble}
\end{figure} 

Now in more detail: decompose the exponent as $\mathcal{I}=iS/\hbar = W + i P,$ where $W$ determines the \emph{weighting} and $P$ is the \emph{phase}. Mathematically, $W$ is also known as the Morse function. Then downward flow is defined by
\begin{equation}
\frac{\mathrm{d}u^i}{\mathrm{d}\lambda} = -g^{ij}\frac{\partial W}{\partial u^j}\,,
\label{eq:dw}
\end{equation}
where $\lambda$ is a parameter along the flow and $g_{ij}$ is a (Riemannian) metric on the complex plane. To verify that $W$ indeed decreases along such a flow, consider $\frac{\mathrm{d}W}{\mathrm{d} \lambda} = \sum_i\frac{\partial W}{\partial u^i}\frac{\mathrm{d}u^i}{\mathrm{d}\lambda} = -\sum_i\left(\frac{\partial W}{\partial u^i}\right)^2<0.$ To proceed, it is useful to choose a metric. One may choose this for convenience, in our case we will simply take it to be the flat Cartesian metric, $\mathrm{d}s^2 = |\mathrm{d}x|^2.$ With complex coordinates, $(u,\bar{u})=\bigl(u^1+i u^2,u^1-i u^2\bigr)$, the metric is $g_{uu}=g_{\bar{u}\bar{u}}=0,\,g_{u\bar{u}}=g_{\bar{u}u}=1/2$. Then $W=(\mathcal{I}+\bar{\mathcal{I}})/2$ and the flow equations \eqref{eq:dw} become 
\begin{equation}
\frac{\mathrm{d}u}{\mathrm{d}\lambda} = - \frac{\partial {\bar{\cal I}}}{\partial \bar{u}}, \quad \frac{\mathrm{d}\bar{u}}{\mathrm{d}\lambda} = - \frac{\partial {{\cal I}}}{\partial {u}}\,.
\end{equation} 
From these a most useful property immediately follows,
\begin{equation}
\label{eq:imh}
\frac{\mathrm{d} P}{\mathrm{d}\lambda} = \frac{1}{2i}\frac{\mathrm{d}({\cal I} - \bar{\cal I})}{\mathrm{d}\lambda} = \frac{1}{2i}\left( \frac{\partial {\cal I}}{\partial u}\frac{\mathrm{d}u}{\mathrm{d}\lambda} - \frac{\partial \bar{\cal I}}{\partial \bar{u}}\frac{\mathrm{d}\bar{u}}{\mathrm{d}\lambda}\right) = 0\,.
\end{equation}
That is to say, the phase $P$ is constant along a steepest descent flow. This provides both a useful way of finding thimbles numerically (just plot the locus of points with the same phases as those of the saddle points), and it leads to a huge simplification of the integral, because now along a thimble it does not oscillate anymore! Rather, it is maximally convergent, see Fig.~\ref{fig:thimble}. In fact, the integral along a thimble is convergent if
\begin{align}
\left| \int_{{\cal J}_\sigma} \mathrm{d} x e^{iS[x]/\hbar} \right| 
\leq \int_{{\cal J}_\sigma} |\mathrm{d}x| \left| e^{iS[x]/\hbar} \right| 
=
\int_ {{\cal J}_\sigma} \mathrm|{d}x| e^{W(x)} 
< \infty\,.
\label{eq:conv}
\end{align}
If we denote the length along the curve as $l= \int |dx|$, then we will get convergence if $W(x(l))< -\ln (l) + A$, for some constant $A$, as $l \rightarrow \infty.$ Thus only fairly weak assumptions must be made to guarantee convergence.

In an analogous manner to downwards flows, one can define upwards flows,
\begin{equation}
\frac{\mathrm{d}u^i}{\mathrm{d}\lambda} = +g^{ij}\frac{\partial W}{\partial u^j}\,,
\label{eq:uw}
\end{equation}
and along such flows the phase $P$ is consequently also fixed. Hence we arrive at the picture that from saddle points $\sigma$ emanate equal numbers of thimbles ${\cal J}_\sigma$ and steepest ascent contours ${\cal K}_\sigma.$ 

We should discuss a subtlety: it can happen that what departs from one saddle point as a steepest descent contour arrives at another saddle point with lower weighting (and thus looks like a steepest ascent contour from the point of view of the second saddle point). Such a degeneracy can arise for example when saddle points occur in complex conjugate pairs. To deal with such a situation unambiguously, one can add a symmetry breaking term to $S[x],$ multiplied by a parameter $\epsilon.$ This will break the degeneracy, and then one can let $\epsilon \to 0$ at the end of the calculation. Note that all thimbles then run as far as they can, meaning $W \to - \infty$ and the integrand always falls off to zero along thimbles. Likewise, steepest ascent contours reach $W \to + \infty.$

One important question remains to be addressed: it is not clear yet which saddles, and thus which thimbles, contribute to the integral and which do not. Do they all contribute, or only some? If possible degeneracies are resolved as described above, then we can associate a single saddle point to every thimble, and to every steepest ascent contour. As an equation, we can express this as an intersection
\begin{equation}
{\rm Int}({\cal J}_\sigma, {\cal K}_{\sigma'})=\delta_{\sigma \sigma'}\,, \label{eq:intersection}
\end{equation}
where we have implicitly chosen a direction for the contours. Our goal is to write the integration contour as a sum of thimbles,
\begin{equation}
\label{eq:contourexp}
{\cal C} = \sum_\sigma n_\sigma {\cal J}_\sigma\,, 
\end{equation}
where the coefficients $n_\sigma$ take the values $0$ or $\pm1$ (the sign depends on the relative orientation of ${\cal C}$ and the thimbles). Now it is easy to determine these coefficients, we just intersect both sides of the above equation with ${\cal K}_{\sigma'},$ to obtain 
\begin{align}
n_\sigma= {\rm Int}(\mathcal{C}, {\cal K}_{\sigma})\,.
\end{align}
In words, this means that a  saddle point contributes if its steepest ascent contour intersects the original integration contour. This makes complete sense: we have an oscillating integral along the original contour; this will contain many cancellations. Now we want to deform this into a non-oscillating integral. Along the new contour, the modulus of the integrand must be smaller, because there will be no cancellations along this new integration domain. Hence, from the original integration contour, we must be able to flow down towards a saddle point, if it is to be relevant. Or, equivalently, from the saddle point we must be able to flow up to the original contour.

Usually, an invariant definition requires that ${\cal C}$ runs between singularities, either at finite locations or at infinity. But then it may happen that the thimbles approach the singularities, or infinity, at different angles than ${\cal C}.$ Hence one must make sure that in addition, the arcs either near a singularity or at infinity, linking ${\cal C}$ to the appropriate thimble, give zero contribution to the integral. In fact, this is easy to verify in simple models. The reason is simply that the weighting runs to minus infinity along the thimbles, and if these are to be relevant then they must directly link up to ${\cal C},$ without an intervening region of divergence. For a more detailed treatment of this point, see \cite{Feldbrugge:2017kzv}.

Using the rewriting of the integration domain, we can now express the original, conditionally convergent integral as a sum over absolutely convergent integrals along thimbles,
\begin{equation}
\label{eq:contour}
I=\int_{\cal C} \mathrm{d} x \, e^{\frac{i}{\hbar}S[x]} = \sum_\sigma n_\sigma \int_{{\cal J}_\sigma} \mathrm{d} x \, e^{\frac{i}{\hbar}S[x]}\,.
\end{equation}
This is the result we wanted to arrive at. It defines the original integral in a precise and unambiguous manner. For higher-dimensional integrals, this is especially important, as it guarantees that one can use Fubini's theorem, which states that the order of performing the integrals does not matter, as long as they are all absolutely convergent. (This justifies the procedure done repeatedly in the main text, namely to perform the integral over the scale factor first, and then that over the lapse.) 

It can occur that when the parameters of the integral are varied, different thimbles become relevant, {\it i.e.} the $n_\sigma$ can change. This important effect is called a Stokes phenomenon, and it can have interesting consequences as saddle points that are unimportant for some parameters can become dominant in other parameter ranges. We will encounter examples of this effect in the main text.

But there is more: the rewriting \eqref{eq:contour} also allows for a useful approximation scheme, called the saddle point approximation. This is because the integral along each thimble is strongly peaked around the saddle point  $x_\sigma$. And this the more so, the smaller $\hbar$ is. Since in physical applications $\hbar$ is indeed very small, this approximation is typically highly accurate. Furthermore, we can pull out the phase from each integral along thimbles, since it is constant there. Putting all this together, we obtain
\begin{align}
I = \int_{\cal C} \mathrm{d}x \, e^{i S[x]/\hbar} 
 & = \sum_\sigma n_\sigma \, e^{i \, P(x_\sigma)}\int_{{\cal J}_\sigma} e^W \mathrm{d}x \\ 
& \approx \sum_\sigma n_\sigma \, e^{i S(x_\sigma)/\hbar}\left[A_\sigma+\mathcal{O}(\hbar)\right]\,.
\end{align} 
Here $A_\sigma$ is a factor that one can get by integrating over fluctuations around the saddle point, {\it i.e.} by performing a Gaussian integral over the action expanded to second order. Further sub-leading terms can be calculated perturbatively, but in fact very often the leading, saddle point contribution is all we will require. 

Thus Picard-Lefschetz theory provides a highly useful tool for defining and evaluating oscillating integrals.

\bibliographystyle{elsarticle-num.bst}
\bibliography{NoBoundaryReview.bib}

\end{document}